\pdfoutput=1
\def\QCCDArxivBuild{1}
\def\QCCDNoNotes{1}
\newif\ifqccdarxiv
\ifdefined\QCCDArxivBuild\qccdarxivtrue\else\qccdarxivfalse\fi
\ifqccdarxiv
\documentclass[sigconf,screen,nonacm]{acmart}
\else
\documentclass[sigconf,screen,review]{acmart}
\fi
\newif\ifqccdreference
\ifdefined\QCCDReferenceBuild\qccdreferencetrue\else\qccdreferencefalse\fi
\newif\ifqccdtutorial
\ifqccdreference\qccdtutorialtrue\else\qccdtutorialfalse\fi
\qccdtutorialfalse
\newif\ifqccdformal
\qccdformalfalse
\ifqccdreference\qccdformaltrue\fi
\newif\ifqccdproof
\qccdprooffalse
\ifqccdformal\qccdprooftrue\fi

\usepackage{amsmath,amsthm,stmaryrd,booktabs,mathpartir,xcolor,listings,tikz,multirow,array,enumitem,xspace,graphicx}
\usetikzlibrary{arrows.meta,positioning,calc,shapes.geometric,shapes.misc,fit,backgrounds,decorations.pathreplacing}

\usepackage{colortbl}
\definecolor{qink}{HTML}{1C2A4A}
\definecolor{qblue}{HTML}{2A78D6}
\definecolor{qyellow}{HTML}{E8B940}
\definecolor{qgreen}{HTML}{0B7A4B}
\definecolor{qorange}{HTML}{B26A00}
\definecolor{qred}{HTML}{C62828}
\definecolor{qrail}{HTML}{CFCECA}
\definecolor{qgrey}{HTML}{8A8985}
\definecolor{qcard}{HTML}{FBFAF7}
\definecolor{qpaper}{HTML}{F7F6F2}
\definecolor{qsoft}{HTML}{EEF3FB}
\definecolor{qwarm}{HTML}{FFF3E0}
\definecolor{qmint}{HTML}{E6F4EC}
\definecolor{qrose}{HTML}{FBE9E7}
\definecolor{qpurple}{HTML}{7B3FA0}
\definecolor{qteal}{HTML}{006C78}

\definecolor{qcodeblue}{HTML}{234F7D}
\definecolor{qcodepurple}{HTML}{603D78}
\definecolor{qcodegreen}{HTML}{35623B}
\definecolor{qcodeorange}{HTML}{865000}
\definecolor{qcodeteal}{HTML}{245D63}

\DeclareRobustCommand{\qccdkeyword}[2]{\ifmmode
  \text{\normalfont\ttfamily\bfseries\color{#1}#2}\else
  {\normalfont\ttfamily\bfseries\color{#1}#2}\fi}
\DeclareRobustCommand{\hwkw}[1]{\qccdkeyword{qcodeblue}{#1}}
\DeclareRobustCommand{\circkw}[1]{\qccdkeyword{qcodepurple}{#1}}
\DeclareRobustCommand{\rulekw}[1]{\qccdkeyword{qcodegreen}{#1}}
\DeclareRobustCommand{\execkw}[1]{\qccdkeyword{qcodeorange}{#1}}
\DeclareRobustCommand{\irkw}[1]{\qccdkeyword{qcodeteal}{#1}}

\tikzset{
  ion/.style={rounded corners=4pt, fill=qink, text=white, font=\sffamily\bfseries\footnotesize, inner xsep=2.5pt, inner ysep=1.2pt},
  ion new/.style={ion, fill=qgreen},
  ion moving/.style={ion, fill=qblue},
  lbl/.style={font=\sffamily\footnotesize, text=qgrey, inner sep=1pt},
  hop/.style={-{Stealth[length=4pt,width=3.5pt]}, qink, line width=1pt},
  hop blue/.style={hop, qblue},
  hop green/.style={hop, qgreen},
  hop orange/.style={hop, qorange},
  hop red/.style={hop, qred, dashed},
  box/.style={draw=qink, fill=white, rounded corners=3pt, line width=0.6pt, font=\sffamily\footnotesize, inner sep=3pt, minimum height=13pt},
  box gate/.style={box, fill=qwarm},
  box one/.style={box, fill=qsoft},
  box meas/.style={box, fill=qpaper},
  box ok/.style={box, fill=qmint, draw=qgreen},
  box bad/.style={box, fill=qrose, draw=qred},
  box pass/.style={box, fill=qcard, minimum width=44pt, align=center},
  box trusted/.style={box, fill=qmint, draw=qgreen, align=center, line width=0.8pt},
  box untrusted/.style={box, fill=qpaper, draw=qgrey, align=center, dashed},
  art/.style={font=\ttfamily\footnotesize, text=qgrey, inner sep=1pt},
  flow/.style={-{Stealth[length=4pt,width=3.5pt]}, qgrey, line width=0.8pt},
  flow ink/.style={-{Stealth[length=4pt,width=3.5pt]}, qink, line width=0.9pt},
  layer/.style={fill=qpaper, rounded corners=4pt, inner sep=3pt},
  dep/.style={-{Stealth[length=3.5pt,width=3pt]}, qgrey, line width=0.7pt},
  sectag/.style={fill=qink, text=white, rounded corners=1.5pt, inner xsep=2pt,
    inner ysep=1.4pt, font=\sffamily\bfseries\fontsize{6.5}{7}\selectfont},
  sechead/.style={inner sep=0pt, anchor=west, text=qink,
    font=\sffamily\bfseries\fontsize{7.5}{9}\selectfont},
  okmark/.style={text=qgreen, font=\sffamily\bfseries\fontsize{7.5}{9}\selectfont, inner sep=0pt},
  badmark/.style={text=qred, font=\sffamily\bfseries\fontsize{7.5}{9}\selectfont, inner sep=0pt},
}
\newcommand{\QCCDFigHead}[3]{%
  \node[sectag,anchor=west] (#1) at (0,0) {#2};
  \node[sechead] at ($(#1.east)+(4pt,0)$) {#3};}
\newcommand{\cmark}{\ensuremath{\checkmark}}
\newcommand{\xmark}{\ensuremath{\times}}

\newcommand{\capa}{\mathrm{cap}}

\newcommand{\pos}{\mathrm{pos}}

\newcommand{\via}{\mathrm{via}}

\newcommand{\rulename}[1]{\textsf{#1}}
\newcommand{\R}[1]{\rulename{R#1}}
\newcommand{\Rule}[1]{{\upshape\textsc{#1}}}

\newcommand{\MS}{\mathrm{MS}}
\newcommand{\CX}{\mathrm{CX}}
\newcommand{\Rp}{\mathrm{R}}
\newcommand{\ket}[1]{|#1\rangle}

\newcommand{\qccdglyphbox}[1]{\tikz[x=.068em,y=.1875ex,baseline=0pt,line cap=round,line join=round]{%
  \path[use as bounding box] (0,0) rectangle (22,8);#1}}
\newcommand{\qgT}[3][4]{\fill[qblue,rounded corners=.25pt] (#2-#3,#1-1.2) rectangle (#2+#3,#1+1.2);}
\newcommand{\qgV}[3]{\fill[qblue,rounded corners=.25pt] (#1-.75,#2-#3) rectangle (#1+.75,#2+#3);}
\newcommand{\qgZ}[3][4]{\fill[qorange,rounded corners=.25pt] (#2-#3,#1-1.2) rectangle (#2+#3,#1+1.2);}
\newcommand{\qgJ}[2]{\fill[qink] (#1,#2) circle[radius=.5pt];}
\newcommand{\qgN}[2]{\path[fill=white,draw=qpurple,line width=.32pt] (#1,#2) circle[radius=.66pt];}
\tikzset{qgseg/.style={qgrey,line width=.45pt}, qgedge/.style={qpurple,line width=.4pt}}
\newcommand{\qccdglyph}[2]{\ifcsname qccd@#1@#2\endcsname\qccdglyphbox{\csname qccd@#1@#2\endcsname}%
  \else\PackageError{qccd-style}{Unknown pictogram key `#2'}{Add a drawing for it in qccd-style.tex.}\fi}
\DeclareRobustCommand{\pglyph}[1]{\qccdglyph{p}{#1}}
\DeclareRobustCommand{\dglyph}[1]{\qccdglyph{d}{#1}}
\DeclareRobustCommand{\Pg}[2]{#1\nobreak\hspace{.18em}\pglyph{#2}}
\DeclareRobustCommand{\Dg}[2]{#1\nobreak\hspace{.18em}\dglyph{#2}}
\makeatletter
\def\qccd@defglyph#1#2#3{\expandafter\def\csname qccd@#1@#2\endcsname{#3}}
\qccd@defglyph{p}{chain}{\draw[qgedge] (2,4)--(20,4);\qgN{2}{4}\qgN{6.5}{4}\qgN{11}{4}\qgN{15.5}{4}\qgN{20}{4}}
\qccd@defglyph{p}{local}{\draw[qgedge] (2,2)--(20,2) (2,2) to[bend left=55] (11,2) (6.5,2) to[bend left=55] (15.5,2)
  (11,2) to[bend left=55] (20,2);\qgN{2}{2}\qgN{6.5}{2}\qgN{11}{2}\qgN{15.5}{2}\qgN{20}{2}}
\qccd@defglyph{p}{star}{\draw[qgedge] (11,4)--(2.5,4) (11,4)--(5.5,7.3) (11,4)--(16.5,7.3) (11,4)--(19.5,4) (11,4)--(16.5,.7)
  (11,4)--(5.5,.7);\qgN{2.5}{4}\qgN{5.5}{7.3}\qgN{16.5}{7.3}\qgN{19.5}{4}\qgN{16.5}{.7}\qgN{5.5}{.7}\fill[qpurple] (11,4) circle[radius=.8pt];}
\qccd@defglyph{p}{sparse}{\draw[qgedge] (2,2) to[bend left=50] (20,2) (6.5,2) to[bend left=45] (15.5,2);
  \qgN{2}{2}\qgN{6.5}{2}\qgN{11}{2}\qgN{15.5}{2}\qgN{20}{2}}
\qccd@defglyph{p}{dense}{\draw[qgedge,line width=.35pt] (11,7.4)--(19.5,4.6)--(16.3,.6)--(5.7,.6)--(2.5,4.6)--cycle
  (11,7.4)--(16.3,.6)--(2.5,4.6)--(19.5,4.6)--(5.7,.6)--cycle;
  \qgN{11}{7.4}\qgN{19.5}{4.6}\qgN{16.3}{.6}\qgN{5.7}{.6}\qgN{2.5}{4.6}}
\qccd@defglyph{d}{line-c4}{\draw[qgseg] (1,4)--(21,4);\qgT{2}{.9}\qgT{4.57}{.9}\qgT{7.14}{.9}\qgT{9.71}{.9}\qgT{12.29}{.9}\qgT{14.86}{.9}\qgT{17.43}{.9}\qgT{20}{.9}}
\qccd@defglyph{d}{line-c6}{\draw[qgseg] (1,4)--(21,4);\qgT{2.3}{1.25}\qgT{5.78}{1.25}\qgT{9.26}{1.25}\qgT{12.74}{1.25}\qgT{16.22}{1.25}\qgT{19.7}{1.25}}
\qccd@defglyph{d}{line-c8}{\draw[qgseg] (1,4)--(21,4);\qgT{2.6}{1.55}\qgT{6.8}{1.55}\qgT{11}{1.55}\qgT{15.2}{1.55}\qgT{19.4}{1.55}}
\qccd@defglyph{d}{line-c12}{\draw[qgseg] (1,4)--(21,4);\qgT{3}{1.9}\qgT{8.33}{1.9}\qgT{13.67}{1.9}\qgT{19}{1.9}}
\qccd@defglyph{d}{line-c16}{\draw[qgseg] (1,4)--(21,4);\qgT{3.8}{2.7}\qgT{11}{2.7}\qgT{18.2}{2.7}}
\qccd@defglyph{d}{line-c24}{\draw[qgseg] (1,4)--(21,4);\qgT{5.6}{4.3}\qgT{16.4}{4.3}}
\qccd@defglyph{d}{line-c32}{\draw[qgseg] (1,4)--(21,4);\qgT{6.6}{5.4}\qgT{17.6}{3.3}}
\qccd@defglyph{d}{line-c48}{\qgT{11}{9.8}}
\qccd@defglyph{d}{line-l6}{\csname qccd@d@line-c6\endcsname}
\qccd@defglyph{d}{line-moveless}{\draw[qgseg] (1,4)--(21,4);\qgT{2}{1}\qgT{5}{1}\qgT{8}{1}\qgT{11}{1}\qgT{14}{1}\qgT{17}{1}\qgT{20}{1}}
\qccd@defglyph{d}{line-chain2}{\draw[qgseg] (1,4)--(21,4);\qgT{1.6}{.55}\qgT{3.8}{.55}\qgT{5.9}{.55}\qgT{8.1}{.55}\qgT{10.3}{.55}
  \qgT{12.4}{.55}\qgT{14.6}{.55}\qgT{16.8}{.55}\qgT{18.9}{.55}\qgT{21}{.55}}
\qccd@defglyph{d}{single-trap}{\qgT{11}{10}}
\qccd@defglyph{d}{ring-c12}{\draw[qgseg] (1.2,1.2) rectangle (20.8,6.8);\qgT[1.2]{7}{3}\qgT[1.2]{15}{3}\qgT[6.8]{7}{3}\qgT[6.8]{15}{3}
  \qgJ{1.2}{1.2}\qgJ{20.8}{1.2}\qgJ{20.8}{6.8}\qgJ{1.2}{6.8}}
\qccd@defglyph{d}{ring-cyclone-c12}{\csname qccd@d@ring-c12\endcsname}
\qccd@defglyph{d}{ring-c6}{\draw[qgseg] (1.2,1.2) rectangle (20.8,6.8);\qgT[1.2]{4.4}{1.5}\qgT[1.2]{8.8}{1.5}\qgT[1.2]{13.2}{1.5}\qgT[1.2]{17.6}{1.5}
  \qgT[6.8]{4.4}{1.5}\qgT[6.8]{8.8}{1.5}\qgT[6.8]{13.2}{1.5}\qgT[6.8]{17.6}{1.5}\qgJ{1.2}{1.2}\qgJ{20.8}{1.2}\qgJ{20.8}{6.8}\qgJ{1.2}{6.8}}
\qccd@defglyph{d}{ring-cyclone-c4}{\draw[qgseg] (1.2,1.2) rectangle (20.8,6.8);\qgT[1.2]{3.6}{.8}\qgT[1.2]{6.4}{.8}\qgT[1.2]{9.6}{.8}\qgT[1.2]{12.4}{.8}
  \qgT[1.2]{15.6}{.8}\qgT[1.2]{18.4}{.8}\qgT[6.8]{3.6}{.8}\qgT[6.8]{6.4}{.8}\qgT[6.8]{9.6}{.8}\qgT[6.8]{12.4}{.8}\qgT[6.8]{15.6}{.8}\qgT[6.8]{18.4}{.8}
  \qgJ{1.2}{1.2}\qgJ{20.8}{1.2}\qgJ{20.8}{6.8}\qgJ{1.2}{6.8}}
\qccd@defglyph{d}{ringcross-c6}{\draw[qgseg] (4,1.2) rectangle (18,6.8) (4,1.2)--(18,6.8) (4,6.8)--(18,1.2);
  \qgT[1.2]{8.6}{1.6}\qgT[1.2]{13.4}{1.6}\qgT[6.8]{8.6}{1.6}\qgT[6.8]{13.4}{1.6}\qgV{4}{4}{1.5}\qgV{18}{4}{1.5}
  \qgJ{4}{1.2}\qgJ{18}{1.2}\qgJ{18}{6.8}\qgJ{4}{6.8}\qgJ{11}{4}}
\qccd@defglyph{d}{ring-docks}{\draw[qgseg] (3,2.6) rectangle (19,5.4) (7,5.4)--(7,6.8) (15,5.4)--(15,6.8) (7,2.6)--(7,1.2) (15,2.6)--(15,1.2);
  \qgT[6.8]{7}{1.7}\qgT[6.8]{15}{1.7}\qgT[1.2]{7}{1.7}\qgT[1.2]{15}{1.7}\qgJ{7}{5.4}\qgJ{15}{5.4}\qgJ{7}{2.6}\qgJ{15}{2.6}}
\qccd@defglyph{d}{racetrack-h2}{\draw[qgseg,rounded corners=1.6pt] (1,1.2) rectangle (21,6.8);
  \qgZ[1.2]{7.5}{1.8}\qgZ[1.2]{14.5}{1.8}\qgZ[6.8]{7.5}{1.8}\qgZ[6.8]{14.5}{1.8}}
\qccd@defglyph{d}{dualloop-coupled}{\draw[qgseg] (1.2,1.2) rectangle (20.8,6.8) (6,3.2) rectangle (16,4.8) (8.5,1.2)--(8.5,3.2) (13.5,4.8)--(13.5,6.8);
  \qgT[1.2]{4}{1.4}\qgT[1.2]{13.5}{1.4}\qgT[6.8]{8.5}{1.4}\qgT[6.8]{18}{1.4}\fill[qblue] (10.2,3.6) rectangle (11.8,4.4);}
\newcommand{\qgM}[2]{\fill[qblue,rounded corners=.2pt] (#1-1.25,#2-.68) rectangle (#1+1.25,#2+.68);}
\qccd@defglyph{d}{mesh-c6}{\draw[qgseg,line width=.35pt] (3.5,.9)--(18.5,.9) (3.5,2.97)--(18.5,2.97) (3.5,5.03)--(18.5,5.03) (3.5,7.1)--(18.5,7.1)
  (3.5,.9)--(3.5,7.1) (8.5,.9)--(8.5,7.1) (13.5,.9)--(13.5,7.1) (18.5,.9)--(18.5,7.1);
  \foreach \qgy in {.9,2.97,5.03,7.1}{\qgM{3.5}{\qgy}\qgM{8.5}{\qgy}\qgM{13.5}{\qgy}\qgM{18.5}{\qgy}}}
\qccd@defglyph{d}{ladder-c12}{\draw[qgseg] (3.2,1.4)--(18.8,1.4) (3.2,6.6)--(18.8,6.6) (3.2,1.4)--(3.2,6.6) (8.4,1.4)--(8.4,6.6) (13.6,1.4)--(13.6,6.6) (18.8,1.4)--(18.8,6.6);
  \qgT[1.4]{3.2}{1.9}\qgT[1.4]{8.4}{1.9}\qgT[1.4]{13.6}{1.9}\qgT[1.4]{18.8}{1.9}\qgT[6.6]{3.2}{1.9}\qgT[6.6]{8.4}{1.9}\qgT[6.6]{13.6}{1.9}\qgT[6.6]{18.8}{1.9}}
\qccd@defglyph{d}{ladder}{\draw[qgseg] (1,1.4)--(21,1.4) (1,6.6)--(21,6.6) (6,1.4)--(6,6.6) (16,1.4)--(16,6.6);
  \qgT[1.4]{3}{1.3}\qgT[1.4]{11}{1.3}\qgT[1.4]{19}{1.3}\qgT[6.6]{3}{1.3}\qgT[6.6]{11}{1.3}\qgT[6.6]{19}{1.3}
  \qgJ{6}{1.4}\qgJ{16}{1.4}\qgJ{6}{6.6}\qgJ{16}{6.6}}
\qccd@defglyph{d}{bus-c6}{\draw[qgseg] (3.5,4)--(18.5,4) (3.5,1.4)--(3.5,6.6) (8.5,1.4)--(8.5,6.6) (13.5,1.4)--(13.5,6.6) (18.5,1.4)--(18.5,6.6);
  \qgV{3.5}{6.6}{1.3}\qgV{8.5}{6.6}{1.3}\qgV{13.5}{6.6}{1.3}\qgV{18.5}{6.6}{1.3}\qgV{3.5}{1.4}{1.3}\qgV{8.5}{1.4}{1.3}\qgV{13.5}{1.4}{1.3}\qgV{18.5}{1.4}{1.3}
  \qgJ{3.5}{4}\qgJ{8.5}{4}\qgJ{13.5}{4}\qgJ{18.5}{4}}
\qccd@defglyph{d}{bus-c12}{\draw[qgseg] (6,4)--(16,4) (6,1.4)--(6,6.6) (16,1.4)--(16,6.6);
  \fill[qblue,rounded corners=.25pt] (4.3,5.4) rectangle (7.7,7.9) (14.3,5.4) rectangle (17.7,7.9) (4.3,.1) rectangle (7.7,2.6) (14.3,.1) rectangle (17.7,2.6);
  \qgJ{6}{4}\qgJ{16}{4}}
\qccd@defglyph{d}{grid-g2x3}{\draw[qgseg] (4,4)--(18,4) (4,1.4)--(4,6.6) (11,1.4)--(11,6.6) (18,1.4)--(18,6.6);
  \qgV{4}{6.6}{1.3}\qgV{11}{6.6}{1.3}\qgV{18}{6.6}{1.3}\qgV{4}{1.4}{1.3}\qgV{11}{1.4}{1.3}\qgV{18}{1.4}{1.3}\qgJ{4}{4}\qgJ{11}{4}\qgJ{18}{4}}
\qccd@defglyph{d}{tree-c6}{\draw[qgseg] (11,7.3)--(3.5,4.4) (11,7.3)--(8.5,4.4) (11,7.3)--(13.5,4.4) (11,7.3)--(18.5,4.4)
  (3.5,4.4)--(2.2,1.4) (3.5,4.4)--(4.8,1.4) (8.5,4.4)--(7.2,1.4) (8.5,4.4)--(9.8,1.4) (13.5,4.4)--(12.2,1.4) (13.5,4.4)--(14.8,1.4)
  (18.5,4.4)--(17.2,1.4) (18.5,4.4)--(19.8,1.4);
  \qgV{2.2}{1.4}{1.3}\qgV{4.8}{1.4}{1.3}\qgV{7.2}{1.4}{1.3}\qgV{9.8}{1.4}{1.3}\qgV{12.2}{1.4}{1.3}\qgV{14.8}{1.4}{1.3}\qgV{17.2}{1.4}{1.3}\qgV{19.8}{1.4}{1.3}
  \qgJ{11}{7.3}\qgJ{3.5}{4.4}\qgJ{8.5}{4.4}\qgJ{13.5}{4.4}\qgJ{18.5}{4.4}}
\qccd@defglyph{d}{tree-c12}{\draw[qgseg] (11,6.8)--(3.5,2) (11,6.8)--(8.5,2) (11,6.8)--(13.5,2) (11,6.8)--(18.5,2);
  \fill[qblue,rounded corners=.25pt] (2.1,.1) rectangle (4.9,3.4) (7.1,.1) rectangle (9.9,3.4) (12.1,.1) rectangle (14.9,3.4) (17.1,.1) rectangle (19.9,3.4);
  \qgJ{11}{6.8}}
\qccd@defglyph{d}{grid-c2}{\draw[qgseg] (2,1.2)--(20,1.2) (2,6.8)--(20,6.8) (2,1.2)--(2,6.8) (11,1.2)--(11,6.8) (20,1.2)--(20,6.8);
  \qgT[1.2]{6.5}{.9}\qgT[1.2]{15.5}{.9}\qgT[6.8]{6.5}{.9}\qgT[6.8]{15.5}{.9}\qgV{2}{4}{.9}\qgV{11}{4}{.9}\qgV{20}{4}{.9}
  \qgJ{2}{1.2}\qgJ{11}{1.2}\qgJ{20}{1.2}\qgJ{2}{6.8}\qgJ{11}{6.8}\qgJ{20}{6.8}}
\qccd@defglyph{d}{grid-zoned}{\draw[qgseg] (2,1.2)--(20,1.2) (2,6.8)--(20,6.8) (2,1.2)--(2,6.8) (11,1.2)--(11,6.8) (20,1.2)--(20,6.8);
  \qgT[1.2]{6.5}{2.2}\qgT[1.2]{15.5}{2.2}\qgT[6.8]{6.5}{2.2}\qgT[6.8]{15.5}{2.2}\qgZ[1.2]{6.5}{.7}\qgZ[1.2]{15.5}{.7}\qgZ[6.8]{6.5}{.7}\qgZ[6.8]{15.5}{.7}
  \qgJ{2}{1.2}\qgJ{11}{1.2}\qgJ{20}{1.2}\qgJ{2}{6.8}\qgJ{11}{6.8}\qgJ{20}{6.8}}
\qccd@defglyph{d}{wise}{\draw[qgseg,line width=.3pt] (1.5,1.2)--(20.5,1.2) (1.5,4)--(20.5,4) (1.5,6.8)--(20.5,6.8);
  \qgT[1.2]{2.4}{.75}\qgT[1.2]{6.2}{.75}\qgT[1.2]{10}{.75}\qgT[1.2]{13.8}{.75}\qgT[1.2]{17.6}{.75}\qgT[4]{4.3}{.75}\qgT[4]{8.1}{.75}\qgT[4]{11.9}{.75}
  \qgT[4]{15.7}{.75}\qgT[4]{19.5}{.75}\qgT[6.8]{2.4}{.75}\qgT[6.8]{6.2}{.75}\qgT[6.8]{10}{.75}\qgT[6.8]{13.8}{.75}\qgT[6.8]{17.6}{.75}}
\qccd@defglyph{d}{grid-pz}{\draw[qgseg] (2,1.4)--(15,1.4) (2,6.6)--(15,6.6) (8.5,1.4)--(8.5,6.6) (15,1.4)--(15,6.6) (15,4)--(17,4);
  \qgT[1.4]{4}{1.6}\qgT[1.4]{11.5}{1.6}\qgT[6.6]{4}{1.6}\qgT[6.6]{11.5}{1.6}\qgZ{19.3}{2.3}\qgJ{8.5}{1.4}\qgJ{8.5}{6.6}\qgJ{15}{1.4}\qgJ{15}{6.6}\qgJ{15}{4}}
\qccd@defglyph{d}{agent}{\draw[qgseg] (2.6,6.4)--(7,4)--(14,5.4)--(19.4,6.6) (7,4)--(4.4,1.2) (14,5.4)--(12.6,1.4) (14,5.4)--(19,2.2);
  \qgT[6.4]{2.6}{1.8}\qgV{4.4}{1.4}{1.3}\qgT[1.4]{12.6}{2.2}\qgT[6.6]{19.4}{1.8}\qgV{19}{2.1}{1.3}\qgJ{7}{4}\qgJ{14}{5.4}}
\makeatother

\lstdefinelanguage{tsir}{
  morekeywords={init,cool,gate,shuttle,rotate,dock,undock,measure,reset,barrier,R,VZ,MS},
  sensitive=true, morecomment=[l]{\#\#}, morestring=[b]",
}
\lstdefinelanguage{qccd-hardware}{
  morekeywords={hardware,architecture,site,trap,junction,path,channel,waveform,
    share,mandatory,optional,pulse,transport,rotation,interaction,readout,cooling,
    primitive,mode,from,to,via,on,at,with,duration,capacity,ports,drive,targets,
    supports,calibration,effect,requires,ensures,amplitude,frequency,phase,
    shape,program,routine,realizes,bind,using,optical,addressed,R,MS,
    broadcast,sites,limit,acquire,pump,controller},
  sensitive=true, morecomment=[l]{//}, morestring=[b]",
}
\lstdefinelanguage{qccd-circuit}{
  morekeywords={OPENQASM,include,qreg,creg,qubit,bit,circuit,qubits,bits,layer,
    H,X,Y,Z,S,T,R,RX,RY,RZ,CX,CZ,VZ,MS,h,x,y,z,s,t,rx,ry,rz,cx,cz,
    measure,reset,barrier,U,if,gate,opaque,pi,sin,cos,tan,exp,ln,sqrt},
  sensitive=true, morecomment=[l]{//}, morestring=[b]",
}
\lstdefinelanguage{qccd-rules}{
  morekeywords={rules,rule,requires,ensures,before,during,after,occ,move,gate,
    pulse,channel,waveform,capacity,forall,in,when,then,and,or,not,
    occ_after,occ_before,occBefore,occDuring,occAfter,start,end,duration,
    moveCount,pulseCount,Rcount,MScount,readoutCount,pumpCount,latchCount,pulses},
  sensitive=true, morecomment=[l]{//}, morestring=[b]",
}
\lstdefinelanguage{qccd-execution}{
  morekeywords={execution,program,place,init,at,emit,pulse,on,with,wait,delay,
    parallel,sequence,barrier,sync,frame,duration,channel,waveform,phase,
    amplitude,frequency,shape,targets,R,MS,VZ,shuttle,rotate,dock,undock,
    cool,measure,reset,tsir,arch,simd,transport,idle,broadcast,readout,pump,
    latch,saved,when,little_endian,via},
  sensitive=true, morecomment=[l]{//}, morestring=[b]",
}
\lstdefinelanguage{qccd-ir}{
  morekeywords={place,move,along,execute,wait,layer,U,CX,R,MS,measure,reset,
    barrier,latch,saved,when,little_endian,candidates,init},
  sensitive=true, morecomment=[l]{//}, morestring=[b]",
}
\lstdefinestyle{qccd-hardware}{language=qccd-hardware,
  keywordstyle=\color{qcodeblue}\bfseries}
\lstdefinestyle{qccd-circuit}{language=qccd-circuit,
  keywordstyle=\color{qcodepurple}\bfseries}
\lstdefinestyle{qccd-rules}{language=qccd-rules,
  keywordstyle=\color{qcodegreen}\bfseries}
\lstdefinestyle{qccd-execution}{language=qccd-execution,
  keywordstyle=\color{qcodeorange}\bfseries}
\lstdefinestyle{qccd-ir}{language=qccd-ir,
  keywordstyle=\color{qcodeteal}\bfseries}
\theoremstyle{acmdefinition}

\AtEndPreamble{\theoremstyle{acmdefinition}\newtheorem{remark}[theorem]{Remark}}
\input{qccd-hardware}
\makeatletter
\newcommand{\QCCDLiteratureGeometry}[1]{%
  \QCCDGeometry{#1}{6}%
  \def\qccd@font{11.4}\def\qccd@texthalo{1.1}%
  \def\qccd@bounds{(0,-2) rectangle (152,142)}%
}
\expandafter\def\csname qccd@architecture@kmw-fragment\endcsname{%
  \QCCDLiteratureGeometry{24}
  \def\qccd@sites{M0,M1,M2,M3,G0,G1}\def\qccd@junctions{J0,J1,J2,J3}%
  \QCCDDeclareSite{M0}{30}{20}{90}{qhwData}{2}{qhwData}
  \QCCDDeclareSite{M1}{30}{122}{90}{qhwData}{2}{qhwData}
  \QCCDDeclareSite{M2}{122}{20}{90}{qhwData}{2}{qhwData}
  \QCCDDeclareSite{M3}{122}{122}{90}{qhwData}{2}{qhwData}
  \QCCDDeclareSite{G0}{76}{50}{0}{qhwTrap}{2}{qhwTrap}
  \QCCDDeclareSite{G1}{76}{94}{0}{qhwTrap}{2}{qhwTrap}
  \QCCDDeclareJunction{J0}{30}{50}{qhwGrid}
  \QCCDDeclareJunction{J1}{122}{50}{qhwGrid}
  \QCCDDeclareJunction{J2}{30}{94}{qhwGrid}
  \QCCDDeclareJunction{J3}{122}{94}{qhwGrid}
  \foreach \a/\b in {M0/J0,J0/J2,J2/M1,M2/J1,J1/J3,J3/M3,
    J0/G0,G0/J1,J2/G1,G1/J3}{\QCCDRail{\a}{\b}}
}
\expandafter\def\csname qccd@scene@kmw-fragment\endcsname{%
  \QCCDHardware{kmw-fragment}
  \QCCDIon{q0}{M0}{0}\QCCDIon{q1}{G0}{1}
  \draw[-{Stealth[length=3.5bp,width=2.8bp]},qblue,line width=.85bp]
    (30,29)--(30,50)--(66,50);
}
\expandafter\def\csname qccd@architecture@pino-fragment\endcsname{%
  \QCCDLiteratureGeometry{32}
  \def\qccd@sites{S0,G0,S1,G1}\def\qccd@junctions{}%
  \QCCDDeclareSite{S0}{18}{70}{0}{qhwData}{2}{qhwData}
  \QCCDDeclareSite{G0}{57}{70}{0}{qhwTrap}{2}{qhwTrap}
  \QCCDDeclareSite{S1}{96}{70}{0}{qhwData}{2}{qhwData}
  \QCCDDeclareSite{G1}{135}{70}{0}{qhwTrap}{2}{qhwTrap}
  \foreach \a/\b in {S0/G0,G0/S1,S1/G1}{\QCCDRail{\a}{\b}}
}
\expandafter\def\csname qccd@scene@pino-fragment\endcsname{%
  \QCCDHardware{pino-fragment}
  \QCCDIon{q0}{S0}{0}\QCCDIon{q1}{G0}{1}
  \draw[-{Stealth[length=3.5bp,width=2.8bp]},qblue,line width=.85bp]
    (29,70)--(46,70);
}
\expandafter\def\csname qccd@architecture@lekitsch-fragment\endcsname{%
  \QCCDLiteratureGeometry{32}
  \def\qccd@sites{W,G,L,D}\def\qccd@junctions{J}%
  \QCCDDeclareSite{W}{76}{120}{90}{qhwData}{2}{qhwData}
  \QCCDDeclareSite{G}{18}{70}{0}{qhwTrap}{2}{qhwTrap}
  \QCCDDeclareSite{L}{134}{70}{0}{qhwData}{2}{qhwData}
  \QCCDDeclareSite{D}{76}{18}{90}{qhwTrap}{2}{qhwTrap}
  \QCCDDeclareJunction{J}{76}{70}{qhwGrid}
  \foreach \a/\b in {W/J,G/J,L/J,D/J}{\QCCDRail{\a}{\b}}
}
\expandafter\def\csname qccd@scene@lekitsch-fragment\endcsname{%
  \QCCDHardware{lekitsch-fragment}
  \QCCDIon{q0}{W}{0}\QCCDIon{q1}{G}{1}
  \draw[-{Stealth[length=3.5bp,width=2.8bp]},qblue,line width=.85bp]
    (76,105)--(76,70)--(34,70);
}
\expandafter\def\csname qccd@architecture@moses-fragment\endcsname{%
  \QCCDLiteratureGeometry{32}
  \def\qccd@sites{U0,U1,R,G1,G0,L}\def\qccd@junctions{}%
  \QCCDDeclareSite{U0}{48}{22}{0}{qhwData}{2}{qhwData}
  \QCCDDeclareSite{U1}{96}{22}{0}{qhwData}{2}{qhwData}
  \QCCDDeclareSite{R}{129}{70}{90}{qhwData}{2}{qhwData}
  \QCCDDeclareSite{G1}{96}{116}{0}{qhwTrap}{2}{qhwTrap}
  \QCCDDeclareSite{G0}{48}{116}{0}{qhwTrap}{2}{qhwTrap}
  \QCCDDeclareSite{L}{15}{70}{90}{qhwData}{2}{qhwData}
  \draw[qhwRail,line width=\qccd@railwidth\QCCDUnit,line cap=round]
    (U0) .. controls (15,22) and (15,42) .. (L)
    .. controls (15,98) and (15,116) .. (G0);
  \draw[qhwRail,line width=\qccd@railwidth\QCCDUnit,line cap=round]
    (U1) .. controls (129,22) and (129,42) .. (R)
    .. controls (129,98) and (129,116) .. (G1);
  \foreach \x/\y/\a in {23/36/135,18/48/110,18/91/70,23/104/45,
     121/36/45,126/48/70,126/91/110,121/104/135}
    {\QCCDElectrodePair{\x}{\y}{\a}}
  \QCCDRail{U0}{U1}\QCCDRail{G0}{G1}
}
\expandafter\def\csname qccd@scene@moses-fragment\endcsname{%
  \QCCDHardware{moses-fragment}
  \QCCDIon{q0}{L}{0}\QCCDIon{q1}{G0}{1}
  \draw[-{Stealth[length=3.5bp,width=2.8bp]},qblue,line width=.85bp]
    (15,84) .. controls (15,107) and (24,116) .. (35,116);
}
\makeatother

\newcommand{\iscasubmissionnumber}{DRAFT}

\setcopyright{none}
\acmConference[ISCA '27]{54th IEEE/ACM International Symposium on Computer Architecture}{June 5--9, 2027}{Atlanta, Georgia, USA}
\acmYear{2027}
\acmISBN{}
\acmDOI{}

\newcommand{\QCCDPhysicalMainText}{%
\paragraph{Cost model}
We compare modeled device execution time under one extension of the
QCCDSim reference model~\cite{murali2020architecting}, the same for all
designs. In microseconds: a one-qubit gate takes 5, a readout 400 and a
reset 50~\cite{jones2026surfacecodes}; a split and a merge 80 each; a
segment 5, a three-way junction 100 and a four-way junction 120. A
transport is one split, travel and merge between two traps and is charged
for every segment, junction and bend (a change of travel direction) on
its path; a bend and a port of a trap with three or more segments take
100, our assumption where the reference model is silent. The two-qubit gate follows the rule for frequency-modulated
gates of Murali et al.~\cite{murali2020architecting} at the declared trap capacity $c$,
$\max(13.33\,c-54,\,100)$: 100 up to 11 ions, 266 at 24 and 586 at 48. A
gate SWAP costs three two-qubit gates of its trap, as in QCCDSim.
}

\newcounter{bodycontentpages}
\newcounter{appendixstartpage}
\ifqccdarxiv
\hypersetup{pdftitle={QCCDCert: Formally Verified Compilation for QCCD Co-Design},pdfsubject={},pdfkeywords={}}
\else
\hypersetup{pdftitle={Verified Compilation for Software-Hardware Co-Design of QCCD Quantum Computers},pdfauthor={},pdfsubject={},pdfkeywords={}}
\fi

\newif\ifqccdnotes
\ifdefined\QCCDNoNotes\qccdnotesfalse\else\qccdnotestrue\fi

\DeclareRobustCommand{\rev}[1]{#1}

\DeclareRobustCommand{\rednote}[1]{\ifqccdnotes{\color{red}#1}\fi}
\begin{document}

\ifqccdarxiv
\title{QCCDCert: Formally Verified Compilation for QCCD Co-Design}
\author{John Zhuoyang Ye}
\affiliation{%
  \institution{University of California, Los Angeles}
  \city{Los Angeles}
  \state{CA}
  \country{USA}}
\email{yezhuoyang@cs.ucla.edu}
\author{Jens Palsberg}
\affiliation{%
  \institution{University of California, Los Angeles}
  \city{Los Angeles}
  \state{CA}
  \country{USA}}
\email{palsberg@cs.ucla.edu}
\author{Ke Sun}
\affiliation{%
  \institution{University of California, Berkeley}
  \city{Berkeley}
  \state{CA}
  \country{USA}}
\email{kesun@berkeley.edu}
\author{Hartmut H\"{a}ffner}
\affiliation{%
  \institution{University of California, Berkeley}
  \city{Berkeley}
  \state{CA}
  \country{USA}}
\email{hhaeffner@berkeley.edu}
\renewcommand{\shortauthors}{Ye, Palsberg, Sun and H\"{a}ffner}

\else
\title{Verified Compilation for Software--Hardware Co-Design of QCCD Quantum Computers}
\author{ISCA 2027 Submission
  \textbf{\#\iscasubmissionnumber} -- Confidential Draft -- Do NOT Distribute!!}
\renewcommand{\shortauthors}{ISCA 2027 Submission \#\iscasubmissionnumber}
\fi

\begin{abstract}
Quantum charge-coupled device (QCCD) architectures enable scalable
trapped-ion quantum computing by dynamically shuttling ions between
specialized processing and storage zones. A QCCD can be built with many
trap sizes and layouts. Comparing them requires compiling every program
under each design's transport, capacity and control constraints, and
trusting the target code of each compilation. We present a co-design framework built
on one Lean-verified compiler. The hardware description and its operating rules
are compiler inputs, so a single theorem covers every candidate design:
the target code of every compilation preserves its program's behavior and obeys the
hardware rules. Across 75 benchmark programs of 11 to 130 qubits, the best
design found takes 66\% less modeled execution time on average than the
worst design of equal capacity. An AI agent that proposes devices based on our compiler's checked results
beats the fastest of these designs on 71 of 75 programs. 
Our verified compiler provably
avoids the 11 defects that we report in three previous compilers. It scales to hundreds of qubits, compiling 512-qubit programs
for a 38-trap device in 1 to 29 minutes.
\end{abstract}
\maketitle

\section{Introduction}
\label{sec:intro}

\begin{figure}[t]
\centering
\begingroup
\makeatletter
\patchcmd{\qccd@architecture@ring}{{L0}}{{}}{}{}
\expandafter\patchcmd\csname qccd@architecture@line4\endcsname{{2}{qhwTrap}}{{4}{qhwTrap}}{}{}
\expandafter\patchcmd\csname qccd@architecture@line4\endcsname{{2}{qhwTrap}}{{4}{qhwTrap}}{}{}
\expandafter\patchcmd\csname qccd@architecture@line4\endcsname{{2}{qhwTrap}}{{4}{qhwTrap}}{}{}
\expandafter\patchcmd\csname qccd@architecture@line4\endcsname{{2}{qhwTrap}}{{4}{qhwTrap}}{}{}
\patchcmd{\qccd@architecture@grid}
  {{T0_1h}{140}{203}{0}{qhwTrap}{2}}
  {{T0_1h}{140}{203}{0}{qhwTrap}{4}}{}{}
\makeatother
\resizebox{\columnwidth}{!}{%
\begin{tikzpicture}[x=1pt,y=1pt,font=\sffamily\fontsize{7.5}{9}\selectfont,
  head/.style={inner sep=0pt,text=black,align=left,anchor=north west,
    font=\sffamily\bfseries\fontsize{8.5}{10}\selectfont},
  txt/.style={inner sep=0pt,text=black,align=center},
  small/.style={txt,font=\sffamily\fontsize{6.8}{8}\selectfont},
  feature/.style={txt,font=\sffamily\bfseries\fontsize{7.4}{9}\selectfont},
  card/.style={draw=qgrey!65,fill=qcard,rounded corners=2.5pt,line width=.65pt},
  trustbadge/.style={circle,fill=white,line width=.9pt,minimum size=14.5pt,
    inner sep=0pt,font=\fontsize{11}{11}\selectfont},
  flow/.style={-{Stealth[length=3.6pt,width=3pt]},line width=.85pt,
    draw=qgreen,rounded corners=2pt}]
\path[use as bounding box] (-6,5) rectangle (248,-294);

\node[head] at (0,4) {(a) Comparison with prior work};
\path[fill=qmint!45] (212,-11) rectangle (248,-91);
\node[small,anchor=west,font=\sffamily\bfseries\fontsize{7.2}{8.5}\selectfont]
  at (3,-22) {Feature};
\node[small,font=\sffamily\fontsize{6.7}{7.5}\selectfont] at (122,-22)
  {QCCDSim\\\cite{murali2020architecting}};
\node[small,font=\sffamily\fontsize{6.7}{7.5}\selectfont] at (158,-22)
  {Muzzle the\\Shuttle~\cite{saki2022muzzle}};
\node[small,font=\sffamily\fontsize{6.7}{7.5}\selectfont] at (194,-22)
  {MQT\\IonShuttler\\\cite{schoenberger2025multizone}};
\node[feature] at (230,-22) {Ours};
\draw[qgrey,line width=.6pt] (0,-34)--(248,-34);
\foreach \y in {-53,-72,-91}
  {\draw[qgrey!35,line width=.35pt] (0,\y)--(248,\y);}
\node[txt,anchor=west,font=\sffamily\fontsize{7.4}{9}\selectfont]
  at (3,-43.5) {Hardware formalization (\S\ref{sec:hardware-model})};
\node[txt,anchor=west,font=\sffamily\fontsize{7.4}{9}\selectfont]
  at (3,-62.5) {Flexible operating rules (\S\ref{sec:model})};
\node[txt,anchor=west,font=\sffamily\fontsize{7.4}{9}\selectfont]
  at (3,-81.5) {Correctness proof (\S\ref{sec:checker})};
\foreach \y in {-43.5,-62.5,-81.5}{
  \foreach \x in {122,158,194}
    {\node[trustbadge,draw=qred,text=qred] at (\x,\y) {\xmark};}
  \node[trustbadge,draw=qgreen,text=qgreen] at (230,\y) {\cmark};
}

\node[head] at (0,-101) {(b) Automatic, large-scale co-design};
\node[small] at (124,-116) {Fixed program, total ion capacity and cost model};
\node[small,font=\sffamily\bfseries\fontsize{7}{8}\selectfont]
  at (58,-128) {1. Propose device + rules};
\node[small,font=\sffamily\bfseries\fontsize{7}{8}\selectfont]
  at (192.5,-128) {2. Compile and check};

\path[card] (0,-137) rectangle (117,-207);
\node[feature] at (58.5,-145)
  {Hardware formalization (\S\ref{sec:hardware-model})};
\node[small,font=\sffamily\fontsize{6.4}{7.5}\selectfont]
  at (58.5,-155) {16 ion slots in every layout};
\node[inner sep=0pt] at (21,-175) {\QCCDThumb{line4}{11pt}};
\node[inner sep=0pt] at (60,-175) {\QCCDThumb{ring}{30pt}};
\node[inner sep=0pt] at (99,-175) {\QCCDThumb{grid}{30pt}};
\foreach \x/\name in {21/Line,60/Ring,99/Grid}
  {\node[small] at (\x,-193) {\name};}
\node[small,font=\sffamily\fontsize{6.2}{7}\selectfont]
  at (58.5,-201) {Capsules: traps; lines: transport paths};

\path[card] (0,-215) rectangle (117,-271);
\node[feature] at (58.5,-223)
  {Flexible operating rules (\S\ref{sec:model})};
\foreach \y in {-237,-251}
  {\path[draw=qgrey!40,fill=white,rounded corners=1.5pt,line width=.4pt]
    (4,\y+5.5) rectangle (113,\y-5.5);}
\node[small] at (58.5,-237) {At most 2 ions per gate zone};
\node[small] at (58.5,-251) {One optical action at a time};
\node[small] at (58.5,-264) {Rules supplied as inputs};

\path[card,draw=qgreen,fill=qmint!30] (137,-137) rectangle (248,-207);
\node[feature] at (192.5,-145) {Correctness proof (\S\ref{sec:checker})};
\node[small] at (192.5,-156) {Verified compiler (\S\ref{sec:compiler})};
\node[small] at (192.5,-166) {Behavior kept; rules obeyed};
\node[trustbadge,minimum size=17pt,font=\fontsize{13}{13}\selectfont,
  draw=qgreen,text=qgreen] at (151,-184) {\cmark};
\foreach \y/\bars in {-176/{0/12,16/16,36/12},-183/{4/16,24/8,36/16},-190/{0/8,12/20,36/8}}{
  \foreach \x/\w in \bars
    {\path[fill=qteal!55] (168+\x,\y) rectangle +(\w,-4);}
}
\draw[-{Stealth[length=2pt]},qgrey,line width=.4pt] (224,-194)--(241,-194);
\node[small,font=\sffamily\fontsize{6.2}{7}\selectfont] at (233,-187) {time};
\node[small] at (192.5,-201) {Checked target code and schedule};

\path[card,draw=qgreen,fill=qmint!30] (137,-215) rectangle (248,-271);
\node[feature] at (192.5,-223) {3. Score: finish time (\S\ref{sec:codesign})};
\foreach \x/\h/\c in {150/20/qgrey,173/11/qgreen,196/24/qgrey}
  {\path[fill=\c!65] (\x,-254) rectangle +(10,\h);}
\draw[qgrey,line width=.4pt] (146,-254)--(210,-254);
\foreach \x/\name in {155/Line,178/Ring,201/Grid}
  {\node[small] at (\x,-261) {\name};}
\node[small,font=\sffamily\itshape\fontsize{6.2}{7.5}\selectfont]
  at (228,-238) {Schematic};
\node[small] at (228,-255) {Lower\\is better};

\draw[flow] (117,-173)--(137,-173);
\draw[flow] (58.5,-207)--(58.5,-215);
\draw[flow] (117,-243)--(127,-243)--(127,-195)--(137,-195);
\draw[flow] (192.5,-207)--(192.5,-215);
\draw[flow] (192.5,-271)--(192.5,-290)--(-5,-290)--(-5,-173)--(0,-173);
\node[small,anchor=south] at (121,-285)
  {4. Use schedule records to refine device + rules};
\end{tikzpicture}}
\endgroup
\caption{Verified QCCD co-design. (a) Hardware formalization means
machine-checked semantics; flexible rules are assertion
inputs; the correctness proof is one reusable compiler theorem. (b) Only
complete, checked target code receives a score, and its records guide the
next proposal. Schedule rows are traps and bars are events, schematic
and not measured.}
\label{fig:codesign-overview}
\Description{A single-column, two-part figure. The comparison table has
separate columns for QCCDSim, Muzzle the Shuttle, MQT IonShuttler and our
work. Its three rows are hardware formalization, flexible operating rules
and correctness proof, each linked to its section. Prior tools have circled
red crosses and our work has circled green checks for these formally defined
features. Below, line, ring and grid QCCD thumbnails and concrete operating
rules feed a verified compiler. A correctness proof and checks
support the accepted target code and its schedule. Schematic finish-time bars score
candidate designs; schedule feedback refines the next device and rules.}
\end{figure}
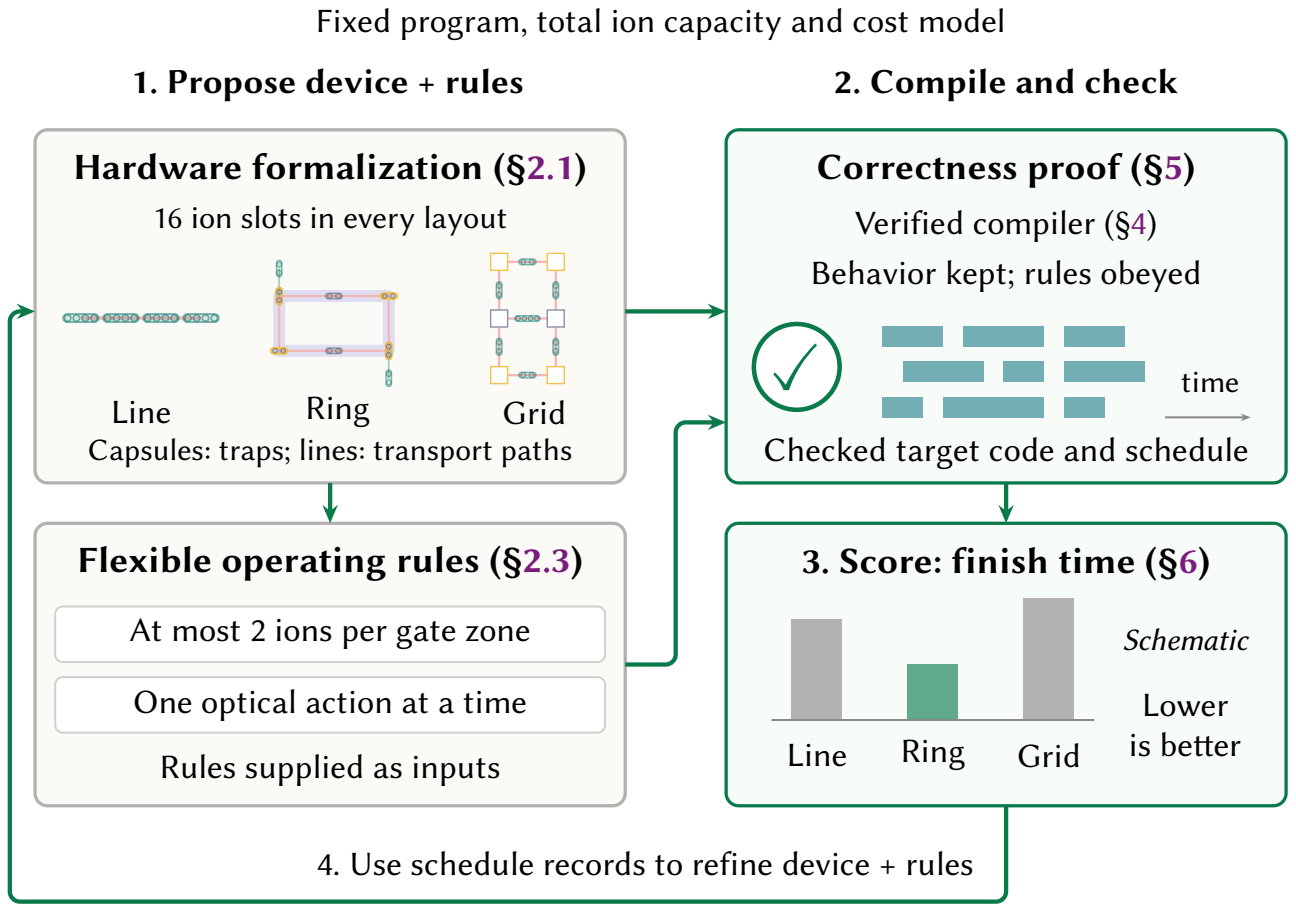

Quantum computers promise large speedups, and the resources that they
need keep falling: factoring a 2048-bit RSA integer with Shor's
algorithm~\cite{shor1994algorithms} was estimated at 20 million noisy
qubits in 2021~\cite{Gidney2021howtofactorbit} and at tens of thousands
today~\cite{cain2026shor,pappalardo2026shorcat}. Every such estimate
assumes an architecture and a compiler that maps the algorithm onto it. A
compiled program of this size cannot be checked by simulation, so a better idea is to verify the compiler itself.

Trapped ions are a leading platform, with the highest gate fidelity and
the longest coherence
time~\cite{pino2021demonstration,moses2023race}. A quantum
charge-coupled device (QCCD)~\cite{kielpinski2002architecture} scales
them by holding the ions in many small traps joined by transport segments
and junctions; a two-qubit gate needs both ions in one trap. Designing a
QCCD means choosing the traps, their sizes, their connections and their
controls.

Prior work proposes and evaluates such designs.
QCCDSim~\cite{murali2020architecting} compares trap capacities and
\Dg{linear}{line-l6} and \Dg{grid}{grid-g2x3} layouts; Muzzle the
Shuttle~\cite{saki2022muzzle} and S-SYNC~\cite{zhu2025ssync} reduce
shuttling on a given device; others propose one architecture, such as a
grid with shared control wiring
(\Dg{WISE}{wise})~\cite{malinowski2023wise}, a
\Dg{race track}{racetrack-h2}~\cite{moses2023race},
\Dg{rings}{ring-cyclone-c12}~\cite{khan2026cyclone} and surface-code
layouts~\cite{jones2026surfacecodes}. Each fixes one or a few designs. In this research, we address the following question: 
\textbf{What if instead we want the best design for a specific
circuit?} Circuits move ions differently: some keep most gates inside one
trap, others, such as the quantum Fourier transform, make ions travel all
the time. Finding the best design for a circuit is a software--hardware
co-design problem: we compile the same circuit for many candidate designs
and compare the results (Fig.~\ref{fig:codesign-overview}). Human
designers or AI agents can propose such
candidates~\cite{krishnan2023archgym}, but an agent can also exploit
mistakes in the simulator that scores
them~\cite{gupta2026archagent}. A co-design search is therefore only as
trustworthy as the compiler behind its scores, which leads to our second
question.

\textbf{What kind of compiler does a co-design search need?} We argue that it must meet two requirements. First, the compiler must
generate good code for every design (\emph{Requirement~1}), or the search
undervalues designs whose features it does not use. Second, the compiler must be
verified (\emph{Requirement~2}), because incorrect target code can make
a design look better than it is. \textbf{To show that this risk is real, we audited three published QCCD
compilers and report 11 defects} (Section~\ref{sec:eval-compiler}). Their compiled
schedules overfill traps, send two ions through one junction at once, or
take an ion from the middle of a chain, so the affected designs look
larger or faster than they are. Existing verified quantum
compilers~\cite{hietala2021voqc,shi2019certiq} cannot exclude such
defects: they verify circuit rewrites and do not model ion movement,
capacity or shared controls.

We therefore introduce a new verified QCCD compiler and a co-design loop built on
it. Three challenging problems had no existing solution, and their solutions are the new ideas of this paper.

\textbf{Challenge 1: How does an architecture specification express that
qubits move?}
A processor with fixed qubits, such as a superconducting chip, is typically specified by a coupling graph: a two-qubit gate is allowed only
on an edge~\cite{li2019tackling,tan2020optimal}. In a QCCD, whether an instruction may run depends on the state of the device throughout the instruction, including intermediate states during transport. Moreover, each laboratory operates its device under policies of its own, such as which zones may run gates and which electrodes share a signal~\cite{pino2021demonstration,moses2023race,malinowski2023wise}. 
The audited
compilers build such rules into their code but sometimes break them (Fig.~\ref{fig:baseline-bugs}). \textbf{Our
idea is to give every rule a single form: an assertion over an \emph{event
frame}}, which is the state of the device before, during and after an
instruction. The structure of a device and its operating rules form a
hardware description that is an input to the compiler
(Section~\ref{sec:background}).

\textbf{Challenge 2: How do we reason about ions that move in parallel
under shared control?}
Scalable designs drive many traps with one control signal, a
\emph{broadcast} channel. Moves of different ions are then no longer
independent: they must make the same motion, and the signal also moves
ions that the program did not name. A compiler that ignores this returns code that the device cannot run. A conservative compiler, which
lets one transport use the channel at a time, can make a 63-qubit Fourier
transform 4.2 times slower. \textbf{Our idea is to make broadcast a
primitive of the target code.} One transport instruction lists all the
moves that happen together and is legal only as a whole, so a parallel
movement that the shared signal cannot produce is one illegal instruction, never
two legal ones. And a channel is
not a resource that a transport holds but a signal that is
\emph{activated}: transports of the same motion share one activation,
and a proved scheduler may
run them ahead of unrelated work. This brings the Fourier transform
within 0.3\% of a device with a channel for every transport
(Section~\ref{sec:broadcast}).

\textbf{Challenge 3: How do we search such a huge design space?}
A
QCCD designer chooses the number of traps and the capacity of each, the
segments and junctions that connect them, the control signals that they
share, and where each qubit starts. As an example, a budget of 96 ion slots splits into traps of
two or more ions in more than 13 million ways,
before any choice of layout, control channels or placement.  \textbf{Our idea is a co-design loop with feedback provided by checked target code.} We compile the benchmark programs for 89 manual designs,
test candidate design principles on the checked results, and give the principles that hold to an AI agent. The agent then proposes a device, the verified compiler scores it, and records of the checked target code guide the next proposal (Section~\ref{sec:codesign}).

\textbf{We develop a single compiler with a single proof that carries these ideas.} The compiler first
fixes \emph{what} is computed, an ordered sequence of native pulses that
is the same for every device, and then chooses \emph{how} a device
computes it: placement, transport, pulse sharing and timing. A \emph{router}
is a procedure that chooses the transports that bring the ions of each
gate into one trap. Routers and
the agent only propose; the compiler checks every proposed
step~\cite{pnueli1998translation,tristan2008formal}
(Section~\ref{sec:compiler}). We prove in Lean~4 that each pass preserves
the behavior of the program and that its output satisfies the hardware
rules, and we compose the proofs into a single theorem, as CompCert does for
C~\cite{leroy2009compcert} (Section~\ref{sec:checker}). Inspired
by architecture-parameterized verified compilation, as in CakeML's
compilation to several machine-code targets~\cite{fox2017verified}, we
treat each QCCD hardware design as an input of the compiler and not as a
separate compilation target. We can therefore explore different hardware
designs with the same compiler and the same correctness theorem.

If the compiler returns target code, that code is correct and
follows all the rules. Moreover, the execution time that the compiler
reports for it, which scores a design, is the time that the stated
hardware model assigns. We contribute:

\begin{itemize}[leftmargin=*,nosep]
\item An architecture-specification language for moving qubits: a device and its
operating rules, each an assertion over event frames
(Section~\ref{sec:background}).
\item A model of parallel movement under shared control: broadcast is
a primitive of the target code, and a broadcast channel is scheduled by its
activations (Section~\ref{sec:broadcast}).
\item A list of 11 defects in three published QCCD compilers that can bias design comparisons, each excluded from our compiler by our theorem
(Section~\ref{sec:eval-compiler}).
\item A verified compiler with checked routing proposals, pulse sharing
and a parallel schedule, which compiles 512-qubit programs in 1 to 29 minutes
(Sections~\ref{sec:compiler}, \ref{sec:checker}
and~\ref{sec:eval-scalability}).
\item A co-design study of 75 programs of 11 to 130 qubits at equal ion
capacity: the best design takes 66.4\% less modeled time than the worst
on average, and an AI agent that proposes devices based
on the checked results is faster than the best design on 71 of the 75 programs, equal on 3 and
slower on 1
(Sections~\ref{sec:codesign} and~\ref{sec:eval}).
\end{itemize}

\section{Architecture specification: device and rules}
\label{sec:background}

What must a designer write down so that a compiler can decide whether target code may run on a device? In a QCCD the answer depends on where the
ions are before, during and after an instruction, and on when it runs
(Challenge~1 of Section~\ref{sec:intro}). \textbf{Our specification language gives every operating rule a single form: an assertion about the
state of one instruction, the \emph{event frame}.}
Section~\ref{sec:hardware-model} describes the structure of a device,
the candidate of Fig.~\ref{fig:codesign-overview}b;
Section~\ref{sec:programs} introduces the event frame, and
Section~\ref{sec:model} the rules that are checked on it, all on a running example.

\subsection{Describing a QCCD device}
\label{sec:hardware-model}

\begin{figure}[t]
\centering
\resizebox{\columnwidth}{!}{%
\begin{tikzpicture}[x=1pt,y=1pt,font=\sffamily\fontsize{7}{8.4}\selectfont,
  call/.style={inner sep=0pt,text=qink,align=center},
  lead/.style={qgrey,line width=.4pt}]
\path[use as bounding box] (-2,6) rectangle (242,-160);
\QCCDFigHead{tag}{\S2.1}{Hardware description: the running-example device}

\begin{scope}
\QCCDUnit=.86bp
\begin{scope}[shift={(28,-40)},x=\QCCDUnit,y=-\QCCDUnit]
  \QCCDDevice{primer}{9}
  \QCCDIon{q0}{S}{0}\QCCDIon{q1}{G}{1}
  \draw[qhwBeam,line width=2.2\QCCDUnit] (200,-36)--(200,-8);
  \draw[qhwBeam,line width=1.2\QCCDUnit] (184,-8)--(216,-8);
  \draw[qblue,line width=1.3\QCCDUnit,-{Stealth[length=4bp,width=3.2bp]}]
    (25,66)--(110,66)--(188,66);
  \fill[qblue] (110,66) circle[radius=2.6\QCCDUnit];
  \draw[qblue,densely dotted,line width=.8\QCCDUnit] (25,58)--(25,66) (188,58)--(188,66);
\end{scope}
\end{scope}

\node[call,anchor=east,text=qhwBeam] at (195,-14) {optical channel 7};
\node[call,anchor=west,text=qhwBeam] at (205,-14) {R, MS,};
\node[call,anchor=west,text=qhwBeam] at (205,-22.5) {readout};
\foreach \x/\s/\c in {49.5/S/1,122.6/J/1,200/G/2}{
  \node[call,anchor=north,fill=white,inner sep=1pt] at (\x,-69.5)
    {\textbf{\s}\quad$\capa(\mathrm{\s})=\c$};}
\node[call,anchor=north,text=qblue] at (120,-100)
  {path $p_0$: S$\to$J$\to$G,\ \ $\via(p_0)=\{\mathrm J\}$,\ \ 2 ticks,\ \ transport channel 0};
\node[call,anchor=north west,text=qgrey] at (0,-112)
  {placement $\pos$: $q_0\mapsto\mathrm S$, $q_1\mapsto\mathrm G$};

\draw[qrail,line width=.5pt] (0,-123)--(240,-123);
\node[inner sep=0pt,anchor=north west] at (0,-127) {\begin{minipage}{240pt}
  \lstinputlisting[style=qccd-hardware,
    basicstyle=\ttfamily\fontsize{6.6}{8}\selectfont\color{black},
    xleftmargin=0pt,aboveskip=0pt,belowskip=0pt,breaklines=false]%
    {paper-listing-primer-hardware.txt}
\end{minipage}};
\end{tikzpicture}}
\caption{The running example of Section~\ref{sec:background}. A device
is a list of declarations: sites (traps, junctions or segments) with
capacities $\capa$, paths between sites with intermediate sites $\via$,
and calibrations that bind native operations to a site and a control
channel. $\pos$ places the ions $q_0,q_1$.}
\label{fig:hardware-model}
\Description{Three sites S, J and G in a row with capacities one, one and
two. Ion q0 is at S and ion q1 at G. A blue arrow marks path p0 from S
through J to G. A laser beam marks the optical channel at G. Below, the
hardware declaration lists the sites, path, calibration and readout.}
\end{figure}

Co-design changes the hardware, so the hardware must be an input to the
compiler, as machine descriptions are for some classical
compilers~\cite{fauth1995nml,halambi1999expression,ramsey1998machine,armstrong2019isa,ikarashi2022exo}.
Ours describes where ions rest, how they move and which controls they
share. A QCCD holds its ions in traps; ions move between traps along
segments, which meet at
junctions~\cite{kielpinski2002architecture,pino2021demonstration}. We
describe all of these places by a single abstraction, as \emph{sites}. A site $s$ is
any place where an ion can stay or pass through, and its capacity
$\capa(s)$ is the largest number of ions it may hold. A \emph{placement}
$\pos$ maps each ion to a site. A \emph{path} $p$ is a declared transport
from a source site to a destination site, with the sites it passes
through, $\via(p)$, and its duration; one path may stand for several
physical steps, such as split, shuttle and
merge~\cite{molavi2026generating,bach2025position}.
Fig.~\ref{fig:hardware-model} shows the running example: a storage site
S with one slot, a transit site J with one slot and a gate zone G with
two slots, joined by a path $p_0$ from S through J to G that takes two
ticks.

An ion can be operated on only at a site that has a control for the
operation. A \emph{calibration} binds one native operation to a site, a
control channel and a duration, like a \texttt{defcal} of
OpenQASM~3~\cite{cross2022openqasm3} or a Jaqal gate
file~\cite{morrison2020jaqal} without the pulse shape. Our native
operations are single-ion rotations $\mathrm R(\theta,\phi)$, two-ion
M\o{}lmer--S\o{}rensen gates $\mathrm{MS}(\theta)$, readout, reset by
optical pumping, and a controller latch that saves the result of a
classical comparison. In Fig.~\ref{fig:hardware-model}, all optical
controls are at G on channel~7. A \emph{channel} is a control signal
that several sites may share. An ion moves when the voltages on the
electrodes beneath it change. A \emph{waveform} is the sequence of
voltages over time that a transport channel applies to carry an ion along
one path: for a programmer, a named \emph{movement pattern} such as
``one trap to the right''. Each path names the transport channel that
drives it and the waveform that the channel applies for it.
A \emph{direct} device gives every path its own channel; a
\emph{broadcast} device lets many paths share a channel
(Section~\ref{sec:broadcast}).

\subsection{Programs and the event frame}
\label{sec:programs}

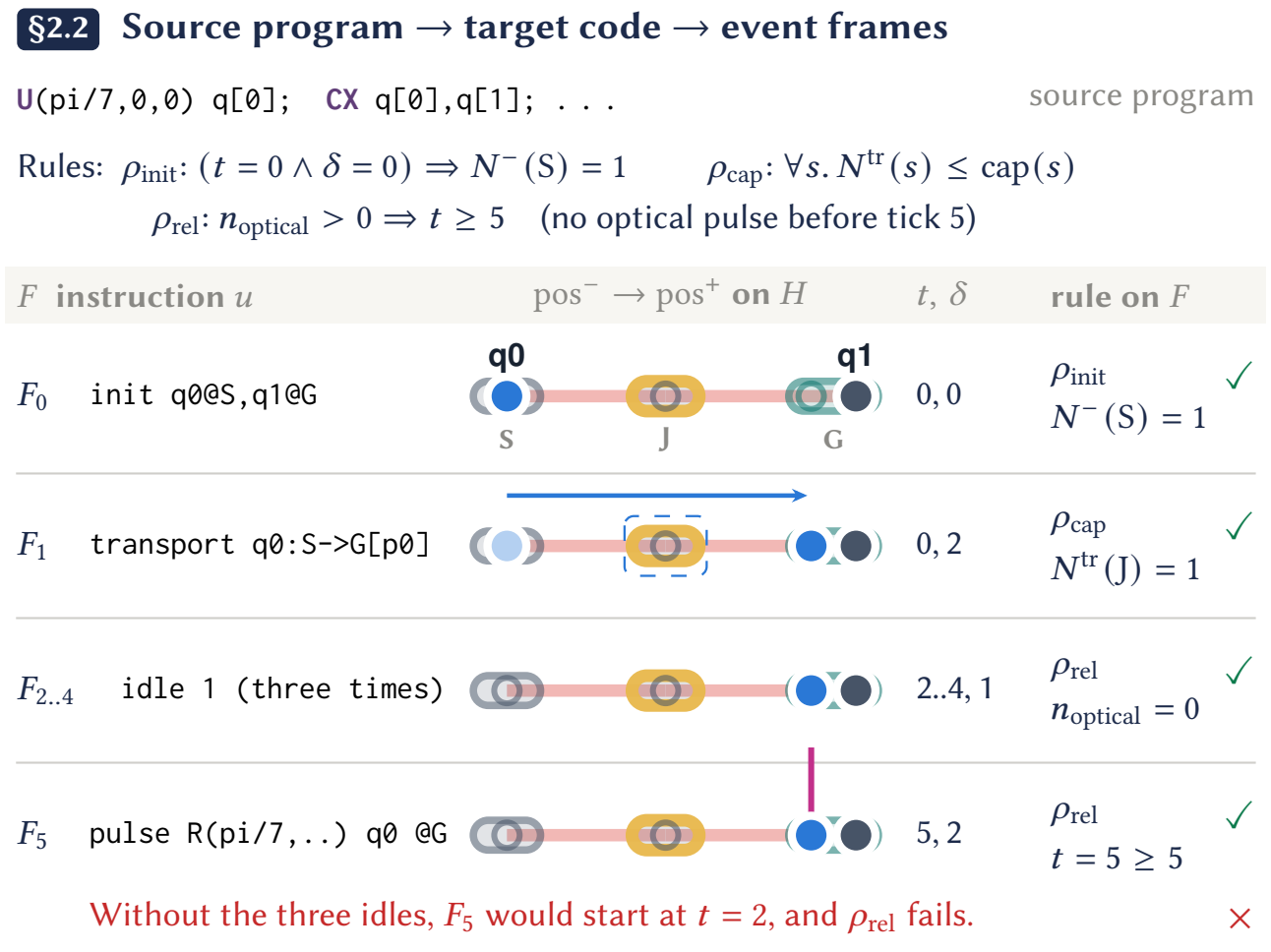
\begin{figure}[t]
\centering
\newcommand{\FrameDevice}[3]{
  \begin{scope}\QCCDUnit=.36bp\qccdpadsfalse#3
  \begin{scope}[shift={(86,#1+9)},x=\QCCDUnit,y=-\QCCDUnit]
    \QCCDDevice{primer}{17}#2
  \end{scope}\end{scope}}
\newcommand{\FrameRule}[3]{
  \node[inner sep=0pt,anchor=west,text=qink] at (200,#1+4) {#2};
  \node[inner sep=0pt,anchor=west,text=qink] at (200,#1-4.5) {#3};
  \node[okmark] at (237,#1+4) {\cmark};}
\resizebox{\columnwidth}{!}{%
\begin{tikzpicture}[x=1pt,y=1pt,font=\sffamily\fontsize{6.8}{8}\selectfont,
  code/.style={inner sep=0pt,anchor=west,text=black,
    font=\ttfamily\fontsize{6.3}{7.6}\selectfont},
  colh/.style={inner sep=0pt,anchor=west,text=qgrey,
    font=\sffamily\bfseries\fontsize{6.6}{8}\selectfont},
  fr/.style={inner sep=0pt,anchor=west,text=qink,
    font=\sffamily\bfseries\fontsize{6.8}{8}\selectfont},
  txt/.style={inner sep=0pt,anchor=west,text=qink},
  sep/.style={qrail,line width=.4pt}]
\path[use as bounding box] (-2,6) rectangle (242,-178);
\QCCDFigHead{tag}{\S2.2}{Source program $\to$ target code $\to$ event frames}

\node[code,anchor=north west] at (0,-11)
  {\color{qcodepurple}\bfseries U\color{black}\mdseries(pi/7,0,0) q[0];\ \
   \color{qcodepurple}\bfseries CX\color{black}\mdseries\ q[0],q[1]; \dots};
\node[txt,anchor=north east,text=qgrey] at (240,-11.5) {source program};
\node[txt,anchor=north west] at (0,-23)
  {Rules:\ \ $\rho_{\rm init}$: $(t=0\land\delta=0)\Rightarrow N^-(\mathrm S)=1$
   \qquad$\rho_{\rm cap}$: $\forall s.\,N^{\rm tr}(s)\leq\capa(s)$};
\node[txt,anchor=north west] at (26,-33.5)
  {$\rho_{\rm rel}$: $n_{\rm optical}>0\Rightarrow t\geq5$\quad (no optical pulse before tick 5)};

\fill[qpaper] (-2,-46) rectangle (242,-57);
\node[colh] at (0,-51.5) {$F$\ \ instruction $u$};
\node[colh] at (100,-51.5) {$\pos^-\to\pos^+$ on $H$};
\node[colh] at (174,-51.5) {$t,\,\delta$};
\node[colh] at (200,-51.5) {rule on $F$};

\node[fr] at (0,-71) {$F_0$};
\node[code] at (14,-71) {init q0@S,q1@G};
\FrameDevice{-71}{\QCCDIon[qblue]{q0}{S}{0}\QCCDIon{q1}{G}{1}
  \foreach \s in {S,J,G}{\node[text=qgrey,font=\sffamily\bfseries\fontsize{5.5}{6}\selectfont,
    inner sep=0pt] at ($(\s)+(0,23)$) {\s};}}{}
\node[txt] at (174,-71) {0,\,0};
\FrameRule{-71}{$\rho_{\rm init}$}{$N^-(\mathrm S)=1$}

\draw[sep] (0,-86)--(240,-86);
\node[fr] at (0,-100) {$F_1$};
\node[code] at (14,-100) {transport q0:S->G[p0]};
\FrameDevice{-100}{\draw[qblue,dashed,line width=1.2\QCCDUnit,rounded corners=5\QCCDUnit]
    (88,9) rectangle (132,41);
  \QCCDIon[qblue!35]{q0}{S}{0}\QCCDIon{q1}{G}{1}\QCCDIon[qblue]{q0n}{G}{0}
  \draw[qblue,line width=2.4\QCCDUnit,-{Stealth[length=3bp,width=2.6bp]}]
    (25,-2)--(110,-2)--(186,-2);}{\qccdionlabelsfalse}
\node[txt] at (174,-100) {0,\,2};
\FrameRule{-100}{$\rho_{\rm cap}$}{$N^{\rm tr}(\mathrm J)=1$}

\draw[sep] (0,-114)--(240,-114);
\node[fr] at (0,-128) {$F_{2..4}$};
\node[code] at (20,-128) {idle 1 (three times)};
\FrameDevice{-128}{\QCCDIon[qblue]{q0}{G}{0}\QCCDIon{q1}{G}{1}}{\qccdionlabelsfalse}
\node[txt] at (174,-128) {2..4,\,1};
\FrameRule{-128}{$\rho_{\rm rel}$}{$n_{\rm optical}=0$}

\draw[sep] (0,-142)--(240,-142);
\node[fr] at (0,-156) {$F_5$};
\node[code] at (14,-156) {pulse R(pi/7,..) q0 @G};
\FrameDevice{-156}{\draw[qhwBeam,line width=3.2\QCCDUnit] (188,-22)--(188,14);
  \QCCDIon[qblue]{q0}{G}{0}\QCCDIon{q1}{G}{1}}{\qccdionlabelsfalse}
\node[txt] at (174,-156) {5,\,2};
\FrameRule{-156}{$\rho_{\rm rel}$}{$t=5\geq5$}
\node[txt,text=qred] at (14,-172)
  {Without the three idles, $F_5$ would start at $t=2$, and $\rho_{\rm rel}$ fails.};
\node[badmark] at (237,-172) {\xmark};
\end{tikzpicture}}
\caption{Target code runs as a sequence of event frames
$F=(H,\pos^-,\pos^+,u,t,\delta)$: the device, the placements before and
after, the instruction, its start time and its duration. A rule must
hold in every frame.}
\label{fig:event-frames}
\Description{A table of event frames for the running example. F0
initializes q0 at S and q1 at G. F1 transports q0 from S through J to G
and reserves J. F2 to F4 are idles. F5 is the first R pulse at tick five.
Each row shows the device, the start time and duration, and a rule that
holds on the frame. Without the idles the release rule fails.}
\end{figure}

The input is a \emph{source program} in
OpenQASM~2~\cite{cross2017openqasm}: single-qubit gates U, controlled-X
gates CX, gate definitions, barriers, measurement, reset and conditions
on a whole register, such as \texttt{if(c==1)}. For one device, the compiler produces
\emph{target code}: an initial placement and a list of instructions
(column $u$ of Fig.~\ref{fig:event-frames}). Each instruction names the declared
resources it uses: \texttt{transport q0:S->G[p0]} uses path $p_0$, and
\texttt{pulse R(..) q0 @G via c0} uses calibration $c_0$ at G. An
instruction prefixed by \texttt{when saved} acts only if the saved
comparison was true.

\paragraph{When is target code legal?}
The device runs each instruction as one \emph{event}: a transport changes
only the placement, gates R and MS apply their ideal unitaries, readout writes
a classical bit, reset prepares $\ket{0}$, and an idle changes only the
time. An event is \emph{legal} if it passes three core checks. (i) Every
ion that it uses is where the event needs it: at the source of its path
or at the site of its calibration. (ii) No site holds more ions than its
capacity, before, during or after the event. (iii) Each shared channel
applies one waveform, and every ion that it drives without a switch is
listed as moving. \textbf{Target code is legal if its initial placement
respects all capacity constraints and every one of its events is legal} and satisfies
the designer's rules (Section~\ref{sec:model}). 
The core checks apply to
every device; the rules that a designer adds can only add restrictions. We assume
that the device can carry out each declared event within its declared
duration whenever the core checks pass (the
\emph{primitive-realization assumption}): our proofs concern this
discrete model.

\paragraph{Event frames}
What does a rule need to know to decide whether an instruction may run? A
capacity rule needs the placements and the sites that ions pass through,
and a timing rule needs the start time and duration. We collect these in an
\emph{event frame} $F=(H,\pos^-,\pos^+,u,t,\delta)$: the hardware $H$,
the placements $\pos^-$ before and $\pos^+$ after, the instruction $u$,
its start time $t$ and its duration $\delta$. A frame is the program state
as one instruction sees it. Initialization is a frame $F_0$ at time zero
with zero duration and no activity.

Fig.~\ref{fig:event-frames} steps through the frames of the running
example under an initialization rule $\rho_{\rm init}$, a capacity rule
$\rho_{\rm cap}$ and a release rule $\rho_{\rm rel}$. Frame $F_1$ moves
$q_0$ to G and reserves J for $[0,2)$. Rule $\rho_{\rm rel}$ forbids
optical pulses before tick~5, so the compiler inserts three idle frames
before the first pulse: changing the rules changes the target code.

\subsection{Operating rules: assertions over one frame}
\label{sec:model}

A designer completes the specification by writing the operating rules of
a device as assertions over event frames. Why a language instead of a
fixed list of checks? Operating rules differ between devices and change
during co-design. A laboratory may limit a gate zone to two ions, allow
one laser pulse per event, or forbid pulses until the optics have
settled. Built into the compiler, every new rule would need a new proof. When written in a single language, rules are inputs and one proof covers them all.
We use first-order assertions with Boolean connectives, natural-number
arithmetic and finite quantification over the declared sites
$\mathcal S$, paths $\mathcal P$ and transport channels $\mathcal C_T$.
They are the assertions of Hoare logic~\cite{hoare1969axiomatic} with one
change of role: a Hoare assertion is proved about one program, while
ours is an input that the compiler enforces: it returns target code only
if every rule is true on every frame of that code. The soundness of this
check is proved once, for all rules and programs.
\ifqccdreference Appendix~\ref{app:syntax-rules} gives the full grammar.\fi

\begin{figure}[t]
\centering
\resizebox{\columnwidth}{!}{%
\begin{tikzpicture}[x=1pt,y=1pt,font=\sffamily\fontsize{6.8}{8}\selectfont,
  txt/.style={inner sep=0pt,anchor=west,text=qink},
  obs/.style={inner sep=1pt,text=qink,fill=white,align=center},
  fam/.style={inner sep=0pt,anchor=west,text=qink,
    font=\sffamily\bfseries\fontsize{6.8}{8}\selectfont},
  elab/.style={draw=qgreen,fill=qmint,rounded corners=2pt,line width=.5pt,
    inner sep=2.5pt,align=center,text=qink},
  flowg/.style={-{Stealth[length=3.5pt,width=3pt]},qgreen,line width=.7pt},
  sep/.style={qrail,line width=.4pt}]
\path[use as bounding box] (-2,6) rectangle (242,-200);
\QCCDFigHead{tag}{\S2.3}{Operating rules: assertions over one event frame}

\node[txt,text=qgrey] at (0,-12) {(a) Observations on $F_1$ = transport q0:\,S$\to$G along $p_0$, $t=0$, $\delta=2$};
\begin{scope}
\QCCDUnit=.62bp\qccdpadsfalse\qccdionlabelsfalse
\begin{scope}[shift={(48,-34)},x=\QCCDUnit,y=-\QCCDUnit]
  \QCCDDevice{primer}{12}
  \draw[qblue,dashed,line width=1.2\QCCDUnit,rounded corners=5\QCCDUnit]
    (88,9) rectangle (132,41);
  \QCCDIon[qblue!35]{q0}{S}{0}\QCCDIon{q1}{G}{1}\QCCDIon[qblue]{q0n}{G}{0}
  \draw[qblue,line width=2\QCCDUnit,-{Stealth[length=3.5bp,width=3bp]}]
    (25,0)--(110,0)--(186,0);
\end{scope}
\end{scope}
\node[obs,anchor=north] at (63.5,-57)  {$N^-(\mathrm S)=1$\\$N^+(\mathrm S)=0$};
\node[obs,anchor=north] at (116,-57) {$N^{\rm tr}(\mathrm J)=1$\\{\color{qgrey}passes via J}};
\node[obs,anchor=north] at (172,-57) {$N^-(\mathrm G)=1$\\$N^+(\mathrm G)=2$};
\node[obs,anchor=south,text=qblue] at (116,-30) {\rev{$M(p_0)=M(0)=1$,\ \ $W(0)=1$}};
\node[txt,anchor=north west,align=left] at (204,-30)
  {$n_{\rm move}=1$\\$n_{\rm optical}=0$\\$P(\mathrm G)=0$};
\node[txt,anchor=north west] at (0,-80) {$N^{\rm tr}(s)=|\{q\mid\pos^-(q)=s\lor\pos^+(q)=s\}|+|\{m\mid s\in\via(m)\}|$};
\node[txt,anchor=north west,text=qgrey] at (0,-90) {residents, departures and arrivals, plus ions passing through $s$};

\draw[sep] (0,-102)--(240,-102);
\node[txt,text=qgrey] at (0,-109) {(b) Three families of rules, evaluated on $F_1$};
\node[fam] at (0,-122) {Occupancy};
\node[txt] at (68,-122) {$\forall s{:}\mathcal S.\;N^{\rm tr}(s)\leq\capa(s)$};
\node[txt] at (176,-122) {$1\le1,\,1\le1,\,2\le2$};
\node[fam] at (0,-135) {Transport, control};
\node[txt] at (68,-135) {$\forall c{:}\mathcal C_T.\;W(c)\leq1$};
\node[txt] at (176,-135) {\rev{$W(0)=1\le1$}};
\node[fam] at (0,-148) {Operation, timing};
\node[txt] at (68,-148) {$n_{\rm optical}>0\Rightarrow t\geq5$};
\node[txt] at (176,-148) {$n_{\rm optical}=0$};
\foreach \y in {-122,-135,-148}{\node[okmark] at (237,\y) {\cmark};}

\draw[sep] (0,-158)--(240,-158);
\node[txt,text=qgrey] at (0,-165) {(c) One proof for every formula: elaborate, then evaluate};
\node[elab,anchor=west] (phi) at (0,-182) {$\forall s{:}\mathcal S.$\\$N^{\rm tr}(s)\leq\capa(s)$};
\node[elab,anchor=west] (core) at (74,-182)
  {$N^{\rm tr}(\mathrm S)\leq1\land N^{\rm tr}(\mathrm J)\leq1$\\$\land\,N^{\rm tr}(\mathrm G)\leq2$};
\node[elab,anchor=west] (ok) at (186,-182) {\cmark\ or \xmark\\on every $F$};
\draw[flowg] (phi)--node[above,inner sep=1.5pt,font=\sffamily\fontsize{6}{7}\selectfont]{\rev{elaborate}}(core);
\draw[flowg] (core)--node[above,inner sep=1.5pt,font=\sffamily\fontsize{6}{7}\selectfont]{eval}(ok);
\node[txt,anchor=north,text=qgreen] at (120,-194)
  {\rev{Same truth value before and after elaboration (proved in Lean)}};
\end{tikzpicture}}
\caption{Operating rules are formulas over the observations of one event
frame. (a) The counts a formula can read on the transport frame of
Fig.~\ref{fig:event-frames}: $M$ counts moving ions and $W$ distinct
waveforms on transport channel 0. (b) One rule of each family. (c)
Elaboration expands quantifiers over the declared resources, and a
verified evaluator decides the result on every frame.}
\label{fig:assertions}
\Description{Top: the S, J, G device during the transport of q0 from S
to G, annotated with the counts N minus, N plus, N transit, M and W.
Middle: an occupancy, a transport and a timing rule, each true on the
frame. Bottom: a quantified formula is elaborated into a conjunction over
the three sites and evaluated.}
\end{figure}
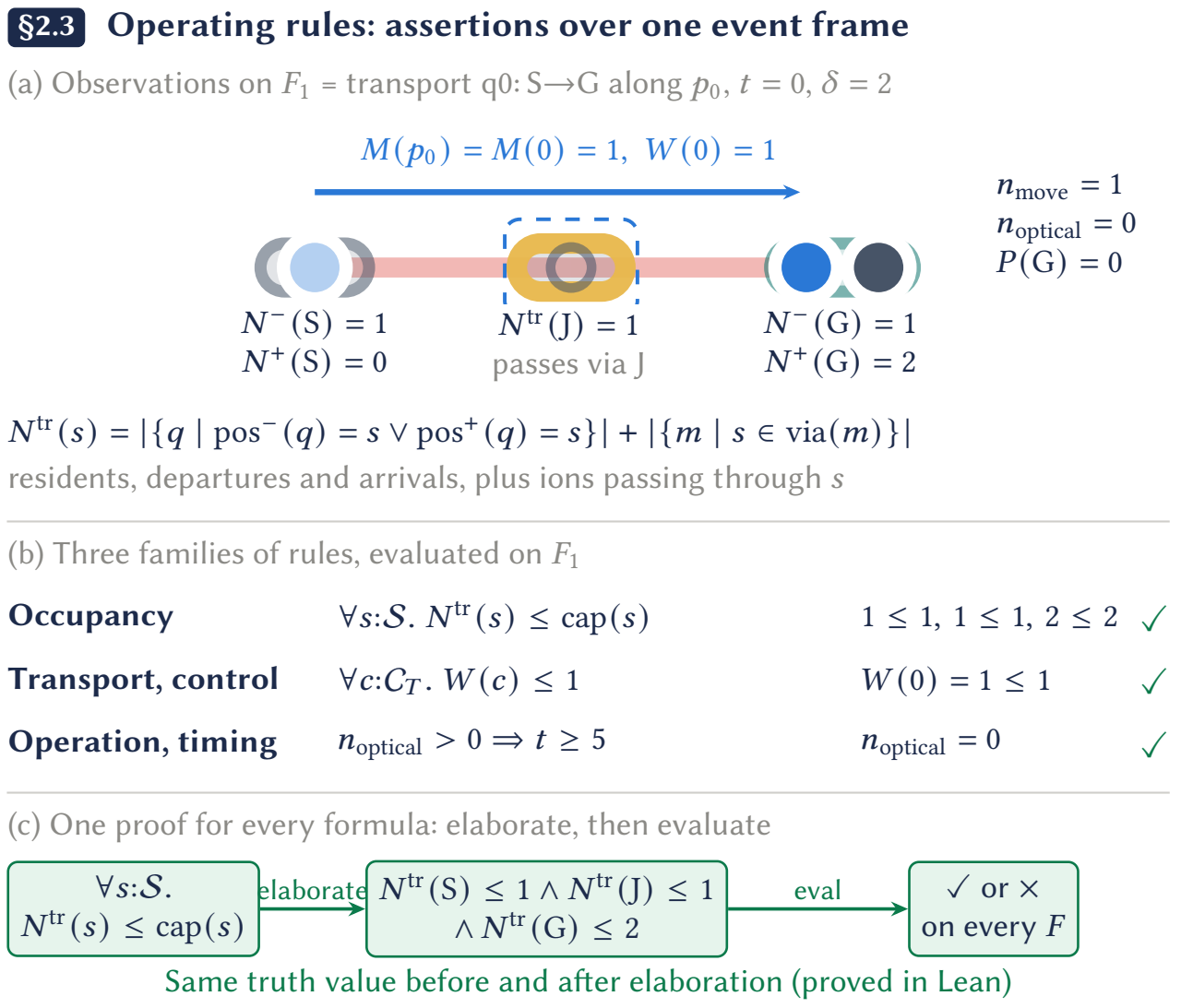

\paragraph{What a rule observes}
A named assertion $\mathsf{assert}\ a:\varphi$ must hold in the initial
frame and in every later frame. The formula $\varphi$ reads counts of one
frame, shown on the transport frame $F_1$ in Fig.~\ref{fig:assertions}a.
$N^-(s)$ and $N^+(s)$ count the ions at site $s$ before and after the
event. The transit reservation $N^{\rm tr}(s)$ counts the ions that use
$s$ during the event: each ion present before or after it, plus every
move of its move list $L$ that passes through $s$,
\[
N^{\mathrm{tr}}(s)=
 |\{q\mid\pos^-(q)=s\lor\pos^+(q)=s\}|
 +|\{m\in L\mid s\in\via(m)\}|.
\]
This conservative count holds for the whole event and detects an ion
that passes through a full trap or through a junction that another ion
uses. $M(p)$ and $M(c)$ count the ions that move
along path $p$ and on transport channel $c$, and $W(c)$ counts the
distinct waveforms requested on $c$. $P(s)$ counts the optical actions at
$s$, and $n_k$ counts the actions of kind $k$: move, optical, R, MS,
readout, pump or latch.

\paragraph{Three families of rules}
Fig.~\ref{fig:assertions}b evaluates one rule of each family on $F_1$.
\emph{Occupancy}: $\forall s{:}\mathcal S.\;N^{\rm tr}(s)\leq\capa(s)$
restates the core capacity check, and $N^{\rm tr}(\mathrm G)\leq1$
operates the two-slot zone G with one ion without changing the device.
\emph{Transport and control}: $\forall p{:}\mathcal P.\;M(p)\leq1$ allows
one ion per path, and $\forall c{:}\mathcal C_T.\;W(c)\leq1$ extends one
waveform per channel from broadcast channels to direct ones.
\emph{Operation and timing}: $n_{\rm optical}\leq1$ allows one optical
action per event, $\forall s{:}\mathcal S.\;P(s)\leq1$ one per site, and
$n_{\rm optical}>0\Rightarrow t\geq\tau$ sets a release time $\tau$.

\paragraph{Why does a new rule need no new proof?}
\label{sec:elaboration}
A designer writes a rule in general terms, such as ``no site exceeds
its capacity''. \emph{Elaboration} compiles this general rule into
explicit checks for one device (Fig.~\ref{fig:assertions}c): on the
device with sites S, J and G, ``for every site'' becomes three
conditions joined by ``and''. Elaboration and the checks run again in
every compilation; the proof is done once, in two general parts.
First, the compiler theorem (Section~\ref{sec:checker}) holds for every
device and every set of explicit checks. Second, elaboration preserves
truth, by induction over formulas: for all $H$, frames $F$ and rules
$\varphi$, with $F\models_H\varphi$ the meaning of the rule as written,
\[
\mathsf{elab}_H(\varphi)\Downarrow\psi\;\Longrightarrow\;
\bigl(\mathsf{eval}(\psi,F)=\mathsf{true}\iff F\models_H\varphi\bigr).
\]
\textbf{Together, for every device, rule and program,
target code that the compiler returns satisfies the rule as the designer
wrote it.} A new rule can make a compilation fail, but it needs no new
proof.

\section{Parallel movement under shared control}
\label{sec:broadcast}

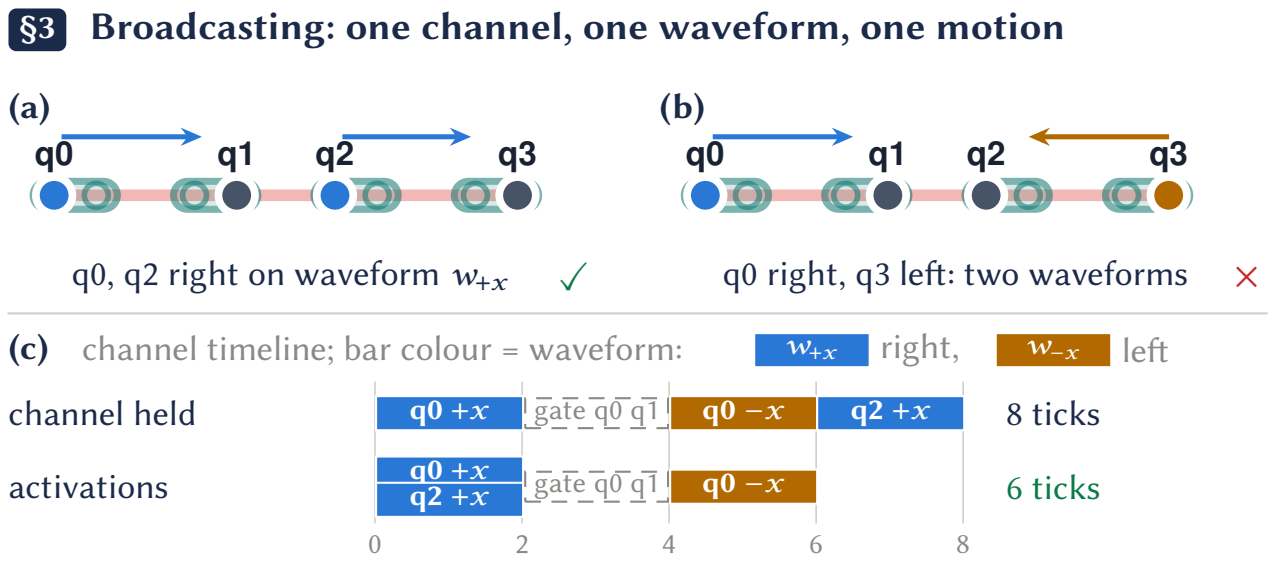
\begin{figure}[t]
\centering
\newcommand{\BcastChain}[3]{
  \begin{scope}\QCCDUnit=.37bp\qccdpadsfalse
  \begin{scope}[shift={(#1-9,#2+8)},x=\QCCDUnit,y=-\QCCDUnit]
    \QCCDDevice{line4}{17}#3
  \end{scope}\end{scope}}
\resizebox{\columnwidth}{!}{%
\begin{tikzpicture}[x=1pt,y=1pt,font=\sffamily\fontsize{6.8}{8}\selectfont,
  txt/.style={inner sep=0pt,anchor=west,text=qink,align=left},
  lab/.style={inner sep=0pt,anchor=west,text=qink,
    font=\sffamily\bfseries\fontsize{6.8}{8}\selectfont},
  mv/.style={-{Stealth[length=4bp,width=3.4bp]},line width=2.6\QCCDUnit},
  bar/.style={draw=white,line width=.4pt,minimum height=6.5pt,inner sep=0pt,
    font=\sffamily\bfseries\fontsize{5.8}{6.5}\selectfont,text=white},
  gate/.style={draw=qgrey,densely dashed,line width=.4pt,minimum height=5.7pt,
    minimum width=27.2pt,inner sep=0pt,anchor=west,fill=white,text=qgrey,
    font=\sffamily\fontsize{5.6}{6.3}\selectfont},
  sep/.style={qrail,line width=.4pt}]
\path[use as bounding box] (-2,6) rectangle (242,-103);
\QCCDFigHead{tag}{\S3}{Broadcasting: one channel, one waveform, one motion}

\node[lab] at (0,-15) {(a)};
\BcastChain{0}{-15}{\QCCDIon[qblue]{q0}{C0}{0}\QCCDIon[qblue]{q2}{C2}{0}
  \QCCDIon{q1}{C1}{1}\QCCDIon{q3}{C3}{1}
  \draw[mv,qblue] (52,36)--(124,36);\draw[mv,qblue] (196,36)--(262,36);}
\node[txt] at (12,-47) {q0, q2 right on waveform $w_{+x}$};
\node[okmark] at (108,-47) {\cmark};

\node[lab] at (124,-15) {(b)};
\BcastChain{124}{-15}{\QCCDIon[qblue]{q0}{C0}{0}\QCCDIon[qorange]{q3}{C3}{1}
  \QCCDIon{q1}{C1}{1}\QCCDIon{q2}{C2}{0}
  \draw[mv,qblue] (52,36)--(124,36);\draw[mv,qorange] (286,36)--(214,36);}
\node[txt] at (136,-47) {q0 right, q3 left: two waveforms};
\node[badmark] at (236,-47) {\xmark};

\draw[sep] (0,-54)--(240,-54);
\node[lab] at (0,-61) {(c)};
\node[txt,text=qgrey] at (14,-61) {channel timeline; bar colour = waveform:};
\node[bar,fill=qblue,minimum width=22pt,anchor=west] at (142,-61) {$w_{+x}$};
\node[txt,text=qgrey] at (166,-61) {right,};
\node[bar,fill=qorange,minimum width=22pt,anchor=west] at (188,-61) {$w_{-x}$};
\node[txt,text=qgrey] at (212,-61) {left};
\foreach \t in {0,2,4,6,8}{
  \draw[qrail,line width=.3pt] ({70+14*\t},-67)--({70+14*\t},-94);
  \node[inner sep=0pt,text=qgrey,font=\sffamily\fontsize{5.6}{6}\selectfont] at ({70+14*\t},-98) {\t};}
\node[txt] at (0,-73) {channel held};
\node[bar,fill=qblue,minimum width=28pt,anchor=west] at (70,-73) {q0 $+x$};
\node[gate] at (98.4,-73) {gate q0 q1};
\node[bar,fill=qorange,minimum width=28pt,anchor=west] at (126,-73) {q0 $-x$};
\node[bar,fill=qblue,minimum width=28pt,anchor=west] at (154,-73) {q2 $+x$};
\node[txt] at (190,-73) {8 ticks};
\node[txt] at (0,-87) {activations};
\node[bar,fill=qblue,minimum width=28pt,anchor=west] at (70,-84.5) {q0 $+x$};
\node[bar,fill=qblue,minimum width=28pt,anchor=west] at (70,-89.5) {q2 $+x$};
\node[gate] at (98.4,-87) {gate q0 q1};
\node[bar,fill=qorange,minimum width=28pt,anchor=west] at (126,-87) {q0 $-x$};
\node[txt,text=qgreen] at (190,-87) {6 ticks};
\end{tikzpicture}}
\caption{Broadcasting: one transport channel with switches drives all four
traps. (a) One waveform moves q0 and q2 one trap right in one event. (b)
Opposite motions would need two waveforms on one channel and are
rejected. (c) Each bar is one transport, coloured by the waveform
that the channel applies for it. The code moves q0 right, runs a gate on
q0 and q1, moves q0 back, then moves q2 right. If each transport holds the
channel, that of q2 waits behind the return of q0. Our scheduler records
each waveform that the channel applies as an \emph{activation}, and lets
q0 and q2, which make the same motion on disjoint ions and sites as in
(a), share one activation.}
\label{fig:broadcast}
\Description{Three panels. (a) A line of four traps holding ions q0, q1,
q2 and q3, one per trap; q0 and q2 both move one trap to the right under
one waveform, marked legal. (b) The same line and placement with q0 moving
right and q3 moving left in one event, marked illegal. (c) Two timelines
of the transport channel for the code that moves q0 right, runs a gate on
q0 and q1 in ticks 2 to 4, moves q0 back left and moves q2 right. If a
transport holds the channel, the three transports take ticks 0 to 2, 4 to
6 and 6 to 8. With activations, the moves of q0 and q2 to the right share
ticks 0 to 2 and the return of q0 takes ticks 4 to 6.}
\end{figure}

A QCCD is fast only if many ions move at once, and scalable designs
move them with few control signals: electrodes with the same role on
many traps are wired to one transport channel~\cite{malinowski2023wise}.
When such a \emph{broadcast} channel applies a waveform, every trap on it
receives the same voltages, so every ion that it moves makes the same
motion (Fig.~\ref{fig:broadcast}a). In this section, we explain how
target code expresses this movement and how the compiler schedules it
(Challenge~2).

\textbf{Our idea is to make broadcast a primitive of the target code.}
Target code represents simultaneous moves with a single transport instruction that lists every move that happens together. The core checks accept the list as a whole: one waveform on a shared channel, and every ion that this waveform drives on the list. Thus, a parallel movement that the shared signal cannot produce is illegal. Verification becomes simpler for the same reason: legality is one check of one instruction on one frame.

\paragraph{Why are two obvious models bad?}
The first model has one instruction per ion and lets moves of
different ions run together. It accepts target code that no device can
run: two moves can cross one junction between their endpoints, one
signal cannot push one ion left and another right
(Fig.~\ref{fig:broadcast}b), and the signal also moves every ion at a
site that the channel drives without a switch, a \emph{spectator}. The second model
treats a shared channel like a trap, a resource that one transport holds
at a time. It is safe but far too strict: a 63-qubit Fourier transform
takes 4.2 times longer than on a direct device
(Section~\ref{sec:eval-broadcast}).

\paragraph{One instruction, many moves, one waveform}
Each move of a transport names an ion and a declared path, and the
whole list is one event with one frame. Besides the checks of
Section~\ref{sec:programs}, no two moves may cross one segment in
opposite directions or pass through a common site, and on a shared
channel the moves of one event must name the same waveform. A name alone is for bookkeeping, so we also
check the device description: paths that share a waveform on a shared channel must have the
same sequence of hops and the same duration, written $\mathsf{Uniform}(H)$. \textbf{All ions that a
channel moves together then make the same motion in the same time}
(Fig.~\ref{fig:broadcast}a). On a channel without switches, declared
\emph{mandatory}, an event must list a move for every ion at the source
of a path with its waveform, so a spectator cannot be forgotten.

\paragraph{Scheduling: a channel is not held, it is activated}
The
compiler's last pass starts every event as
soon as everything that it holds is free (Section~\ref{sec:compiler}): a
transport holds its ions, its source, destination and transit sites, and
its direct channels. A broadcast channel cannot be on this list, or we
are back to the exclusive model. But a channel keeps
no state between transports, so we record only its
\emph{activations}: an activation is one waveform that the channel
applies from a start time to an end time. A transport may start when every activation
on its channel has ended, begins only after the transport ends, or is the
same waveform over the same interval. Transports of the same motion on
disjoint ions and sites thus start together on one activation: in
Fig.~\ref{fig:broadcast}c, the transports of q0 and q2 to the right share one activation and the schedule
shrinks from 8 to 6 ticks. This is safe: the transport of q2 runs before the gate and the return of
q0, which precede it in the code, but shares no ion and no site with either, so either order gives the same
result, and every event is checked again on its frame at its new start
time (Theorem~\ref{thm:compile-correct}).

\section{A verified compilation engine for co-design}
\label{sec:compiler}
\label{sec:overview}

With the hardware, the programs and the rules defined, we now explain how
a single compiler builds target code for any candidate device, the
\emph{verified compiler} box of Fig.~\ref{fig:codesign-overview}; no pass
has a device-specific case. The pipeline has CompCert's shape, a chain of intermediate languages with one proved pass between
neighbours~\cite{leroy2009compcert,kumar2014cakeml}; where a pass must
search, a proved checker validates the
result~\cite{pnueli1998translation,necula2000translation,tristan2008formal,six2020certified}.

\input{fig-compiler-passes}

\paragraph{Example}
Fig.~\ref{fig:compiler-passes} follows one compilation through this
section and the next: two CX gates on disjoint qubit pairs, for a device
with the layout of QCCDSim's \Dg{grid}{grid-g2x3}, six two-ion traps
around three junctions. One transport channel without switches drives the three column
segments: when it applies the waveform $w_{\downarrow}$, every ion in the
top row moves down through its junction. The rules $R$ are transit
capacity and one waveform per channel, and the initial placement is given
(Fig.~\ref{fig:compiler-passes}a).

\paragraph{Source passes (stages 1--4)}
They fix one ordered native computation that every candidate must
implement: they expand includes, gate definitions and registers into U,
CX, measurement, reset and barrier effects, and implement U and CX by R
and MS pulses with exact angles such as $\pi/7$. Each CX of the example
becomes one MS pulse and four rotations
(Fig.~\ref{fig:compiler-passes}b).

\paragraph{Binding, placement and routing (stages 5--7)}
Control binding keeps every declared realization of each
native effect, placement assigns each ion an initial site, and routing
moves operands. In the
example q0 must join q1 in B0; the channel that moves it also moves q2,
so the router emits one transport with two moves. \textbf{Throughout
routing, every physical prefix reaches a legal placement and has the same
quantum and classical effect as its native prefix.} \textbf{Every routing proposal is checked:} where the bounded search fails, the compiler accepts a complete
proposal from an external router and admits each proposed transport only
with the legality proof that the search
requires~\cite{rideau2010validating}.
\ifqccdreference
Table~\ref{tab:irs} lists the intermediate languages.
\fi

\paragraph{A parallel move is checked as a whole}
The transport of the example is one event with one frame, and the core
checks accept its whole list of moves
(Fig.~\ref{fig:compiler-passes}c). We proposed
three other moves to the compiler, and these checks rejected each
(Fig.~\ref{fig:compiler-passes}e).

\paragraph{Pulse sharing and time (stages 8--9)}
Stage 8 replaces adjacent R operations by one declared optical broadcast
when they have the same waveform and guard on distinct ions and the
rules permit the packed event; its simultaneous action is proved equal
to theirs. Stage 9 assigns declared durations and inserts waits; a
guarded operation reserves the same duration on both branches, so
legality and time do not depend on measurement outcomes. The result is
the target code $K$: nine instructions, 22 ticks one after another. \textbf{A
parallel schedule then lets independent events overlap:} each event
starts as early as every resource that it uses is free (its ions, sites,
direct channels, classical bits and the controller; broadcast channels
follow Section~\ref{sec:broadcast}), and the
rules are rechecked on every frame at its new start time. After the transport, B0 and B2
hold different ions under different lasers, so their gates overlap and
the timeline $T$ ends at tick 14 (Fig.~\ref{fig:compiler-passes}d). We
prove once that events sharing no resource commute, so every timeline
equals the program order up to swaps of independent
neighbours~\cite{mazurkiewicz1987trace,lipton1975reduction}. \textbf{Exchanges restore chain order:} a
trap releases only the ion at the end that faces the segment, so a last
pass inserts and charges an exchange before each transport whose ion is
elsewhere in the chain.
\label{sec:impl}
When a placement is unroutable, the driver revisits earlier choices
and then tries a fallback router; the theorem holds for every search
configuration.

\section{End-to-end correctness}
\label{sec:checker}

We now state what returned target code guarantees, the
verified-compilation step of Fig.~\ref{fig:codesign-overview}b, as one
implication with three clauses; Fig.~\ref{fig:compiler-passes}b shows
each clause on the example of Section~\ref{sec:compiler}.
\label{sec:correctness}

\paragraph{Notation}
We write $\mathsf{compile}(H,R,P,\kappa)\Downarrow(K,T)$ when the
compiler, given a hardware description $H$, rules $R$, a source program
$P$ and a search configuration $\kappa$, returns target code $K$ with a
timeline $T$; this premise states only what the procedure
returned. $T$ lists steps $e_1,\dots,e_m$, the instructions of $K$ with
chain exchanges inserted, their start times $t_i$ and a finish time
$t_{\mathrm{end}}$. $H$ declares the duration $\delta_i$ of $e_i$; let
$\tau_i=[t_i,t_i+\delta_i)$. The frame of $e_i$ is
$F_i=(H,\pos_{i-1},\pos_i,e_i,t_i,\delta_i)$, where the placements
$\pos_0,\dots,\pos_m$ follow the steps in program order, and $F_0$ is the
initial frame.

\begin{theorem}[Verified compilation]
\label{thm:compile-correct}\label{thm:reported}\label{thm:broadcast}
For all $H$, $R$, $P$, $\kappa$, $K$ and $T$, under the declared
ideal primitive semantics,
\[
\mathsf{compile}(H,R,P,\kappa)\Downarrow(K,T)\;\Longrightarrow\;
\mathsf{Meaning}\land\mathsf{Legality}\land\mathsf{Time}.
\]
\end{theorem}

The clauses are the following conjunctions, numbered for later
reference.
\begin{align}
\mathsf{Meaning}:\;\;&\mathsf{erase}(e_1..e_m)=K,\quad
  \mathsf{proj}(K)=\mathsf{native}(P),\label{eq:meaning}\\
&\forall\eta,\sigma.\,\exists\sigma'.\;
  \langle P,\mathsf{obs}\,\sigma\rangle\Downarrow_{\eta}\sigma'\,\land\,
  \mathsf{obs}\,\llbracket K^{T}\rrbracket_{H,\eta}\,\sigma\simeq\sigma'\notag\\
\mathsf{Legality}:\;\;&\forall i.\;F_i\models\mathsf{Core}_H\,\land\,
  \forall\varphi\in R.\;F_i\models_H\varphi,\label{eq:legality}\\
&\mathsf{chain}_H(e_1..e_m)\Downarrow\chi\,\land\,\chi\sim\pos_m,\notag\\
&\forall i<j.\;e_i\mathrel{\#}e_j\Rightarrow t_i+\delta_i\le t_j,\notag\\
&\forall i,j,c.\;\tau_i\cap\tau_j=\emptyset\,\lor\,
  (w_i(c)=w_j(c)\land\tau_i=\tau_j),\notag\\
&\mathsf{Uniform}(H)\Rightarrow\forall a,b\in e_i.\;
  \mu(a)=\mu(b)\land d(a)=d(b)\notag\\
\mathsf{Time}:\;\;&t_{\mathrm{end}}=\max\nolimits_i(t_i+\delta_i)\le
  \textstyle\sum_i\delta_i,\label{eq:time}\\
&\forall\,\text{pulse }e_i.\;\delta_i=d(k_i)\notag
\end{align}

\emph{Meaning.} $\mathsf{erase}$ deletes the exchanges, $\mathsf{proj}$
deletes transports and waits and unpacks shared pulses, and
$\mathsf{native}(P)$ is the ordered
native program of stages 1--4: nothing is dropped, merged or reordered.
$\eta$ assigns the symbolic angles, $\sigma$ is any initial quantum and
classical state, $\Downarrow_\eta$ is the source semantics,
$\llbracket K^{T}\rrbracket$ runs $K$ in order of the start times of
$T$, $\mathsf{obs}$ hides the compiler's saved bit, and $\simeq$ is
equality up to the listing order of measurement branches.
\emph{Legality.} $\mathsf{Core}_H$ are the core checks of
Section~\ref{sec:programs} for an instruction of $K$, and
$F\models_H\varphi$ is satisfaction of the rule as the designer wrote
it. $\mathsf{chain}_H$
replays the steps on the declared ion chains,
where an ion leaves a trap only at its end; its result $\chi$ holds
the ions of $\pos_m$. $e_i\mathrel{\#}e_j$ if the steps share an ion, a
site, a direct channel, a classical bit or the controller, or if one is
a global wait. On a
broadcast channel $c$, where $e_i$ applies waveform $w_i(c)$, two steps
are disjoint in time or form one activation.
$\mathsf{Uniform}(H)$ is the device condition of
Section~\ref{sec:broadcast}, and $a,b$ are moves of one transport on one
broadcast channel, with motion $\mu$ and duration $d$. \emph{Time.}
$k_i$ is the calibration of a pulse at the trap that holds its ions.

\paragraph{One proof for every device}
The theorem quantifies over $H$ and $R$
(Section~\ref{sec:elaboration}), and it is a soundness theorem: it
constrains every result that the compiler returns. In Lean it is one declaration,
\nolinkurl{verified_compilation}, whose conclusion is the conjunction of
the three clauses. We prove it by composing the contracts of the passes
of Fig.~\ref{fig:compiler-passes}: source passes preserve the semantics
of their languages, and physical passes also establish legality. Our
programs are finite, so preservation is an equality of
results, as in QWIRE and VOQC~\cite{paykin2017qwire,hietala2021voqc}.
\ifqccdformal
Appendix~\ref{app:formal} proves the theorem pass by pass and replays
the proof on one program.
\fi

\ifqccdreference
\paragraph{Artifact proof map}
The following declarations connect each pass to the invariants above.
\begin{center}
\begingroup\small
\begin{tabular}{@{}p{.25\columnwidth}p{.71\columnwidth}@{}}
Theorem~\ref{thm:compile-correct} & \nolinkurl{verified_compilation}\\
Success contract & \nolinkurl{compileQASM_sound}\\
Observation & \nolinkurl{QASMCompilation.source_correct}\\
Source expansion & \nolinkurl{compileFrontend_source_correct}\\
Native selection & \nolinkurl{selectProgram_correct}\\
Binding/placement & \nolinkurl{Binding.stage_correct}, \nolinkurl{Placed.stage_correct}\\
Routing prefixes & \nolinkurl{CoreExecution.prefix_invariant}\\
Optical broadcast & \nolinkurl{Packed.stage_correct}\\
Timing, legality & \nolinkurl{Timed.stage_correct}\\
Assertion meaning & \nolinkurl{Assertions.elaborate_correct}\\
Assertion legality & \nolinkurl{Assertions.initial_satisfies}, \nolinkurl{Assertions.run_satisfies}\\
Routing proposals & \nolinkurl{RoutePlan.compileQASM_sound}\\
Reported timeline & \nolinkurl{Reported.reported_sound}, \nolinkurl{TimeOrder.reported_time_order_sound}\\
Independent events & \nolinkurl{Independence.hardware_comm}, \nolinkurl{TimeOrder.resource_order}\\
Reported time & \nolinkurl{ReportedTime.finish_latest_end}, \nolinkurl{ReportedTime.gate_durations}\\
One per defect & \nolinkurl{Defects.QS1_ion_order_is_replayed}, \ldots\\
Input acceptance & \nolinkurl{Acceptance.compileFrontend_accepts}\\
Text layout & \nolinkurl{parseSourceUnit_leading}, \nolinkurl{parse_layout}\\
Progress & \nolinkurl{compileQASMTotal_complete}, \nolinkurl{Fallback.routeEither_complete}\\
Heat count & \nolinkurl{Evaluation.reported_heat}, \nolinkurl{Heat.move_charged}\\
Heat model & \nolinkurl{printed_heat}, \nolinkurl{Strict.heatModel_file}\\
Input files & \nolinkurl{Strict.hardwareFile_file}, \nolinkurl{Strict.rulesProgram_file}\\
Chain length & \nolinkurl{Defects.M7_gate_charged_for_its_chain_length}\\
Exact minimum & \nolinkurl{Exact.compile_minimal}, \nolinkurl{Exact.compile_complete}\\
Chain order & \nolinkurl{Chain.check_agrees}, \nolinkurl{Chain.timeline_exclusive}\\
Broadcast transport & \nolinkurl{Broadcast.execution_same_motion}, \nolinkurl{Shared.timeline_exclusive}\\
\end{tabular}
\endgroup
\end{center}
\fi

\paragraph{What is proved, checked and assumed}
\label{sec:validation}\label{sec:trust}
\textbf{Our compiler proof is general:} Lean's kernel checks it
once, and no compilation needs simulation or a proof certificate;
external routing proposals pass a
proved checker~\cite{tristan2008formal,necula1997proof}. \textbf{The rest is
assumed,} which is the primitive-realization assumption of
Section~\ref{sec:background}, and Lean's compiler and runtime. An audit of the dependencies of 739 headline
declarations admits only the axioms \nolinkurl{propext},
\nolinkurl{Classical.choice} and \nolinkurl{Quot.sound}.

\section{Co-design Loop}
\label{sec:codesign}
Sections~\ref{sec:background}--\ref{sec:checker} make one candidate
design cheap to judge. In this section, we explain how we choose the
candidates, the third challenge of Section~\ref{sec:intro}. The
choices interact: a larger trap saves transports and slows every gate in
it, and the same traps declared in another order can be four times
slower (Section~\ref{sec:eval-search}).

\paragraph{The loop}
Fig.~\ref{fig:codesign-overview}b repeats four steps for a fixed source
program, total ion capacity and cost model. A proposer, a person or an AI
agent, describes a device. The verified compiler produces target code;
only complete, checked target code receives a score, its parallel finish
time. Per-trap gate, transport and
exchange records of that code then guide the next proposal. The
evaluator never changes while the proposals do, and a proposal is only a
compiler input, so it cannot win with target code that breaks a rule.

\paragraph{Learning principles from manual designs}
What should a proposer know before its first proposal? We compile the benchmark programs for the manual designs of
Section~\ref{sec:eval}, which are 15 published families, and we test 67
candidate \emph{design principles} on the checked target code
(Section~\ref{sec:eval-families}). Each is a quantity defined on the
hardware description, on the program's interaction graph or on both,
with a predicted effect on the finish time. We evaluate each principle by measuring the rank correlation between the corresponding architectural property and program completion time. We consider that a principle is validated if the correlation has the predicted sign, is statistically significant (\(p < 0.05\), after correction for multiple tests), and also has the predicted sign on held-out devices where the principle can be tested. Fewer than one-third of the principles are validated. The most important ones concern matching the device architecture to the program's interaction graph and choosing trap sizes that preserve fast gate execution.

\paragraph{The agent}
Section~\ref{sec:eval-search} instantiates the loop with an AI agent. A
proposal is a planar layout on an integer lattice: traps with their
capacities and controls, junctions, segments, and the order in which
traps are declared, which sets the initial placement. Layout rules are
checked before any compilation. In the co-design loop, we fix the
operating rules and the control channels, and the agent varies the
layout, the trap sizes and the placement.

\section{Evaluation}
\label{sec:eval}

Seven research questions (RQs) test the requirements and challenges of
Section~\ref{sec:intro}.

\begin{description}[leftmargin=1em,nosep]
\item[RQ1] \emph{Is the compiler's code competitive (Requirement~1)?}
Yes: fewer transports than QCCDSim on all six runs.
\item[RQ2] \emph{Can a search trust existing compilers (Requirement~2)?}
No: we report 11 defects, each excluded from ours by our theorem.
\item[RQ3] \emph{How much does the design matter?} At equal capacity the
best design takes 66.4\% less time than the worst on average.
\item[RQ4] \emph{What does the broadcast model buy (Challenge~2)?} One
transport channel in place of up to 23 costs at most 8\%.
\item[RQ5] \emph{Can we learn guiding design principles
(Challenge~3)?} Yes: nine hold; the main one is to fit the device to
the circuit.
\item[RQ6] \emph{Can a search find better devices (Challenge~3)?} Yes:
an AI agent beats the fastest design of equal capacity on 71 of 75
programs.
\item[RQ7] \emph{Does verified compilation scale?} Yes: 512-qubit
programs compile in 67 seconds to 29 minutes.
\end{description}

Every reported design is the best observed within stated
candidates and budgets.

\subsection{Setup}

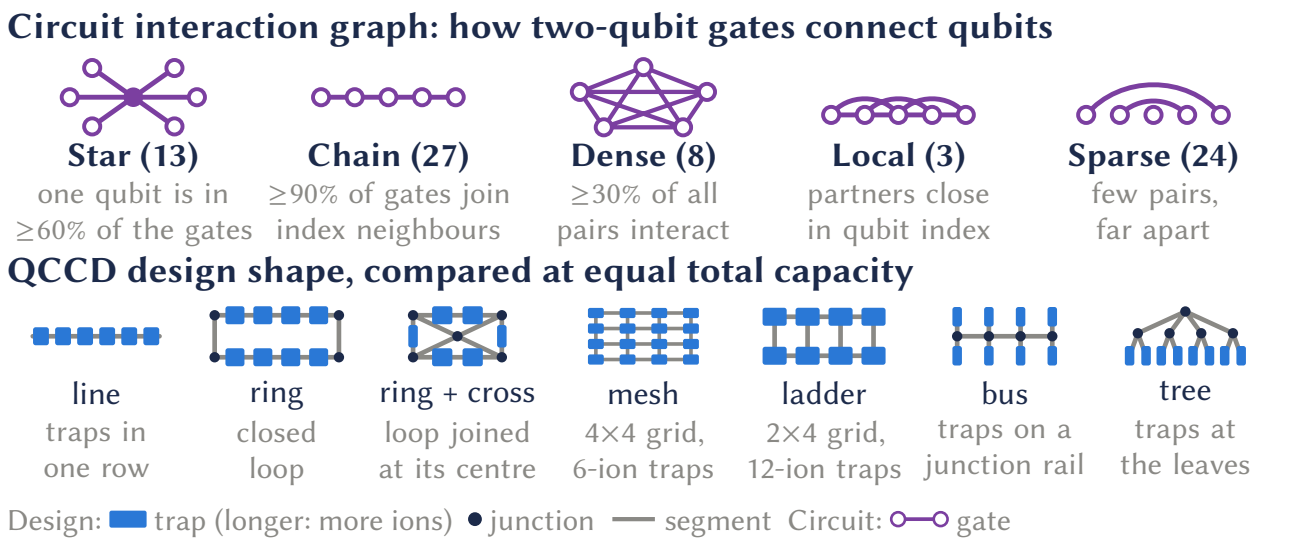
\begin{figure}[t]
\centering
\begin{tikzpicture}[x=1pt,y=1pt,
  hd/.style={anchor=west,inner sep=0pt,text=qink,font=\sffamily\bfseries\fontsize{7}{8}\selectfont},
  nm/.style={anchor=north,inner sep=0pt,text=qink,align=center,font=\sffamily\bfseries\fontsize{6.5}{7.5}\selectfont},
  rl/.style={anchor=north,inner sep=0pt,text=qgrey,align=center,text width=46pt,font=\sffamily\fontsize{5.8}{6.8}\selectfont}]
  \path[use as bounding box] (0,5) rectangle (240,-100);
  \node[hd] at (0,0) {Circuit interaction graph: how two-qubit gates connect qubits};
  \foreach \x/\k/\n/\c/\r in {
      24/star/Star/13/{one qubit is in\\${\ge}60\%$ of the gates},
      72/chain/Chain/27/{${\ge}90\%$ of gates join\\index neighbours},
      120/dense/Dense/8/{${\ge}30\%$ of all\\pairs interact},
      168/local/Local/3/{partners close\\in qubit index},
      216/sparse/Sparse/24/{few pairs,\\far apart}}{
    \node[anchor=north,inner sep=0pt] at (\x,-6) {\scalebox{2.3}{\pglyph{\k}}};
    \node[nm] at (\x,-21) {\n\ (\c)};
    \node[rl] at (\x,-28.5) {\r};
  }
  \node[hd] at (0,-46) {QCCD design shape, compared at equal total capacity};
  \foreach \x/\k/\n/\r in {17/line-c6/line/{traps in\\one row},51/ring-c6/ring/{closed\\loop},
      85/ringcross-c6/ring + cross/{loop joined\\at its centre},
      120/mesh-c6/mesh/{4$\times$4 grid,\\6-ion traps},154/ladder-c12/ladder/{2$\times$4 grid,\\12-ion traps},
      188/bus-c6/bus/{traps on a\\junction rail},222/tree-c6/tree/{traps at\\the leaves}}{
    \node[anchor=north,inner sep=0pt] at (\x,-52) {\scalebox{1.95}{\dglyph{\k}}};
    \node[nm,font=\sffamily\fontsize{6.2}{7}\selectfont] at (\x,-66) {\n};
    \node[rl,text width=34pt] at (\x,-73.5) {\r};
  }
  \node[anchor=north west,inner sep=0pt,text=qgrey,font=\sffamily\fontsize{5.8}{6.8}\selectfont]
    at (0,-90) {Design:\ \tikz[baseline=-2pt]{\fill[qblue,rounded corners=.4pt] (0,-1.6) rectangle (7,1.6);}~trap (longer: more ions)\enspace
      \tikz[baseline=-2pt]{\fill[qink] (0,0) circle[radius=1.3pt];}~junction\enspace
      \tikz[baseline=-2pt]{\draw[qgrey,line width=.8pt] (0,0)--(8,0);}~segment\enspace
      Circuit:\ \tikz[baseline=-2pt]{\draw[qpurple,line width=.7pt] (0,0)--(8,0);
      \path[fill=white,draw=qpurple,line width=.5pt] (0,0) circle[radius=1.3pt] (8,0) circle[radius=1.3pt];}~gate};
\end{tikzpicture}
\caption{Circuit interaction graphs and QCCD design shapes. A circuit's
pattern is the first rule, left to right, that its gates match;
``local'': partners are on average a quarter as far apart in index as
random pairs, or closer. A ladder is a grid with two rows; grid sizes
are those at 96 ions.}
\label{fig:workloads}
\end{figure}

\paragraph{Workloads}
We take every original circuit of the medium and large tiers of
QASMBench~\cite{li2022qasmbench} with 11 to 134 qubits, QCCDSim's
benchmark circuits~\cite{murali2020architecting}, and generated
circuits of 16 to 128 qubits: Greenberger--Horne--Zeilinger (GHZ)
states, Bernstein--Vazirani, and the quantum approximate optimization
algorithm (QAOA) on a ring. A rule
fixed in advance admits a program if source preparation succeeds and it
expands to at most 30,000 operations: 75 of 89 programs, of 11 to 130
qubits. By their \emph{interaction graph}, the qubit pairs that two-qubit
gates join (Fig.~\ref{fig:workloads}), 13 are \Pg{star}{star}, 27
\Pg{chain}{chain}, 8 \Pg{dense}{dense},
3 \Pg{local}{local}, 24 \Pg{sparse}{sparse}.

\paragraph{Devices of equal capacity}
We compare designs of equal size, because a design with more ion
slots could be faster only for being larger.
A program with $n$ qubits is therefore compiled only for designs
whose traps hold exactly $C$ ions in total, the smallest multiple of 24
with $n\le 0.7\,C$, so no device is much larger than its program.


\QCCDPhysicalMainText

\paragraph{Compilation and acceptance}
All experiments use pulse sharing, the parallel schedule and chain
order. Qubits fill the traps in declared order in
proportion to capacity. A design's time for a program is
$t_{\mathrm{end}}$. We count a compilation only after
three independent re-checks: a replay of placement, capacity and pulses,
a recomputation of the parallel start times, and a replay of the chain
order. Experiments run on one Intel i5-12400 workstation with 64\,GiB of
memory.

\subsection{RQ1: Is the compiler's code competitive?}
\label{sec:eval-quality}

\begin{figure}[t]
\centering
\begin{tikzpicture}[x=1pt,y=.88pt,font=\sffamily\fontsize{6.4}{7.6}\selectfont,
 lab/.style={inner sep=0pt,text=qink},val/.style={inner sep=0.6pt,text=qink,font=\sffamily\fontsize{5.4}{6.2}\selectfont}]
\path[use as bounding box] (0,108) rectangle (240,-28);
\draw[qrail,line width=.3pt] (22.0,15.36)--(240.0,15.36);
\node[lab,anchor=east] at (20.0,15.36) {0.5};
\draw[qink,line width=.5pt] (22.0,30.71)--(240.0,30.71);
\node[lab,anchor=east] at (20.0,30.71) {1};
\draw[qrail,line width=.3pt] (22.0,46.07)--(240.0,46.07);
\node[lab,anchor=east] at (20.0,46.07) {1.5};
\draw[qrail,line width=.3pt] (22.0,61.43)--(240.0,61.43);
\node[lab,anchor=east] at (20.0,61.43) {2};
\draw[qrail,line width=.3pt] (22.0,76.79)--(240.0,76.79);
\node[lab,anchor=east] at (20.0,76.79) {2.5};
\draw[qink,line width=.5pt] (22.0,0.0)--(240.0,0.0);
\node[lab,rotate=90] at (4,43.0) {ratio to QCCDSim};
\fill[qteal!70] (25.27,0.0) rectangle (32.27,17.55);
\node[val,anchor=south] at (28.77,17.55) {0.57};
\fill[qorange!80] (33.47,0.0) rectangle (40.47,32.73);
\node[val,anchor=south] at (36.97,32.73) {1.07};
\fill[qblue!70] (41.67,0.0) rectangle (48.67,29.72);
\node[val,anchor=south] at (45.17,29.72) {0.97};
\node[lab,anchor=north] at (37.57,-3) {BV-64};
\node[lab,anchor=north] at (37.57,-12) {\dglyph{line-l6}\,L6};
\node[val,anchor=north] at (37.57,-20.5) {13.27\,ms};
\fill[qteal!70] (56.41,0.0) rectangle (63.41,10.24);
\node[val,anchor=south] at (59.91,10.24) {0.33};
\fill[qorange!80] (64.61,0.0) rectangle (71.61,35.95);
\node[val,anchor=south] at (68.11,35.95) {1.17};
\fill[qblue!70] (72.81,0.0) rectangle (79.81,29.97);
\node[val,anchor=south] at (76.31,29.97) {0.98};
\node[lab,anchor=north] at (68.71,-3) {Adder};
\node[lab,anchor=north] at (68.71,-12) {\dglyph{line-l6}\,L6};
\node[val,anchor=north] at (68.71,-20.5) {90.14\,ms};
\fill[qteal!70] (87.56,0.0) rectangle (94.56,24.57);
\node[val,anchor=south] at (91.06,24.57) {0.80};
\fill[qorange!80] (95.76,0.0) rectangle (102.76,20.68);
\node[val,anchor=south] at (99.26,20.68) {0.67};
\fill[qblue!70] (103.96,0.0) rectangle (110.96,17.43);
\node[val,anchor=south] at (107.46,17.43) {0.57};
\node[lab,anchor=north] at (99.86,-3) {QAOA-64};
\node[lab,anchor=north] at (99.86,-12) {\dglyph{line-l6}\,L6};
\node[val,anchor=north] at (99.86,-20.5) {162.36\,ms};
\fill[qteal!70] (118.70,0.0) rectangle (125.70,30.35);
\node[val,anchor=south] at (122.20,30.35) {0.99};
\fill[qorange!80] (126.90,0.0) rectangle (133.90,81.31);
\node[val,anchor=south] at (130.40,81.31) {2.65};
\fill[qblue!70] (135.10,0.0) rectangle (142.10,69.38);
\node[val,anchor=south] at (138.60,69.38) {2.26};
\node[lab,anchor=north] at (131.00,-3) {QFT-64};
\node[lab,anchor=north] at (131.00,-12) {\dglyph{line-l6}\,L6};
\node[val,anchor=north] at (131.00,-20.5) {333.07\,ms};
\fill[qteal!70] (149.84,0.0) rectangle (156.84,17.55);
\node[val,anchor=south] at (153.34,17.55) {0.57};
\fill[qorange!80] (158.04,0.0) rectangle (165.04,28.62);
\node[val,anchor=south] at (161.54,28.62) {0.93};
\fill[qblue!70] (166.24,0.0) rectangle (173.24,25.83);
\node[val,anchor=south] at (169.74,25.83) {0.84};
\node[lab,anchor=north] at (162.14,-3) {BV-64};
\node[lab,anchor=north] at (162.14,-12) {\dglyph{grid-g2x3}\,$G2\mathord\times3$};
\node[val,anchor=north] at (162.14,-20.5) {14.35\,ms};
\fill[qteal!70] (180.99,0.0) rectangle (187.99,24.57);
\node[val,anchor=south] at (184.49,24.57) {0.80};
\fill[qorange!80] (189.19,0.0) rectangle (196.19,35.87);
\node[val,anchor=south] at (192.69,35.87) {1.17};
\fill[qblue!70] (197.39,0.0) rectangle (204.39,33.52);
\node[val,anchor=south] at (200.89,33.52) {1.09};
\node[lab,anchor=north] at (193.29,-3) {QAOA-64};
\node[lab,anchor=north] at (193.29,-12) {\dglyph{grid-g2x3}\,$G2\mathord\times3$};
\node[val,anchor=north] at (193.29,-20.5) {191.81\,ms};
\fill[qteal!70] (212.13,0.0) rectangle (219.13,30.35);
\node[val,anchor=south] at (215.63,30.35) {0.99};
\fill[qorange!80] (220.33,0.0) rectangle (227.33,53.27);
\node[val,anchor=south] at (223.83,53.27) {1.73};
\fill[qblue!70] (228.53,0.0) rectangle (235.53,45.26);
\node[val,anchor=south] at (232.03,45.26) {1.47};
\node[lab,anchor=north] at (224.43,-3) {QFT-64$^{\dagger}$};
\node[lab,anchor=north] at (224.43,-12) {\dglyph{line-l6}\,L6};
\node[val,anchor=north] at (224.43,-20.5) {333.07\,ms};
\fill[qteal!70] (28.0,97.0) rectangle (36.0,103.0);
\node[lab,anchor=west] at (39.0,100.0) {transports};
\fill[qorange!80] (98.0,97.0) rectangle (106.0,103.0);
\node[lab,anchor=west] at (109.0,100.0) {finish};
\fill[qblue!70] (168.0,97.0) rectangle (176.0,103.0);
\node[lab,anchor=west] at (179.0,100.0) {finish, Ours$^{\ast}$};
\end{tikzpicture}
\caption{The verified compiler against QCCDSim on QCCDSim's benchmark runs;
below 1 is better than QCCDSim, whose finish time is under each run.
Ours$^{\ast}$ sets single-qubit gate durations to zero, as QCCDSim assumes. $\dagger$: barriers removed.
QCCDSim's schedules for QFT-64 and for QAOA-64 on $G2\mathord\times3$ fail the chain-order check.}
\label{fig:qccdsim-baseline}
\Description{Grouped bars for seven runs: the ratio of transports and of finish time of the verified compiler to QCCDSim.}
\end{figure}
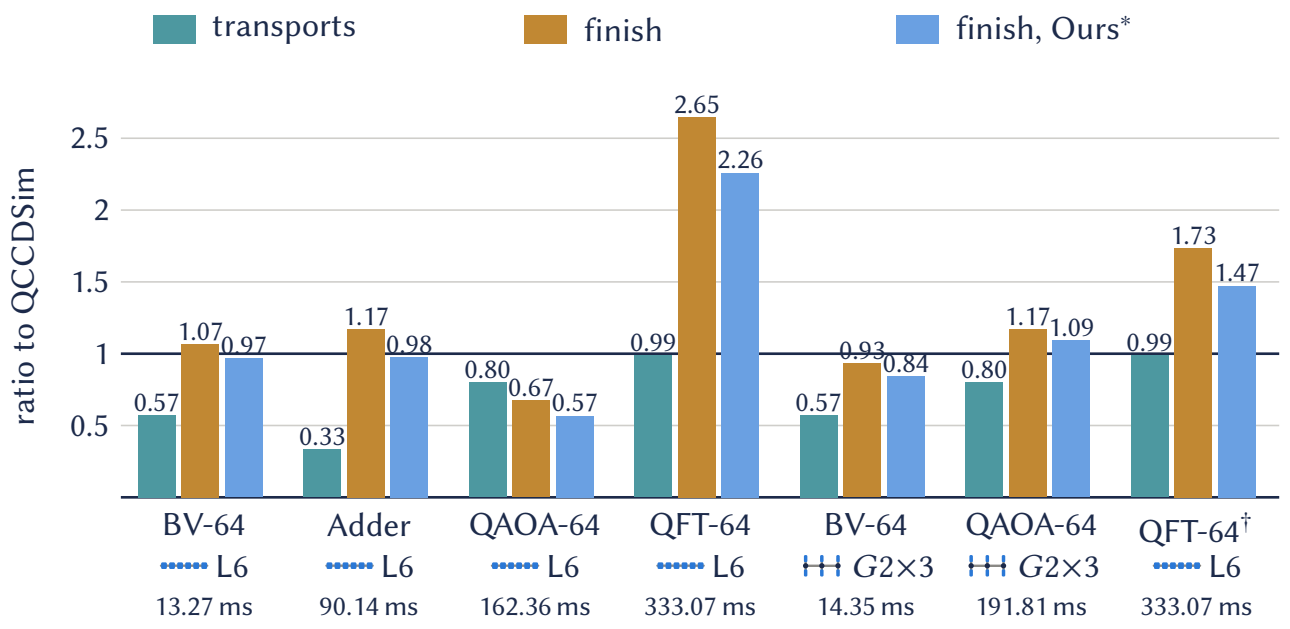

Requirement~1
of Section~\ref{sec:intro} is good code for every design, because a
compiler that wastes transports undervalues its designs. Here we test whether our compiler meets
this requirement. We compare with QCCDSim on its
benchmark circuits and its two six-trap devices, published in the same
paper~\cite{murali2020architecting}, a \Dg{line}{line-l6} and
a \Dg{bus}{grid-g2x3}, from the initial
placement and chain order that QCCDSim chose
(Fig.~\ref{fig:qccdsim-baseline}).

\textbf{Fig.~\ref{fig:qccdsim-baseline}
shows that our verified compiler needs fewer transports than QCCDSim on
all six runs,} down to
one third on the adder. When every gate is charged, our target code
finishes earlier on two runs (by 33\% and 7\%), and on four runs under
QCCDSim's timing convention. We lose time on the Fourier transform, whose
source contains barriers that our compiler obeys. Overall, the verified
compiler generates code that is competitive with QCCDSim's, with fewer
transports on every run.

\subsection{RQ2: Can a search trust existing compilers?}
\label{sec:eval-compiler}

\begin{figure*}[t]
\centering
\newcommand{\BugDevice}[5][.5bp]{%
  \begin{scope}\QCCDUnit=#1\qccdpadsfalse\qccdionlabelsfalse
  \begin{scope}[shift={(#2,#3)},x=\QCCDUnit,y=-\QCCDUnit]
    \QCCDDevice{#4}{12}#5
  \end{scope}\end{scope}}
\newcommand{\BugIon}[5]{\QCCDIon[#1]{#2}{#3}{#4}\BugNumber{($(#2)+(0,-20)$)}{qhwInk}{#5}}
\newcommand{\BugNumber}[3]{\begingroup\contourlength{.8bp}%
  \node[inner sep=0pt,text=#2,font=\fontencoding{T1}\fontfamily{phv}\fontseries{b}%
    \fontsize{5.8bp}{5.8bp}\selectfont] at #1 {\contour{white}{#3}};\endgroup}
\newcommand{\BugName}[3]{\node[inner sep=0pt,text=qgrey,
  font=\sffamily\fontsize{6.2}{7}\selectfont,anchor=north] at (#1,#2) {#3};}
\resizebox{\textwidth}{!}{%
\begin{tikzpicture}[x=1pt,y=1pt,font=\sffamily\fontsize{7}{8.4}\selectfont,
  head/.style={inner sep=0pt,anchor=north west,text=qink,
    font=\sffamily\bfseries\fontsize{7.6}{9}\selectfont},
  txt/.style={inner sep=0pt,anchor=north west,text=qink,align=left},
  rule/.style={inner sep=0pt,anchor=north west,text=qgrey,align=left,
    font=\sffamily\itshape\fontsize{6.8}{8}\selectfont},
  bad/.style={inner sep=0pt,anchor=north west,text=qred,align=left,
    font=\sffamily\bfseries\fontsize{6.8}{8}\selectfont},
  good/.style={inner sep=0pt,anchor=north west,text=qgreen,align=left,
    font=\sffamily\bfseries\fontsize{6.8}{8}\selectfont},
  move/.style={-{Stealth[length=3pt,width=2.6pt]},line width=.8pt},
  sep/.style={qrail,line width=.5pt}]
\path[use as bounding box] (0,2) rectangle (516,-146);
\draw[sep] (170,0)--(170,-144) (344,0)--(344,-144);
\foreach \x in {0,176,350}{\draw[sep] (\x,-110)--(\x+164,-110);}

\node[head] at (0,0) {(a) QCCDSim~\cite{murali2020architecting}: chain order};
\node[rule] at (0,-11) {Own rule: only the end ion leaves; else gate swap.};
\node[txt] at (0,-23) {\rev{States 1\,2\,3: swap 1 and 3, then state 1 leaves right}};
\newcommand{\BugLeft}{\path[fill=qhwData!40,draw=white,line width=3.9\QCCDUnit] (47,0) circle[radius=8.8\QCCDUnit];
  \BugNumber{(47,-20)}{qhwInk!55}{1}\draw[move,qgrey] (59,0)--(73,0);}
\BugDevice{40}{-48}{bugtrap}{\BugIon{qhwData}{i3}{T}{0}{3}\BugIon{qblue}{i2}{T}{1}{2}\BugLeft
  \draw[move,qred,rounded corners=2pt] ($(i2)+(0,10)$)--($(i2)+(0,15)$)--(-64,15);}
\BugDevice{124}{-48}{bugtrap}{\BugIon{qblue}{i2}{T}{0}{2}\BugIon{qhwData}{i3}{T}{1}{3}\BugLeft}
\BugName{40}{-58}{\rev{states after the exit}}\BugName{124}{-58}{QCCDSim record}
\node[inner sep=0pt,text=qred,font=\sffamily\bfseries\fontsize{11}{12}\selectfont]
  at (82,-48) {$\neq$};
\node[bad] at (0,-70) {\rev{Next, state 2 leaves left with no swap,}\\\rev{although state 3 is at the left end}};
\node[good] at (0,-89) {\rev{Ours: update the state order; exchange}\\\rev{states 2 and 3 before the left exit}};
\node[txt] at (0,-114) {4 of 11 schedules fail the chain check;\\\rev{QFT-64: 69 missing exchanges (32.9\,ms)}};
\node[rule] at (0,-134) {Excluded by the chain machine (\S5)};

\node[head] at (176,0) {(b) Muzzle the Shuttle~\cite{saki2022muzzle}: capacity};
\node[rule] at (176,-11) {Own rule: a full trap cannot receive an ion.};
\node[txt] at (176,-23) {Re-balancing moves an ion T0$\to$T2 via full T1:};
\BugDevice{203}{-50}{bugline}{%
  \BugIon{qhwData}{i0}{T0}{0}{0}\BugIon{qhwData}{i2}{T0}{1}{2}\BugIon{qhwData!35}{i3}{T0}{2}{3}
  \BugIon{qhwData}{i1}{T1}{0}{1}\BugIon{qhwData}{i4}{T1}{1}{4}\BugIon{qhwData}{i5}{T1}{2}{5}
  \BugIon{qhwData}{i6}{T2}{0}{6}
  \path[fill=qred,draw=white,line width=3\QCCDUnit] (70,0) circle[radius=8.8\QCCDUnit];
  \BugNumber{(70,-20)}{qred}{3}
  \draw[move,qred,rounded corners=2pt] ($(i3)+(0,-28)$)--($(i3)+(0,-35)$)
    --($(T2-slot-2)+(0,-35)$)--($(T2-slot-2)+(0,-11)$);}
\BugName{203}{-60}{T0 (full)}\BugName{258}{-60}{T1 (cap.\ 3)}\BugName{313}{-60}{T2}
\node[bad] at (176,-70) {T1 holds 4 ions in 3 places during\\the merge};
\node[good] at (176,-89) {\rev{Ours: transit capacity rejects four ions}\\\rev{in three slots; first make room in T1}};
\node[txt] at (176,-114) {27 of 76 schedules, all five programs of\\its own evaluation; the authors' fix removes it};
\node[rule] at (176,-134) {Excluded by transit capacity (Thm.~\ref{thm:compile-correct})};

\node[head] at (350,0) {(c) MQT IonShuttler~\cite{schoenberger2025multizone}: junctions};
\node[rule] at (350,-11) {Own rule: moves in one step share no junction.};
\node[txt] at (350,-23) {Step 18, one corner junction J of a 2$\times$2 grid:};
\BugDevice{410}{-53}{bugcross}{\QCCDIon[qblue]{ib}{W}{0}\QCCDIon[qorange]{io}{N}{0}
  \BugNumber{(-8,8)}{qhwInk}{J}
  \draw[move,qblue] (-46,-9)--(64,-9);
  \draw[move,qorange] (9,-32)--(9,30);}
\node[inner sep=0pt,text=qgrey,font=\sffamily\fontsize{6.2}{7}\selectfont,anchor=west,
  align=left] at (455,-48) {blue: leaves a gate zone,\\exempt from the test};
\node[bad] at (350,-70) {Two ions cross J in the same step};
\node[good] at (350,-89) {Ours: J is a site of capacity 1; events\\that share a site never overlap};
\node[txt] at (350,-114) {69 of 498 schedules, all on grids with three or\\four gate zones joined by one-site paths};
\node[rule] at (350,-134) {Excluded by site exclusion (\S5)};
\end{tikzpicture}}
\caption{One rule violation in each of three published QCCD compilers,
run unchanged; each breaks a rule that the tool itself states (italic).
Numbers in (a) name quantum states. A gate SWAP exchanges states,
not ion positions: the rightmost ion leaves, giving state order 3,\,2;
the stale record 2,\,3 omits a later exchange. In (c), a gate zone is a trap
where gates run.}
\label{fig:baseline-bugs}
\Description{Three panels. (a) A QCCDSim trap whose true chain is 3, 2
while QCCDSim records 2, 3, so state 2 later leaves from the middle. (b) A
Muzzle-the-Shuttle re-balancing move that merges a fourth ion into a full
trap of capacity three. (c) Two ions crossing one junction of an
MQT-IonShuttler grid in the same time step. Each panel states the extent
and the property of our compiler that excludes it.}
\end{figure*}
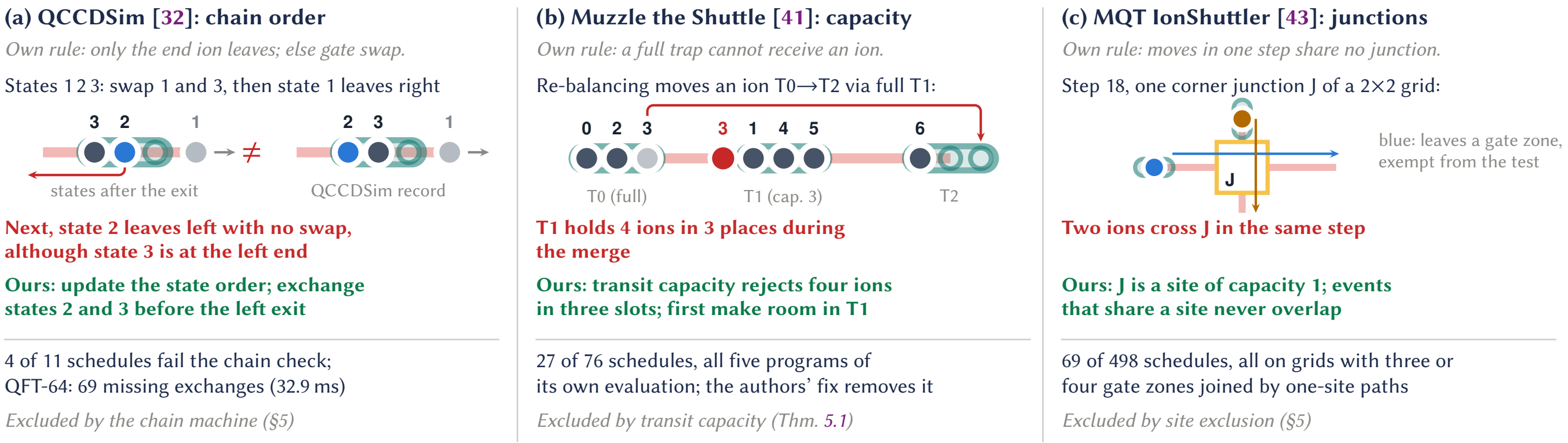

Requirement~2 of Section~\ref{sec:intro} is a verified compiler,
because incorrect target code can make a design look better than it is.
Here we test whether published compilers can be trusted. We ran
three published compilers unchanged, replayed every recorded schedule on
independent checkers and traced each finding to the upstream source. We used
Fable~5.2 to help explore the open-source code and find these defects.
\textbf{We report 11 defects, counting shared code once, and each is
excluded from our compiler by 
Theorem~\ref{thm:compile-correct}.}

\emph{Legality} (Eq.~\ref{eq:legality}) excludes the five defects that break a rule which the
tool itself states (Fig.~\ref{fig:baseline-bugs}): a stale record of ion
order in QCCDSim, traps above capacity at the start and in transit in
Muzzle the Shuttle and QCCDSim, and two ions in one junction in two
modes of MQT IonShuttler. \emph{Meaning} (Eq.~\ref{eq:meaning}) excludes three: two that
silently drop gates or merge registers, and a run that stops at its step
limit but is reported as finished. \emph{Time} (Eq.~\ref{eq:time}) excludes three step
counts that are off by one, because it fixes $t_{\mathrm{end}}$ as the
latest end of a step.

\textbf{These defects bias design comparisons in one direction.}
QCCDSim's stale chain record omits 69 of the 251 exchanges of a
64-qubit Fourier transform on six 16-slot traps, 9.9\% of its reported
finish time, which favours exchange-heavy layouts. Muzzle the Shuttle~\cite{saki2022muzzle} overfills traps
in all five programs of its own evaluation, which grants an extra trap
slot. QCCDSim's parser
drops two-qubit gates in 13 of 47 QASMBench programs, one reason we
compare with it only on its own benchmarks. In short, each audited
compiler returns target code that breaks its own rules, in ways that
favour some designs, so a design search cannot rely on it.

\subsection{RQ3: How much does the design matter?}
\label{sec:eval-selection}

\input{fig-eval-bars}

\paragraph{Designs}
A search pays only if designs differ for the same program. We fix
capacity and vary trap size and layout; each design is one hardware
description for the same compiler (Challenge~1). The designs are a
\Dg{line}{line-c8} of $C/c$ equal traps for every trap capacity
$c\in\{4,6,8,12,16,24,32,48\}$ that divides $C$ and, from 48 ions on, a
\Dg{ring}{ring-c12} and a two-row \Dg{ladder}{ladder-c12} of 12-ion
traps: 64 designs over the eight classes from 24 to 192 ions. Of 1,017
compilations under a 600-second limit, 918 pass every check.

\paragraph{Which design wins}
\textbf{The design decides most of the execution time: the best
design takes 66.4\% less modeled time on average than the worst}
(median 70.9\%, range 36.8\% to 92.1\%) and 22.4\% less than the program's
median design (Fig.~\ref{fig:eval-bars}). \textbf{The best design
depends on the program:} the
\Dg{ladder}{ladder-c12} wins 20 of the 75 programs, the \Dg{line}{line-c8} of 8-ion traps
17, and the \Dg{line}{line-c4} of 4-ion traps and the
\Dg{ring}{ring-c12} 12 each. Under a constant 100\,$\mu$s gate the mean reduction falls to
39.4\%, and \Dg{lines}{line-c24} of 24- or 48-ion traps become the
fastest design for 25 of 75 programs. The design therefore changes the
execution time by two thirds on average, and each program has its own
best design.

\subsection{\rev{RQ4: What does the broadcast model buy?}}
\label{sec:eval-broadcast}

Scalable
devices drive many traps with one broadcast channel
(Section~\ref{sec:broadcast}, Challenge~2). Here we test whether our
model of such a channel lets programs run as fast as on a device with a
channel for every transport. We take ten designs of RQ3 and compile each
program three ways: (i) on the design as it is, with a channel for every
transport; (ii) with one broadcast channel for all transports of the
device, 10 channels in place of 86, held by one transport at a time; and
(iii) on the same broadcast device with the activation schedule of
Section~\ref{sec:broadcast}. All three complete for 34
program--design pairs.

\textbf{With our broadcast model, one shared channel is as fast as
a channel for every transport:} the median ratio of (iii) to (i) is
1.00 and the worst is 1.08. \textbf{Holding the channel costs up to a
factor of 4.2:} under (ii) these programs finish up to 2.0 times later
(median 4\%), the 29-qubit Fourier transform on twelve 4-ion
traps~\dglyph{line-c4} takes 176.1\,ms against 87.5\,ms under (i) and
87.8\,ms under (iii), and a 63-qubit one on 24 such traps takes 4.2
times longer. Every broadcast channel of these runs has switches.
Theorem~\ref{thm:compile-correct} covers such channels, and it also
covers channels without switches (Section~\ref{sec:broadcast}).
The broadcast model thus lets a design with a ninth of the transport
channels run within 8\% of the fully connected one.

\subsection{RQ5: Can we learn guiding design principles?}
\label{sec:eval-families}

A search over devices needs guidance (Challenge~3). Every candidate is
a compiler input and every time is checked, so the 67 \emph{design
principles} of Section~\ref{sec:codesign} can be tested at scale: on
more than 3,000 checked compilations, for the designs of RQ3, those below
and held-out random devices. Those that hold brief the agent of RQ6.

\paragraph{Designs}
We instantiate 15 published
QCCD designs~\cite{khan2026cyclone,murali2020architecting,malinowski2023wise,schoenberger2025multizone,moses2023race}
at every capacity class by one parameter rule per family,
fixed before compilation, and add ten generic shapes whose gate traps
all have one size, so that only the layout differs
(Fig.~\ref{fig:workloads}). Each design takes
its fastest checked target code over two initial placements and two
routers: checked proposals (Section~\ref{sec:compiler}) and the transport
search for 60\,s. The 65 programs have up to 98 qubits.

\input{fig-eval-graphs}
\begin{table}[t]
\caption{Design rule from the comparison of 25 designs: the QCCD design to choose for a circuit interaction graph. Within 10\%: programs for which the chosen design is within 10\% of the fastest of all 25. Fastest: programs won outright.}
\label{tab:design-choice}
\centering\footnotesize\setlength{\tabcolsep}{1.9pt}
\begin{tabular}{@{}lrlrl@{}}
\toprule
Interaction graph & Progr. & QCCD design to choose & Within 10\% & Fastest\\
\midrule
\pglyph{chain}\,Chain & 22 & \dglyph{line-moveless}\,line, 7-ion traps & 13 & small-trap lines: 20\\
\pglyph{star}\,Star & 12 & \dglyph{bus-c12}\,bus, 12-ion traps & 11 & buses: 11\\
\pglyph{sparse}\,Sparse & 20 & \dglyph{ladder-c12}\,ladder, 12-ion traps & 18 & ladder: 15\\
\pglyph{dense}\,Dense & 8 & \dglyph{line-moveless}\,line, 7-ion traps & 3 & none above 3\\
\pglyph{local}\,Local & 3 & \dglyph{line-moveless}\,line, 7-ion traps & 1 & 12-ion line: 2\\
\midrule
\textbf{By this rule} & 65 & three designs & \textbf{46} & \\
Best single design & 65 & \dglyph{line-moveless}\,line, 7-ion traps & 27 & \\
\bottomrule
\end{tabular}
\end{table}

\paragraph{Which principles hold}
Of the 67 candidates, here we examine the 31 that state one quantity as
advice. \textbf{Nine hold, on the analysed designs and on held-out
designs} (Fig.~\ref{fig:principles}). Four are known from the device alone,
four are measured on the checked target code, and the ninth, the fit of
the device to the program, relates the two before any compilation.

\paragraph{The main principle: fit the device to the interaction graph}
No design is good everywhere: the best, a \Dg{line}{line-moveless} of
7-ion traps, is within 10\% of the fastest design for 27 of 65 programs.
The interaction graph decides
(Table~\ref{tab:design-choice}). \Pg{Chain}{chain} programs join neighbouring
qubits, and \Dg{lines}{line-c6} of small traps win 20 of 22.
\Pg{Star}{star} programs go to a junction \Dg{bus}{bus-c12} on 11 of 12,
because the central qubit reaches every trap without passing through another.
\Pg{Sparse}{sparse} programs go to the \Dg{ladder}{ladder-c12} of 12-ion
traps on 15 of 20: its second rail lets ions pass. \Pg{Dense}{dense}
programs have no winner. Choosing among these three designs by
interaction graph is within 10\% of the fastest of all 25 for 46 programs.

\paragraph{A metric for the fit}
To state the principle as a number, the \emph{fit distance} $d(G,H)$ of
a device $H$ to a program with interaction graph $G$ is the transport
time per two-qubit gate under the best placement of the qubits in the
traps, within their capacities; we take the best placement that
simulated annealing finds. Among generic designs of one
trap size, which differ in layout only, \textbf{a smaller fit
distance goes with a shorter time for 47 of 65 programs} and for 21 of 22 \Pg{chain}{chain}
programs (Fig.~\ref{fig:fit}). On the random devices of RQ6 it does for
all six programs.
\Pg{Star}{star} programs are the exception (4 of 12): a bus has longer
routes, but no ion passes through a trap. In summary, nine principles
hold on held-out designs, and the strongest is to fit the device to the
circuit's interaction graph.

\subsection{RQ6: Can a search find better devices?}
\label{sec:eval-search}

\paragraph{Workflow}
An AI agent runs the loop of Section~\ref{sec:codesign} (Challenge~3)
for one fixed program at a time. Each of six programs, chosen before the search with
one or two per interaction pattern, has a budget of 30 proposals (4 to 12
used) and a second, briefed round, each compiled with both routers for
60 to 150\,s. The second round adds the principles of
Section~\ref{sec:codesign} and 20 more proposals at once, so it
tests the briefed loop as a whole.

\paragraph{References}
Each of the six programs has 25 manual reference designs of its capacity
class, from RQ3 and RQ5; later searches keep the five best. The declared
trap order of a proposal decides where qubits
start, so an agent also chooses a placement; every manual design is
therefore evaluated under three placement rules, and the best counts.

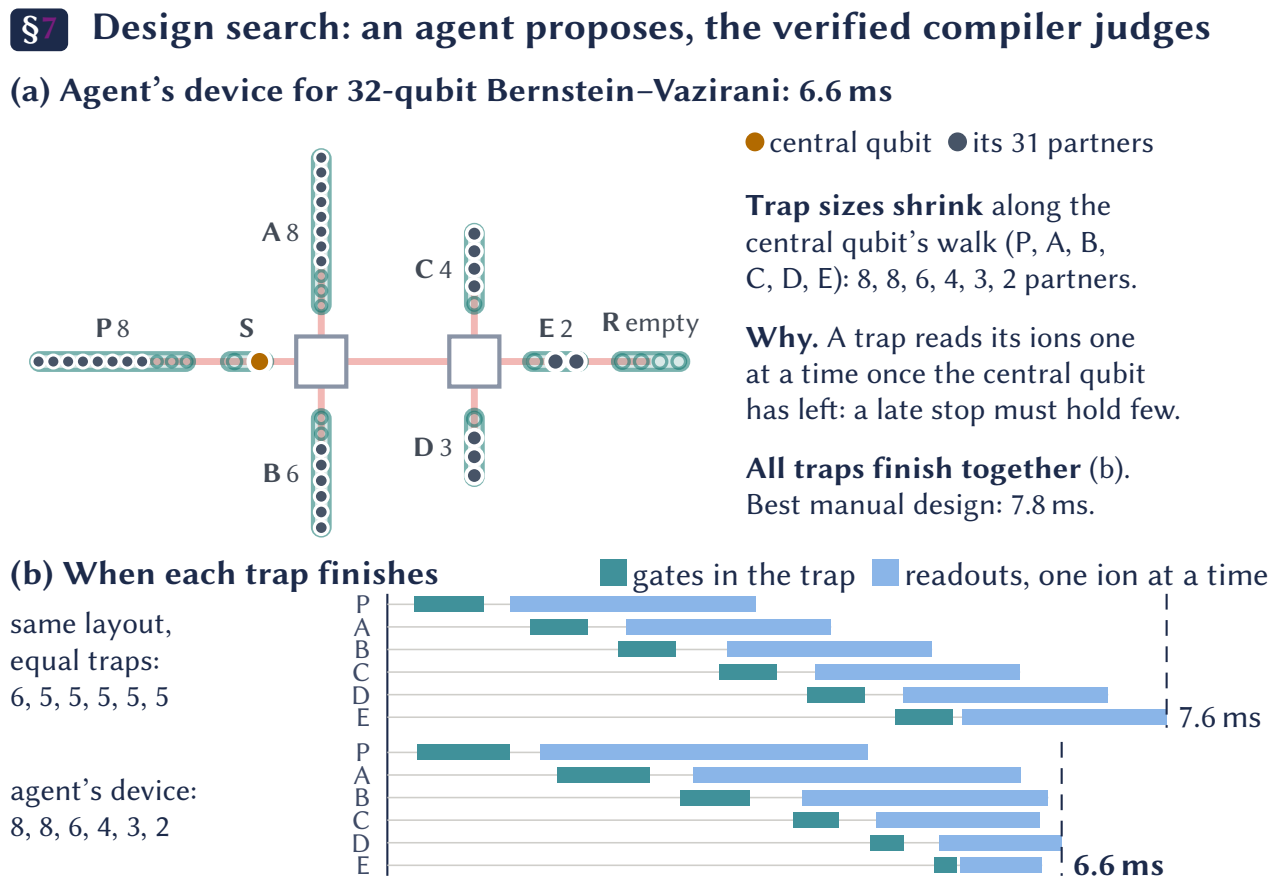
\begin{figure}[t]
\centering
\newcommand{\SpineName}[4]{\node[inner sep=0pt,anchor=#3,text=qhwInk!82,
  font=\sffamily\fontsize{5.4}{6}\selectfont] at ($(#1)+#2$) {\textbf{#1}#4};}
\resizebox{\columnwidth}{!}{%
\begin{tikzpicture}[x=1pt,y=1pt,font=\sffamily\fontsize{6.8}{8}\selectfont,
  lab/.style={inner sep=0pt,text=qink,font=\sffamily\fontsize{6.3}{7.4}\selectfont},
  small/.style={inner sep=0pt,text=qink,anchor=north west,align=left,
    font=\sffamily\fontsize{5.9}{6.9}\selectfont},
  tiny/.style={inner sep=0pt,text=qink!85,font=\sffamily\fontsize{5.2}{6}\selectfont},
  head/.style={lab,anchor=west,font=\sffamily\bfseries\fontsize{6.5}{7.5}\selectfont}]
\path[use as bounding box] (-2,6) rectangle (242,-161.2);
\QCCDFigHead{tag}{\rev{\S\ref{sec:eval}}}{Design search: an agent proposes, the verified compiler judges}

\node[head] at (0,-12) {(a) Agent's device for 32-qubit Bernstein--Vazirani: 6.6\,ms};
\begin{scope}\QCCDUnit=.44bp\qccdpadsfalse\qccdionlabelsfalse\QCCDLongTraps
\begin{scope}[shift={(2,-19)},x=\QCCDUnit,y=-\QCCDUnit]
  \QCCDDevice{agenttaper}{12}
  \QCCDIon[qorange]{centre}{S}{1}
  \foreach \i in {0,...,7}{\QCCDIon{p\i}{P}{\i}\QCCDIon{a\i}{A}{\i}}
  \foreach \i in {2,...,7}{\QCCDIon{b\i}{B}{\i}}
  \foreach \i in {0,...,3}{\QCCDIon{c\i}{C}{\i}}
  \foreach \i in {1,...,3}{\QCCDIon{d\i}{D}{\i}}
  \foreach \i in {1,2}{\QCCDIon{e\i}{E}{\i}}
  \foreach \s/\n in {A/8,B/6,C/4,D/3}{\SpineName{\s}{(-9,0)}{east}{\,\n}}
  \foreach \s/\n in {P/8,E/2}{\SpineName{\s}{(0,-10)}{south}{\,\n}}
  \SpineName{S}{(0,-10)}{south}{}\SpineName{R}{(0,-10)}{south}{\,empty}
\end{scope}\end{scope}
\node[small] at (140,-19)
  {\tikz[baseline=-2pt]{\fill[qorange] (0,0) circle[radius=1.8pt];}\,central qubit\enspace
   \tikz[baseline=-2pt]{\fill[qhwData] (0,0) circle[radius=1.8pt];}\,its 31 partners};
\node[small] at (140,-31)
  {\textbf{Trap sizes shrink} along the\\central qubit's walk (P, A, B,\\C, D, E): 8, 8, 6, 4, 3, 2 partners.};
\node[small] at (140,-56)
  {\textbf{Why.} A trap reads its ions one\\at a time once the central qubit\\has left: a late stop must hold few.};
\node[small] at (140,-81)
  {\textbf{All traps finish together} (b).\\Best manual design: 7.8\,ms.};

\node[head] at (0,-104) {(b) When each trap finishes};
\node[lab,anchor=east] at (240,-104)
  {\tikz[baseline=.1pt]{\path[fill=qteal!75] (0,0) rectangle (5pt,5pt);}\,gates in the trap\enspace
   \tikz[baseline=.1pt]{\path[fill=qblue!55] (0,0) rectangle (5pt,5pt);}\,readouts, one ion at a time};
\def\T#1{{72+19.5*(#1)}}
\newcommand{\TrapRow}[6]{%
  \node[tiny,anchor=east] at (69,#1) {#2};
  \draw[qrail,line width=.3pt] (\T{0},#1)--(\T{#6},#1);
  \path[fill=qteal!75] (\T{#3},#1-1.5) rectangle (\T{#4},#1+1.5);
  \path[fill=qblue!55] (\T{#5},#1-1.5) rectangle (\T{#6},#1+1.5);}
\draw[qink,line width=.4pt] (72,-107.4)--(72,-161.1);
\node[small,anchor=west] at (0,-120.2) {same layout,\\equal traps:\\6, 5, 5, 5, 5, 5};
\TrapRow{-109.4}{P}{0.260}{0.935}{1.200}{3.600}
\TrapRow{-113.7}{A}{1.400}{1.960}{2.335}{4.335}
\TrapRow{-118}{B}{2.260}{2.820}{3.320}{5.320}
\TrapRow{-122.3}{C}{3.245}{3.805}{4.180}{6.180}
\TrapRow{-126.6}{D}{4.105}{4.665}{5.040}{7.040}
\TrapRow{-130.9}{E}{4.965}{5.525}{5.610}{7.610}
\draw[qink,densely dashed,line width=.4pt] (\T{7.610},-107.4)--(\T{7.610},-132.9);
\node[lab,anchor=west] at ({74+19.5*7.61},-130.9) {7.6\,ms};
\node[small,anchor=west] at (0,-148.4) {agent's device:\\8, 8, 6, 4, 3, 2};
\TrapRow{-137.6}{P}{0.290}{1.195}{1.490}{4.690}
\TrapRow{-141.9}{A}{1.660}{2.565}{2.985}{6.185}
\TrapRow{-146.2}{B}{2.865}{3.540}{4.055}{6.455}
\TrapRow{-150.5}{C}{3.965}{4.410}{4.770}{6.370}
\TrapRow{-154.8}{D}{4.710}{5.040}{5.385}{6.585}
\TrapRow{-159.1}{E}{5.340}{5.555}{5.595}{6.395}
\draw[qink,densely dashed,line width=.4pt] (\T{6.585},-135.6)--(\T{6.585},-161.1);
\node[lab,anchor=west] at ({74+19.5*6.585},-159.1) {\textbf{6.6\,ms}};
\end{tikzpicture}}
\caption{(a) An agent's device: traps shrink along the central qubit's path. (b)
Their readouts then end together; with equal traps they end one after
another.}
\label{fig:design-search}
\Description{The agent's device in the hardware drawing style of the
running example. A row holds a trap P with eight ions, a small trap S with
the central qubit alone, two four-way junctions, a trap E with two ions and an
empty spare trap R. The first junction carries a trap A with eight ions
above and a trap B with six ions below; the second carries a trap C with
four ions above and a trap D with three ions below. Below, two timelines
with one row per gate trap, each showing the central qubit's gates in the trap and
then the readouts of its ions. With traps of equal size, holding 6, 5, 5,
5, 5 and 5 ions, the readouts end at 3.6, 4.3, 5.3, 6.2, 7.0 and 7.6 ms,
one after another. On the agent's device, holding 8, 8, 6, 4, 3 and 2
ions, they end at 4.7, 6.2, 6.5, 6.4, 6.6 and 6.4 ms.}
\end{figure}

\begin{figure}[t]
\centering
\resizebox{\columnwidth}{!}{%
\begin{tikzpicture}[x=1pt,y=1pt,font=\sffamily\fontsize{6.2}{7.4}\selectfont,
  lab/.style={inner sep=0pt,text=qink},
  ms/.style={inner sep=0pt,text=qink!75,font=\sffamily\fontsize{5.8}{7}\selectfont},
  pct/.style={inner sep=0pt,text=qred,font=\sffamily\bfseries\fontsize{6.4}{7.6}\selectfont}]
\path[use as bounding box] (0,12.0) rectangle (240,-95.0);
\node[lab,anchor=west] at (0,7.0) {\tikz[baseline=.1pt]{\path[fill=qgrey!45,draw=qgrey,line width=.4pt] (0pt,0pt) rectangle (7pt,5pt);}~best manual design\hspace{9pt}\tikz[baseline=.1pt]{\path[fill=qteal!70,draw=qteal,line width=.4pt] (0pt,0pt) rectangle (7pt,5pt);}~device proposed by the agent};
\node[ms,anchor=east] at (240,7.0) {finish time};
\node[lab,anchor=east] at (54.0,-3.50) {qec9xz-17\,\pglyph{sparse}};
\path[fill=qgrey!45,draw=qgrey,line width=.4pt] (58.0,-5.80) rectangle (162.0,-1.20);
\path[fill=qteal!70,draw=qteal,line width=.4pt] (58.0,-5.80) rectangle (117.99,-1.20);
\node[pct,anchor=east] at (190.0,-3.50) {$-$42.3\%};
\node[ms,anchor=east] at (240,-3.50) {8.56\,$\to$\,4.94\,ms};
\node[lab,anchor=east] at (54.0,-10.50) {dnn-51\,\pglyph{sparse}};
\path[fill=qgrey!45,draw=qgrey,line width=.4pt] (58.0,-12.80) rectangle (162.0,-8.20);
\path[fill=qteal!70,draw=qteal,line width=.4pt] (58.0,-12.80) rectangle (122.32,-8.20);
\node[pct,anchor=east] at (190.0,-10.50) {$-$38.2\%};
\node[ms,anchor=east] at (240,-10.50) {61.3\,$\to$\,37.9\,ms};
\node[lab,anchor=east] at (54.0,-20.00) {dnn-16\,\pglyph{chain}};
\path[fill=qgrey!45,draw=qgrey,line width=.4pt] (58.0,-22.30) rectangle (162.0,-17.70);
\path[fill=qteal!70,draw=qteal,line width=.4pt] (58.0,-22.30) rectangle (129.86,-17.70);
\node[pct,anchor=east] at (190.0,-20.00) {$-$30.9\%};
\node[ms,anchor=east] at (240,-20.00) {21.1\,$\to$\,14.6\,ms};
\node[lab,anchor=east] at (54.0,-27.00) {ghz-32\,\pglyph{chain}};
\path[fill=qgrey!45,draw=qgrey,line width=.4pt] (58.0,-29.30) rectangle (162.0,-24.70);
\path[fill=qteal!70,draw=qteal,line width=.4pt] (58.0,-29.30) rectangle (146.02,-24.70);
\node[pct,anchor=east] at (190.0,-27.00) {$-$15.4\%};
\node[ms,anchor=east] at (240,-27.00) {6.35\,$\to$\,5.37\,ms};
\node[lab,anchor=east] at (54.0,-36.50) {multiplier-75\,\pglyph{local}};
\path[fill=qgrey!45,draw=qgrey,line width=.4pt] (58.0,-38.80) rectangle (162.0,-34.20);
\path[fill=qteal!70,draw=qteal,line width=.4pt] (58.0,-38.80) rectangle (141.17,-34.20);
\node[pct,anchor=east] at (190.0,-36.50) {$-$20.0\%};
\node[ms,anchor=east] at (240,-36.50) {1167\,$\to$\,933\,ms};
\node[lab,anchor=east] at (54.0,-43.50) {sup-64\,\pglyph{local}};
\path[fill=qgrey!45,draw=qgrey,line width=.4pt] (58.0,-45.80) rectangle (162.0,-41.20);
\path[fill=qteal!70,draw=qteal,line width=.4pt] (58.0,-45.80) rectangle (151.58,-41.20);
\node[pct,anchor=east] at (190.0,-43.50) {$-$10.0\%};
\node[ms,anchor=east] at (240,-43.50) {75.4\,$\to$\,67.9\,ms};
\node[lab,anchor=east] at (54.0,-53.00) {bv-32\,\pglyph{star}};
\path[fill=qgrey!45,draw=qgrey,line width=.4pt] (58.0,-55.30) rectangle (162.0,-50.70);
\path[fill=qteal!70,draw=qteal,line width=.4pt] (58.0,-55.30) rectangle (145.41,-50.70);
\node[pct,anchor=east] at (190.0,-53.00) {$-$16.0\%};
\node[ms,anchor=east] at (240,-53.00) {7.84\,$\to$\,6.59\,ms};
\node[lab,anchor=east] at (54.0,-60.00) {bv-128\,\pglyph{star}};
\path[fill=qgrey!45,draw=qgrey,line width=.4pt] (58.0,-62.30) rectangle (162.0,-57.70);
\path[fill=qteal!70,draw=qteal,line width=.4pt] (58.0,-62.30) rectangle (146.25,-57.70);
\node[pct,anchor=east] at (190.0,-60.00) {$-$15.1\%};
\node[ms,anchor=east] at (240,-60.00) {25.7\,$\to$\,21.8\,ms};
\node[lab,anchor=east] at (54.0,-69.50) {qft-18\,\pglyph{dense}};
\path[fill=qgrey!45,draw=qgrey,line width=.4pt] (58.0,-71.80) rectangle (162.0,-67.20);
\path[fill=qteal!70,draw=qteal,line width=.4pt] (58.0,-71.80) rectangle (145.97,-67.20);
\node[pct,anchor=east] at (190.0,-69.50) {$-$15.4\%};
\node[ms,anchor=east] at (240,-69.50) {32.4\,$\to$\,27.4\,ms};
\node[lab,anchor=east] at (54.0,-76.50) {qft-64 nb\,\pglyph{dense}};
\path[fill=qgrey!45,draw=qgrey,line width=.4pt] (58.0,-78.80) rectangle (162.0,-74.20);
\path[fill=qteal!70,draw=qteal,line width=.4pt] (58.0,-78.80) rectangle (150.31,-74.20);
\node[pct,anchor=east] at (190.0,-76.50) {$-$11.2\%};
\node[ms,anchor=east] at (240,-76.50) {422\,$\to$\,375\,ms};
\draw[qink,line width=.5pt] (58.0,-81.50)--(166.2,-81.50);
\draw[qink,line width=.5pt] (58.0,-81.50)--(58.0,-83.50);
\node[ms,anchor=north] at (58.0,-84.50) {0};
\draw[qink,line width=.5pt] (110.0,-81.50)--(110.0,-83.50);
\node[ms,anchor=north] at (110.0,-84.50) {50\%};
\draw[qink,line width=.5pt] (162.0,-81.50)--(162.0,-83.50);
\node[ms,anchor=north] at (162.0,-84.50) {100\%};
\node[ms,anchor=north west] at (171.0,-84.50) {of the best manual time};
\end{tikzpicture}}
\caption{The agent's device against the best manual design under three placement rules:
per interaction graph (purple), the two largest reductions among the
67 programs with five or more manual designs compared under each rule.}
\label{fig:agent-bars}
\Description{Horizontal bars for 10 programs, two per circuit interaction graph. Each grey bar is
the best manual design at 100 percent and the teal bar over it is the device proposed by the
agent. qec9xz-17: minus 42.3 percent. dnn-51: minus 38.2 percent. dnn-16: minus 30.9 percent. ghz-32: minus 15.4 percent. multiplier-75: minus 20.0 percent. sup-64: minus 10.0 percent. bv-32: minus 16.0 percent. bv-128: minus 15.1 percent. qft-18: minus 15.4 percent. qft-64 nb: minus 11.2 percent.}
\end{figure}
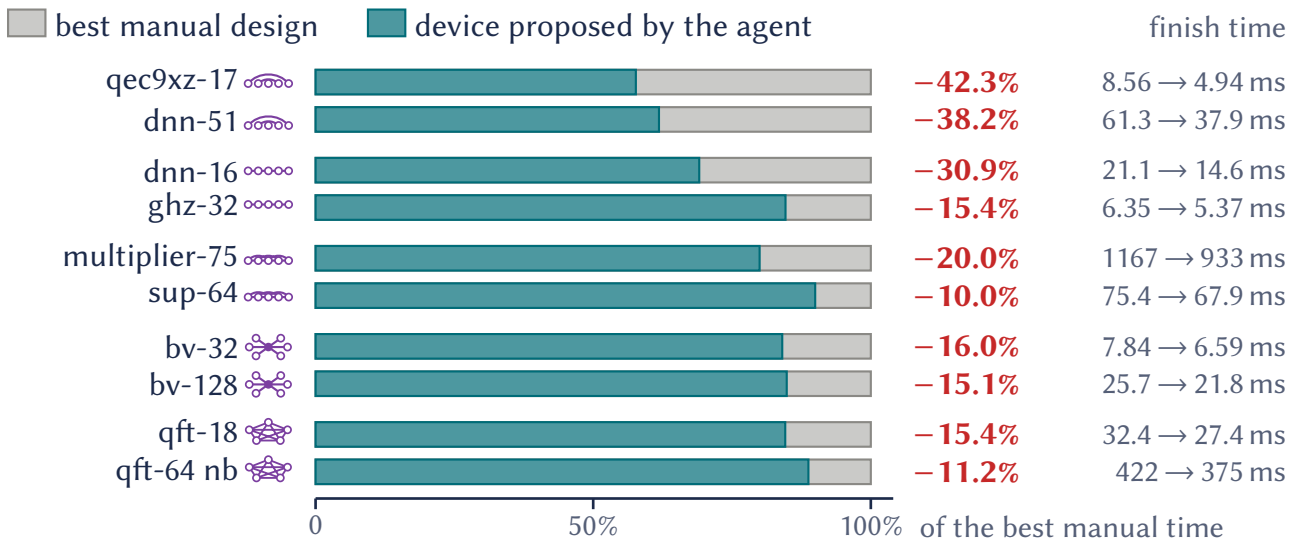

\paragraph{Results}
For each of the six programs the agent proposes a device that is faster
than every manual design, by 1.4\% to
33.1\%, most for the \Pg{sparse}{sparse} programs, whose qubit groups a
proposed device can place side by side. We then ran the
loop on the other 69 programs: seven against their three best manual
designs, then 62 with 15 proposals each and the lessons of the earlier
searches as a brief. \textbf{Over all 75 programs the agent's
design is faster than the fastest design of
Fig.~\ref{fig:eval-bars} on 71,} equal on 3 and slower on 1, Quantum Volume, by 7\% (yellow bars). The
median reduction
is 15\%, and 26\% on the 24 sparse programs. Against further manual
designs and all three placements the agent still wins 67
(largest reductions: Fig.~\ref{fig:agent-bars}). Blind
search falls short: all 30 random designs of each of the six programs
are slower than the program's best manual design. The agents' devices use more
parts; the model prices only time.

\paragraph{What the agents found}
Fig.~\ref{fig:design-search}a shows the device for Bernstein--Vazirani
on 32 qubits, whose 31 gates share one central qubit. It visits
six gate traps, and their sizes shrink along its walk: they hold 8, 8, 6,
4, 3 and 2 partners. The
checked schedule shows why. A trap reads its ions one at a time, and only
after the central qubit has left, so a trap visited late has little time and should
hold few ions. With equal traps on the same layout the readouts end one
after another, at 7.61\,ms; on the agent's device five of the six traps
finish within 0.4\,ms (Fig.~\ref{fig:design-search}b), at 6.59\,ms, 16\%
below the best manual design. The sizes carry most of the gain: the best
manual layout with traps sized this way takes 6.77\,ms. Agents found the
same taper for all eight generated
GHZ and Bernstein--Vazirani programs, with gains of 7\% to 16\%. The declared order is part of a design: the
28-qubit adder's traps declared left to right take 107.9\,ms, and
23.7\,ms in the agent's order. With checked feedback, an AI agent
therefore finds a faster device than the fastest design of
Fig.~\ref{fig:eval-bars} for 71 of 75 programs.

\subsection{RQ7: Does verified compilation scale?}
\label{sec:eval-scalability}

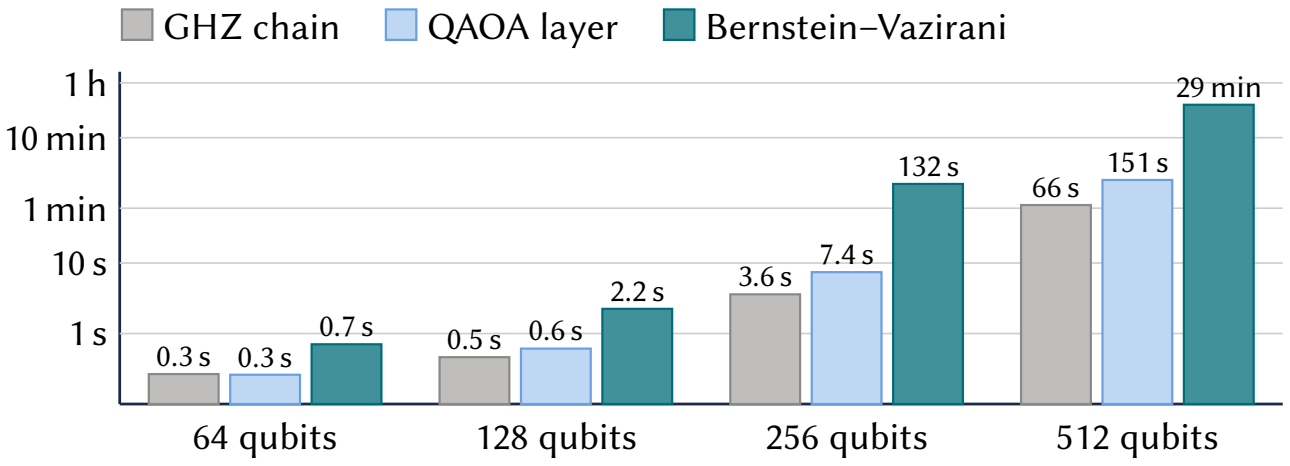
\begin{figure}[t]
\centering
\resizebox{\columnwidth}{!}{%
\begin{tikzpicture}[x=1pt,y=1pt,font=\sffamily\fontsize{7}{8.4}\selectfont,
  lab/.style={inner sep=0pt,text=black},
  val/.style={inner sep=0pt,text=black,font=\sffamily\fontsize{5.8}{6.8}\selectfont}]
\draw[qink,line width=.5pt] (30.0,0)--(240.0,0);
\draw[qink,line width=.5pt] (30.0,0)--(30.0,60.0);
\draw[qrail,line width=.3pt] (30.0,12.73)--(240.0,12.73);
\node[lab,anchor=east] at (28.0,12.73) {1\,s};
\draw[qrail,line width=.3pt] (30.0,25.46)--(240.0,25.46);
\node[lab,anchor=east] at (28.0,25.46) {10\,s};
\draw[qrail,line width=.3pt] (30.0,35.36)--(240.0,35.36);
\node[lab,anchor=east] at (28.0,35.36) {1\,min};
\draw[qrail,line width=.3pt] (30.0,48.09)--(240.0,48.09);
\node[lab,anchor=east] at (28.0,48.09) {10\,min};
\draw[qrail,line width=.3pt] (30.0,58.00)--(240.0,58.00);
\node[lab,anchor=east] at (28.0,58.00) {1\,h};
\path[fill=qgrey!55,draw=qgrey,line width=.4pt] (35.25,0) rectangle (47.85,5.41);
\node[val,anchor=south] at (41.55,6.61) {0.3\,s};
\path[fill=qblue!30,draw=qblue!70,line width=.4pt] (49.95,0) rectangle (62.55,5.29);
\node[val,anchor=south] at (56.25,6.49) {0.3\,s};
\path[fill=qteal!75,draw=qteal,line width=.4pt] (64.65,0) rectangle (77.25,10.79);
\node[val,anchor=south] at (70.95,11.99) {0.7\,s};
\node[lab,anchor=north] at (56.25,-3) {64 qubits};
\path[fill=qgrey!55,draw=qgrey,line width=.4pt] (87.75,0) rectangle (100.35,8.46);
\node[val,anchor=south] at (94.05,9.66) {0.5\,s};
\path[fill=qblue!30,draw=qblue!70,line width=.4pt] (102.45,0) rectangle (115.05,10.01);
\node[val,anchor=south] at (108.75,11.21) {0.6\,s};
\path[fill=qteal!75,draw=qteal,line width=.4pt] (117.15,0) rectangle (129.75,17.19);
\node[val,anchor=south] at (123.45,18.39) {2.2\,s};
\node[lab,anchor=north] at (108.75,-3) {128 qubits};
\path[fill=qgrey!55,draw=qgrey,line width=.4pt] (140.25,0) rectangle (152.85,19.81);
\node[val,anchor=south] at (146.55,21.01) {3.6\,s};
\path[fill=qblue!30,draw=qblue!70,line width=.4pt] (154.95,0) rectangle (167.55,23.81);
\node[val,anchor=south] at (161.25,25.01) {7.4\,s};
\path[fill=qteal!75,draw=qteal,line width=.4pt] (169.65,0) rectangle (182.25,39.74);
\node[val,anchor=south] at (175.95,40.94) {132\,s};
\node[lab,anchor=north] at (161.25,-3) {256 qubits};
\path[fill=qgrey!55,draw=qgrey,line width=.4pt] (192.75,0) rectangle (205.35,35.93);
\node[val,anchor=south] at (199.05,37.13) {66\,s};
\path[fill=qblue!30,draw=qblue!70,line width=.4pt] (207.45,0) rectangle (220.05,40.46);
\node[val,anchor=south] at (213.75,41.66) {151\,s};
\path[fill=qteal!75,draw=qteal,line width=.4pt] (222.15,0) rectangle (234.75,54.04);
\node[val,anchor=south] at (228.45,55.24) {29\,min};
\node[lab,anchor=north] at (213.75,-3) {512 qubits};
\node[lab,anchor=south west] at (30.0,64.00) {\tikz[baseline=.1pt]{\path[fill=qgrey!55,draw=qgrey,line width=.4pt] (0pt,0pt) rectangle (5.5pt,5.5pt);}~GHZ chain\hspace{8pt}\tikz[baseline=.1pt]{\path[fill=qblue!30,draw=qblue!70,line width=.4pt] (0pt,0pt) rectangle (5.5pt,5.5pt);}~QAOA layer\hspace{8pt}\tikz[baseline=.1pt]{\path[fill=qteal!75,draw=qteal,line width=.4pt] (0pt,0pt) rectangle (5.5pt,5.5pt);}~Bernstein--Vazirani};
\end{tikzpicture}}
\caption{Compile time of the verified compiler on lines of 16-ion traps (log scale). All three programs compile at 512 qubits.}
\label{fig:eval-scale}
\Description{Bar chart of compile time on a logarithmic axis for a GHZ chain, one QAOA layer and Bernstein-Vazirani at 64, 128, 256 and 512 qubits. Times grow from under a second to 67 seconds, 151 seconds and 29 minutes at 512 qubits.}
\end{figure}

To test widths beyond the design studies, we compile a GHZ chain,
Bernstein--Vazirani and one QAOA layer on a ring for \Dg{lines}{line-c16} of capacity-16
traps under a 30-minute limit (Fig.~\ref{fig:eval-scale}). \textbf{All three compile at 512 qubits:} the GHZ chain in 67\,s, the QAOA layer in 151\,s and
Bernstein--Vazirani in 29 minutes. A transport-legality check that was quadratic in the number of ions
dominated compile time; we replaced it by a linear computation proved
equal to the original definition, which keeps every theorem
statement as it is and takes the GHZ chain from 57 minutes to 67\,s.
The verified compiler thus handles 512-qubit programs within half an
hour on one workstation.
\section{Related work}
\label{sec:related}

\paragraph{QCCD architecture and compilation}
Prior QCCD work studies trap capacity and
connectivity~\cite{murali2020architecting}, moving-chain and laser-head
co-design~\cite{wu2021tilt}, linear and junction
shuttling~\cite{durandau2023automated,schoenberger2024shuttling}, transport
fidelity~\cite{saki2022muzzle}, multi-zone
compilation~\cite{schoenberger2025multizone} and joint placement,
shuttling and SWAP scheduling~\cite{zhu2025ssync}. These works fix the
device or the compiler; our
compiler takes device and rules as inputs. \rev{WISE proposes shared
transport control with switches~\cite{malinowski2023wise}; we give such
wiring a model, a scheduler and a proof
(Section~\ref{sec:broadcast}).}

\paragraph{Declarative retargeting and verified compilation}
Amaro generates mapping compilers from device state machines, including
trapped-ion models~\cite{molavi2026generating}\rev{; our description adds
durations, shared controls and operating rules, and our theorem is
machine-checked for every description}. VOQC proves circuit
transformations and SWAP mapping in Coq~\cite{hietala2021voqc}, CertiQ
\rev{and Giallar verify} Qiskit passes with symbolic reasoning and
SMT~\cite{shi2019certiq,tao2022giallar}, Weaver validates neutral-atom motion and circuit
equivalence~\cite{kirmemis2025weaver}, and Vonk proposes languages with
physical and timing contracts~\cite{vonk2025formal}.

\paragraph{Quantum architecture specification languages}
The two architecture specification languages closest to ours are
that of MQT IonShuttler~\cite{schoenberger2025multizone}, with which we
compare in Section~\ref{sec:eval-compiler}, and
OpenQL~\cite{khammassi2022openql}, which has many strengths but does not
support all the QCCD features that we consider in this paper.
\section{Conclusion}
\label{sec:concl}

We presented a verified compiler for QCCD co-design. Hardware descriptions
and operating rules are compiler inputs, and one Lean theorem guarantees
that all accepted target code preserves the source program and obeys the
device and its rules, relative to the stated hardware model. A design search can therefore trust
its comparisons of compiled designs: at equal capacity the best
design takes 66.4\% less modeled time than the worst on average, and an
AI agent beats that best design on 71 of the 75 programs. Our framework paves the way to large-scale
software--hardware co-design of QCCD devices.

\paragraph{Artifact availability}
The compiler, its Lean proofs, examples, evaluation tools and recorded
results are at
\url{https://anonymous.4open.science/r/ISCA27ArtifactQCCD-129C/}; one
command regenerates the evaluation figures from them.

\begin{acks}
Generative AI tools assisted with manuscript revision, code, proofs and
analysis scripts. Lean's kernel checks the mechanized theorems.
\end{acks}

\clearpage
\setcounter{bodycontentpages}{\numexpr\value{page}-1\relax}
\typeout{QCCD-MAIN-PAGES: \arabic{bodycontentpages}}
\bibliography{refs}
\ifqccdarxiv\qccdreferencetrue\qccdformaltrue\qccdtutorialtrue\qccdprooftrue\fi
\newif\ifqccdanyappendix
\qccdanyappendixfalse
\ifqccdreference\qccdanyappendixtrue\fi
\ifqccdformal\qccdanyappendixtrue\fi
\ifqccdtutorial\qccdanyappendixtrue\fi
\ifqccdanyappendix
  \clearpage
  \onecolumn
  \setcounter{appendixstartpage}{\value{page}}
  \appendix
  \ifqccdreference\begin{center}
{\large\bfseries Appendix: examples, languages, proofs and reference material}\\[4pt]
{\normalsize From standard OpenQASM to legal QCCD target code}
\end{center}

The appendices go from simple to complex, and each assumes only the main
text and the appendices before it.
\ifqccdtutorial
A tutorial first follows three small programs through every stage of the
compiler and explains the proof on them.
\fi
A complete example then compiles one larger quantum/classical program. The
formal appendices define every language, judgment and compilation rule that
the correctness proof uses, replay the proof on one program, and prove
Theorem~\ref{thm:compile-correct} in full. The last appendices are reference
material: the complete concrete grammars, hardware and intermediate-language
detail, the catalogue of constraints with ion examples, and the full design
comparison of the evaluation.

\noindent\textbf{Table of contents}\par\smallskip
\begingroup\small
\renewcommand{\arraystretch}{1.16}
\noindent\begin{minipage}[t]{.48\linewidth}
\begin{tabular}[t]{@{}p{.86\linewidth}r@{}}
\ifqccdtutorial
\hyperref[app:tutorial]{\textbf{Tutorial: three compilations, stage by stage}} & \pageref{app:tutorial}\\
\quad\hyperref[app:tut-one]{Example 1: one CX through every stage} & \pageref{app:tut-one}\\
\quad\hyperref[app:tut-two]{Example 2: feedback and a changed rule} & \pageref{app:tut-two}\\
\quad\hyperref[app:tut-three]{Example 3: broadcast, parallel time, chain order} & \pageref{app:tut-three}\\
\quad\hyperref[app:tut-proof]{How the verification works} & \pageref{app:tut-proof}\\[3pt]
\fi
\hyperref[app:native]{\textbf{Complete quantum/classical example}} & \pageref{app:native}\\
\quad\hyperref[app:native-hardware]{Two-zone hardware and calibration} & \pageref{app:native-hardware}\\
\quad\hyperref[app:native-source]{Input OpenQASM program} & \pageref{app:native-source}\\
\quad\hyperref[app:native-target]{Complete target code} & \pageref{app:native-target}\\
\quad\hyperref[app:native-rules]{Rules and timing of the example} & \pageref{app:native-rules}\\
\quad\hyperref[app:native-bridge]{Why the saved condition matters} & \pageref{app:native-bridge}\\[3pt]
\hyperref[app:formal]{\textbf{Languages and their meaning}} & \pageref{app:formal}\\
\quad\hyperref[app:f-syntax]{Syntax} & \pageref{app:f-syntax}\\
\quad\hyperref[app:f-static]{Static semantics: well-formedness} & \pageref{app:f-static}\\
\quad\hyperref[app:f-semantics]{Dynamic semantics: what a program computes} & \pageref{app:f-semantics}\\
\quad\hyperref[app:f-physical]{Physical states, frames and core legality} & \pageref{app:f-physical}\\
\quad\hyperref[app:f-assertions]{Rules over frames, and legal runs} & \pageref{app:f-assertions}\\[3pt]
\hyperref[app:f-compile]{\textbf{Compilation rules}} & \pageref{app:f-compile}\\[3pt]
\hyperref[app:hoare-example]{\textbf{The proof on one program}} & \pageref{app:hoare-example}\\
\quad\hyperref[app:replay-stages]{Every stage of one compilation} & \pageref{app:replay-stages}\\
\quad\hyperref[app:hoare]{Hoare logic for target code} & \pageref{app:hoare}\\
\quad\hyperref[app:hoare-derivation]{The same proof as a Hoare derivation} & \pageref{app:hoare-derivation}\\[3pt]
\hyperref[app:proof]{\textbf{Complete proof of the main theorem}} & \pageref{app:proof}\\
\quad\hyperref[app:pf-statement]{The theorem, clause by clause} & \pageref{app:pf-statement}\\
\quad\hyperref[app:pf-objects]{Objects of the reported timeline} & \pageref{app:pf-objects}\\
\quad\hyperref[app:pf-sequential]{The sequential contract} & \pageref{app:pf-sequential}\\
\quad\hyperref[app:pf-chain]{Chain order} & \pageref{app:pf-chain}\\
\quad\hyperref[app:pf-schedule]{The schedules} & \pageref{app:pf-schedule}\\
\quad\hyperref[app:pf-commute]{Events on disjoint resources commute} & \pageref{app:pf-commute}\\
\quad\hyperref[app:pf-order]{The timeline in order of time} & \pageref{app:pf-order}\\
\quad\hyperref[app:pf-assembly]{Assembling the proof} & \pageref{app:pf-assembly}\\[3pt]
\end{tabular}
\end{minipage}\hfill
\begin{minipage}[t]{.48\linewidth}
\begin{tabular}[t]{@{}p{.86\linewidth}r@{}}
\hyperref[app:syntax]{\textbf{Complete language syntax}} & \pageref{app:syntax}\\
\quad\hyperref[app:syntax-names]{Names, expressions and sequences} & \pageref{app:syntax-names}\\
\quad\hyperref[app:syntax-hardware]{Physical hardware programs} & \pageref{app:syntax-hardware}\\
\quad\hyperref[app:syntax-circuit]{Standard OpenQASM source} & \pageref{app:syntax-circuit}\\
\quad\hyperref[app:syntax-ir]{Intermediate programs at all nine stages} & \pageref{app:syntax-ir}\\
\quad\hyperref[app:syntax-execution]{Target code} & \pageref{app:syntax-execution}\\
\quad\hyperref[app:syntax-rules]{Concrete and quantified rule grammars} & \pageref{app:syntax-rules}\\
\quad\hyperref[app:syntax-mapping]{Representation, readers and printers} & \pageref{app:syntax-mapping}\\[3pt]
\hyperref[sec:native-languages]{\textbf{Hardware and intermediate-language detail}} & \pageref{sec:native-languages}\\[3pt]
\hyperref[app:rules]{\textbf{Complete constraint catalogue}} & \pageref{app:rules}\\
\quad\hyperref[app:rule-hardware]{Hardware declarations: H1--H12} & \pageref{app:rule-hardware}\\
\quad\hyperref[app:rule-circuit]{Source programs: C1--C8} & \pageref{app:rule-circuit}\\
\quad\hyperref[app:rule-placement]{Placement and initialization: P1--P4} & \pageref{app:rule-placement}\\
\quad\hyperref[app:rule-transport]{Shuttling and shared controls: T1--T7} & \pageref{app:rule-transport}\\
\quad\hyperref[app:rule-frames]{Physical effects and timing: E1--E10} & \pageref{app:rule-frames}\\[3pt]
\hyperref[app:rule-language]{\textbf{Expressing and extending policies}} & \pageref{app:rule-language}\\
\quad\hyperref[app:rule-observations]{Every observation on a concrete frame} & \pageref{app:rule-observations}\\
\quad\hyperref[app:rule-policies]{Six concrete policies} & \pageref{app:rule-policies}\\
\quad\hyperref[app:rule-extension]{Expressibility and extensions} & \pageref{app:rule-extension}\\[3pt]
\hyperref[app:design-comparison]{\textbf{Full design comparison}} & \pageref{app:design-comparison}\\
\end{tabular}
\end{minipage}
\endgroup\medskip
\fi
  \ifqccdtutorial

\newcommand{\TutCBarsParallel}{\TutBar{0}{0}{2}{gate}\TutBar{0}{2}{2}{move}\TutBar{1}{2}{2}{move}\TutBar{4}{2}{2}{wave0}\TutBar{1}{4}{4}{gate}\TutBar{1}{8}{2}{gate}\TutBar{1}{10}{2}{gate}\TutBar{2}{0}{2}{gate}\TutBar{2}{2}{2}{move}\TutBar{3}{2}{2}{move}\TutBar{4}{2}{2}{wave0}\TutBar{3}{4}{4}{gate}\TutBar{3}{8}{2}{gate}\TutBar{3}{10}{2}{gate}\TutBar{1}{12}{2}{gate}\TutBar{1}{14}{2}{move}\TutBar{2}{14}{2}{move}\TutBar{4}{14}{2}{wave0}\TutBar{3}{16}{2}{move}\TutBar{2}{16}{2}{move}\TutBar{4}{16}{2}{wave1}\TutBar{2}{18}{4}{gate}\TutBar{2}{22}{2}{gate}\TutBar{2}{24}{2}{gate}\TutBar{2}{26}{3}{read}\TutBar{2}{29}{3}{read}}
\newcommand{\TutCBarsChain}{\TutBar{0}{0}{2}{gate}\TutBar{0}{2}{2}{move}\TutBar{1}{2}{2}{move}\TutBar{4}{2}{2}{wave0}\TutBar{1}{4}{4}{gate}\TutBar{1}{8}{2}{gate}\TutBar{1}{10}{2}{gate}\TutBar{2}{0}{2}{gate}\TutBar{2}{2}{2}{move}\TutBar{3}{2}{2}{move}\TutBar{4}{2}{2}{wave0}\TutBar{3}{4}{4}{gate}\TutBar{3}{8}{2}{gate}\TutBar{3}{10}{2}{gate}\TutBar{1}{12}{2}{gate}\TutBar{1}{14}{12}{swap}\TutBar{1}{26}{2}{move}\TutBar{2}{26}{2}{move}\TutBar{4}{26}{2}{wave0}\TutBar{3}{12}{12}{swap}\TutBar{3}{28}{2}{move}\TutBar{2}{28}{2}{move}\TutBar{4}{28}{2}{wave1}\TutBar{2}{30}{4}{gate}\TutBar{2}{34}{2}{gate}\TutBar{2}{36}{2}{gate}\TutBar{2}{38}{3}{read}\TutBar{2}{41}{3}{read}}

\tikzset{
  twire/.style={qink,line width=.5pt},
  tcwire/.style={qink,double,double distance=.9pt,line width=.35pt},
  tgate/.style={draw=qink,fill=white,line width=.5pt,inner sep=1.6pt,minimum size=9pt,
    font=\sffamily\fontsize{5.8}{6.8}\selectfont},
  tlab/.style={inner sep=0pt,text=qink,font=\sffamily\fontsize{6.3}{7.5}\selectfont},
  thead/.style={inner sep=0pt,anchor=west,text=qink,font=\sffamily\bfseries\fontsize{7}{8.4}\selectfont},
  tnote/.style={inner sep=0pt,anchor=north west,text=qink!70,align=left,
    font=\sffamily\itshape\fontsize{5.8}{6.8}\selectfont},
  tcard/.style={draw=qrail,fill=qpaper,rounded corners=2pt,line width=.4pt},
  tproved/.style={draw=qgreen,fill=qmint,rounded corners=1.5pt,line width=.4pt},
  tbgate/.style={fill=qteal!70,draw=qteal,line width=.3pt},
  tbmove/.style={fill=qblue!45,draw=qblue,line width=.3pt},
  tbwave0/.style={fill=qblue!75,draw=qblue,line width=.3pt},
  tbwave1/.style={fill=qorange!70,draw=qorange,line width=.3pt},
  tbread/.style={fill=qgrey!45,draw=qgrey,line width=.3pt},
  tbswap/.style={fill=qred!55,draw=qred,line width=.3pt},
  tbwait/.style={draw=none},
}
\newcommand{\TutCtrl}[2]{\fill[qink] (#1,#2) circle[radius=1.5pt];}
\newcommand{\TutTarg}[2]{\draw[qink,line width=.5pt,fill=white] (#1,#2) circle[radius=3.3pt];
  \draw[qink,line width=.5pt] (#1-3.3,#2)--(#1+3.3,#2) (#1,#2-3.3)--(#1,#2+3.3);}
\newcommand{\TutCX}[3]{\draw[twire] (#1,#2)--(#1,#3);\TutCtrl{#1}{#2}\TutTarg{#1}{#3}}
\newcommand{\TutMeter}[2]{\node[tgate,minimum width=11pt,minimum height=9pt] at (#1,#2) {};
  \draw[qink,line width=.4pt] (#1-3.2,#2-1.9) arc[start angle=180,end angle=0,radius=3.2pt];
  \draw[qink,line width=.4pt] (#1,#2-1.9)--(#1+2.6,#2+2.9);}
\newcommand{\TutListing}[5][qccd-execution]{%
  \node[inner sep=0pt,anchor=north west] at (#2,#3) {\begin{minipage}{#4}
    \lstinputlisting[style=#1,basicstyle=\ttfamily\fontsize{5.6}{7}\selectfont\color{black},
      xleftmargin=0pt,aboveskip=0pt,belowskip=0pt,breaklines=false]{#5}\end{minipage}};}
\newcommand{\TutStage}[6]{%
  \node[thead] at (#1,#2) {#4};
  \node[tnote,text width=#3] (why) at (#1,#2-6) {#5};
  \node[inner sep=0pt,anchor=north west] at ($(why.south west)+(0,-3)$) {\begin{minipage}{#3}
    \lstinputlisting[style=qccd-execution,basicstyle=\ttfamily\fontsize{5.6}{7}\selectfont\color{black},
      xleftmargin=0pt,aboveskip=0pt,belowskip=0pt,breaklines=false]{#6}\end{minipage}};}
\newcommand{\TutBar}[4]{\path[tb#4] (#2,-#1) rectangle ++(#3,-.78);}
\newcommand{\TutFigure}[3]{\begin{center}#1\par\vspace{1pt}\captionof{figure}{#2}\label{#3}\end{center}}
\newcommand{\TutIR}[5]{%
  \node[draw=qink,fill=white,rounded corners=2pt,line width=.5pt,minimum width=46pt,
    minimum height=19pt,inner sep=0pt,anchor=north west,align=center] at (#1*58,-16)
    {{\bfseries #2}\\[-1pt]{\fontsize{5.6}{6.4}\selectfont #3}};
  \node[tnote,anchor=north,align=center,text width=54pt] at (#1*58+23,-38) {#4};
  \node[fill=qsoft,rounded corners=1pt,inner sep=1.3pt,anchor=north,text=black,
    font=\ttfamily\fontsize{4.9}{5.6}\selectfont] at (#1*58+23,-56) {#5};}
\newcommand{\TutTee}[4]{%
  \begin{scope}\QCCDUnit=.3bp\qccdpadsfalse
  \begin{scope}[shift={(#1,#2)},x=\QCCDUnit,y=-\QCCDUnit]
    \QCCDDevice{parity}{20}#4
    \node[tlab,anchor=north,font=\sffamily\bfseries\fontsize{5.6}{6}\selectfont] at (55,84) {S};
    \node[tlab,anchor=north,font=\sffamily\bfseries\fontsize{5.6}{6}\selectfont] at (245,84) {G};
    \node[tlab,anchor=west,font=\sffamily\bfseries\fontsize{5.6}{6}\selectfont] at (176,150) {P};
    \node[tlab,anchor=south west,font=\sffamily\bfseries\fontsize{5.6}{6}\selectfont] at (172,22) {J};
    \node[tlab,anchor=north west,font=\sffamily\fontsize{5.8}{6.6}\selectfont] at (10,190) {#3};
  \end{scope}\end{scope}}
\newcommand{\TutTeeArrow}[2]{\draw[qblue,line width=1.4\QCCDUnit,-{Stealth[length=3.4bp,width=2.8bp]}] (#1,8)--(#2,8);}

\section{Tutorial: three compilations, stage by stage}
\label{app:tutorial}
This first appendix needs only
Sections~\ref{sec:background} to~\ref{sec:checker} of the main text. It names
the programs that the compiler passes through (Fig.~\ref{fig:tut-pipeline}),
follows three small inputs from OpenQASM to target code, and explains the
proof on the same examples.\ifqccdreference\ Appendix~\ref{app:native} then
compiles a larger program, Appendices~\ref{app:formal} and~\ref{app:f-compile}
define every language and compilation rule, Appendix~\ref{app:hoare-example}
replays the proof on one program, and Appendix~\ref{app:proof} proves
Theorem~\ref{thm:compile-correct} in full.\fi\ Every listing below is the
compiler's output for the printed inputs; none is edited by hand. The
listings name rules \texttt{init}, \texttt{cap}, \texttt{rel}, \ldots; the
text writes them $\rho_{\rm init}$, $\rho_{\rm cap}$, $\rho_{\rm rel}$, as the
main text does, and prints calibrations as \texttt{c0}, \texttt{c1}, \ldots.

\TutFigure{\resizebox{\textwidth}{!}{%
\begin{tikzpicture}[x=1pt,y=1pt,font=\sffamily\fontsize{6.5}{7.8}\selectfont]
\path[use as bounding box] (-2,6) rectangle (518,-166);
\node[thead] at (0,0) {The programs of one compilation, what each stage adds, and the line it produces in Example~1};
\TutIR{0}{Source}{OpenQASM 2}{the program\\as written}{CX q[0],q[1];}
\TutIR{1}{Effects}{$P_E$}{gates expanded,\\conditions saved}{CX q0,q1;}
\TutIR{2}{Pulses}{$P_P$}{U, CX become\\R, MS pulses}{R; MS; R; R; R;}
\TutIR{3}{Bound}{$P_B$}{candidate\\controls}{R q0 \{c3\};}
\TutIR{4}{Placed}{$\langle\pos_0,P_P\rangle$}{initial\\sites}{init q0@S,q1@G;}
\TutIR{5}{Routed}{}{transports,\\chosen controls}{transport q0:S->G;}
\TutIR{6}{Packed}{}{shared\\pulses}{broadcast R q0,q1;}
\TutIR{7}{Timed}{target code $K$}{durations\\and waits}{idle 1; (t=2..4)}
\TutIR{8}{Timeline}{}{overlap,\\exchanges}{starts 0,2,3,...}
\foreach \i/\s in {0/1--3,1/4,2/5,3/6,4/7,5/8,6/9,7/+}{
  \draw[flow ink] (\i*58+46.5,-25.5)--(\i*58+57.5,-25.5);
  \node[inner sep=0pt,text=qteal,font=\sffamily\bfseries\fontsize{5.6}{6}\selectfont] at (\i*58+52,-20.5) {\s};}
\draw[qgrey,line width=.5pt] (0,-72)--(0,-75)--(162,-75)--(162,-72);
\node[tlab,anchor=north] at (81,-77) {source semantics $\llbracket\cdot\rrbracket$: input ensemble $\mapsto$ output ensemble};
\draw[qgrey,line width=.5pt] (232,-72)--(232,-75)--(452,-75)--(452,-72);
\node[tlab,anchor=north] at (342,-77) {hardware semantics $\llbracket\cdot\rrbracket_H$: state $\sigma\mapsto\sigma'$, one event at a time};
\path[tproved] (0,-90) rectangle (510,-100);
\node[tlab,anchor=west] at (4,-95) {\textbf{Meaning.} The observable result of $K$, also in the order of the timeline, is the meaning of the source, for every input state and angle value.};
\path[tproved] (116,-103) rectangle (510,-113);
\node[tlab,anchor=west] at (120,-108) {\textbf{Meaning.} Erasing transports, waits and sharing from $K$ gives back $P_P$ exactly.};
\path[tproved] (232,-116) rectangle (510,-126);
\node[tlab,anchor=west] at (236,-121) {\textbf{Legality.} Every event frame passes the core checks and the rules $R$, at its start time.};
\path[tproved] (406,-129) rectangle (510,-139);
\node[tlab,anchor=east] at (402,-134) {\textbf{Legality and Time} on the timeline: steps that share a resource never overlap; only the trap-end ion departs.};
\node[tlab,anchor=west] at (410,-134) {finish = latest end};
\foreach \x/\t/\b in {
  0/{Configuration}/{$(\pos,\,t,\,\sigma)$: placement, time, and an ensemble $\sigma$ with one branch $(C,b,\rho)$ per outcome},
  132/{Observation}/{$\mathrm{obs}(C,b,\rho)=(C,\rho)$ hides only the compiler's saved flag $b$},
  264/{Event frame}/{$F=(H,\pos^-,\pos^+,u,t,\delta)$: the state as one instruction sees it},
  396/{Rule}/{a formula over $N^-,N^+,N^{\rm tr},M,W,P,n_k,t,\delta$ of one frame}}{
  \path[tcard] (\x,-144) rectangle (\x+120,-165);
  \node[tlab,anchor=north west,font=\sffamily\bfseries\fontsize{6.2}{7.2}\selectfont] at (\x+3,-146.5) {\t};
  \node[tlab,anchor=north west,text width=114pt,font=\sffamily\fontsize{5.7}{6.5}\selectfont] at (\x+3,-153) {\b};}
\end{tikzpicture}}}
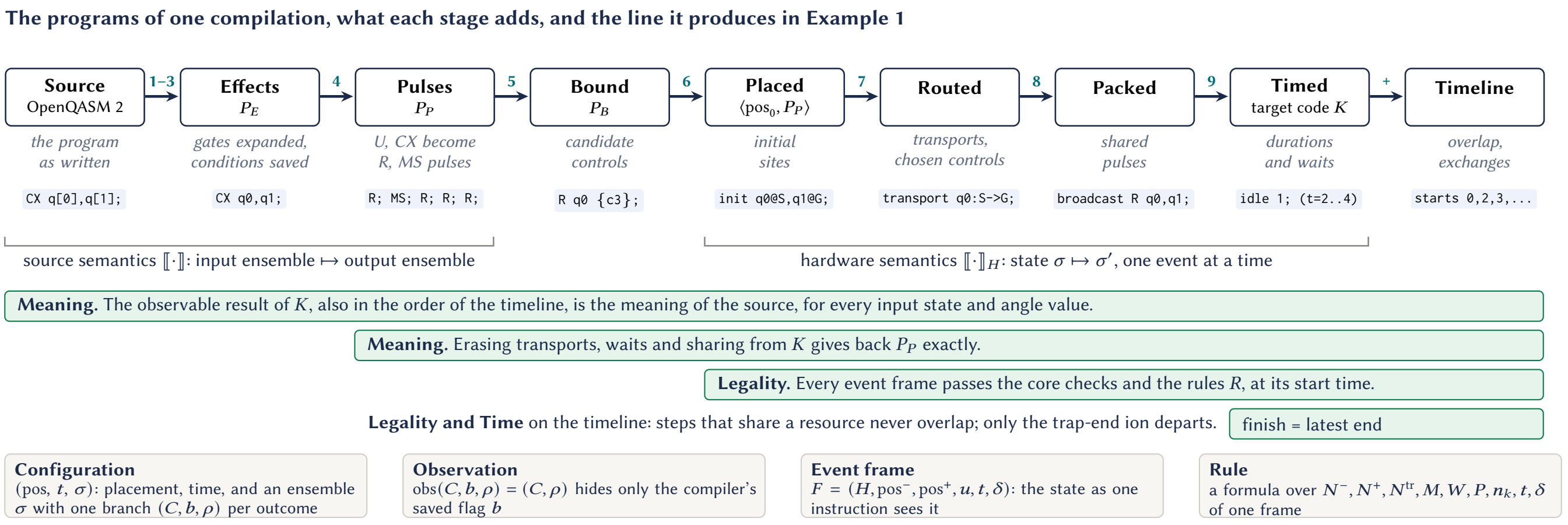
{The abstractions of the compiler. Nine stages turn the source into timed
target code $K$; teal numbers are stage numbers, and ``+'' is the reported
timeline, with the parallel schedule and chain order. Green bars show which
programs each clause of Theorem~\ref{thm:compile-correct} (Meaning,
Legality, Time) relates.}{fig:tut-pipeline}

\TutFigure{\resizebox{\textwidth}{!}{%
\begin{tikzpicture}[x=1pt,y=1pt,font=\sffamily\fontsize{6.5}{7.8}\selectfont]
\path[use as bounding box] (-2,6) rectangle (518,-112);
\node[thead] at (0,0) {(a) Source program and its circuit};
\TutListing[qccd-circuit]{0}{-8}{80pt}{tut-1-source.txt}
\foreach \y/\l in {-16/$q_0$,-30/$q_1$}{\draw[twire] (104,\y)--(156,\y);\node[tlab,anchor=east] at (102,\y) {\l};}
\draw[tcwire] (104,-44)--(156,-44);\node[tlab,anchor=east] at (102,-44) {$c_0$};
\TutCX{120}{-16}{-30}
\draw[tcwire] (140,-30)--(140,-44);\TutMeter{140}{-30}
\node[thead] at (0,-60) {(c) Rule program};
\TutListing[qccd-rules]{0}{-68}{160pt}{tut-1-rules.txt}
\node[tnote,text width=160pt] at (0,-93) {$N^-$: ions before the event; $N^{\rm tr}$: ions that use the
  site during it; $n_{\rm optical}$: optical actions; $t,\delta$: start and duration.};
\node[thead] at (176,0) {(b) Device and initial placement};
\begin{scope}
\QCCDUnit=.6bp
\begin{scope}[shift={(184,-34)},x=\QCCDUnit,y=-\QCCDUnit]
  \QCCDDevice{primer}{11}
  \QCCDIon{q0}{S}{0}\QCCDIon{q1}{G}{1}
  \draw[qhwBeam,line width=2.2\QCCDUnit] (200,-36)--(200,-8);
  \draw[qhwBeam,line width=1.2\QCCDUnit] (184,-8)--(216,-8);
  \draw[qblue,line width=1.3\QCCDUnit,-{Stealth[length=4bp,width=3.2bp]}] (25,66)--(110,66)--(188,66);
  \fill[qblue] (110,66) circle[radius=2.6\QCCDUnit];
  \node[tlab,anchor=north] at (25,72) {\textbf{S}\ \ cap.\ 1};
  \node[tlab,anchor=north] at (110,72) {\textbf{J}\ \ cap.\ 1};
  \node[tlab,anchor=north] at (200,72) {\textbf{G}\ \ cap.\ 2};
  \node[tlab,anchor=east,text=qhwBeam] at (192,-26) {optical channel 7};
\end{scope}
\end{scope}
\node[tlab,anchor=north west,text=qblue] at (184,-92) {path $p_0$: S$\to$J$\to$G, two ticks; J is reserved in transit};
\node[thead] at (346,0) {(d) Hardware program};
\TutListing[qccd-hardware]{346}{-8}{172pt}{tut-1-hardware.txt}
\end{tikzpicture}}}
{Example~1, the inputs: an OpenQASM program, a rule program and a hardware
program for the device S, J, G of the main text. Its calibrations $c_0$ to
$c_2$ serve the U gate there and are unused here.}{fig:tut-one-inputs}

\subsection{Example 1: one CX, stage by stage}
\label{app:tut-one}

\paragraph{Inputs (Fig.~\ref{fig:tut-one-inputs})}
The hardware program declares what the device can do: three sites with
capacities, one path, and one calibration per pulse that the laser at G can
play. It does not say where ions start. The rule program adds three
operating rules. The source is ordinary OpenQASM.

\paragraph{Stages (Fig.~\ref{fig:tut-one-stages})}
Read the seven listings in order; each is the complete program after that
stage, and the note above it says what the stage did and why that is
correct. Stages 1 to 4 fix \emph{what} is computed and never look at the
device. Stages 5 to 9 decide \emph{how}: which control plays each pulse,
where ions start, how they travel, which pulses share a laser shot, and
when everything happens.

\paragraph{Frames (Table~\ref{tab:tut-frames})}
The target code is a list of events. Each event has a frame, and every rule
is evaluated on every frame. The transport $F_1$ shows the transit count:
$q_0$ counts at S, where it leaves, at J, which it crosses, and at G, where
it joins $q_1$. The rule $\rho_{\rm rel}$ is why three waits appear.

\begin{center}\begin{minipage}{\linewidth}\centering
\captionof{table}{The frames of Example~1. $N^{\rm tr}$ is given for S, J, G.}
\label{tab:tut-frames}
\footnotesize\setlength{\tabcolsep}{3pt}
\begin{tabular}{@{}llrcrl@{}}
\toprule
$F$ & instruction $u$ & $t,\delta$ & $N^{\rm tr}$ & $n_{\rm opt}$ & rule that decides\\
\midrule
$F_0$ & init $q_0$@S, $q_1$@G & 0, 0 & 1, 0, 1 & 0 & $\rho_{\rm init}$: $N^-(\mathrm S)=1$\\
$F_1$ & transport $q_0$: S$\to$G & 0, 2 & 1, 1, 2 & 0 & $\rho_{\rm cap}$: $1\le1$, $1\le1$, $2\le2$\\
$F_{2..4}$ & idle 1 & 2..4, 1 & 0, 0, 2 & 0 & $\rho_{\rm rel}$: no optical action\\
$F_5$ & R on $q_0$ & 5, 2 & 0, 0, 2 & 1 & $\rho_{\rm rel}$: $5\ge5$\\
$F_6$ & MS on $q_0,q_1$ & 7, 4 & 0, 0, 2 & 1 & \\
$F_7$ & broadcast R on $q_0,q_1$ & 11, 2 & 0, 0, 2 & 2 & \\
$F_8$ & R on $q_0$ & 13, 2 & 0, 0, 2 & 1 & \\
$F_9$ & readout $q_1\to c_0$ & 15, 3 & 0, 0, 2 & 1 & finish at tick 18\\
\bottomrule
\end{tabular}
\end{minipage}\end{center}

\TutFigure{\resizebox{\textwidth}{!}{%
\begin{tikzpicture}[x=1pt,y=1pt,font=\sffamily\fontsize{6.5}{7.8}\selectfont]
\path[use as bounding box] (-2,6) rectangle (518,-198);
\TutStage{0}{0}{94pt}{Stages 1--3: Effects}{Includes are resolved, angles
  checked, gate definitions and registers expanded. \emph{Correct because}
  each operation is kept, in order.}{tut-1-effects.txt}
\TutStage{106}{0}{92pt}{Stage 4: Pulses}{The CX becomes five native
  pulses. \emph{Correct because} the five act as CX on every state, up to a
  global phase.}{tut-1-pulses.txt}
\TutStage{210}{0}{104pt}{Stage 5: Bound}{Each pulse gets the declared
  controls with exactly its waveform. \emph{Correct because} a missing
  control fails here, not later.}{tut-1-bound.txt}
\TutStage{326}{0}{92pt}{Stage 6: Placed}{An initial site per ion.
  \emph{Correct because} it fits every capacity and the rule
  $\rho_{\rm init}$: one ion at S.}{tut-1-placed.txt}
\node[tcard,anchor=north west,text width=84pt,inner sep=3pt,align=left,
  font=\sffamily\fontsize{5.8}{6.9}\selectfont] at (428,2)
  {\textbf{How to read.} Each box is the whole program after that stage.
   Nothing before stage 6 mentions the device, so every candidate design
   starts from the same pulses.};
\TutStage{0}{-94}{150pt}{Stage 7: Routed}{A transport brings $q_0$ to G,
  and each pulse picks one candidate there. \emph{Correct because} the
  transport is legal at its placement, and a transport changes where ions
  are, never the quantum state.}{tut-1-routed.txt}
\TutStage{166}{-94}{150pt}{Stage 8: Packed}{Two equal R pulses on distinct
  ions become one declared broadcast $b_0$. \emph{Correct because} the
  broadcast acts as the two pulses, and expanding it gives them
  back.}{tut-1-packed.txt}
\TutStage{332}{-94}{184pt}{Stage 9: Timed}{Declared durations, and waits
  where a rule needs them: $\rho_{\rm rel}$ forbids optical actions before tick
  5. \emph{Correct because} a wait changes only the time, and every frame is
  checked at its own $t$ and $\delta$.}{tut-1-timed.txt}
\end{tikzpicture}}}
{Example~1, the program after every stage. The last listing is the
target code: nine events, 18 ticks.}{fig:tut-one-stages}

\TutFigure{\resizebox{\textwidth}{!}{%
\begin{tikzpicture}[x=1pt,y=1pt,font=\sffamily\fontsize{6.5}{7.8}\selectfont]
\path[use as bounding box] (-2,6) rectangle (518,-250);
\node[thead] at (0,0) {(a) Source program and its circuit};
\TutListing[qccd-circuit]{0}{-8}{80pt}{tut-2-source.txt}
\foreach \y/\l in {-74/$q_0$,-88/$q_1$,-102/$q_2$}{\draw[twire] (16,\y)--(150,\y);\node[tlab,anchor=east] at (14,\y) {\l};}
\draw[tcwire] (16,-118)--(150,-118);\node[tlab,anchor=east] at (14,-118) {$c_0$};
\TutCX{36}{-74}{-102}\TutCX{58}{-88}{-102}
\draw[tcwire] (84,-102)--(84,-118);\TutMeter{84}{-102}
\draw[tcwire] (118,-118)--(118,-107);\fill[qink] (118,-118) circle[radius=1.5pt];
\node[tgate] at (118,-102) {$\lvert0\rangle$};
\node[tlab,anchor=north,font=\sffamily\fontsize{5.6}{6.4}\selectfont] at (118,-121) {if $c=1$};
\node[thead] at (0,-138) {(d) Hardware program};
\TutListing[qccd-hardware]{0}{-146}{172pt}{tut-2-hardware.txt}
\node[thead] at (180,0) {(b) The three transports and the readout};
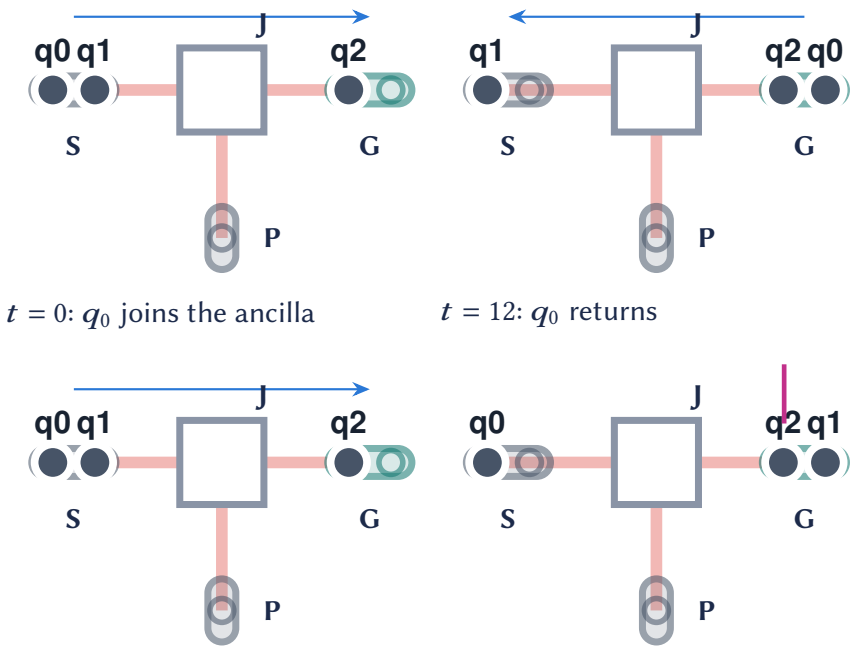
\TutTee{184}{-12}{$t=0$: $q_0$ joins the ancilla}{\QCCDIon{q0}{S}{0}\QCCDIon{q1}{S}{1}\QCCDIon{q2}{G}{0}\TutTeeArrow{55}{245}}
\TutTee{268}{-12}{$t=12$: $q_0$ returns}{\QCCDIon{q1}{S}{0}\QCCDIon{q2}{G}{0}\QCCDIon{q0}{G}{1}\TutTeeArrow{245}{55}}
\TutTee{184}{-84}{$t=14$: $q_1$ joins}{\QCCDIon{q0}{S}{0}\QCCDIon{q1}{S}{1}\QCCDIon{q2}{G}{0}\TutTeeArrow{55}{245}}
\TutTee{268}{-84}{$t=26$: readout of $q_2$}{\QCCDIon{q0}{S}{0}\QCCDIon{q2}{G}{0}\QCCDIon{q1}{G}{1}
  \draw[qhwBeam,line width=3\QCCDUnit] (232,-8)--(232,30);}
\node[thead] at (180,-162) {(c) Two rule programs for the same device};
\TutListing[qccd-rules]{180}{-170}{168pt}{tut-2-rules-one-optical.txt}
\node[tnote,text width=166pt] at (180,-202) {Rule program A is the first three
  assertions: 14 events, 34 ticks. Program B adds $\rho_{\rm one}$, one optical
  action per event: each broadcast is two pulses again, 16 events, 38 ticks.};
\node[thead] at (362,0) {(e) Target code under rule program A};
\TutListing{362}{-8}{156pt}{tut-2-timed.txt}
\node[tnote,text width=152pt] at (362,-136) {\textbf{Feedback.} The latch reads
  the register once and saves the comparison; \texttt{when saved} guards
  the pump. If the flag is false the event is an idle of the same four
  ticks, so both outcomes have one placement and one finish time.};
\node[tnote,text width=152pt] at (362,-178) {\textbf{Routing.} G holds two ions,
  so $q_0$ must leave before $q_1$ enters. Each transport reserves J:
  $N^{\rm tr}(\mathrm J)=1$. Rule $\rho_{\rm traffic}$ allows one ion per path
  per event.};
\end{tikzpicture}}}
{Example~2: a parity check with feedback on a junction device. Changing
only the rule program changes the target code from 34 to 38 ticks; the proof
is the same.}{fig:tut-two}

\subsection{Example 2: feedback, and a rule that changes the target code}
\label{app:tut-two}

\paragraph{The program}
Two data ions each interact with one ancilla. The ancilla is read, and it
is reset only if it read 1 (Fig.~\ref{fig:tut-two}a). The device has
storage S for two ions, a gate zone G for two, a parking site P and a
junction J between them.

\paragraph{Feedback}
The source tests the register in \texttt{if(c==1)}. The compiler turns the
test into a \emph{latch} that samples it once, and marks the guarded
operations \texttt{when saved}. The physical time and placement of a
guarded event do not depend on the outcome, so one target code covers both
branches, and the rules are checked for both.

\paragraph{Rules are inputs}
The same hardware under rule program B gives different target code
(Fig.~\ref{fig:tut-two}c): the rule $\rho_{\rm one}$ rejects the broadcast,
whose frame has $n_{\rm optical}=2$, so stage 8 keeps the two pulses and
the target code takes four more ticks. Nothing in the compiler or its proof
names this rule.

\TutFigure{\resizebox{\textwidth}{!}{%
\begin{tikzpicture}[x=1pt,y=1pt,font=\sffamily\fontsize{6.5}{7.8}\selectfont]
\path[use as bounding box] (-2,6) rectangle (518,-322);
\node[thead] at (0,0) {(a) Source program and its circuit};
\TutListing[qccd-circuit]{0}{-8}{80pt}{tut-3-source.txt}
\foreach \y/\l in {-82/$q_0$,-96/$q_1$,-110/$q_2$,-124/$q_3$}{\draw[twire] (16,\y)--(146,\y);\node[tlab,anchor=east] at (14,\y) {\l};}
\draw[tcwire] (16,-140)--(146,-140);\node[tlab,anchor=east] at (14,-140) {$c$};
\TutCX{34}{-82}{-96}\TutCX{34}{-110}{-124}\TutCX{62}{-82}{-124}
\draw[tcwire] (92,-82)--(92,-140);\TutMeter{92}{-82}
\draw[tcwire] (118,-124)--(118,-140);\TutMeter{118}{-124}
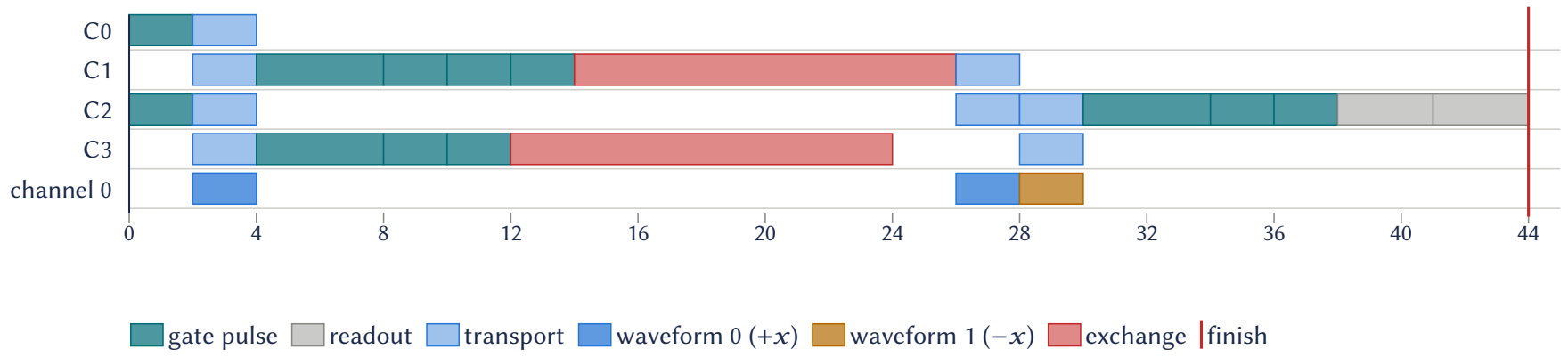
\node[thead] at (0,-160) {(b) Device, initial placement, first two transports};
\begin{scope}
\QCCDUnit=.5bp\qccdpadsfalse
\begin{scope}[shift={(-8,-168)},x=\QCCDUnit,y=-\QCCDUnit]
  \QCCDDevice{line4}{12}
  \QCCDIon[qblue]{q0}{C0}{0}\QCCDIon{q1}{C1}{0}\QCCDIon[qblue]{q2}{C2}{0}\QCCDIon{q3}{C3}{0}
  \foreach \s in {C0,C1,C2,C3}{\node[tlab,anchor=north,font=\sffamily\bfseries\fontsize{5.6}{6}\selectfont] at ($(\s)+(0,22)$) {\s};}
  \draw[qblue,line width=1.4\QCCDUnit,-{Stealth[length=3.4bp,width=2.8bp]}] (52,28)--(124,28);
  \draw[qblue,line width=1.4\QCCDUnit,-{Stealth[length=3.4bp,width=2.8bp]}] (196,28)--(268,28);
\end{scope}
\end{scope}
\node[tnote,text width=158pt] at (0,-224) {Both moves make the motion $+x$ on waveform 0
  of the shared channel: one play of the waveform moves both ions.};
\node[thead] at (0,-250) {(c) Rule program};
\TutListing[qccd-rules]{0}{-258}{180pt}{tut-3-rules.txt}
\node[thead] at (176,0) {(d) Parallel schedule: 32 ticks instead of 44 in sequence};
\foreach \y/\macro/\len in {-14/\TutCBarsParallel/32,-92/\TutCBarsChain/44}{
  \begin{scope}[shift={(206,\y)},x=6.9pt,y=8.6pt]
    \foreach \r/\l in {0/C0,1/C1,2/C2,3/C3,4/channel 0}{
      \node[tlab,anchor=east,font=\sffamily\fontsize{5.6}{6.4}\selectfont] at (-.5,-\r-.4) {\l};
      \draw[qrail,line width=.3pt] (0,-\r-.89)--(45,-\r-.89);}
    \macro
    \draw[qink,line width=.4pt] (0,0)--(0,-5);
    \foreach \t in {0,4,...,44}{\draw[qgrey,line width=.3pt] (\t,-5)--(\t,-5.25);
      \node[tlab,anchor=north,font=\sffamily\fontsize{5.2}{6}\selectfont] at (\t,-5.3) {\t};}
    \draw[qred,line width=.6pt] (\len,.2)--(\len,-5.1);
  \end{scope}}
\node[thead] at (176,-80) {(e) With chain order: two exchanges are inserted, 44 ticks};
\node[tlab,anchor=north west,font=\sffamily\fontsize{5.7}{6.6}\selectfont] at (206,-158)
  {\tikz[baseline=-2pt]{\path[tbgate] (0,-2.4) rectangle (7,2.4);}\,gate pulse\enspace
   \tikz[baseline=-2pt]{\path[tbread] (0,-2.4) rectangle (7,2.4);}\,readout\enspace
   \tikz[baseline=-2pt]{\path[tbmove] (0,-2.4) rectangle (7,2.4);}\,transport\enspace
   \tikz[baseline=-2pt]{\path[tbwave0] (0,-2.4) rectangle (7,2.4);}\,waveform 0 ($+x$)\enspace
   \tikz[baseline=-2pt]{\path[tbwave1] (0,-2.4) rectangle (7,2.4);}\,waveform 1 ($-x$)\enspace
   \tikz[baseline=-2pt]{\path[tbswap] (0,-2.4) rectangle (7,2.4);}\,exchange\enspace
   \tikz[baseline=-2pt]{\draw[qred,line width=.6pt] (0,-3)--(0,3);}\,finish};
\node[thead] at (176,-178) {(f) Why the exchanges: the chain machine};
\node[tnote,text width=160pt,font=\sffamily\fontsize{5.9}{7.1}\selectfont,text=qink] at (176,-186)
  {\textbf{C1.} $q_0$ entered from the left: chain $[q_0\;q_1]$.\\
   It must leave to the right, where $q_1$ sits.\\
   exchange $q_0,q_1$ gives $[q_1\;q_0]$; now $q_0$ departs.\\[3pt]
   \textbf{C3.} Chain $[q_2\;q_3]$; $q_3$ must leave to the left.\\
   exchange $q_3,q_2$ gives $[q_3\;q_2]$; now $q_3$ departs.\\[3pt]
   An exchange costs three two-qubit gates (12 ticks) and holds its trap
   and both ions.\\ Final chains: C1 $[q_1]$, C2 $[q_0\;q_3]$, C3 $[q_2]$.};
\node[thead] at (338,-178) {(g) Hardware program};
\TutListing[qccd-hardware]{338}{-186}{180pt}{tut-3-hardware.txt}
\end{tikzpicture}}}
{Example~3: a line of four traps with one broadcast transport channel. The
first two transports share one play of a waveform; gates in different traps
overlap; with chain order the compiler inserts two exchanges before the
ions that are not at a trap end can leave.}{fig:tut-three}

\subsection{Example 3: broadcast transport, parallel time, chain order}
\label{app:tut-three}

\paragraph{The device}
Four traps in a line share one transport channel with two waveforms, one
per direction; each trap has its own laser channel
(Fig.~\ref{fig:tut-three}b,g). The rules place one ion per trap and allow
one waveform per channel per event.

\paragraph{Parallel time}
Stage 9 produces one sequence of 18 events, 44 ticks. The parallel schedule
keeps this sequence and gives each event the earliest start at which its
ions, sites and channels are free (Fig.~\ref{fig:tut-three}d). The pulses
in C1 and C3 overlap. The two $+x$ transports start together at tick 2 on
one play of waveform 0. The two later transports go in opposite directions,
need two waveforms, and follow one another.

\paragraph{Chain order}
Stages 1 to 9 track which trap holds each ion. A trap also orders its ions,
and only the ion at the end that faces a path can leave through it. The
last pass replays the target code on a chain machine and inserts an exchange
wherever the departing ion is not at that end
(Fig.~\ref{fig:tut-three}e,f). The two exchanges overlap in time in
different traps; the result takes 44 ticks.

\paragraph{What is proved about them}
Both passes are covered by Theorem~\ref{thm:compile-correct}. The schedule
keeps the instruction sequence and only assigns start times. Two events
that share an ion, a site, a direct channel, a classical bit or the
controller never overlap, and two events on one broadcast channel are apart
in time or play one waveform together (Legality). Running the events in
order of their start times gives the same result as in program order,
because events that share no resource act in either order with the same
result (Meaning). The rules are evaluated again at the new start times, with
the sequential timing kept if one fails. The chain machine accepts a
departure only of the ion at the trap end; the compiler reports a timeline
only if its replay is accepted, and an accepted replay keeps in every chain
exactly the ions that the target code places in that trap.

\TutFigure{\resizebox{\textwidth}{!}{%
\begin{tikzpicture}[x=1pt,y=1pt,font=\sffamily\fontsize{6.5}{7.8}\selectfont,
  prog/.style={draw=qink,fill=white,rounded corners=2pt,line width=.5pt,minimum width=40pt,minimum height=13pt,inner sep=1pt},
  mean/.style={draw=qgreen,fill=qmint,rounded corners=2pt,line width=.5pt,minimum width=56pt,minimum height=13pt,inner sep=1pt},
  lem/.style={tnote,anchor=north,align=center,text width=82pt}]
\path[use as bounding box] (-2,6) rectangle (518,-236);
\node[thead] at (0,0) {(a) One equation per pass; their composition is the Meaning clause};
\foreach \x/\p/\m in {30/{$P$}/{$\llbracket P\rrbracket(\rho)$},
    126/{$P_E$}/{$\llbracket P_E\rrbracket(\rho)$},
    222/{$P_P$}/{$\llbracket P_P\rrbracket(\rho)$},
    318/{routed $A$}/{$\mathrm{obs}\,\llbracket A\rrbracket_H(\sigma_0)$},
    402/{packed $A'$}/{$\mathrm{obs}\,\llbracket A'\rrbracket_H(\sigma_0)$},
    486/{target code $K$}/{$\mathrm{obs}\,\llbracket K\rrbracket_H(\sigma_0)$}}{
  \node[prog] (p\x) at (\x,-18) {\p};
  \node[mean] (m\x) at (\x,-48) {\m};
  \draw[qgrey,densely dashed,line width=.4pt] (p\x)--(m\x);}
\foreach \a/\b/\s in {30/126/1--3,126/222/4,222/318/5--7,318/402/8,402/486/9}{
  \draw[flow ink] (p\a)--(p\b) node[midway,above=0pt,inner sep=1pt,text=qteal,font=\sffamily\bfseries\fontsize{5.6}{6}\selectfont] {\s};
  \node[inner sep=0pt,text=qgreen,font=\sffamily\bfseries\fontsize{9}{9}\selectfont] at ($(m\a)!.5!(m\b)$) {$=$};}
\node[lem] at (78,-58) {expansion keeps every operation, its order, and the moment a condition is read};
\node[lem] at (174,-58) {U and CX act as their R/MS pulses on every state, up to a global phase};
\node[lem] at (270,-58) {a transport changes the placement only; a pulse acts on the ions at its site};
\node[lem,text width=70pt] at (360,-58) {a broadcast acts as the pulses it replaces};
\node[lem,text width=70pt] at (444,-58) {a wait changes the time only};
\node[tlab,anchor=west] at (0,-90) {$\sigma_0$ is the input ensemble and $\mathrm{obs}$ hides the saved flag. The Legality clause is an
  invariant of the same induction: every prefix of $A$, $A'$ and $K$ ends in a legal placement, and every event of $K$ is legal on its frame.};

\node[thead] at (0,-108) {(b) Stage 4 on Example 1: five pulses for one CX};
\foreach \y in {-130,-150}{\draw[twire] (6,\y)--(30,\y);\draw[twire] (50,\y)--(206,\y);}
\TutCX{18}{-130}{-150}
\node[tlab,font=\sffamily\bfseries\fontsize{8}{8}\selectfont] at (40,-140) {$\simeq$};
\node[tgate] at (72,-130) {R$(\pi/2,\pi/2)$};
\node[tgate,minimum height=29pt] at (112,-140) {MS$(\pi/2)$};
\node[tgate] at (150,-130) {R$(-\pi/2,0)$};
\node[tgate] at (150,-150) {R$(-\pi/2,0)$};
\node[tgate] at (190,-130) {R$(-\pi/2,\pi/2)$};
\node[tnote,text width=204pt] at (0,-162) {Proved for the operators, so it holds
  when the two ions are entangled with others. $\simeq$: equal up to a global phase, which no
  measurement observes.};

\node[thead] at (0,-186) {(c) Stage 7 on Example 1: the invariant at every prefix};
\node[inner sep=0pt,anchor=north west,font=\sffamily\fontsize{5.8}{7.2}\selectfont] at (0,-194) {%
  \begin{tabular}{@{}lll@{}}
  prefix $A$ of the routed program & placement & pulses done, $\pi(A)$\\\hline
  (empty) & $q_0$@S, $q_1$@G & none\\
  transport & $q_0$@G, $q_1$@G & none\\
  transport; R & $q_0$@G, $q_1$@G & R\\
  transport; R; MS & $q_0$@G, $q_1$@G & R; MS
  \end{tabular}};
\node[tnote,text width=140pt] at (214,-196) {$\llbracket A\rrbracket_H$ and
  $\llbracket\pi(A)\rrbracket$ agree on the logical state at every row.
  The next step is either a transport that is legal here, or the next pulse
  with its ions already at a site that has the control.};

\node[thead] at (222,-108) {(d) Rules are compiler inputs};
\node[tlab,anchor=north west,align=left,text width=132pt,font=\sffamily\fontsize{5.8}{7}\selectfont] at (222,-116)
  {$\rho_{\rm cap}$: $\forall s.\;N^{\rm tr}(s)\le\capa(s)$\\
   elaborates on S, J, G to\\
   $N^{\rm tr}(\mathrm S)\le1\wedge N^{\rm tr}(\mathrm J)\le1\wedge N^{\rm tr}(\mathrm G)\le2$\\[2pt]
   One lemma, proved for all formulas: the expansion holds on a frame
   exactly when the formula does. The evaluator then decides it on each
   frame. A new rule needs no new proof.};

\node[thead] at (366,-108) {(e) What rests on what};
\path[tproved] (366,-116) rectangle (516,-150);
\node[tlab,anchor=north west,text width=144pt,font=\sffamily\fontsize{5.7}{6.7}\selectfont] at (369,-118)
  {\textbf{Proved once} (Lean kernel): stages 1 to 9 for every program,
   device and rule set; rule elaboration; the parallel schedule; that
   events on disjoint resources commute; what the chain machine accepts.};
\path[draw=qorange,fill=qwarm,rounded corners=1.5pt,line width=.4pt,densely dashed] (366,-153) rectangle (516,-194);
\node[tlab,anchor=north west,text width=144pt,font=\sffamily\fontsize{5.7}{6.7}\selectfont] at (369,-155)
  {\textbf{Decided in each compilation} by a proved procedure inside the
   compiler: legality of every transport, including proposed ones; the
   rules on every frame; the rules again at parallel start times; the
   chain replay.};
\path[tcard] (366,-197) rectangle (516,-234);
\node[tlab,anchor=north west,text width=144pt,font=\sffamily\fontsize{5.7}{6.7}\selectfont] at (369,-199)
  {\textbf{Assumed}: the device performs each declared event in its
   declared time (the primitive-realization assumption); Lean's compiler
   and runtime. The listings are displays, outside the theorem.};
\end{tikzpicture}}}
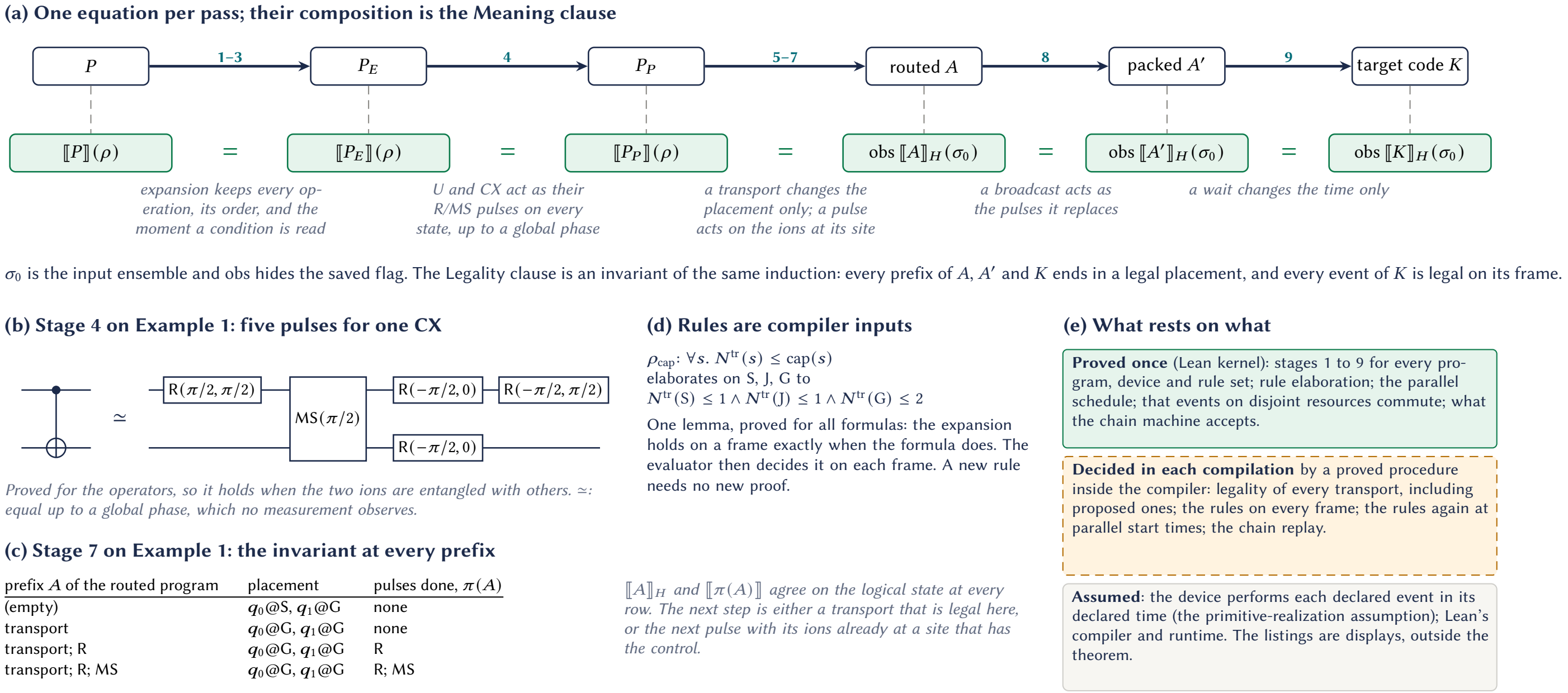
{The proof in one picture. (a) Each pass has its own equation between
meanings; chaining them gives the Meaning clause of
Theorem~\ref{thm:compile-correct}. (b) to (d) show three of the steps on
Example~1. (e) The parts are separated by how they are
established.}{fig:tut-proof}

\subsection{How the verification works}
\label{app:tut-proof}

\paragraph{Statement}
For all programs $P$, devices $H$, rule programs $R$ and search settings
$\kappa$, if compilation returns target code $K$ with a timeline $T$, then
the three clauses of Theorem~\ref{thm:compile-correct}, Meaning, Legality and
Time, hold (Fig.~\ref{fig:tut-pipeline}). The compiler may fail, for example
when a control is missing or the search bound is reached; a failure returns
no target code.

\paragraph{Why one proof covers every device}
$H$ and $R$ are ordinary arguments of the compiler and of the theorem. No
stage inspects which device it is given: stages 1 to 4 do not see $H$ at
all, and stages 5 to 9 read it only through its declarations and through
one evaluator for rules. The three devices of this appendix, the designs
of the evaluation and every drawing of the design agent are instances of
the same statement.

\paragraph{Meaning: the same result}
Each pass has a lemma that relates the meaning of its input to the meaning
of its output (Fig.~\ref{fig:tut-proof}a), proved by induction over the
program. Stage 4 rests on operator identities such as
Fig.~\ref{fig:tut-proof}b. Stage 7 rests on the invariant of
Fig.~\ref{fig:tut-proof}c: after every prefix, the physical state is the
native state at the placement reached. Stages 8 and 9 add sharing and
waits, which do not change the logical state. Stages 5 to 9 only add
annotations, transports, waits and sharing, so erasing them from $K$ gives
back the pulses of stage 4, in order, with their operands and guards; a
dropped or reordered gate would make the two differ. On the timeline, events
that overlap share no resource, and such events act in either order with
the same result, so running $K$ in order of time gives the same result too.

\paragraph{Legality: every frame is legal}
Legality is carried as an invariant of the construction. The router admits
a transport only together with a proof that it passes the core checks at
the current placement; a transport proposed by an external router is
admitted the same way or rejected. The timing stage evaluates every rule
on every frame, for both outcomes of a guarded event
(Fig.~\ref{fig:tut-proof}d), and the timeline reevaluates them at the
reported start times. The scheduler never overlaps two events that share a
resource, and the chain machine accepts only departures from a trap end.

\paragraph{Time: the declared durations}
Every event is charged the duration that the device declares for it, a
pulse that of the calibration at the trap that holds its ions, and the
reported finish is the latest end of an event, never more than the sum of
the durations (Example~3: 32 ticks against 44).

\paragraph{What a reader has to trust}
Fig.~\ref{fig:tut-proof}e separates the three kinds of evidence. The
proofs use no axiom beyond the three standard ones of Lean.
\ifqccdformal
Appendix~\ref{app:formal} defines every rule that this proof uses,
Appendix~\ref{app:hoare-example} replays the proof on Example~1, and
Appendix~\ref{app:proof} proves it for every input.
\fi
\clearpage\fi
  \ifqccdreference\providecommand{\lean}[1]{{\upshape\urlstyle{tt}\nolinkurl{#1}}}
\section{Complete quantum/classical example}
\label{app:native}

The tutorial compiled three small programs. This appendix compiles one larger
program that uses every kind of instruction at once: a U gate, CX gates, a
conditional CX, a register measurement under a condition that the
measurement overwrites, reset and a barrier, compiled to calibrated R/MS
pulses, shuttling, optical broadcasts, acquisition, pumping and saved
conditions on a two-zone device. In this appendix S and G abbreviate the
artifact site names \texttt{storage} and \texttt{gate}, and the device is
named \texttt{TwoZone} (\texttt{two-zone-exact-angles} in the artifact). The
listings are the compiler's input and output with these names and with
redundant parentheses removed; the expression trees are unchanged.

\begin{figure}[t]
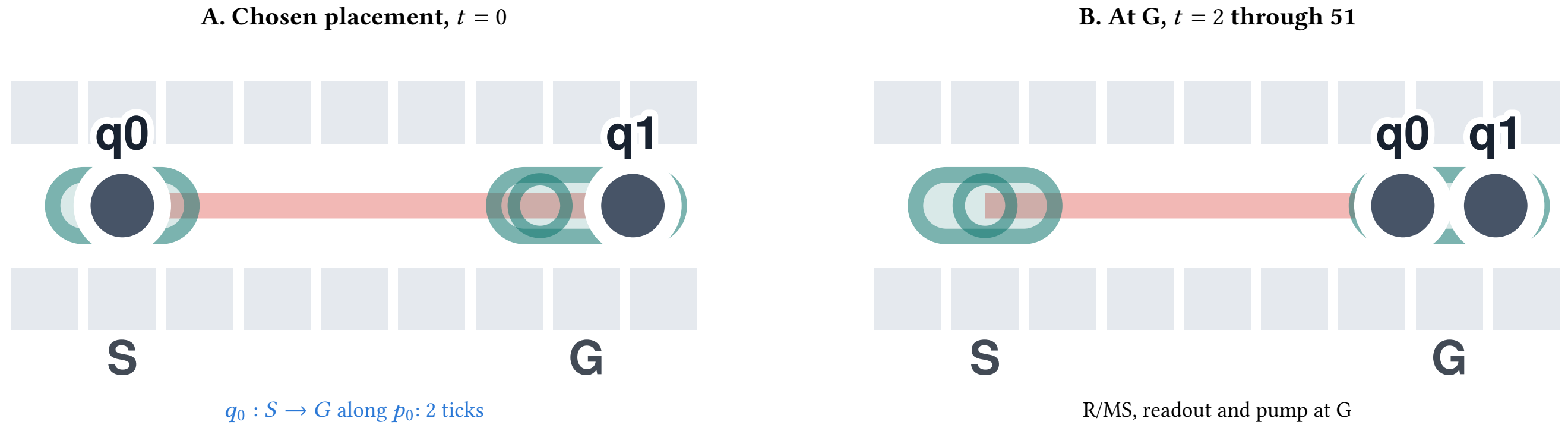

\centering
\begin{minipage}[t]{.46\columnwidth}
\centering\small\textbf{A. Chosen placement, $t=0$}\\[3pt]
\resizebox{\linewidth}{!}{\QCCDTwoZoneSnapshot{0}}\\[-2pt]
{\footnotesize\textcolor{qblue}{$q_0:S\to G$ along $p_0$: 2 ticks}}
\end{minipage}\hfill
\begin{minipage}[t]{.46\columnwidth}
\centering\small\textbf{B. At G, $t=2$ through 51}\\[3pt]
\resizebox{\linewidth}{!}{\QCCDTwoZoneSnapshot{1}}\\[-2pt]
{\footnotesize R/MS, readout and pump at G}
\end{minipage}
\caption{A compiler-chosen placement and its reached placement, drawn with
the website's trap, electrode, slot and ion primitives. All logical outcomes
share this same route and finish at tick 51. The empty storage site stays
available while both ions undergo the declared physical effects at G.}
\label{fig:appendix-dynamic}
\end{figure}

\subsection{Complete hardware program}
\label{app:native-hardware}

\begin{lstlisting}[style=qccd-hardware,basicstyle=\ttfamily\footnotesize\color{black},
  morekeywords={broadcast,sites,limit,acquire,pump,controller},xleftmargin=0pt]
hardware TwoZone;
site S[1];
site G[2];
path p0:S->G channel 0 waveform 0 duration 2;
path p1:G->S channel 0 waveform 1 duration 2;
optical channel 7;
calibration c0:R(pi/7,pi/2-0) @G channel 7 duration 2;
calibration c1:R(pi,0) @G channel 7 duration 2;
calibration c2:R(pi,(0+0)/2) @G channel 7 duration 2;
calibration c3:R(pi/2,pi/2) @G channel 7 duration 2;
calibration c4:MS(pi/2) @G channel 7 duration 4;
calibration c5:R(-(pi/2),0) @G channel 7 duration 2;
calibration c6:R(-(pi/2),pi/2) @G channel 7 duration 2;
broadcast b0:R(-(pi/2),0) sites G
  channel 7 duration 2 limit 2;
acquire a0 @G channel 7 duration 3;
pump p0 @G channel 7 duration 4;
controller duration 1;
\end{lstlisting}

The broken line in the broadcast declaration is only typographic wrapping.
There are no nested scopes. Transport path \texttt{p0} and pump \texttt{p0}
belong to different reference tables. The expressions $\pi/2-0$ and
$(0+0)/2$ intentionally retain the trees produced by Pulse Selection.
Although c1 and c2 have equal numerical phase, binding retains both exact
declarations. The broadcast uses c5 for each addressed ion.

S has no optical capability, while G has every needed capability. Initial
placement is absent from this input: the compiler must choose it using
capacity and the rules below. Neither an optical broadcast nor a scalar pulse
implicitly affects unlisted spectator ions.

\subsection{Complete source program}
\label{app:native-source}

\noindent\begin{minipage}{\columnwidth}
\begin{lstlisting}[style=qccd-circuit,basicstyle=\ttfamily\footnotesize\color{black},
  morekeywords={U,if,pi},xleftmargin=0pt]
OPENQASM 2.0;
qreg q[2];
creg c[2];
U(pi/7,0,0) q[0];
CX q[0],q[1];
measure q[0] -> c[0];
if(c==1) CX q[0],q[1];
if(c==1) measure q -> c;
reset q[0];
if(c==1) reset q[1];
barrier q;
\end{lstlisting}
\end{minipage}\par

No source operation is moved across another or across the barrier.
During execution, the register measurement samples its condition before either
readout writes a bit. Effect Expansion records this behavior in the following
fragment. Capability Binding later attaches eligible acquisition declarations:
\par\noindent\begin{minipage}{\columnwidth}
\begin{lstlisting}[style=qccd-ir,basicstyle=\ttfamily\footnotesize\color{black},
  morekeywords={latch,saved,little_endian,when,measure},xleftmargin=0pt]
latch saved <- little_endian(c0,c1) == 1;
when saved measure q0 -> c0;
when saved measure q1 -> c1;
\end{lstlisting}
\end{minipage}\par
Here the lowest-addressed bit is least significant. Each line retains its
source origin, which is omitted from the display.

\subsection{Complete target code}
\label{app:native-target}

The following is the compiler's output, with the site names abbreviated and
redundant parentheses removed (for example, the compiler prints
\texttt{R((pi/7),((pi/2)-0))} for \texttt{R(pi/7,(pi/2-0))}); every expression
tree is unchanged. Lines wrap only for column width.
The target contains physical effects and controller operations:
\begin{lstlisting}[style=qccd-execution,basicstyle=\ttfamily\footnotesize\color{black},
  morekeywords={transport,idle,broadcast,readout,pump,latch,saved,when,little_endian,via},
  xleftmargin=0pt]
init q0@S,q1@G;
simd transport q0:S->G[p0];
idle 1;
idle 1;
idle 1;
pulse R(pi/7,(pi/2-0)) q0 @G via c0;
pulse R(pi,0) q0 @G via c1;
pulse R(pi,(0+0)/2) q0 @G via c2;
pulse R(pi/2,pi/2) q0 @G via c3;
pulse MS(pi/2) q0,q1 @G via c4;
broadcast R(-(pi/2),0) q0@G, R(-(pi/2),0) q1@G via b0;
pulse R(-(pi/2),pi/2) q0 @G via c6;
readout q0 -> c0 @G via a0;
latch saved <- little_endian(c0,c1) == 1;
when saved pulse R(pi/2,pi/2) q0 @G via c3;
when saved pulse MS(pi/2) q0,q1 @G via c4;
when saved broadcast R(-(pi/2),0) q0@G, R(-(pi/2),0) q1@G via b0;
when saved pulse R(-(pi/2),pi/2) q0 @G via c6;
latch saved <- little_endian(c0,c1) == 1;
when saved readout q0 -> c0 @G via a0;
when saved readout q1 -> c1 @G via a0;
pump q0 @G via p0;
latch saved <- little_endian(c0,c1) == 1;
when saved pump q1 @G via p0;
barrier q0,q1;
\end{lstlisting}

The first three R pulses implement the source U. The following R pulse,
MS pulse, optical broadcast and final R pulse implement CX.
Broadcasting combines its two adjacent disjoint R
occurrences. The guarded CX uses the same physical sequence with one saved
condition. Each disabled pulse, acquisition or pump becomes an idle with its
own declared duration. The final barrier is retained and takes zero time.

\subsection{Rules and the complete timing calculation}
\label{app:native-rules}

The two supplied rules are $\rho_{\rm init}$ and $\rho_{\rm rel}$ of the main
text:
\[
(t=0\land\delta=0)\Rightarrow N^-(\mathrm S)=1,\qquad
1\le n_{\rm optical}\Rightarrow5\le t.
\]
The first admits the chosen placement $(S,G)$, while the second requires the three
waits after the shuttle. Readout and pump contribute to $n_{\rm optical}$. Latching
does not. Timing is consecutive and follows the declarations:

\par\smallskip
\begingroup\small
\setlength{\tabcolsep}{3pt}\renewcommand{\arraystretch}{1.12}
\noindent\begin{tabular}{@{}p{.64\columnwidth}rr@{}}
\toprule
Physical fragment & Start & End\\
\midrule
Shuttle $q_0:S\to G$ & 0 & 2\\
Three unit waits & 2 & 5\\
Three R pulses for U & 5 & 11\\
R/MS/broadcast/R for CX & 11 & 21\\
Readout $q_0\to c_0$ & 21 & 24\\
Latch first condition & 24 & 25\\
Guarded CX pulse sequence & 25 & 35\\
Latch register-measurement condition & 35 & 36\\
Two guarded readouts & 36 & 42\\
Pump $q_0$ & 42 & 46\\
Latch reset condition & 46 & 47\\
Guarded pump $q_1$ & 47 & 51\\
Barrier $q_0,q_1$ & 51 & 51\\
\bottomrule
\end{tabular}
\endgroup\par\smallskip

Every outcome ends at $(G,G)$ and tick 51. In Lean, kernel evaluation checks
that the compiler returns exactly this target code for this source and device
(\lean{Examples.compile_qasm}), and the general theorem then gives its legal
51-tick run and its projection to the pulse program, for every two-ion input
ensemble (\lean{Examples.expected_execution_legal},
\lean{Examples.expected_execution_preserves}).

\subsection{Why the saved condition matters}
\label{app:native-bridge}

Consider $q_0=0,q_1=1$ and classical register $c=1$ at the source statement
\texttt{\circkw{if}(c==1) \circkw{measure} q -> c;}.
The first acquisition sets $c_0=0$. The second must still execute and set
$c_1=1$, yielding $c=2$. Retesting the source comparison between acquisitions
would incorrectly skip the second. The latch samples the condition once, and
classical writes preserve that sampled value until the next latch.
The artifact's separate self-overwrite example reaches this situation from
all-zero input by inserting a reset before the conditional measurement.

The proof lifts this argument by induction over the complete expanded
operation and incoming branch ensemble. For a false sampled condition the
whole word is identity on observable state. For a true condition every
effect executes, including effects that overwrite the condition register.
Source states contain no saved flag, so equivalence erases that flag and
retains every quantum outcome and classical write.

The theorem assumes the declared ideal R/MS, acquisition, pumping and
coherent-transport effects (Section~\ref{sec:trust}). It begins with decoded
source, hardware and rule values; Appendix~\ref{app:syntax-mapping} states
what is proved about the readers. Search may reject a feasible input because
binding is exact or route/timing search is bounded. These limits do not
weaken the universal correctness statement for successful compilation.

\begin{figure*}[p]
\centering
\resizebox{\linewidth}{!}{\input{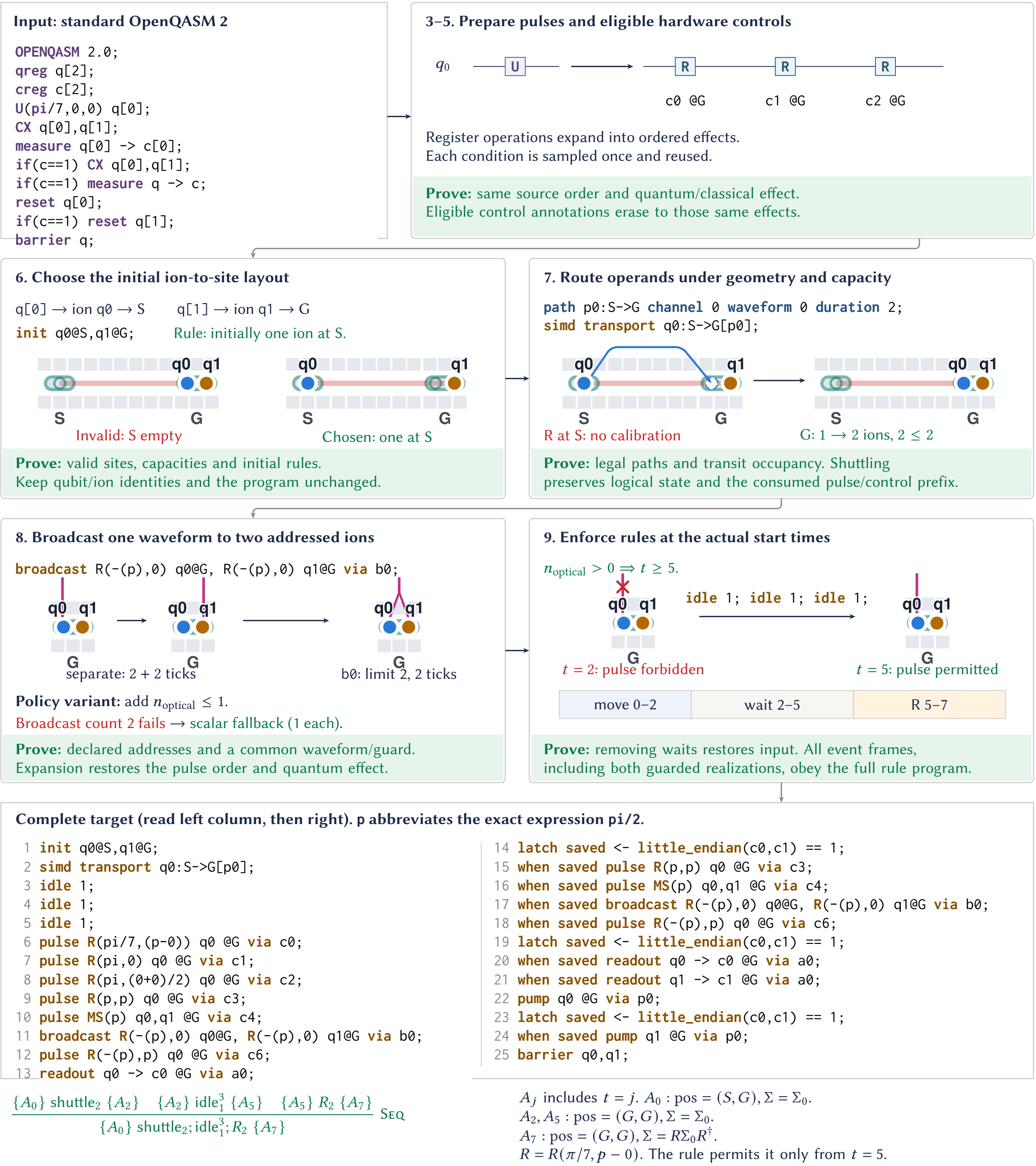}}
\caption{The complete two-zone compilation and its local proof obligations.
The initial rule requires one ion at S; the release rule forbids optical
activity before tick 5.
The baseline target finishes at tick 51; the separately marked one-pulse
policy requires scalar fallback. Read target lines 1--13, then 14--25.
\texttt{p} abbreviates the exact expression \texttt{pi/2}. The Hoare derivation
composes target lines 2--6; subscripts are durations.}
\label{fig:worked-example}
\label{app:native-walkthrough}
\end{figure*}
\clearpage\fi
  \ifqccdformal
\providecommand{\jd}[1]{\mathsf{#1}}
\providecommand{\sem}[1]{\llbracket#1\rrbracket}
\providecommand{\fseq}[1]{\overline{#1}}
\providecommand{\gm}{\gamma}
\providecommand{\nxt}{\jd{next}}
\providecommand{\dur}{\jd{dur}}
\providecommand{\rlz}{\jd{realize}}
\providecommand{\Ntr}{N^{\mathrm{tr}}}
\providecommand{\lean}[1]{{\upshape\urlstyle{tt}\nolinkurl{#1}}}
\providecommand{\wsaved}{\irkw{when}\;\irkw{saved}}
\providecommand{\hsaved}{\execkw{when}\;\execkw{saved}}
\providecommand{\runs}[1]{\stackrel{#1}{\Longrightarrow}}
\providecommand{\steps}[1]{\stackrel{#1}{\longrightarrow}}
\providecommand{\FormalBox}[1]{\par\smallskip\noindent\textbf{#1}\quad}
\raggedbottom
\newcounter{fexample}
\providecommand{\FormalExample}[2]{\par\medskip\noindent\refstepcounter{fexample}\label{#1}%
  \textbf{{\sffamily\color{qblue}Example~\thefexample}\quad#2}\quad}
\providecommand{\FxFigure}[3]{\begin{center}#1\par\captionof{figure}{#2}\label{#3}\end{center}}
\providecommand{\fxdm}[1]{[#1]}
\providecommand{\fxtt}{\ensuremath{\mathsf{tt}}}
\providecommand{\fxff}{\ensuremath{\mathsf{ff}}}
\providecommand{\fxhalf}{\tfrac12}
\tikzset{
  fxwire/.style={qink,line width=.5pt},
  fxcwire/.style={qink,double,double distance=.9pt,line width=.35pt},
  fxgate/.style={draw=qink,fill=white,line width=.5pt,inner sep=1.6pt,minimum size=10pt,
    font=\sffamily\scriptsize},
  fxlab/.style={inner sep=0pt,text=qink,font=\sffamily\scriptsize},
  fxhead/.style={inner sep=0pt,anchor=west,text=qink,font=\sffamily\bfseries\footnotesize},
  fxnote/.style={inner sep=0pt,text=qink!75,align=left,font=\sffamily\scriptsize},
  fxbox/.style={draw=qink,fill=white,rounded corners=2pt,line width=.45pt,inner sep=2.6pt,
    align=center,font=\scriptsize},
  fxin/.style={fxbox,fill=qsoft},
  fxok/.style={fxbox,draw=qgreen,fill=qmint},
  fxbad/.style={fxbox,draw=qred,fill=qrose},
  fxarr/.style={-{Stealth[length=3.6pt,width=3pt]},qink,line width=.5pt},
  fxrule/.style={inner sep=1.2pt,text=qblue,font=\scshape\scriptsize},
  fxmove/.style={fill=qblue!35,draw=qblue,line width=.35pt},
  fxidle/.style={fill=white,draw=qgrey,densely dashed,line width=.35pt},
  fxopt/.style={fill=qteal!30,draw=qteal,line width=.35pt},
  fxread/.style={fill=qgrey!35,draw=qgrey,line width=.35pt},
  fxlatch/.style={fill=qpurple!25,draw=qpurple,line width=.35pt},
  fxpump/.style={fill=qorange!30,draw=qorange,line width=.35pt},
}
\providecommand{\FxDev}[3]{\begin{scope}\QCCDUnit=.4bp\qccdpadsfalse
  \begin{scope}[shift={(#1,#2)},x=\QCCDUnit,y=-\QCCDUnit]
    \QCCDDevice{primer}{15}#3
    \foreach \fxs/\fxx in {S/25,J/110,G/200}{%
      \node[fxlab,anchor=north,font=\sffamily\bfseries\scriptsize] at (\fxx,44) {\fxs};}
  \end{scope}\end{scope}}
\providecommand{\FxTee}[3]{\begin{scope}\QCCDUnit=.3bp\qccdpadsfalse
  \begin{scope}[shift={(#1,#2)},x=\QCCDUnit,y=-\QCCDUnit]
    \QCCDDevice{parity}{20}#3
    \node[fxlab,anchor=north,font=\sffamily\bfseries\scriptsize] at (55,84) {S};
    \node[fxlab,anchor=north,font=\sffamily\bfseries\scriptsize] at (245,84) {G};
    \node[fxlab,anchor=west,font=\sffamily\bfseries\scriptsize] at (176,150) {P};
    \node[fxlab,anchor=west,font=\sffamily\bfseries\scriptsize] at (181,74) {J};
  \end{scope}\end{scope}}
\providecommand{\FxEval}[6]{\node[#6] (#1) at (#2,#3) {#4\\[-1pt]{\sffamily\bfseries #5}};}

\section{Languages and their meaning}
\label{app:formal}
\label{app:lean}

The previous appendices showed compilations by example. This appendix
defines every object that the correctness proof manipulates, in the order in
which the proof needs them: the syntax of each language
(\S\ref{app:f-syntax}), when an input is well formed (\S\ref{app:f-static}),
what a program computes (\S\ref{app:f-semantics}), how an instruction changes
the physical state and when it passes the core checks
(\S\ref{app:f-physical}), and the rules over event frames together with legal
runs (\S\ref{app:f-assertions}). Appendix~\ref{app:f-compile} then states the
nine compilation stages as rules, Appendix~\ref{app:hoare-example} replays the
correctness proof on one program, and Appendix~\ref{app:proof} proves
Theorem~\ref{thm:compile-correct} in general. Every definition, rule and
lemma below is a definition or theorem of the Lean development; its Lean name
is given in typewriter, relative to the namespace \lean{QCCD.Dynamic} where it
lies there. Some names, such as \lean{CoreLegal}, \lean{Run} or
\lean{compileQASM}, also exist in the older namespaces \lean{QCCD.General}
and \lean{QCCD.Native}; the appendices always mean the \lean{QCCD.Dynamic}
version. Each subsection ends with worked examples on one small
program and one small device (Examples~\ref{ex:f-languages}
to~\ref{ex:f-one}). Every step of an example names the definition or rule
that it applies, and the compiled programs and times are the compiler's output
for the stated inputs.

\paragraph{Fonts and notation}
Text of a program is set in typewriter, with keywords in the colour of
their language: \circkw{CX} (source), \irkw{MS} (intermediate programs),
\execkw{pulse} (target code), \hwkw{site} (hardware) and \rulekw{assert}
(rules). Metavariables that range over syntax are italic ($P$, $h$,
$\varphi$). Judgments and semantic functions are sans serif
($\jd{core}$, $\nxt$). Rule names are in small capitals. A bar denotes a
finite list: $\fseq{x}=x_1\ldots x_m$, with $\epsilon$ the empty list and
$x;\fseq{x}$ a list with first element $x$. All programs are finite lists
of instructions; no language has a loop. We keep the notation of the main
text: $H$ is a hardware description, $R$ the rules as the designer writes
them, $P$ a source program, $K$ the target code, $\pos$ a placement, $F$ an
event frame, $\sigma$ an ensemble of quantum and classical branches, $\delta$
the duration of an instruction and $d(x)$ the declared duration of a
declaration $x$.
\subsection{Syntax}
\label{app:f-syntax}

Let $q$ range over ions, $c$ over classical bits, $s$ over sites, $p$ over
paths, $a$ over names, $n$, $d$ and $t$ over natural numbers, and $\alpha$
over angle expressions (exact expression trees over rationals,
\circkw{pi}, names and arithmetic).
\ifqccdreference Appendix~\ref{app:syntax} gives the complete concrete
grammar of every language; the abstract syntax below is what the proof
manipulates.\fi

\FormalBox{Source} (OpenQASM~2.0 after parsing; $u$ is $a$ or $a[n]$)
\[
\begin{array}{@{}r@{\;}c@{\;}l@{}}
S&::=&\circkw{include}\;\textit{file}\mid
      \circkw{qreg}\;a[n]\mid\circkw{creg}\;a[n]\mid
      \circkw{gate}\;a(\fseq{a})\;\fseq{a}\;\{\fseq{B}\}\mid
      \circkw{opaque}\;a(\fseq{a})\;\fseq{a}\\
&\mid&O\mid\circkw{if}\,(a\,\texttt{==}\,n)\;O\mid\circkw{barrier}\;\fseq{u}\\
O&::=&\circkw{U}(\alpha,\alpha,\alpha)\;u\mid\circkw{CX}\;u,u\mid
      a(\fseq{\alpha})\;\fseq{u}\mid\circkw{measure}\;u\to u\mid
      \circkw{reset}\;u\\
B&::=&\circkw{U}(\alpha,\alpha,\alpha)\;a\mid\circkw{CX}\;a,a\mid
      a(\fseq{\alpha})\;\fseq{a}\mid\circkw{barrier}\;\fseq{a}
\qquad\qquad P\;::=\;\fseq{S}
\end{array}
\]
A gate body $\fseq B$ contains unitary operations and barriers on its
formal operands. An \circkw{opaque} gate is declared without a body; calling
it is rejected. Includes are resolved by stage~1.

\FormalBox{Effect and pulse programs} (flat; a guard mark $\gm$ is empty
or \wsaved)
\[
\begin{array}{@{}r@{\;}c@{\;}l@{}}
\gm&::=&\epsilon\mid\wsaved\\
e&::=&\gm\;\irkw{U}(\alpha,\alpha,\alpha)\;q\mid\gm\;\irkw{CX}\;q,q\mid
      \gm\;\irkw{measure}\;q\to c\mid\gm\;\irkw{reset}\;q\mid
      \gm\;\irkw{barrier}\;\fseq{q}\mid
      \irkw{latch}\;\fseq{c}\,\texttt{==}\,n\\
w&::=&\irkw{R}(\alpha,\alpha)\;q\mid\irkw{MS}(\alpha)\;q,q\\
i&::=&\gm\;w\mid\gm\;\irkw{measure}\;q\to c\mid\gm\;\irkw{reset}\;q\mid
      \gm\;\irkw{barrier}\;\fseq{q}\mid
      \irkw{latch}\;\fseq{c}\,\texttt{==}\,n\\
P_E&::=&\langle n_q,n_c,\fseq{e}\rangle\qquad
P_P\;::=\;\langle n_q,n_c,\fseq{i}\rangle
\end{array}
\]
$P_E$ is an effect program and $P_P$ a pulse program on $n_q$ ions and
$n_c$ classical bits (\lean{Dynamic.Program}, \lean{Dynamic.PulseProgram}).

\FormalBox{Hardware description} $H$ is a list of declarations
(\lean{Dynamic.Device}):
\[
\begin{array}{@{}r@{\;}c@{\;}l@{}}
D&::=&\hwkw{site}\;s[n]\mid
      \hwkw{path}\;p:s\to\fseq{s}\to s\;\hwkw{channel}\;n\;
        \hwkw{waveform}\;n\;\hwkw{duration}\;d\mid
      \hwkw{share}\;\hwkw{channel}\;n\;(\hwkw{mandatory}\mid\hwkw{optional})\\
&\mid&\hwkw{optical}\;\hwkw{channel}\;n\mid
      \hwkw{calibration}\;k:W\,@s\;\hwkw{channel}\;n\;\hwkw{duration}\;d\;
        \rev{[\hwkw{ions}\;n]}\mid
      \hwkw{acquire}\;a\,@s\;\hwkw{channel}\;n\;\hwkw{duration}\;d\\
&\mid&\hwkw{pump}\;a\,@s\;\hwkw{channel}\;n\;\hwkw{duration}\;d\mid
      \hwkw{broadcast}\;b:\hwkw{R}(\alpha,\alpha)\;\hwkw{sites}\;\fseq{s}\;
        \hwkw{channel}\;n\;\hwkw{duration}\;d\;\hwkw{limit}\;n\\
&\mid&\hwkw{controller}\;\hwkw{duration}\;d
\qquad\qquad W\;::=\;\hwkw{R}(\alpha,\alpha)\mid\hwkw{MS}(\alpha)
\end{array}
\]
We write $\capa(s)$ for the capacity of site $s$; $\jd{src}(p)$,
$\jd{dst}(p)$ and $\via(p)$ for the first, last and intermediate sites of
path $p$; and $\jd{site}(k)$, $\jd{ch}(k)$, $\jd{wave}(k)$ for the fields
of a declaration $k$. \rev{The optional $\hwkw{ions}\;n$ of a calibration
declares the number of ions that its site must hold when the calibration
is used (\S\ref{app:f-physical}).} For a pulse, $\jd{wave}(\irkw{R}(\theta,\varphi)\,q)
=\hwkw{R}(\theta,\varphi)$ and $\jd{ions}(\irkw{R}(\theta,\varphi)\,q)=q$,
and likewise for \irkw{MS}.

\FormalBox{Target code} (\lean{Dynamic.HardwareInstruction},
\lean{Dynamic.HardwareExecution})
\[
\begin{array}{@{}r@{\;}c@{\;}l@{}}
h&::=&\execkw{transport}\;\fseq{m}\mid
      \gm\;\execkw{pulse}\;w\;\execkw{via}\;k\mid
      \gm\;\execkw{broadcast}\;\fseq{w}\;\execkw{via}\;b[\fseq{k}]\mid
      \gm\;\execkw{readout}\;q\to c\;\execkw{via}\;a\mid
      \gm\;\execkw{pump}\;q\;\execkw{via}\;a\\
&\mid&\execkw{latch}\;\fseq{c}\,\texttt{==}\,n\mid\execkw{idle}\;n\mid
      \gm\;\execkw{barrier}\;\fseq{q}
\qquad m\;::=\;q:p
\qquad K\;::=\;\langle\pos_0,\fseq{h}\rangle
\end{array}
\]
A placement $\pos$ is a list of sites, one per ion: $\pos(q)$ is the site
of ion $q$. Target code is an initial placement and a list of
instructions; the Lean development calls it an execution. Printed target
code also shows the site of each
declaration, as in \texttt{@G}; that site is $\jd{site}(k)$ and is not
part of the syntax.

\FormalBox{Assertions} (\lean{Assertions.Formula}; core rules
\lean{Dynamic.Rule})
\[
\begin{array}{@{}r@{\;}c@{\;}l@{}}
x&::=&a\mid\textit{variable}\qquad
\mathcal D\;::=\;\mathcal S\mid\mathcal P\mid\mathcal C_T\\
o&::=&t\mid\delta\mid t{+}\delta\mid n_{\rm move}\mid n_{\rm optical}\mid
      n_{\rm R}\mid n_{\rm MS}\mid n_{\rm readout}\mid n_{\rm pump}\mid
      n_{\rm latch}\\
T&::=&n\mid o\mid N^-(x)\mid N^+(x)\mid\Ntr(x)\mid\capa(x)\mid P(x)\mid
      M(x)\mid W(x)\mid T+T\mid T\mathbin{\dot-}T\\
\varphi&::=&\top\mid\bot\mid T\le T\mid T=T\mid\varphi\land\varphi\mid
      \varphi\lor\varphi\mid\lnot\varphi\mid\varphi\Rightarrow\varphi\mid
      \forall x{:}\mathcal D.\,\varphi\mid\exists x{:}\mathcal D.\,\varphi\\
r&::=&T_0\le T_0\mid T_0=T_0\mid r\land r\mid r\lor r\mid\lnot r\mid
      r\Rightarrow r\qquad
\text{where $T_0$ is a term without variables and without $\capa$}
\end{array}
\]
A rule program $R$ is a list of named formulas
$\rulekw{assert}\;a:\varphi$, the rules as the designer writes them. The
domains $\mathcal S$, $\mathcal P$ and $\mathcal C_T$ are the declared sites,
paths and transport channels, written \texttt{Sites}, \texttt{Paths} and
\texttt{Channels} in rule text. The compiler checks \emph{core rules} $r$,
the explicit checks of Section~\ref{sec:elaboration}, which have no quantifier
and no variable; $\hat R$ denotes a list of core rules, and elaboration
(\S\ref{app:f-assertions}) turns $R$ into $\hat R$ for one device.

\FormalExample{ex:f-languages}{One computation in every language.}
The \emph{feedback example} is the program of
Examples~\ref{ex:f-languages} to~\ref{ex:f-one}. It applies a CX to two
ions, reads the second, and resets it if the result was 1
(Fig.~\ref{fig:f-example}a). The device is the device S, J, G of the main
text (Fig.~\ref{fig:f-example}b). The source, the hardware description
and the rules are inputs. The effect program, the pulse program and the
target code are what the compiler returns for these inputs
(Fig.~\ref{fig:f-example}c). Table~\ref{tab:f-example} prints all six and
gives, for each line, the production of this section that derives it.

\FxFigure{%
\begin{tikzpicture}[x=1pt,y=1pt]
\path[use as bounding box] (0,7) rectangle (500,-84);
\node[fxhead] at (0,0) {(a) The circuit of the source $P$};
\foreach \y/\l in {-22/$q_0$,-40/$q_1$}{\draw[fxwire] (18,\y)--(132,\y);
  \node[fxlab,anchor=east] at (15,\y) {\l};}
\draw[fxcwire] (18,-62)--(132,-62);\node[fxlab,anchor=east] at (15,-62) {$c_0$};
\draw[fxwire] (42,-22)--(42,-40);\fill[qink] (42,-22) circle[radius=1.6pt];
\draw[fxwire,fill=white] (42,-40) circle[radius=3.6pt];
\draw[fxwire] (38.4,-40)--(45.6,-40) (42,-36.4)--(42,-43.6);
\draw[fxcwire] (74,-45)--(74,-62);
\node[fxgate,minimum width=13pt] at (74,-40) {};
\draw[qink,line width=.4pt] (70.4,-42) arc[start angle=180,end angle=0,radius=3.6pt];
\draw[qink,line width=.4pt] (74,-42)--(77,-36.6);
\draw[fxcwire] (110,-62)--(110,-46);\fill[qink] (110,-62) circle[radius=1.6pt];
\node[fxgate] at (110,-40) {$\ket0$};
\node[fxlab,anchor=north] at (110,-67) {if $c=1$};
\node[fxlab,anchor=north] at (42,-67) {CX};
\node[fxlab,anchor=north] at (74,-67) {measure};
\node[fxhead] at (158,0) {(b) The device $H$, placement $\pos_0$};
\FxDev{170}{-24}{\QCCDIon{q0}{S}{0}\QCCDIon{q1}{G}{1}
  \draw[qblue,line width=1.2\QCCDUnit,-{Stealth[length=3.4bp,width=2.8bp]}] (25,72)--(194,72);
  \fill[qblue] (110,72) circle[radius=2.4\QCCDUnit];}
\node[fxlab,anchor=north west,text=qblue] at (160,-57) {path $p_0$: S$\to$J$\to$G, 2 ticks};
\node[fxlab,anchor=north west] at (160,-67)
  {$\capa(\mathrm S)=1$,\ \ $\capa(\mathrm J)=1$,\ \ $\capa(\mathrm G)=2$};
\node[fxlab,anchor=north west] at (160,-77) {optical channel 7 at G};
\node[fxhead] at (300,0) {(c) Inputs (shaded) and compiler outputs};
\node[fxin,minimum width=42pt] (p) at (323,-26) {$P$\\source\\5 statements};
\node[fxbox,minimum width=42pt] (pe) at (379,-26) {$P_E$\\effects\\4 instructions};
\node[fxbox,minimum width=42pt] (pp) at (439,-26) {$P_P$\\pulses\\8 instructions};
\node[fxbox,minimum width=64pt] (e) at (439,-68) {$K$:\ \ target code\\11 events, 23 ticks};
\node[fxin,minimum width=42pt] (h) at (323,-60) {$H$\\device};
\node[fxin,minimum width=42pt] (r) at (323,-79) {rules};
\draw[fxarr] (p)--(pe);\draw[fxarr] (pe)--(pp);\draw[fxarr] (pp)--(e);
\draw[fxarr] (h.east)--(e.west|-h.east);
\draw[fxarr] (r.east)--(e.west|-r.east);
\end{tikzpicture}}
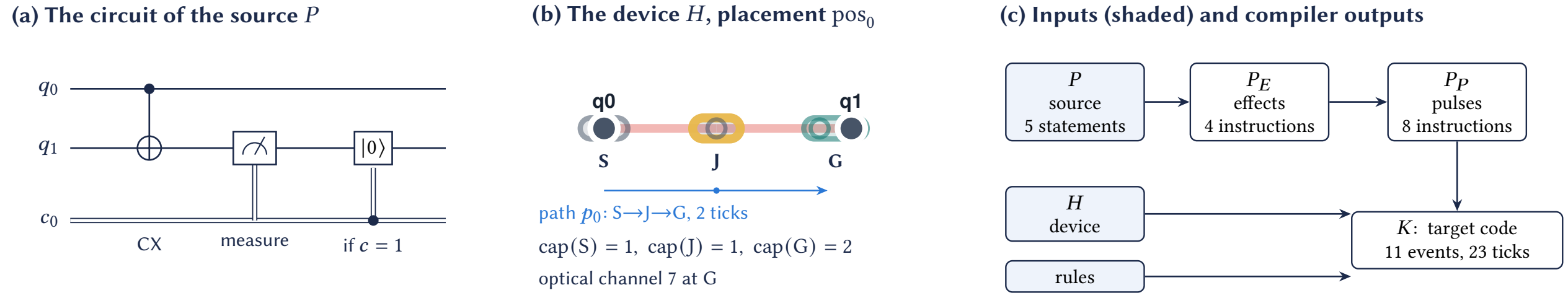
{The feedback example. (a)~The source entangles $q_0$ and $q_1$, reads
$q_1$ into $c_0$ and resets $q_1$ if $c=1$. (b)~The device and the initial
placement $\pos_0$: $q_0$ at S and $q_1$ at G. (c)~The six programs of
Table~\ref{tab:f-example}.}{fig:f-example}

\begin{center}\begin{minipage}{\linewidth}\centering
\captionof{table}{The feedback example in the languages of
\S\ref{app:f-syntax}. The right column names the production that derives
each line. Ions and bits are numbers: $q_j$ is ion $j$ and $c_j$ is bit
$j$.}
\label{tab:f-example}
\footnotesize\setlength{\tabcolsep}{3pt}\renewcommand{\arraystretch}{1.08}
\begin{tabular}{@{}l@{\ \ }l@{\qquad}l@{}}
\toprule
& program text & derived by\\
\midrule
\multicolumn{3}{@{}l}{\emph{Source} $P$\quad(input)}\\
& $\circkw{qreg}\;\texttt{q[2]};\ \ \circkw{creg}\;\texttt{c[1]};$
  & $S::=\circkw{qreg}\;a[n]$\ \ and\ \ $S::=\circkw{creg}\;a[n]$\\
& $\circkw{CX}\;\texttt{q[0]},\texttt{q[1]};$
  & $S::=O$,\ \ $O::=\circkw{CX}\;u,u$,\ \ each $u$ of the form $a[n]$\\
& $\circkw{measure}\;\texttt{q[1]}\to\texttt{c[0]};$
  & $O::=\circkw{measure}\;u\to u$\\
& $\circkw{if}\,(\texttt{c}\,\texttt{==}\,1)\;\circkw{reset}\;\texttt{q[1]};$
  & $S::=\circkw{if}\,(a\,\texttt{==}\,n)\;O$\ \ with\ \ $O::=\circkw{reset}\;u$
    \ \ (Fig.~\ref{fig:f-trees}a)\\[2pt]
\multicolumn{3}{@{}l}{\emph{Effect program} $P_E=\langle2,1,\ e_1\ldots e_4\rangle$\quad(output of stage 3)}\\
$e_1$ & $\irkw{CX}\;q_0,q_1$ & $e::=\gm\;\irkw{CX}\;q,q$\ \ with\ \ $\gm=\epsilon$\\
$e_2$ & $\irkw{measure}\;q_1\to c_0$ & $e::=\gm\;\irkw{measure}\;q\to c$\\
$e_3$ & $\irkw{latch}\;c_0\,\texttt{==}\,1$ & $e::=\irkw{latch}\;\fseq c\,\texttt{==}\,n$\\
$e_4$ & $\wsaved\;\irkw{reset}\;q_1$ & $e::=\gm\;\irkw{reset}\;q$\ \ with\ \ $\gm=\wsaved$\\[2pt]
\multicolumn{3}{@{}l}{\emph{Pulse program} $P_P=\langle2,1,\ i_1\ldots i_8\rangle$\quad(output of stage 4)}\\
$i_1,i_2,i_3$ & $\irkw{R}(\tfrac\pi2,\tfrac\pi2)\;q_0;\ \ \irkw{MS}(\tfrac\pi2)\;q_0,q_1;\ \
  \irkw{R}(-\tfrac\pi2,0)\;q_0$
  & $i::=\gm\;w$,\ \ $w::=\irkw{R}(\alpha,\alpha)\;q\mid\irkw{MS}(\alpha)\;q,q$\\
$i_4,i_5$ & $\irkw{R}(-\tfrac\pi2,0)\;q_1;\ \ \irkw{R}(-\tfrac\pi2,\tfrac\pi2)\;q_0$ & the same\\
$i_6,i_7,i_8$ & $\irkw{measure}\;q_1\to c_0;\ \ \irkw{latch}\;c_0\,\texttt{==}\,1;\ \
  \wsaved\;\irkw{reset}\;q_1$ & as $e_2$, $e_3$, $e_4$\\[2pt]
\multicolumn{3}{@{}l}{\emph{Hardware description} $H$\quad(input; the device also declares
  calibrations $k_0$ to $k_2$, which this program does not use)}\\
& $\hwkw{site}\;\mathrm S[1];\ \ \hwkw{site}\;\mathrm J[1];\ \ \hwkw{site}\;\mathrm G[2];$
  & $D::=\hwkw{site}\;s[n]$;\ \ $\capa(\mathrm S)=\capa(\mathrm J)=1$, $\capa(\mathrm G)=2$\\
& $\hwkw{path}\;p_0:\mathrm S\to\mathrm J\to\mathrm G\;\hwkw{channel}\;0\;
  \hwkw{waveform}\;0\;\hwkw{duration}\;2;$
  & $D::=\hwkw{path}\ldots$;\ \ $\jd{src}(p_0)=\mathrm S$, $\via(p_0)=\mathrm J$,
    $\jd{dst}(p_0)=\mathrm G$\\
& $\hwkw{optical}\;\hwkw{channel}\;7;$ & $D::=\hwkw{optical}\;\hwkw{channel}\;n$\\
& $\hwkw{calibration}\;k_3:\hwkw{R}(\tfrac\pi2,\tfrac\pi2)\,@\mathrm G\;
  \hwkw{channel}\;7\;\hwkw{duration}\;2;$
  & $D::=\hwkw{calibration}\;k:W\,@s\ldots$;\ \ $\jd{site}(k_3)=\mathrm G$, $\jd{ch}(k_3)=7$\\
& $k_4:\hwkw{MS}(\tfrac\pi2)$, duration 4;\ \ $k_5:\hwkw{R}(-\tfrac\pi2,0)$\ \ and\ \
  $k_6:\hwkw{R}(-\tfrac\pi2,\tfrac\pi2)$, duration 2
  & the same, all at G on channel 7\\
& $\hwkw{broadcast}\;b_0:\hwkw{R}(-\tfrac\pi2,0)\;\hwkw{sites}\;\mathrm G\;
  \hwkw{channel}\;7\;\hwkw{duration}\;2\;\hwkw{limit}\;2;$
  & $D::=\hwkw{broadcast}\;b:\hwkw{R}(\alpha,\alpha)\ldots$\\
& $\hwkw{acquire}\;a_0\,@\mathrm G\;\hwkw{channel}\;7\;\hwkw{duration}\;3;\ \
  \hwkw{pump}\;a_1\,@\mathrm G\;\hwkw{channel}\;7\;\hwkw{duration}\;4;$
  & $D::=\hwkw{acquire}\ldots$\ \ and\ \ $D::=\hwkw{pump}\ldots$\\
& $\hwkw{controller}\;\hwkw{duration}\;1;$ & $D::=\hwkw{controller}\;\hwkw{duration}\;d$\\[2pt]
\multicolumn{3}{@{}l}{\emph{Target code} $K=\langle\pos_0,\ h_1\ldots h_{11}\rangle$ with
  $\pos_0=(\mathrm S,\mathrm G)$\quad(output of stage 9)}\\
$h_1$ & $\execkw{transport}\;q_0{:}p_0$ & $h::=\execkw{transport}\;\fseq m$,\ \ $m::=q:p$\\
$h_2,h_3,h_4$ & $\execkw{idle}\;1$ & $h::=\execkw{idle}\;n$\\
$h_5,h_6$ & $\execkw{pulse}\;\irkw{R}(\tfrac\pi2,\tfrac\pi2)\;q_0\;\execkw{via}\;k_3;\ \
  \execkw{pulse}\;\irkw{MS}(\tfrac\pi2)\;q_0,q_1\;\execkw{via}\;k_4$
  & $h::=\gm\;\execkw{pulse}\;w\;\execkw{via}\;k$\\
$h_7$ & $\execkw{broadcast}\;\irkw{R}(-\tfrac\pi2,0)\;q_0,\ \irkw{R}(-\tfrac\pi2,0)\;q_1\;
  \execkw{via}\;b_0[k_5,k_5]$
  & $h::=\gm\;\execkw{broadcast}\;\fseq w\;\execkw{via}\;b[\fseq k]$\\
$h_8$ & $\execkw{pulse}\;\irkw{R}(-\tfrac\pi2,\tfrac\pi2)\;q_0\;\execkw{via}\;k_6$ & as $h_5$\\
$h_9$ & $\execkw{readout}\;q_1\to c_0\;\execkw{via}\;a_0$
  & $h::=\gm\;\execkw{readout}\;q\to c\;\execkw{via}\;a$\\
$h_{10}$ & $\execkw{latch}\;c_0\,\texttt{==}\,1$ & $h::=\execkw{latch}\;\fseq c\,\texttt{==}\,n$\\
$h_{11}$ & $\hsaved\;\execkw{pump}\;q_1\;\execkw{via}\;a_1$
  & $h::=\gm\;\execkw{pump}\;q\;\execkw{via}\;a$\ \ with\ \ $\gm=\hsaved$\\[2pt]
\multicolumn{3}{@{}l}{\emph{Rule program}\quad(input)}\\
& $\rulekw{assert}\;\ensuremath{\rho_{\rm init}}:\ (t=0\land\delta=0)\Rightarrow N^-(\mathrm S)=1$
  & $\varphi::=\varphi\Rightarrow\varphi\mid\varphi\land\varphi\mid T=T$,\ \
    $T::=o\mid n\mid N^-(x)$,\ \ $x::=a$\\
& $\rulekw{assert}\;\ensuremath{\rho_{\rm cap}}:\ \forall s{:}\mathcal S.\ \Ntr(s)\le\capa(s)$
  & $\varphi::=\forall x{:}\mathcal D.\,\varphi\mid T\le T$
    \ \ (Fig.~\ref{fig:f-trees}b)\\
& $\rulekw{assert}\;\ensuremath{\rho_{\rm rel}}:\ n_{\rm optical}>0\Rightarrow t\ge5$
  & $T>0$ abbreviates $1\le T$, and $t\ge5$ abbreviates $5\le t$\\
\bottomrule
\end{tabular}
\end{minipage}\end{center}

\FormalExample{ex:f-trees}{Two derivations in the grammar.}
Fig.~\ref{fig:f-trees} derives one source statement and one assertion
from the start symbol of their grammar; each inner node is one production.
The abstract syntax keeps what the proof needs and drops what parsing
resolves. In $P$ the operand is the text \texttt{q[1]}, a register name
and an index; from $P_E$ on it is the ion $q_1$. The pump is written
$a_1$, to keep it apart from the path $p_0$.

\FxFigure{%
\begin{tikzpicture}[x=1pt,y=1pt]
\path[use as bounding box] (0,7) rectangle (500,-100);
\node[fxhead] at (0,0) {(a) $\circkw{if}\,(\texttt{c}\,\texttt{==}\,1)\;\circkw{reset}\;\texttt{q[1]};$
  \ \ is a statement $S$};
\node[fxbox] (s) at (60,-20) {$S$};
\node[fxbox] (a) at (20,-46) {$a$:\ \texttt{c}};
\node[fxbox] (n) at (60,-46) {$n$:\ 1};
\node[fxbox] (o) at (104,-46) {$O$};
\node[fxbox] (u) at (104,-70) {$u$};
\node[fxbox] (a2) at (82,-94) {$a$:\ \texttt{q}};
\node[fxbox] (n2) at (126,-94) {$n$:\ 1};
\draw[fxwire] (s)--(a) (s)--(n) (s)--(o) (o)--(u) (u)--(a2) (u)--(n2);
\node[fxnote,anchor=west] at (76,-20) {$S::=\circkw{if}\,(a\,\texttt{==}\,n)\;O$};
\node[fxnote,anchor=west] at (120,-58) {$O::=\circkw{reset}\;u$};
\node[fxnote,anchor=west] at (140,-82) {$u::=a[n]$};
\node[fxhead] at (262,0) {(b) $\forall s{:}\mathcal S.\ \Ntr(s)\le\capa(s)$\ \ is a formula $\varphi$};
\node[fxbox] (f) at (330,-20) {$\varphi$};
\node[fxbox] (x) at (280,-46) {$x$:\ $s$};
\node[fxbox] (d) at (330,-46) {$\mathcal D$:\ $\mathcal S$};
\node[fxbox] (g) at (390,-46) {$\varphi$};
\node[fxbox] (t1) at (360,-70) {$T$};
\node[fxbox] (t2) at (420,-70) {$T$};
\node[fxbox] (x1) at (360,-94) {$x$:\ $s$};
\node[fxbox] (x2) at (420,-94) {$x$:\ $s$};
\draw[fxwire] (f)--(x) (f)--(d) (f)--(g) (g)--(t1) (g)--(t2) (t1)--(x1) (t2)--(x2);
\node[fxnote,anchor=west] at (344,-20) {$\varphi::=\forall x{:}\mathcal D.\,\varphi$};
\node[fxnote,anchor=west] at (414,-52) {$\varphi::=T\le T$};
\node[fxnote,anchor=east] at (350,-82) {$T::=\Ntr(x)$};
\node[fxnote,anchor=west] at (430,-82) {$T::=\capa(x)$};
\end{tikzpicture}}
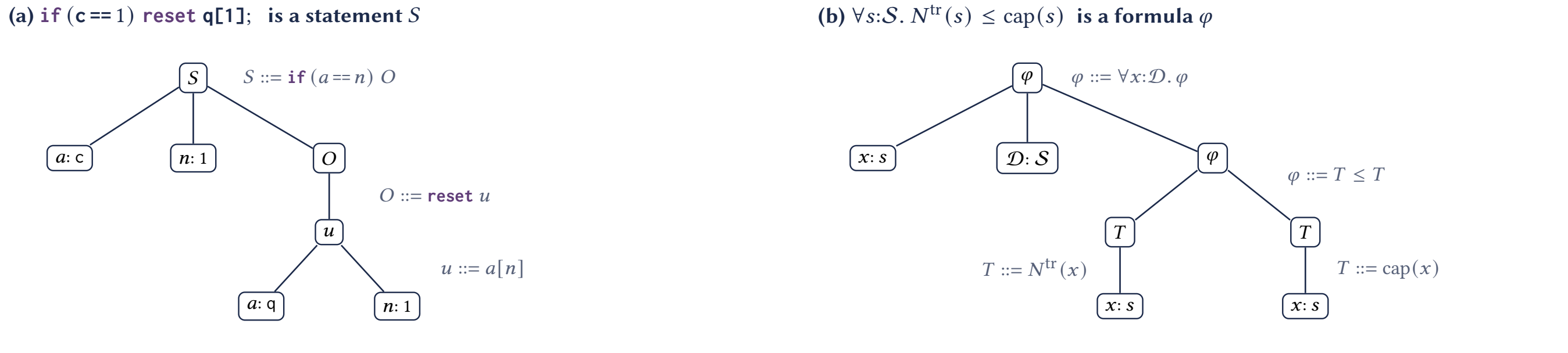
{Derivations in the grammars of \S\ref{app:f-syntax}. (a)~The conditional
reset of the source. (b)~The rule \ensuremath{\rho_{\rm cap}}. A leaf gives the name,
number or variable that the production leaves open.}{fig:f-trees}

\subsection{Static semantics: well-formedness}
\label{app:f-static}

The following judgments say when an input makes sense. They are decidable, and
the compiler evaluates them before it uses an input.

\FormalBox{Devices} $\vdash H\;\jd{ok}$ holds when all of the following
hold (\lean{DeviceWellFormed}):
(W1)~there is a site, site names are distinct, every capacity is
positive, and the legacy operation list of each site has no duplicate (it
grants no capability);
(W2)~the endpoints of every path are distinct sites, its intermediate
sites exist, are distinct and exclude the endpoints, and its duration is
positive;
(W3)~every shared channel is declared once and drives a path;
(W4)~on a mandatory channel, at most one path leaves a given site with a
given waveform;
(W5)~optical channels are distinct;
(W6)~every calibration names a declared optical channel, a positive
duration and a site whose capacity is at least the number of ions of its
waveform;
(W7)~every broadcast profile names a declared optical channel, a
positive duration and limit, and a nonempty list of distinct declared sites,
each of positive capacity;
(W8)~every acquisition and pump names a declared site of positive
capacity, a declared optical channel and a positive duration;
(W9)~a controller, if declared, has a positive duration.

\FormalBox{Placements and operands} Let
$N_{\pos}(s)=|\{q\mid\pos(q)=s\}|$.
\begin{mathpar}
\inferrule*[right=WF-Place]
  {|\pos|=n_q\\ \forall q.\ \pos(q)\in\jd{sites}(H)\\
   \forall s.\ N_{\pos}(s)\le\capa(s)}
  {H\vdash\pos:n_q}
\and
\inferrule*[right=Ok-R]{q<n_q}
  {n_q;n_c\vdash\gm\;\irkw{R}(\theta,\varphi)\;q\;\jd{ok}}
\and
\inferrule*[right=Ok-MS]{q<n_q\\ q'<n_q\\ q\ne q'}
  {n_q;n_c\vdash\gm\;\irkw{MS}(\theta)\;q,q'\;\jd{ok}}
\and
\inferrule*[right=Ok-Measure]{q<n_q\\ c<n_c}
  {n_q;n_c\vdash\gm\;\irkw{measure}\;q\to c\;\jd{ok}}
\and
\inferrule*[right=Ok-Reset]{q<n_q}
  {n_q;n_c\vdash\gm\;\irkw{reset}\;q\;\jd{ok}}
\and
\inferrule*[right=Ok-Barrier]{\forall q\in\fseq{q}.\ q<n_q}
  {n_q;n_c\vdash\gm\;\irkw{barrier}\;\fseq{q}\;\jd{ok}}
\and
\inferrule*[right=Ok-Latch]
  {\fseq{c}\ne\epsilon\\ \fseq{c}\text{ distinct}\\
   \forall c\in\fseq{c}.\ c<n_c}
  {n_q;n_c\vdash\irkw{latch}\;\fseq{c}\,\texttt{==}\,n\;\jd{ok}}
\end{mathpar}
(\lean{PlacementValid}, \lean{PulseOperandsValid}). The rules for
\irkw{U} and \irkw{CX} are those for \irkw{R} and \irkw{MS}. These are the
rules for pulse programs; for effect programs Lean checks the same operand
bounds but does not require the bits of a latch to be nonempty and distinct,
so the judgment above is the stronger of the two.

\FormalBox{Assertions are sorted} A context $\Gamma$ maps variables to
domains. The observations $N^-$, $N^+$, $\Ntr$, $\capa$ and $P$ take a
site and $W$ takes a channel. $M$ stands for two observations, as in the
main text: path use $M(p)$ takes a path and channel use $M(c)$ a channel
(\lean{pathUse}, \lean{channelUse}); the sort of the argument tells them
apart. Write $\jd{dom}(f)$ for the domain of observation $f$.
\begin{mathpar}
\inferrule*[right=A-Var]{\Gamma(x)=\jd{dom}(f)}
  {H;\Gamma\vdash f(x)\;\jd{ok}}
\and
\inferrule*[right=A-Name]{a\in\jd{dom}(f)\text{ of }H}
  {H;\Gamma\vdash f(a)\;\jd{ok}}
\and
\inferrule*[right=A-Quant]{H;\Gamma,x{:}\mathcal D\vdash\varphi\;\jd{ok}}
  {H;\Gamma\vdash\forall x{:}\mathcal D.\,\varphi\;\jd{ok}\\
   H;\Gamma\vdash\exists x{:}\mathcal D.\,\varphi\;\jd{ok}}
\end{mathpar}
The remaining rules are structural: a constant and an observation $o$
are well formed, and a compound term or formula is well formed when its
parts are (\lean{Formula.valid}). The body of a quantifier is checked
even when its domain is empty. A rule program is well sorted,
$H;\cdot\vdash R\;\jd{ok}$, when each of its formulas is. A core rule is
well formed, $H\vdash r\;\jd{ok}$, when every site, path and channel that it
names is declared in $H$ (\lean{ruleWellFormed}), and
$H\vdash\hat R\;\jd{ok}$ when every core rule of $\hat R$ is.

\FormalExample{ex:f-device}{The device and its placements are well formed.}
We check $\vdash H\;\jd{ok}$ for the device of Table~\ref{tab:f-example},
one condition at a time.
\begin{center}\small
\begin{tabular}{@{}l@{\quad}p{.86\linewidth}@{}}
(W1) & three sites with the distinct names S, J, G and capacities $1,1,2>0$\\
(W2) & $p_0$: its endpoints S and G are distinct sites; its one intermediate
  site J is declared and is not an endpoint; its duration is $2>0$\\
(W3), (W4) & $H$ declares no shared channel, so nothing is required\\
(W5) & one optical channel, 7\\
(W6) & $k_0$ to $k_6$ name channel 7, a duration of 2 or 4, and site G; an
  \hwkw{R} has one ion and an \hwkw{MS} two, and $\capa(\mathrm G)=2$\\
(W7) & $b_0$ names channel 7, duration 2, limit 2 and the one site G\\
(W8) & $a_0$ and $a_1$ name site G, channel 7 and the durations 3 and 4\\
(W9) & the controller has duration $1>0$
\end{tabular}
\end{center}
Hence $\vdash H\;\jd{ok}$. A device that fails a condition is rejected
before any program is read: with $\hwkw{site}\;\mathrm G[1]$ in place of
$\hwkw{site}\;\mathrm G[2]$, condition (W6) fails for $k_4$, whose
\hwkw{MS} needs two ions at G, and the compiler reports a malformed
device. For the placements of Fig.~\ref{fig:f-place}, with $n_q=2$:
\begin{mathpar}
\inferrule*[right=WF-Place]
  {|\pos_0|=2\\ \pos_0(q_0)=\mathrm S\\ \pos_0(q_1)=\mathrm G\\
   N_{\pos_0}(\mathrm S)=1\le1\\ N_{\pos_0}(\mathrm J)=0\le1\\
   N_{\pos_0}(\mathrm G)=1\le2}
  {H\vdash\pos_0:2}
\end{mathpar}
The same rule derives $H\vdash\pos_1:2$ for
$\pos_1=(\mathrm G,\mathrm G)$, because $N_{\pos_1}(\mathrm G)=2\le2$.
It derives nothing for $\pos=(\mathrm S,\mathrm S)$: its premise for S
would be $N_{\pos}(\mathrm S)=2\le1$.

\FxFigure{%
\begin{tikzpicture}[x=1pt,y=1pt]
\path[use as bounding box] (0,7) rectangle (500,-62);
\node[fxhead] at (0,0) {(a) $\pos_0=(\mathrm S,\mathrm G)$};
\FxDev{12}{-22}{\QCCDIon{q0}{S}{0}\QCCDIon{q1}{G}{1}}
\node[fxlab,anchor=north west,text=qgreen] at (0,-53) {$N=(1,0,1)\le(1,1,2)$:\ \ \Rule{WF-Place} applies};
\node[fxhead] at (176,0) {(b) $\pos_1=(\mathrm G,\mathrm G)$};
\FxDev{188}{-22}{\QCCDIon{q0}{G}{0}\QCCDIon{q1}{G}{1}}
\node[fxlab,anchor=north west,text=qgreen] at (176,-53) {$N=(0,0,2)\le(1,1,2)$:\ \ \Rule{WF-Place} applies};
\node[fxhead] at (352,0) {(c) $\pos=(\mathrm S,\mathrm S)$};
\FxDev{364}{-22}{\qccdionlabelsfalse\QCCDIon[qred]{qa}{S}{0}
  \path[fill=qred,draw=white,line width=.8\QCCDUnit] ($(S)+(0,-23)$) circle[radius=10.4\QCCDUnit];
  \node[fxlab,anchor=west,text=qred] at ($(S)+(16,-14)$) {$q_0,q_1$};}
\node[fxlab,anchor=north west,text=qred] at (352,-53) {$N(\mathrm S)=2>1=\capa(\mathrm S)$:\ \ no derivation};
\end{tikzpicture}}
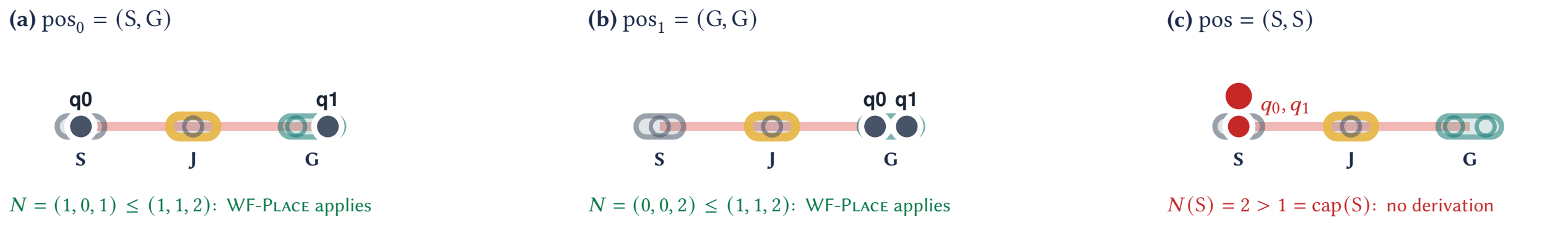
{Placements of two ions on the device S, J, G, with capacities
$(1,1,2)$. $N$ lists the number of ions at S, J and G.}{fig:f-place}

\FormalExample{ex:f-operands}{Operands and sorts.}
The pulse program $P_P$ of Table~\ref{tab:f-example} has $n_q=2$ and
$n_c=1$. Four of its instructions, each with the rule that accepts it:
\begin{mathpar}
\inferrule*[right=Ok-MS]{0<2\\ 1<2\\ q_0\ne q_1}
  {2;1\vdash\irkw{MS}(\tfrac\pi2)\;q_0,q_1\;\jd{ok}}
\and
\inferrule*[right=Ok-Measure]{1<2\\ 0<1}
  {2;1\vdash\irkw{measure}\;q_1\to c_0\;\jd{ok}}
\and
\inferrule*[right=Ok-Latch]{c_0\ne\epsilon\\ c_0\text{ distinct}\\ 0<1}
  {2;1\vdash\irkw{latch}\;c_0\,\texttt{==}\,1\;\jd{ok}}
\and
\inferrule*[right=Ok-Reset]{1<2}
  {2;1\vdash\wsaved\;\irkw{reset}\;q_1\;\jd{ok}}
\end{mathpar}
No rule accepts $\irkw{measure}\;q_1\to c_1$ here, because
\Rule{Ok-Measure} would need $1<n_c=1$, and none accepts
$\irkw{MS}(\tfrac\pi2)\;q_0,q_0$, because \Rule{Ok-MS} needs two
different ions. The rule \ensuremath{\rho_{\rm cap}} is well sorted, with
$\Gamma=s{:}\mathcal S$:
\begin{mathpar}
\inferrule*[right=A-Quant]
 {\inferrule*[right=structural]
   {\inferrule*[right=A-Var]
      {\Gamma(s)=\mathcal S=\jd{dom}(\Ntr)}
      {H;\Gamma\vdash\Ntr(s)\;\jd{ok}}\\
    \inferrule*[right=A-Var]
      {\Gamma(s)=\mathcal S=\jd{dom}(\capa)}
      {H;\Gamma\vdash\capa(s)\;\jd{ok}}}
   {H;\Gamma\vdash\Ntr(s)\le\capa(s)\;\jd{ok}}}
 {H;\cdot\vdash\forall s{:}\mathcal S.\ \Ntr(s)\le\capa(s)\;\jd{ok}}
\and
\inferrule*[right=A-Name]
 {\mathrm S\text{ is a site of }H}
 {H;\cdot\vdash N^-(\mathrm S)\;\jd{ok}}
\end{mathpar}
The second derivation is the one that \ensuremath{\rho_{\rm init}} needs; its other
terms are constants and observations. A formula that applies an
observation to the wrong kind of object has no derivation: in
$\forall p{:}\mathcal P.\ \Ntr(p)\le1$, \Rule{A-Var} would need
$\Gamma(p)=\jd{dom}(\Ntr)$, but $\Gamma(p)=\mathcal P$ and
$\jd{dom}(\Ntr)=\mathcal S$. Elaboration rejects a rule program that
contains this formula (\S\ref{app:f-assertions}).

\subsection{Dynamic semantics: what a program computes}
\label{app:f-semantics}

\FormalBox{Semantic domains} (\lean{Quantum.Branch}, \lean{Quantum.State})
\[
\begin{array}{@{}l@{\qquad}l@{}}
C\in\jd{Store}=\jd{Bit}\to\{0,1\} &
\rho\in\jd{Mat}_{n_q}=\mathbb C^{2^{n_q}\times2^{n_q}}\\
\beta\in\jd{Branch}::=(C,b,\rho),\ b\in\{\jd{tt},\jd{ff}\} &
\sigma\in\jd{Ens}=\jd{Branch}^{*}
\end{array}
\]
A branch is one sequence of measurement outcomes: its classical store
$C$, the compiler's saved flag $b$, and a matrix $\rho$ whose trace is
the probability of the branch; $c$ always names a classical bit and $C(c)$
is its value. An ensemble is an ordered list of branches, and the
\emph{ensemble} $\sigma$ is the quantum and classical state that
Theorem~\ref{thm:compile-correct} quantifies over. The proofs hold for every matrix $\rho$, so they hold for
every density matrix. We write $U\cdot\rho=U\rho U^{\dagger}$ for the
action of an operator, $A_q$ for a one-ion matrix $A$ acting on ion $q$
of the register, and $\eta$ for an environment that gives a real value
$\sem{\alpha}_\eta$ to every angle expression. The auxiliary definitions
are
\[
\begin{array}{@{}r@{\;}c@{\;}l@{\qquad}l@{}}
f^{\dagger}(\beta_1\ldots\beta_m)&=&f(\beta_1)\ldots f(\beta_m)
  &\text{(lifting a branch function to ensembles)}\\
(\gm\,?\,f)(C,b,\rho)&=&
  \begin{cases}(C,b,\rho)&\text{if }\gm=\wsaved\text{ and }b=\jd{ff}\\
               f(C,b,\rho)&\text{otherwise}\end{cases}
  &\text{(guard)}\\
\jd{obs}(C,b,\rho)&=&(C,\rho)
  &\text{(observation, pointwise on ensembles)}\\
\jd{val}_C(c_0\ldots c_m)&=&\textstyle\sum_j C(c_j)\,2^{j}
  &\text{(register value, first bit least significant)}
\end{array}
\]

\FormalBox{Operators} (namespace \lean{Native.Quantum}) With Pauli matrices $X$,
$Y$ and $\Pi^0=\ket0\!\langle0|$, $\Pi^1=\ket1\!\langle1|$,
$K^0=\ket0\!\langle0|$, $K^1=\ket0\!\langle1|$:
\[
\begin{array}{@{}l@{}}
\Rp(\theta,\varphi)=\cos\tfrac\theta2\,I
  -i\sin\tfrac\theta2\,(\cos\varphi\,X+\sin\varphi\,Y)\qquad
\MS(\theta)=\cos\tfrac\theta2\,I-i\sin\tfrac\theta2\,X\!\otimes\!X\\[2pt]
\mathrm U(\theta,\varphi,\lambda)=
 \begin{pmatrix}e^{-i(\varphi+\lambda)/2}\cos\frac\theta2&-e^{-i(\varphi-\lambda)/2}\sin\frac\theta2\\
 e^{i(\varphi-\lambda)/2}\sin\frac\theta2&e^{i(\varphi+\lambda)/2}\cos\frac\theta2
 \end{pmatrix}\qquad
\CX=\ket0\!\langle0|\otimes I+\ket1\!\langle1|\otimes X
\end{array}
\]

\FormalBox{Denotation of an instruction} $\sem{x}_\eta:\jd{Branch}\to
\jd{Ens}$, for effect and pulse instructions alike
(\lean{instructionBranch}, \lean{pulseBranch}). Angles are evaluated in
$\eta$.
\[
\begin{array}{@{}r@{\;}c@{\;}l@{}}
\sem{\gm\;\irkw{U}(\theta,\varphi,\lambda)\;q}_\eta&=&
  \gm\,?\,\bigl((C,b,\rho)\mapsto(C,b,\mathrm U(\theta,\varphi,\lambda)_q\cdot\rho)\bigr)\\
\sem{\gm\;\irkw{CX}\;q,q'}_\eta&=&
  \gm\,?\,\bigl((C,b,\rho)\mapsto(C,b,\CX_{q,q'}\cdot\rho)\bigr)\\
\sem{\gm\;\irkw{R}(\theta,\varphi)\;q}_\eta&=&
  \gm\,?\,\bigl((C,b,\rho)\mapsto(C,b,\Rp(\theta,\varphi)_q\cdot\rho)\bigr)\\
\sem{\gm\;\irkw{MS}(\theta)\;q,q'}_\eta&=&
  \gm\,?\,\bigl((C,b,\rho)\mapsto(C,b,\MS(\theta)_{q,q'}\cdot\rho)\bigr)\\
\sem{\gm\;\irkw{measure}\;q\to c'}_\eta&=&
  \gm\,?\,\bigl((C,b,\rho)\mapsto(C[c'\mapsto0],b,\Pi^0_q\cdot\rho)\;
   (C[c'\mapsto1],b,\Pi^1_q\cdot\rho)\bigr)\\
\sem{\gm\;\irkw{reset}\;q}_\eta&=&
  \gm\,?\,\bigl((C,b,\rho)\mapsto(C,b,K^0_q\cdot\rho+K^1_q\cdot\rho)\bigr)\\
\sem{\gm\;\irkw{barrier}\;\fseq q}_\eta&=&\beta\mapsto\beta\\
\sem{\irkw{latch}\;\fseq c\,\texttt{==}\,n}_\eta&=&
  (C,b,\rho)\mapsto(C,\ \jd{val}_C(\fseq c)=n,\ \rho)
\end{array}
\]
A measurement keeps both outcomes and never renormalizes. A latch reads
the register once and saves the comparison in $b$; no other instruction
changes $b$.

\FormalBox{Denotation of a program} For a list $X$ of effect or pulse
instructions (\lean{interpretCode}, \lean{interpretPulseCode}):
\[
\sem{\epsilon}_\eta\,\sigma=\sigma\qquad\qquad
\sem{x;X}_\eta\,\sigma=\sem{X}_\eta\bigl(\sem{x}_\eta^{\dagger}\,\sigma\bigr)
\]

\FormalBox{Denotation of target code} A guarded event whose flag is
false is realized as a wait of the same duration
(\lean{realize}; $\delta_H$ is defined in \S\ref{app:f-physical}):
\[
\rlz_H(b,h)=\begin{cases}\execkw{idle}\;\delta_H(h)&
  \text{if $h$ is a pulse, broadcast, readout or pump marked \hsaved{}
  and }b=\jd{ff}\\ h&\text{otherwise}\end{cases}
\]
The meaning of an instruction is the action of its realization
(\lean{interpretHardwareBranch}, \lean{activeHardwareBranch}),
$\sem{h}_{H,\eta}(C,b,\rho)=\jd{act}_{H,\eta}(\rlz_H(b,h))(C,b,\rho)$,
where the action of each instruction is
\[
\begin{array}{@{}r@{\;}c@{\;}l@{}}
\jd{act}(\gm\;\execkw{pulse}\;w\;\execkw{via}\;k)&=&
  (C,b,\rho)\mapsto(C,b,W_w\cdot\rho)
  \quad\text{with $W_w$ the matrix of $w$ on its ions}\\
\jd{act}(\gm\;\execkw{broadcast}\;\fseq w\;\execkw{via}\;b'[\fseq k])&=&
  (C,b,\rho)\mapsto\bigl(C,b,\bigl(\textstyle\prod_{q\in\jd{ions}(\fseq w)}
  \Rp(\theta_{b'},\varphi_{b'})_q\bigr)\cdot\rho\bigr)
  \quad\text{with the angles that $b'$ declares}\\
\jd{act}(\gm\;\execkw{readout}\;q\to c'\;\execkw{via}\;a)&=&
  \sem{\irkw{measure}\;q\to c'}\qquad
\jd{act}(\gm\;\execkw{pump}\;q\;\execkw{via}\;a)=\sem{\irkw{reset}\;q}\\
\jd{act}(\execkw{latch}\;\fseq c\,\texttt{==}\,n)&=&
  \sem{\irkw{latch}\;\fseq c\,\texttt{==}\,n}\qquad
\jd{act}(\execkw{transport}\;\fseq m)=\jd{act}(\execkw{idle}\;n)
  =\jd{act}(\gm\;\execkw{barrier}\;\fseq q)=\beta\mapsto\beta
\end{array}
\]
and $\sem{\epsilon}_{H,\eta}\sigma=\sigma$,
$\sem{h;K}_{H,\eta}\sigma=\sem{K}_{H,\eta}(\sem{h}_{H,\eta}^{\dagger}\sigma)$
(\lean{interpretHardwareCode}). A transport and a wait do not change the
ensemble; they change the placement and the clock
(\S\ref{app:f-physical}). This is the declared ideal semantics of the
primitives, the assumption named in Theorem~\ref{thm:compile-correct}.

\FormalBox{Projection} $\jd{proj}$ maps target code to the pulse program
that it performs (\lean{hardwareProject}); Theorem~\ref{thm:compile-correct}
writes $\jd{proj}(K)=\jd{native}(P)$:
\[
\begin{array}{@{}l@{}}
\jd{proj}(\gm\;\execkw{pulse}\;w\;\execkw{via}\;k)=\gm\;w\qquad
\jd{proj}(\gm\;\execkw{broadcast}\;w_1\ldots w_m\;\execkw{via}\;b[\fseq k])
  =\gm\;w_1;\ldots;\gm\;w_m\qquad
\jd{proj}(\execkw{transport}\;\fseq m)=\jd{proj}(\execkw{idle}\;n)=\epsilon\\
\jd{proj}(\gm\;\execkw{readout}\;q\to c\;\execkw{via}\;a)=\gm\;\irkw{measure}\;q\to c
\qquad\jd{proj}(\gm\;\execkw{pump}\;q\;\execkw{via}\;a)=\gm\;\irkw{reset}\;q\\
\jd{proj}(h)=h\ \text{for a latch or a barrier}
\qquad\jd{proj}(\epsilon)=\epsilon\qquad\jd{proj}(h;K)=\jd{proj}(h);\jd{proj}(K)
\end{array}
\]

\FormalBox{Source semantics} The meaning of the source is defined
independently of the compiler, as a big-step relation
$\Delta\vdash\langle S,\sigma\rangle\Downarrow\langle\Delta',\sigma'\rangle$
on observable ensembles: branches $(C,\rho)$ without a saved flag
(\lean{Source.StatementExecutes}). $\Delta$ records the declared
registers and gates. $\Delta\vdash u\mapsto\fseq q$ resolves an operand
to its ions (one ion for $a[n]$, all ions of the register for $a$), and
$\Delta\vdash O:f$ gives a unitary operation its action $f$ on matrices:
$\rho\mapsto U\cdot\rho$ with the matrix of \circkw{U} or \circkw{CX} at
the resolved ions, and for a call of a defined gate the composition of
the actions of its body with the arguments substituted. An operation with
register operands acts as one instance per index, in index order, as
OpenQASM prescribes (\lean{Instances}); a barrier acts as the
identity.
\begin{mathpar}
\inferrule*[right=S-Decl]
  {S\text{ declares a register or gate}\\ \Delta'=\Delta\oplus S}
  {\Delta\vdash\langle S,\sigma\rangle\Downarrow\langle\Delta',\sigma\rangle}
\and
\inferrule*[right=S-Unitary]
  {\Delta\vdash O:f}
  {\Delta\vdash\langle O,\sigma\rangle\Downarrow
   \langle\Delta,\ ((C,\rho)\mapsto(c,f(\rho)))^{\dagger}\sigma\rangle}
\and
\inferrule*[right=S-Measure]
  {\Delta\vdash u\mapsto q_1\ldots q_m\\ \Delta\vdash u'\mapsto c_1\ldots c_m}
  {\Delta\vdash\langle\circkw{measure}\;u\to u',\sigma\rangle\Downarrow
   \langle\Delta,\ \jd{meas}_{q_m,c_m}^{\dagger}(\cdots
   \jd{meas}_{q_1,c_1}^{\dagger}\sigma)\rangle}
\and
\inferrule*[right=S-Reset]
  {\Delta\vdash u\mapsto q_1\ldots q_m}
  {\Delta\vdash\langle\circkw{reset}\;u,\sigma\rangle\Downarrow
   \langle\Delta,\ \jd{rst}_{q_m}^{\dagger}(\cdots\jd{rst}_{q_1}^{\dagger}\sigma)\rangle}
\and
\inferrule*[right=S-If]
  {\Delta\vdash a\mapsto\fseq c\\
   \forall j.\ \bigl(\jd{val}_{C_j}(\fseq c)=n\ \land\
     \Delta\vdash\langle O,\beta_j\rangle\Downarrow\langle\Delta,\sigma_j\rangle\bigr)
     \ \lor\ \bigl(\jd{val}_{C_j}(\fseq c)\ne n\ \land\ \sigma_j=\beta_j\bigr)}
  {\Delta\vdash\langle\circkw{if}\,(a\,\texttt{==}\,n)\;O,\ \beta_1\ldots\beta_m\rangle
   \Downarrow\langle\Delta,\ \sigma_1\ldots\sigma_m\rangle}
\and
\inferrule*[right=S-Seq]
  {\Delta\vdash\langle S,\sigma\rangle\Downarrow\langle\Delta_1,\sigma_1\rangle\\
   \Delta_1\vdash\langle\fseq S,\sigma_1\rangle\Downarrow\langle\Delta',\sigma'\rangle}
  {\Delta\vdash\langle S;\fseq S,\sigma\rangle\Downarrow\langle\Delta',\sigma'\rangle}
\end{mathpar}
Here $\beta_j=(C_j,\rho_j)$, and $\jd{meas}_{q,c}$ and $\jd{rst}_q$ are the
branch functions of \irkw{measure} and \irkw{reset} above without the
flag. In \Rule{S-Measure}, $u$ and $u'$ are both registers or both
single elements. In \Rule{S-If} each incoming branch reads its own
register once, before the operation runs. A barrier leaves the ensemble unchanged, and
$\Delta\vdash\langle\epsilon,\sigma\rangle\Downarrow\langle\Delta,\sigma\rangle$.
A source bundle $P$ with root $r$ executes as its statements with every
\circkw{include} replaced by the statements of the included unit; we write
$P\vdash\sigma\Downarrow\sigma'$ (\lean{Source.BundleExecutes}). The
relation is deterministic (\lean{StatementsExecute.deterministic}).

\FormalExample{ex:f-source}{The source, branch by branch.}
We run the source $P$ of Table~\ref{tab:f-example} on one input and name
the rule of every step (Fig.~\ref{fig:f-branches}a). Notation: a ket
lists $q_0$ and then $q_1$; $\fxdm\psi$ is the matrix
$\ket\psi\!\langle\psi|$; $\ket{+}=(\ket0+\ket1)/\sqrt2$ and
$\ket\Phi=(\ket{00}+\ket{11})/\sqrt2$; a store is written as the value of
its one bit $c_0$. The input is the ensemble of one branch,
$(0,\fxdm{{+}0})$.
\begin{center}\small\setlength{\tabcolsep}{4pt}
\begin{tabular}{@{}lll>{\raggedright\arraybackslash}p{.36\linewidth}l@{}}
\toprule
& statement & rule & what the rule computes & ensemble after\\
\midrule
1 & $\circkw{qreg}\;\texttt{q[2]};\ \circkw{creg}\;\texttt{c[1]};$ & \Rule{S-Decl}, twice
  & $\Delta=\{\texttt q\mapsto q_0q_1,\ \texttt c\mapsto c_0\}$; no change
  & $(0,\fxdm{{+}0})$\\
2 & $\circkw{CX}\;\texttt{q[0]},\texttt{q[1]};$ & \Rule{S-Unitary}
  & $\Delta\vdash O:\rho\mapsto\CX_{q_0,q_1}\cdot\rho$,\ \ and\ \
    $\CX\ket{{+}0}=\ket\Phi$
  & $(0,\fxdm\Phi)$\\
3 & $\circkw{measure}\;\texttt{q[1]}\to\texttt{c[0]};$ & \Rule{S-Measure}
  & $\Delta\vdash\texttt{q[1]}\mapsto q_1$,\ \ $\Delta\vdash\texttt{c[0]}\mapsto c_0$;\ \
    $\Pi^0_{q_1}\cdot\fxdm\Phi=\fxhalf\fxdm{00}$,\ \
    $\Pi^1_{q_1}\cdot\fxdm\Phi=\fxhalf\fxdm{11}$
  & $(0,\fxhalf\fxdm{00})\ \ (1,\fxhalf\fxdm{11})$\\
4 & $\circkw{if}\,(\texttt c\,\texttt{==}\,1)\;\circkw{reset}\;\texttt{q[1]};$ & \Rule{S-If}
  & $\Delta\vdash\texttt c\mapsto c_0$. Branch 1: $\jd{val}=0\ne1$, so it is
    kept. Branch 2: $\jd{val}=1$, so \Rule{S-Reset} runs on it:
    $K^0_{q_1}\cdot\fxhalf\fxdm{11}+K^1_{q_1}\cdot\fxhalf\fxdm{11}
    =0+\fxhalf\fxdm{10}$
  & $(0,\fxhalf\fxdm{00})\ \ (1,\fxhalf\fxdm{10})$\\
\bottomrule
\end{tabular}
\end{center}
\Rule{S-Seq} joins the statements:
$P\vdash(0,\fxdm{{+}0})\Downarrow(0,\fxhalf\fxdm{00})\;(1,\fxhalf\fxdm{10})$.
Each outcome has probability $\fxhalf$, the trace of its matrix. Ion
$q_1$ ends in $\ket0$ in both branches, and $c_0$ records which branch
was taken.

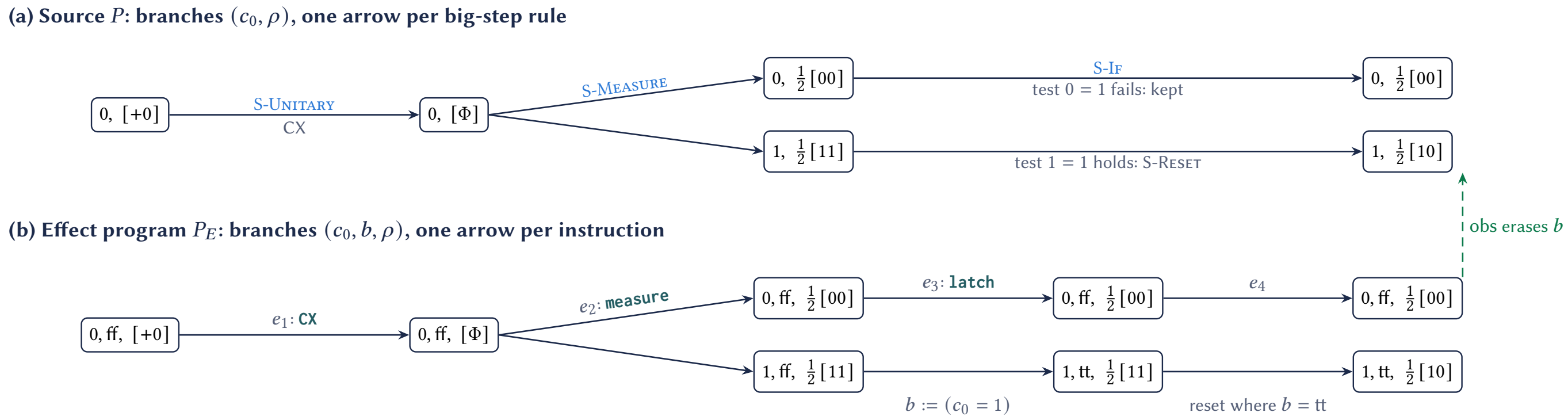
\FxFigure{%
\begin{tikzpicture}[x=1pt,y=1pt]
\path[use as bounding box] (0,7) rectangle (500,-150);
\node[fxhead] at (0,0) {(a) Source $P$: branches $(c_0,\rho)$, one arrow per big-step rule};
\node[fxbox] (a0) at (40,-32) {$0,\ \fxdm{{+}0}$};
\node[fxbox] (a1) at (146,-32) {$0,\ \fxdm\Phi$};
\node[fxbox] (a2a) at (262,-20) {$0,\ \fxhalf\fxdm{00}$};
\node[fxbox] (a2b) at (262,-44) {$1,\ \fxhalf\fxdm{11}$};
\node[fxbox] (a3a) at (458,-20) {$0,\ \fxhalf\fxdm{00}$};
\node[fxbox] (a3b) at (458,-44) {$1,\ \fxhalf\fxdm{10}$};
\draw[fxarr] (a0)--node[fxrule,above]{S-Unitary}node[fxnote,below=2pt]{CX} (a1);
\draw[fxarr] (a1.east)--node[fxrule,above,sloped]{S-Measure} (a2a.west);
\draw[fxarr] (a1.east)--(a2b.west);
\draw[fxarr] (a2a)--node[fxrule,above]{S-If}node[fxnote,below=1pt]{test $0=1$ fails: kept} (a3a);
\draw[fxarr] (a2b)--node[fxnote,below=1pt]{test $1=1$ holds: \Rule{S-Reset}} (a3b);
\node[fxhead] at (0,-70) {(b) Effect program $P_E$: branches $(c_0,b,\rho)$, one arrow per instruction};
\node[fxbox] (b0) at (40,-104) {$0,\fxff,\ \fxdm{{+}0}$};
\node[fxbox] (b1) at (146,-104) {$0,\fxff,\ \fxdm\Phi$};
\node[fxbox] (b2a) at (262,-92) {$0,\fxff,\ \fxhalf\fxdm{00}$};
\node[fxbox] (b2b) at (262,-116) {$1,\fxff,\ \fxhalf\fxdm{11}$};
\node[fxbox] (b3a) at (360,-92) {$0,\fxff,\ \fxhalf\fxdm{00}$};
\node[fxbox] (b3b) at (360,-116) {$1,\fxtt,\ \fxhalf\fxdm{11}$};
\node[fxbox] (b4a) at (458,-92) {$0,\fxff,\ \fxhalf\fxdm{00}$};
\node[fxbox] (b4b) at (458,-116) {$1,\fxtt,\ \fxhalf\fxdm{10}$};
\draw[fxarr] (b0)--node[fxnote,above=2pt]{$e_1$: \irkw{CX}} (b1);
\draw[fxarr] (b1.east)--node[fxnote,above=2pt,sloped]{$e_2$: \irkw{measure}} (b2a.west);
\draw[fxarr] (b1.east)--(b2b.west);
\draw[fxarr] (b2a)--node[fxnote,above=2pt]{$e_3$: \irkw{latch}} (b3a);
\draw[fxarr] (b2b)--(b3b);
\draw[fxarr] (b3a)--node[fxnote,above=2pt]{$e_4$} (b4a);
\draw[fxarr] (b3b)--(b4b);
\node[fxnote,anchor=north] at (311,-124) {$b:=(c_0=1)$};
\node[fxnote,anchor=north] at (409,-124) {reset where $b=\fxtt$};
\draw[fxarr,qgreen,densely dashed] (b4a.north)++(18,0)--node[fxnote,right=2pt,text=qgreen]{$\jd{obs}$ erases $b$}
  ($(a3b.south)+(18,0)$);
\node[fxnote,anchor=north west] at (0,-134) {The observed result of (b) is the result of (a):
  $(0,\fxhalf\fxdm{00})\ (1,\fxhalf\fxdm{10})$. Each box is one branch; its matrix has
  trace $\fxhalf$, the probability of the branch.};
\end{tikzpicture}}
{The feedback example on the input $(0,\fxdm{{+}0})$, with
$\fxdm\psi=\ket\psi\!\langle\psi|$ and $\ket\Phi$ the Bell state.
(a)~Example~\ref{ex:f-source}. (b)~Example~\ref{ex:f-effects}: the latch
saves the test in the flag $b$, and the guarded reset acts only where
$b=\jd{tt}$.}{fig:f-branches}

\FormalExample{ex:f-effects}{The effect program and the target code compute the same branches.}
Run the effect program $P_E$ on $\sigma=(0,\fxff,\fxdm{{+}0})$, using the
denotation of instructions (Fig.~\ref{fig:f-branches}b):
\begin{center}\small\setlength{\tabcolsep}{4pt}
\begin{tabular}{@{}ll>{\raggedright\arraybackslash}p{.40\linewidth}l@{}}
\toprule
& instruction & its denotation on the branches & ensemble after\\
\midrule
$e_1$ & $\irkw{CX}\;q_0,q_1$ & $\gm=\epsilon$, so the guard applies the
  function: $\rho\mapsto\CX_{q_0,q_1}\cdot\rho$ & $(0,\fxff,\fxdm\Phi)$\\
$e_2$ & $\irkw{measure}\;q_1\to c_0$ & both outcomes are kept, each with
  its projected matrix
  & $(0,\fxff,\fxhalf\fxdm{00})\ \ (1,\fxff,\fxhalf\fxdm{11})$\\
$e_3$ & $\irkw{latch}\;c_0\,\texttt{==}\,1$ & each branch sets
  $b:=(\jd{val}_C(c_0)=1)$
  & $(0,\fxff,\fxhalf\fxdm{00})\ \ (1,\fxtt,\fxhalf\fxdm{11})$\\
$e_4$ & $\wsaved\;\irkw{reset}\;q_1$ & guard: branch 1 has $b=\fxff$ and is
  returned unchanged; branch 2 has $b=\fxtt$ and is reset as in
  Example~\ref{ex:f-source}
  & $(0,\fxff,\fxhalf\fxdm{00})\ \ (1,\fxtt,\fxhalf\fxdm{10})$\\
\bottomrule
\end{tabular}
\end{center}
By the two equations for the denotation of a program, the last line is
\[
\sem{P_E}\sigma=\sem{e_4}^{\dagger}(\sem{e_3}^{\dagger}(\sem{e_2}^{\dagger}
(\sem{e_1}^{\dagger}\sigma))),
\qquad\text{and}\qquad
\jd{obs}(\sem{P_E}\sigma)=(0,\fxhalf\fxdm{00})\;(1,\fxhalf\fxdm{10})
\]
is the result of Example~\ref{ex:f-source} from $\jd{obs}(\sigma)$.

The target code $K$ of Table~\ref{tab:f-example} gives the same ensemble,
by the denotation of target code:
\begin{center}\small\setlength{\tabcolsep}{4pt}
\begin{tabular}{@{}l>{\raggedright\arraybackslash}p{.62\linewidth}l@{}}
\toprule
events & realization and action & ensemble after\\
\midrule
$h_1$ to $h_4$ & a transport and three waits: $\jd{act}$ is the identity
  & $(0,\fxff,\fxdm{{+}0})$\\
$h_5$ to $h_8$ & not guarded, so $\rlz_H(b,h)=h$; $\jd{act}$ applies the
  matrices of the five pulses in order, the broadcast $h_7$ applying the
  product of its two. Their product acts as $\CX_{q_0,q_1}$ on every
  matrix (the operator identity in the proof of Lemma~\ref{lem:select})
  & $(0,\fxff,\fxdm\Phi)$\\
$h_9$ & $\jd{act}(\execkw{readout}\;q_1\to c_0\;\execkw{via}\;a_0)
  =\sem{\irkw{measure}\;q_1\to c_0}$
  & $(0,\fxff,\fxhalf\fxdm{00})\ \ (1,\fxff,\fxhalf\fxdm{11})$\\
$h_{10}$ & $\jd{act}(\execkw{latch}\;c_0\,\texttt{==}\,1)
  =\sem{\irkw{latch}\;c_0\,\texttt{==}\,1}$
  & $(0,\fxff,\fxhalf\fxdm{00})\ \ (1,\fxtt,\fxhalf\fxdm{11})$\\
$h_{11}$ & branch 1 has $b=\fxff$: $\rlz_H(\fxff,h_{11})=\execkw{idle}\;4$,
  the identity. Branch 2 has $b=\fxtt$: $\rlz_H(\fxtt,h_{11})=h_{11}$, and
  $\jd{act}(\execkw{pump}\;q_1\;\execkw{via}\;a_1)=\sem{\irkw{reset}\;q_1}$
  & $(0,\fxff,\fxhalf\fxdm{00})\ \ (1,\fxtt,\fxhalf\fxdm{10})$\\
\bottomrule
\end{tabular}
\end{center}
Hence $\sem{K}_{H}\sigma=\sem{P_E}\sigma$ on this input. The wait
$\execkw{idle}\;4$ has the duration of the pump, so both branches take
the same time (\S\ref{app:f-physical}). Projection reads the pulse
program off the target code: $\jd{proj}(h_1)=\cdots=\jd{proj}(h_4)=\epsilon$,
$\jd{proj}(h_5)=i_1$, $\jd{proj}(h_6)=i_2$, $\jd{proj}(h_7)=i_3;i_4$, $\jd{proj}(h_8)=i_5$,
$\jd{proj}(h_9)=i_6$, $\jd{proj}(h_{10})=i_7$ and $\jd{proj}(h_{11})=i_8$, so $\jd{proj}(K)$ is
the code of $P_P$. One operator by hand: $\Rp(\pi,0)=-iX$, so
$\sem{\irkw{R}(\pi,0)\;q_0}(C,b,\fxdm{00})=(C,b,\fxdm{10})$; the phase
$-i$ cancels in $U\cdot\rho=U\rho U^{\dagger}$.

\FormalExample{ex:f-latch}{Why the test is saved.}
A guarded operation may overwrite the register that its condition reads.
Take the source
$\circkw{qreg}\;\texttt{q[2]};\ \circkw{creg}\;\texttt{c[2]};\
\circkw{if}\,(\texttt c\,\texttt{==}\,1)\;\circkw{measure}\;\texttt q\to\texttt c;$
on the input branch with store $(c_0,c_1)=(1,0)$ and matrix $\fxdm{01}$.
The register value is $\jd{val}=1\cdot2^0+0\cdot2^1=1$, so \Rule{S-If}
runs the operation, and \Rule{S-Measure} measures both ions:
$\jd{meas}^{\dagger}_{q_1,c_1}(\jd{meas}^{\dagger}_{q_0,c_0}\,\beta)$.
The effect program for this source is
$\irkw{latch}\;c_0c_1\,\texttt{==}\,1;\ \wsaved\;\irkw{measure}\;q_0\to c_0;\
\wsaved\;\irkw{measure}\;q_1\to c_1$. Its denotation, with $0$ for the
zero matrix:
\begin{center}\small\setlength{\tabcolsep}{4pt}
\begin{tabular}{@{}ll@{}}
\toprule
instruction & ensemble after, as branches $((c_0,c_1),b,\rho)$\\
\midrule
(input) & $((1,0),\fxff,\fxdm{01})$\\
$\irkw{latch}\;c_0c_1\,\texttt{==}\,1$ & $((1,0),\fxtt,\fxdm{01})$\\
$\wsaved\;\irkw{measure}\;q_0\to c_0$
  & $((0,0),\fxtt,\fxdm{01})\ \ ((1,0),\fxtt,0)$\\
$\wsaved\;\irkw{measure}\;q_1\to c_1$
  & $((0,0),\fxtt,0)\ \ ((0,1),\fxtt,\fxdm{01})\ \ ((1,0),\fxtt,0)\ \ ((1,1),\fxtt,0)$\\
\bottomrule
\end{tabular}
\end{center}
After the first measurement the register of the branch with probability
1 holds $(0,0)$, whose value is 0. A guard that tested the register
again would skip the second measurement and leave $c_1=0$. The guard
reads the flag, which is still \fxtt, so $q_1$ is measured and the
branch ends with $(c_0,c_1)=(0,1)$: the four observed branches are those
that \Rule{S-If} and \Rule{S-Measure} give the source. Branches of
probability 0 are kept; the denotation never removes a branch.

\subsection{Physical states, frames and core legality}
\label{app:f-physical}

A physical state is a placement and a clock. An instruction changes them
as follows (\lean{nextPlacement}, \lean{instructionDuration}):
\[
\begin{array}{@{}l@{\qquad}l@{}}
\nxt_H(\pos,\execkw{transport}\;\fseq m)(q)=
 \begin{cases}\jd{dst}(p)&\text{if }(q:p)\in\fseq m\\ \pos(q)&\text{otherwise}\end{cases}
&\nxt_H(\pos,h)=\pos\ \text{ for every other }h\\[8pt]
\delta_H(\execkw{transport}\;\fseq m)=\max_{(q:p)\in\fseq m}d(p)
&\delta_H(\gm\;\execkw{pulse}\;w\;\execkw{via}\;k)=d(k)\\
\delta_H(\gm\;\execkw{broadcast}\;\fseq w\;\execkw{via}\;b[\fseq k])=d(b)
&\delta_H(\gm\;\execkw{readout}\ldots\execkw{via}\;a)=
 \delta_H(\gm\;\execkw{pump}\ldots\execkw{via}\;a)=d(a)\\
\delta_H(\execkw{latch}\ldots)=d(\hwkw{controller})
&\delta_H(\execkw{idle}\;n)=n\qquad\delta_H(\gm\;\execkw{barrier}\;\fseq q)=0
\end{array}
\]
All moves of one transport happen at once.

\FormalBox{Frames} A frame is what one event shows to the rules
(\lean{Dynamic.Frame}): $F=(H,\pos^-,\pos^+,u,t,\delta)$ with the placement
before and after, the instruction $u$ (or $\bot$ for initialization), the
start time and the duration.
\[
\jd{frame}_H(\pos,h,t)=(H,\ \pos,\ \nxt_H(\pos,h),\ h,\ t,\ \delta_H(h))
\qquad\qquad
\jd{init}_H(\pos)=(H,\pos,\pos,\bot,0,0)
\]
The observations of a frame are (\lean{evalNat}), with $L$ the moves of
$u$ ($L=\epsilon$ unless $u$ is a transport) and $\jd{ch}(m)$,
$\jd{wf}(m)$ and $\via(m)$ the channel, waveform and intermediate sites
of the path of move $m$:
\[
\begin{array}{@{}l@{\qquad}l@{}}
N^-(s)=N_{\pos^-}(s)\qquad N^+(s)=N_{\pos^+}(s)
& \Ntr(s)=|\{q\mid\pos^-(q)=s\lor\pos^+(q)=s\}|+|\{m\in L\mid s\in\via(m)\}|\\
M(p)=|\{(q:p')\in L\mid p'=p\}|
& M(c)=|\{m\in L\mid\jd{ch}(m)=c\}|\qquad
  W(c)=|\{\jd{wf}(m)\mid m\in L,\ \jd{ch}(m)=c\}|\\
n_{\rm move}=|L|\qquad n_{\rm optical}=\text{number of optical actions of }u
& P(s)=\text{number of optical actions of $u$ at }s
\end{array}
\]
A pulse, readout or pump is one optical action, and a broadcast of $m$
pulses is $m$. The counts $n_{\rm R}$, $n_{\rm MS}$, $n_{\rm readout}$,
$n_{\rm pump}$ and $n_{\rm latch}$ are the numbers of actions of $u$ of
that kind. Every count is $0$ in the initial frame.

\FormalBox{Core legality} $H;\pos\vdash_{\!a}h$ says that instruction $h$
may fire at placement $\pos$ (\lean{ActiveCoreLegal}). Every rule has the
premise $H\vdash\pos:n_q$, which we do not repeat.
\begin{mathpar}
\inferrule*[right=C-Transport]
  {\fseq m\ne\epsilon\\ \text{the ions of }\fseq m\text{ are distinct}\\
   \forall(q:p)\in\fseq m.\ q<n_q\land\pos(q)=\jd{src}(p)\\
   \forall m\ne m'\in\fseq m.\ \jd{compat}_H(m,m')\\
   \jd{complete}_H(\pos,\fseq m)\\
   H\vdash\nxt_H(\pos,\execkw{transport}\;\fseq m):n_q\\
   \forall s.\ \Ntr(s)\le\capa(s)\\ \delta_H(\execkw{transport}\;\fseq m)>0}
  {H;\pos\vdash_{\!a}\execkw{transport}\;\fseq m}
\and
\inferrule*[right=C-Pulse]
  {n_q;n_c\vdash w\;\jd{ok}\\ k\in H\\ \jd{wave}(k)=\jd{wave}(w)\\
   \forall q\in\jd{ions}(w).\ \pos(q)=\jd{site}(k)\\ \rev{\jd{fits}_H(\pos,k)}}
  {H;\pos\vdash_{\!a}\gm\;\execkw{pulse}\;w\;\execkw{via}\;k}
\and
\inferrule*[right=C-Broadcast]
  {m\ge2\\ \text{the ions of }w_1\ldots w_m\text{ are distinct}\\ b\in H\\
   \forall j.\ H;\pos\vdash_{\!a}\execkw{pulse}\;w_j\;\execkw{via}\;k_j\\
   \forall j.\ \jd{wave}(w_j)=\jd{wave}(b)\land\jd{ch}(k_j)=\jd{ch}(b)\land
     d(k_j)=d(b)\land\jd{site}(k_j)\in\jd{sites}(b)\\
   \forall s\in\jd{sites}(b).\ |\{j\mid\jd{site}(k_j)=s\}|\le\jd{limit}(b)}
  {H;\pos\vdash_{\!a}\gm\;\execkw{broadcast}\;w_1\ldots w_m\;\execkw{via}\;b[k_1\ldots k_m]}
\and
\inferrule*[right=C-Readout]
  {q<n_q\\ c<n_c\\ a\in H\\ \pos(q)=\jd{site}(a)}
  {H;\pos\vdash_{\!a}\gm\;\execkw{readout}\;q\to c\;\execkw{via}\;a}
\and
\inferrule*[right=C-Pump]
  {q<n_q\\ a\in H\\ \pos(q)=\jd{site}(a)}
  {H;\pos\vdash_{\!a}\gm\;\execkw{pump}\;q\;\execkw{via}\;a}
\and
\inferrule*[right=C-Latch]
  {n_q;n_c\vdash\irkw{latch}\;\fseq c\,\texttt{==}\,n\;\jd{ok}\\
   \hwkw{controller}\in H}
  {H;\pos\vdash_{\!a}\execkw{latch}\;\fseq c\,\texttt{==}\,n}
\and
\inferrule*[right=C-Idle]{n>0}{H;\pos\vdash_{\!a}\execkw{idle}\;n}
\and
\inferrule*[right=C-Barrier]{\forall q\in\fseq q.\ q<n_q}
  {H;\pos\vdash_{\!a}\gm\;\execkw{barrier}\;\fseq q}
\end{mathpar}
In \Rule{C-Transport}, $\Ntr$ is taken in the frame of the transport;
$\jd{compat}_H(m,m')$ holds when the two paths traverse no link in
opposite directions, share no intermediate site, and use one waveform if
they are on the same shared channel; $\jd{complete}_H(\pos,\fseq m)$ holds
when, for every move on a mandatory channel with waveform $w$, every ion
at the source of a path on that channel with waveform $w$ is moved along
that path (\lean{TransportLegal}). In \Rule{C-Pulse} to \Rule{C-Pump}, a
declaration in $H$ is well formed because $\vdash H\;\jd{ok}$.
\rev{A device may declare, for a calibration $k$, the number $n$ of ions
that its site must hold when $k$ is used, so that a gate duration can
depend on the length of the chain. $\jd{fits}_H(\pos,k)$ holds when $k$
declares no such number, or when it declares $n$ and
$|\{q\mid\pos(q)=\jd{site}(k)\}|=n$ (\lean{ChainLengthFits}). Each member
of a broadcast satisfies it through \Rule{C-Pulse}. No device in the
examples of this appendix declares a chain length, so the premise holds
trivially in every derivation below.}

An instruction is \emph{core legal} when both of its realizations may
fire (\lean{CoreLegal}):
\[
H;\pos\vdash h\;\jd{core}\quad\text{iff}\quad
H;\pos\vdash_{\!a}h\ \text{ and }\ H;\pos\vdash_{\!a}\rlz_H(\jd{ff},h).
\]
This is the judgment that Theorem~\ref{thm:compile-correct} writes
$F\models\jd{Core}_H$. Its premises contain the three core checks of
Section~\ref{sec:programs}: (i)~every ion that an event uses is where the
event needs it, at the source of its path or at the site of its declaration
($\pos(q)=\jd{src}(p)$, $\pos(q)=\jd{site}(k)$, $\pos(q)=\jd{site}(a)$);
(ii)~no site holds more ions than its capacity before, during or after the
event ($H\vdash\pos:n_q$, $H\vdash\nxt_H(\pos,h):n_q$ and
$\Ntr(s)\le\capa(s)$); and (iii)~each shared channel applies one waveform,
and every ion that a mandatory channel drives is listed as moving
($\jd{compat}_H$ and $\jd{complete}_H$). The remaining premises check
operands, declarations and durations.

\FormalExample{ex:f-run}{The target code, frame by frame.}
The target code $K$ of Table~\ref{tab:f-example} starts in the physical
state $\langle\pos_0,0\rangle$ with $\pos_0=(\mathrm S,\mathrm G)$. The
functions $\nxt_H$ and $\delta_H$ give the states that it passes through
(Fig.~\ref{fig:f-run}). Only the transport changes the placement:
$\nxt_H(\pos_0,h_1)(q_0)=\jd{dst}(p_0)=\mathrm G$ because
$(q_0{:}p_0)$ is a move of $h_1$, and
$\nxt_H(\pos_0,h_1)(q_1)=\pos_0(q_1)=\mathrm G$; call this placement
$\pos_1=(\mathrm G,\mathrm G)$. The durations are those that $H$
declares: $d(p_0)=2$ for $h_1$; $1$ for each wait; $d(k_3)=2$,
$d(k_4)=4$, $d(b_0)=2$ and $d(k_6)=2$ for $h_5$ to $h_8$;
$d(a_0)=3$ for the readout; the controller's $1$ for the latch; and
$d(a_1)=4$ for the pump. The clock therefore reads
$0,2,3,4,5,7,11,13,15,18,19$ at the starts of $h_1$ to $h_{11}$ and $23$
at the end. Table~\ref{tab:f-frames} lists the frames
$F_j=\jd{frame}_H(\pos,h_j,t)$ and their observations.

\FxFigure{%
\begin{tikzpicture}[x=1pt,y=1pt]
\path[use as bounding box] (0,7) rectangle (500,-158);
\node[fxhead] at (0,0) {(a) Placements: before, during and after the transport $h_1$};
\FxDev{14}{-26}{\QCCDIon{q0}{S}{0}\QCCDIon{q1}{G}{1}}
\node[fxlab,anchor=north west] at (0,-54) {$\langle\pos_0,0\rangle$,\ \ $N^-=(1,0,1)$};
\FxDev{192}{-26}{\QCCDIon{q1}{G}{1}
  \path[fill=qblue,draw=white,line width=.8\QCCDUnit] (J) circle[radius=10.4\QCCDUnit];
  \node[fxlab,text=qblue,font=\sffamily\bfseries\tiny] at ($(J)+(0,-18)$) {q0};
  \draw[qblue,line width=1.2\QCCDUnit,-{Stealth[length=3.4bp,width=2.8bp]}] (25,-22)--(188,-22);
  \draw[qblue,densely dotted,line width=.8\QCCDUnit] (S) circle[radius=10.4\QCCDUnit];
  \draw[qblue,densely dotted,line width=.8\QCCDUnit] (G-slot-0) circle[radius=10.4\QCCDUnit];}
\node[fxlab,anchor=north west] at (178,-54) {frame $F_1$, ticks $[0,2)$:\ \ $\Ntr=(1,1,2)$};
\FxDev{374}{-26}{\QCCDIon{q0}{G}{0}\QCCDIon{q1}{G}{1}}
\node[fxlab,anchor=north west] at (360,-54) {$\langle\pos_1,2\rangle$,\ \ $N^+=(0,0,2)$};
\draw[fxarr] (122,-34)--node[fxrule,above]{C-Transport} (176,-34);
\draw[fxarr] (300,-34)--node[fxnote,above=2pt]{$t+\delta_H(h_1)$} (356,-34);
\node[fxhead] at (0,-72) {(b) The clock: each event starts when the one before it ends};
\begin{scope}[shift={(34,-96)},x=19.6pt,y=1pt]
\path[fill=qred!10] (0,10) rectangle (5,-26);
\node[fxnote,anchor=south west,text=qred] at (0,11) {rule \ensuremath{\rho_{\rm rel}}: no optical action before tick 5};
\path[fxmove] (0,0) rectangle (2,-18);\node[fxlab] at (1,-9) {transport};
\foreach \t in {2,3,4}{\path[fxidle] (\t,0) rectangle (\t+1,-18);\node[fxlab,font=\sffamily\tiny] at (\t+.5,-9) {idle};}
\path[fxopt] (5,0) rectangle (7,-18);\node[fxlab] at (6,-9) {R};
\path[fxopt] (7,0) rectangle (11,-18);\node[fxlab] at (9,-9) {MS};
\path[fxopt] (11,0) rectangle (13,-18);\node[fxlab] at (12,-9) {R, R};
\path[fxopt] (13,0) rectangle (15,-18);\node[fxlab] at (14,-9) {R};
\path[fxread] (15,0) rectangle (18,-18);\node[fxlab] at (16.5,-9) {readout};
\path[fxlatch] (18,0) rectangle (19,-18);\node[fxlab,font=\sffamily\tiny] at (18.5,-9) {latch};
\path[fxpump] (19,0) rectangle (23,-9);\node[fxlab,font=\sffamily\tiny] at (21,-4.5) {pump, if $b=\fxtt$};
\path[fxidle] (19,-9) rectangle (23,-18);\node[fxlab,font=\sffamily\tiny] at (21,-13.5) {idle 4, if $b=\fxff$};
\foreach \x/\l in {1/1,2.5/2,3.5/3,4.5/4,6/5,9/6,12/7,14/8,16.5/9,18.5/10,21/11}{
  \node[fxlab,anchor=south] at (\x,1.5) {$h_{\l}$};}
\draw[qink,line width=.4pt] (0,-26)--(23,-26);
\foreach \t in {0,2,3,4,5,7,11,13,15,18,19,23}{\draw[qink,line width=.4pt] (\t,-26)--(\t,-29);
  \node[fxlab,anchor=north] at (\t,-31) {\t};}
\node[fxlab,anchor=east] at (-.3,-31) {$t$};
\end{scope}
\node[fxnote,anchor=north west] at (0,-140) {Placement $\pos_0$ during $[0,2)$ and $\pos_1$ from
  tick 2 on. The guarded pump $h_{11}$ reserves four ticks in both of its realizations, so the
  finish time 23 does not depend on the outcome.};
\end{tikzpicture}}
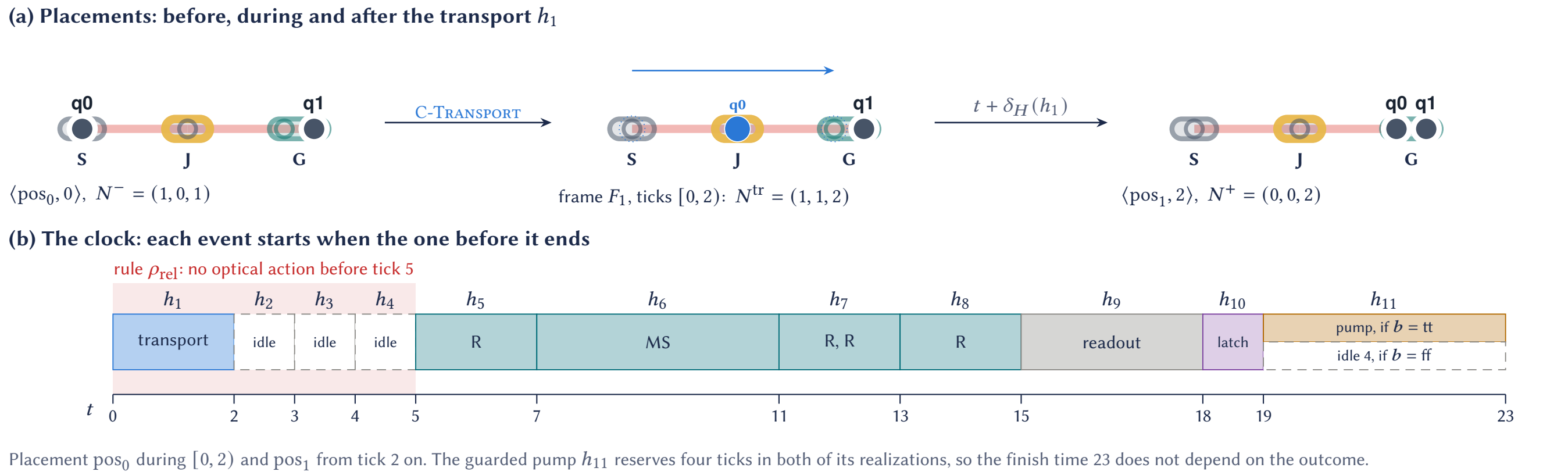
{The run of the feedback example. (a)~The one change of placement. During
the transport, $q_0$ counts at S, which it leaves, at J, which it
crosses, and at G, where it joins $q_1$. (b)~The eleven events on the
clock, with the start times that the compiler reports.}{fig:f-run}

\begin{center}\begin{minipage}{\linewidth}\centering
\captionof{table}{The frames of the feedback example. Triples list S, J,
G. $F_{11}^{\jd{ff}}$ is the frame of the second realization of $h_{11}$,
$\rlz_H(\jd{ff},h_{11})=\execkw{idle}\;4$.}
\label{tab:f-frames}
\footnotesize\setlength{\tabcolsep}{4.5pt}
\begin{tabular}{@{}lllcccccccl@{}}
\toprule
frame & instruction $u$ & $\pos^-\to\pos^+$ & $t$ & $\delta$ & $N^-$ & $N^+$ & $\Ntr$ &
$n_{\rm move}$ & $n_{\rm optical}$ & other counts that are not 0\\
\midrule
$F_0=\jd{init}_H(\pos_0)$ & $\bot$ & $\pos_0\to\pos_0$ & 0 & 0 & 1, 0, 1 & 1, 0, 1 & 1, 0, 1 & 0 & 0 & \\
$F_1$ & $h_1$: \execkw{transport} $q_0{:}p_0$ & $\pos_0\to\pos_1$ & 0 & 2 & 1, 0, 1 & 0, 0, 2 & 1, 1, 2 & 1 & 0 &
  $M(p_0)=1$,\ channel 0: $M=1$, $W=1$\\
$F_2,F_3,F_4$ & $h_2,h_3,h_4$: \execkw{idle} 1 & $\pos_1\to\pos_1$ & 2, 3, 4 & 1 & 0, 0, 2 & 0, 0, 2 & 0, 0, 2 & 0 & 0 & \\
$F_5$ & $h_5$: \execkw{pulse} \irkw{R} $q_0$ \execkw{via} $k_3$ & $\pos_1\to\pos_1$ & 5 & 2 & 0, 0, 2 & 0, 0, 2 & 0, 0, 2 & 0 & 1 & $n_{\rm R}=1$,\ $P(\mathrm G)=1$\\
$F_6$ & $h_6$: \execkw{pulse} \irkw{MS} $q_0,q_1$ \execkw{via} $k_4$ & $\pos_1\to\pos_1$ & 7 & 4 & 0, 0, 2 & 0, 0, 2 & 0, 0, 2 & 0 & 1 & $n_{\rm MS}=1$,\ $P(\mathrm G)=1$\\
$F_7$ & $h_7$: \execkw{broadcast} \irkw{R} $q_0,q_1$ \execkw{via} $b_0$ & $\pos_1\to\pos_1$ & 11 & 2 & 0, 0, 2 & 0, 0, 2 & 0, 0, 2 & 0 & 2 & $n_{\rm R}=2$,\ $P(\mathrm G)=2$\\
$F_8$ & $h_8$: \execkw{pulse} \irkw{R} $q_0$ \execkw{via} $k_6$ & $\pos_1\to\pos_1$ & 13 & 2 & 0, 0, 2 & 0, 0, 2 & 0, 0, 2 & 0 & 1 & $n_{\rm R}=1$,\ $P(\mathrm G)=1$\\
$F_9$ & $h_9$: \execkw{readout} $q_1\to c_0$ \execkw{via} $a_0$ & $\pos_1\to\pos_1$ & 15 & 3 & 0, 0, 2 & 0, 0, 2 & 0, 0, 2 & 0 & 1 & $n_{\rm readout}=1$,\ $P(\mathrm G)=1$\\
$F_{10}$ & $h_{10}$: \execkw{latch} $c_0\,\texttt{==}\,1$ & $\pos_1\to\pos_1$ & 18 & 1 & 0, 0, 2 & 0, 0, 2 & 0, 0, 2 & 0 & 0 & $n_{\rm latch}=1$\\
$F_{11}$ & $h_{11}$: \hsaved{} \execkw{pump} $q_1$ \execkw{via} $a_1$ & $\pos_1\to\pos_1$ & 19 & 4 & 0, 0, 2 & 0, 0, 2 & 0, 0, 2 & 0 & 1 & $n_{\rm pump}=1$,\ $P(\mathrm G)=1$\\
$F_{11}^{\jd{ff}}$ & \execkw{idle} 4 & $\pos_1\to\pos_1$ & 19 & 4 & 0, 0, 2 & 0, 0, 2 & 0, 0, 2 & 0 & 0 & \\
\bottomrule
\end{tabular}
\end{minipage}\end{center}

\noindent In $F_1$ the list of moves is $L=(q_0{:}p_0)$. By the
definition of $\Ntr$: at S, one ion is there before or after ($q_0$,
before) and no move crosses S, so $\Ntr(\mathrm S)=1+0$; at J, no ion
rests there and one move has $\mathrm J\in\via(p_0)$, so
$\Ntr(\mathrm J)=0+1$; at G, $q_1$ is there before and $q_0$ after, so
$\Ntr(\mathrm G)=2+0$. Every instruction of $K$ is core legal:
\begin{mathpar}
\inferrule*[right=C-Transport]
  {(q_0{:}p_0)\ne\epsilon\\ \text{one ion}\\
   q_0<2\\ \pos_0(q_0)=\mathrm S=\jd{src}(p_0)\\
   \text{one move: no pair to compare}\\\\
   \text{channel 0 is not mandatory: }\jd{complete}_H\text{ holds}\\
   H\vdash\pos_1:2\\
   \Ntr=(1,1,2)\le(1,1,2)\\ \delta_H(h_1)=2>0}
  {H;\pos_0\vdash_{\!a}\execkw{transport}\;q_0{:}p_0}
\end{mathpar}
\begin{center}\small\setlength{\tabcolsep}{4pt}
\begin{tabular}{@{}ll>{\raggedright\arraybackslash}p{.74\linewidth}@{}}
\toprule
events & rule & premises, at placement $\pos_1$\\
\midrule
$h_2,h_3,h_4$ & \Rule{C-Idle} & $1>0$\\
$h_5$, $h_8$ & \Rule{C-Pulse} & the pulse is well formed (\Rule{Ok-R}: $0<2$);
  $k_3\in H$ and $\jd{wave}(k_3)=\hwkw{R}(\tfrac\pi2,\tfrac\pi2)$ is the waveform of
  the pulse; $\pos_1(q_0)=\mathrm G=\jd{site}(k_3)$\rev{; $\jd{fits}_H(\pos_1,k_3)$
  holds because $H$ declares no chain length}. Likewise for $h_8$ and $k_6$\\
$h_6$ & \Rule{C-Pulse} & \Rule{Ok-MS} (Example~\ref{ex:f-operands}); $k_4\in H$ with
  waveform $\hwkw{MS}(\tfrac\pi2)$; $\pos_1(q_0)=\pos_1(q_1)=\mathrm G=\jd{site}(k_4)$\\
$h_7$ & \Rule{C-Broadcast} & $m=2$; the ions $q_0,q_1$ are distinct; $b_0\in H$; each
  member is legal by \Rule{C-Pulse} via $k_5$; the waveform
  $\hwkw{R}(-\tfrac\pi2,0)$, channel 7 and duration 2 of $k_5$ are those of $b_0$;
  $\jd{site}(k_5)=\mathrm G\in\jd{sites}(b_0)$; two members at G do not exceed the
  limit 2\\
$h_9$ & \Rule{C-Readout} & $1<2$; $0<1$; $a_0\in H$; $\pos_1(q_1)=\mathrm G=\jd{site}(a_0)$\\
$h_{10}$ & \Rule{C-Latch} & \Rule{Ok-Latch} (Example~\ref{ex:f-operands}); $H$ declares a
  controller\\
$h_{11}$ & \Rule{C-Pump} & $1<2$; $a_1\in H$; $\pos_1(q_1)=\mathrm G=\jd{site}(a_1)$\\
$\rlz_H(\fxff,h_{11})$ & \Rule{C-Idle} & the second realization is $\execkw{idle}\;4$, and $4>0$\\
\bottomrule
\end{tabular}
\end{center}
For $h_1$ to $h_{10}$, which are not guarded, $\rlz_H(\fxff,h)=h$, so
$H;\pos\vdash h\;\jd{core}$ is the line above. For $h_{11}$ it is the
last two lines together. These are the premises of a core run:
$H\vdash\pos_0\runs{K}_{\jd{core}}\pos_1$. Example~\ref{ex:f-rules} adds
the rules and obtains the run.

\FormalExample{ex:f-rejected}{Steps that no rule accepts.}
Fig.~\ref{fig:f-rejected} shows three steps and, for each, the premise
that fails.
(a)~At $\pos_0$, the instruction $\execkw{pulse}\;\irkw{MS}(\tfrac\pi2)\;q_0,q_1\;
\execkw{via}\;k_4$ needs $\pos_0(q_0)=\jd{site}(k_4)$ in
\Rule{C-Pulse}, but $\pos_0(q_0)=\mathrm S$ and
$\jd{site}(k_4)=\mathrm G$. This is why $K$ starts with a transport.
(b)~With three ions at $\pos=(\mathrm S,\mathrm G,\mathrm G)$, the
transport $q_0{:}p_0$ would give
$\nxt_H=(\mathrm G,\mathrm G,\mathrm G)$. The premise
$H\vdash\nxt_H:3$ of \Rule{C-Transport} fails, because \Rule{WF-Place}
would need $N(\mathrm G)=3\le2$; so does the premise
$\Ntr(\mathrm G)=3\le2$.
(c)~Take a device $H'$ with a junction: sites
$\mathrm S[2]$, $\mathrm J[1]$, $\mathrm G[2]$, $\mathrm P[1]$ and paths
$p_0:\mathrm S\to\mathrm J\to\mathrm G$ and
$p_1:\mathrm G\to\mathrm J\to\mathrm S$ on two unshared channels. At
$\pos=(\mathrm S,\mathrm S,\mathrm G)$ the transport
$q_0{:}p_0,\ q_2{:}p_1$ exchanges two ions through J. The placement
after it, $(\mathrm G,\mathrm S,\mathrm S)$, is valid:
$N=(2,0,1,0)\le(2,1,2,1)$ on S, J, G, P. \Rule{C-Transport} still does
not apply. $\jd{compat}_{H'}$ fails, because the two paths traverse the
links S--J and J--G in opposite directions and share the intermediate
site J. The transit premise fails at J, $\Ntr(\mathrm J)=0+2>1$, and at
S, where $q_0$ is before and $q_1,q_2$ are after:
$\Ntr(\mathrm S)=3>2$. A check of the destination alone would accept
this step; the junction defects of Fig.~\ref{fig:baseline-bugs} are of
this kind.

\FxFigure{%
\begin{tikzpicture}[x=1pt,y=1pt]
\path[use as bounding box] (0,7) rectangle (500,-104);
\node[fxhead] at (0,0) {(a) A gate away from its control};
\FxDev{14}{-26}{\QCCDIon[qred]{q0}{S}{0}\QCCDIon{q1}{G}{1}
  \draw[qhwBeam,line width=2.2\QCCDUnit] (200,-40)--(200,-6);
  \node[fxlab,anchor=east,text=qhwBeam] at (192,-30) {laser of $k_4$};}
\node[fxnote,anchor=north west,text=qred,text width=150pt] at (0,-52)
  {\Rule{C-Pulse}: $\pos_0(q_0)=\mathrm S\ne\mathrm G=\jd{site}(k_4)$};
\node[fxhead] at (170,0) {(b) A third ion into a full gate zone};
\FxDev{184}{-26}{\QCCDIon[qred]{q0}{S}{0}\QCCDIon{q1}{G}{0}\QCCDIon{q2}{G}{1}
  \draw[qred,line width=1.2\QCCDUnit,-{Stealth[length=3.4bp,width=2.8bp]}] (25,72)--(194,72);}
\node[fxnote,anchor=north west,text=qred,text width=150pt] at (170,-58)
  {\Rule{C-Transport}: after the step $N(\mathrm G)=3>2$,\\and $\Ntr(\mathrm G)=3>2$};
\node[fxhead] at (340,0) {(c) Two ions exchanged through a junction};
\FxTee{352}{-14}{\QCCDIon[qred]{q0}{S}{0}\QCCDIon{q1}{S}{1}\QCCDIon[qred]{q2}{G}{0}
  \draw[qred,line width=1.4\QCCDUnit,-{Stealth[length=3.4bp,width=2.8bp]}] (75,6)--(225,6);
  \draw[qred,line width=1.4\QCCDUnit,-{Stealth[length=3.4bp,width=2.8bp]}] (225,98)--(75,98);}
\node[fxnote,anchor=north west,text=qred,text width=156pt] at (340,-78)
  {\Rule{C-Transport}: $\jd{compat}$ fails;\\$\Ntr(\mathrm J)=2>1$ and $\Ntr(\mathrm S)=3>2$.\\The placement after the step is valid.};
\end{tikzpicture}}
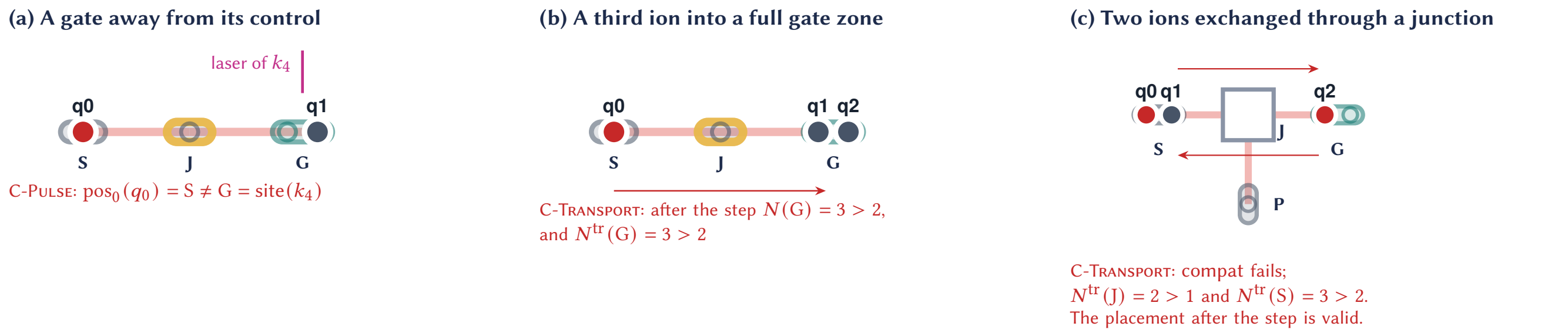
{Three steps that are not core legal, with the failing premise. Red ions
are the operands of the step.}{fig:f-rejected}

\subsection{Rules over frames, and legal runs}
\label{app:f-assertions}

\FormalBox{Meaning of terms and formulas} A variable environment $\xi$
maps variables to sites, paths or channels. A term denotes a natural
number on a frame, $\sem{T}^{F}_{\xi}$: a constant denotes itself, an
observation its value in $F$ (\S\ref{app:f-physical}), $\capa(x)$ the
declared capacity, $+$ addition and $\dot-$ truncated subtraction.
Satisfaction $F;\xi\models_H\varphi$ is the usual one
(\lean{Formula.Holds}):
\[
\begin{array}{@{}l@{\quad}c@{\quad}l@{\qquad\qquad}l@{\quad}c@{\quad}l@{}}
F;\xi\models_H T_1\le T_2&\text{iff}&\sem{T_1}^F_\xi\le\sem{T_2}^F_\xi
&F;\xi\models_H\varphi_1\land\varphi_2&\text{iff}&
  F;\xi\models_H\varphi_1\text{ and }F;\xi\models_H\varphi_2\\
F;\xi\models_H T_1=T_2&\text{iff}&\sem{T_1}^F_\xi=\sem{T_2}^F_\xi
&F;\xi\models_H\varphi_1\lor\varphi_2&\text{iff}&
  F;\xi\models_H\varphi_1\text{ or }F;\xi\models_H\varphi_2\\
F;\xi\models_H\lnot\varphi&\text{iff}&\text{not }F;\xi\models_H\varphi
&F;\xi\models_H\varphi_1\Rightarrow\varphi_2&\text{iff}&
  F;\xi\models_H\varphi_1\text{ implies }F;\xi\models_H\varphi_2\\
F;\xi\models_H\forall x{:}\mathcal D.\,\varphi&\text{iff}&
  \forall v\in\mathcal D_H.\ F;\xi[x\mapsto v]\models_H\varphi
&F;\xi\models_H\exists x{:}\mathcal D.\,\varphi&\text{iff}&
  \exists v\in\mathcal D_H.\ F;\xi[x\mapsto v]\models_H\varphi
\end{array}
\]
with $F;\xi\models_H\top$ always and $F;\xi\models_H\bot$ never, and
$\mathcal D_H$ the sites, paths or transport channels that $H$ declares.
For a closed formula we write $F\models_H\varphi$, and $F\models_H R$ when
every formula of the rule program $R$ holds on $F$: this is satisfaction of
the rules as the designer wrote them, as in
Theorem~\ref{thm:compile-correct}. Core rules have the same clauses without
quantifiers (\lean{RuleHolds}); we write $F\models r$ for a core rule and
$F\models\hat R$ when every core rule of $\hat R$ holds.

\FormalBox{Elaboration} $\jd{elab}_H^{\xi}(\varphi)$ translates a formula to
a core rule (\lean{Formula.lower}); Section~\ref{sec:elaboration} writes
$\jd{elab}_H(\varphi)\Downarrow\psi$ for the closed case. It replaces each
variable by its value in $\xi$ and each $\capa(s)$ by the declared constant,
is the identity on the connectives, and expands quantifiers over the
declarations of $H$:
\[
\jd{elab}_H^{\xi}(\forall x{:}\mathcal D.\,\varphi)
 =\bigwedge_{v\in\mathcal D_H}\jd{elab}_H^{\xi[x\mapsto v]}(\varphi)
\qquad\qquad
\jd{elab}_H^{\xi}(\exists x{:}\mathcal D.\,\varphi)
 =\bigvee_{v\in\mathcal D_H}\jd{elab}_H^{\xi[x\mapsto v]}(\varphi)
\]
An empty conjunction is $0=0$ and an empty disjunction is $\lnot(0=0)$.
A rule program $R$ elaborates to the list $\hat R=\jd{elab}_H(R)$ of its
elaborated formulas when every formula satisfies
$H;\cdot\vdash\varphi\;\jd{ok}$, and is rejected otherwise
(\lean{Assertions.elaborate}).

\begin{lemma}[Elaboration preserves meaning]
\label{lem:elaborate}
For every $H$, $\xi$, $F$ and $\varphi$:\quad
$F\models\jd{elab}_H^{\xi}(\varphi)$ iff $F;\xi\models_H\varphi$.
Consequently, if $R$ elaborates to $\hat R$, then $F\models\hat R$ iff
$F\models_H R$, for every frame $F$.\quad (\lean{Formula.lower_correct},
\lean{Assertions.elaborate_correct})
\end{lemma}
\begin{proof}
By induction on $\varphi$, for all $\xi$. Terms: the elaborated term has
the value $\sem{T}^F_\xi$, by induction on $T$; a variable is replaced by its
value in $\xi$ and $\capa(s)$ by the declared constant, which is the meaning
of both. Connectives: by the induction hypotheses. Quantifiers: a finite
conjunction holds iff each conjunct holds, and each conjunct is the
induction hypothesis at $\xi[x\mapsto v]$; dually for disjunction, and the
empty cases are $0=0$ (true) and $\lnot(0=0)$ (false), as the empty
quantifiers are. The consequence applies the first part with the empty
environment to each formula of $R$.
\end{proof}

\begin{lemma}[Rules are decided by evaluation]
\label{lem:decide}
There is a function $\jd{eval}(r,F)\in\{\jd{tt},\jd{ff}\}$ with
$\jd{eval}(r,F)=\jd{tt}$ iff $F\models r$.\quad
(\lean{evalRule_correct})
\end{lemma}
\begin{proof}
By induction on $r$: a comparison evaluates both terms to numbers, which
are computed from the finite frame, and each connective combines the
results of its parts as its truth table does.
\end{proof}

\noindent By the two lemmas, $\jd{eval}(\psi,F)=\jd{tt}$ for every
$\psi\in\hat R$ iff $F\models_H R$, which is the equation of
Section~\ref{sec:elaboration}. Hence \Rule{Step}, \Rule{Init} and core
legality below are decidable, and a step that the compiler accepts
satisfies the designer's quantified formulas, not only their expansions
(\lean{Assertions.run_satisfies}).

\FormalBox{Legal steps and runs} With the rule program $R$
(\lean{Dynamic.Legal}, \lean{Dynamic.Run}, \lean{Dynamic.InitialLegal}). In
Lean the rule premises are stated for the elaborated rules, $F\models\hat R$;
by Lemma~\ref{lem:elaborate} they are equivalent to the premises
$F\models_H R$ written here.
\begin{mathpar}
\inferrule*[right=Step]
  {H;\pos\vdash h\;\jd{core}\\
   \jd{frame}_H(\pos,h,t)\models_H R\\
   \jd{frame}_H(\pos,\rlz_H(\jd{ff},h),t)\models_H R}
  {H;R\vdash\langle\pos,t\rangle\steps{h}
   \langle\nxt_H(\pos,h),\ t+\delta_H(h)\rangle}
\and
\inferrule*[right=Run-Nil]{ }
  {H;R\vdash\langle\pos,t\rangle\runs{\epsilon}\langle\pos,t\rangle}
\and
\inferrule*[right=Run-Cons]
  {H;R\vdash\langle\pos,t\rangle\steps{h}\langle\pos_1,t_1\rangle\\
   H;R\vdash\langle\pos_1,t_1\rangle\runs{K}\langle\pos',t'\rangle}
  {H;R\vdash\langle\pos,t\rangle\runs{h;K}\langle\pos',t'\rangle}
\and
\inferrule*[right=Init]
  {H\vdash\pos:n_q\\ H\vdash\hat R\;\jd{ok}\\ \jd{init}_H(\pos)\models_H R}
  {H;R\vdash\pos\;\jd{init}}
\end{mathpar}
Here $\hat R=\jd{elab}_H(R)$, and $H\vdash\hat R\;\jd{ok}$ says that every
site, path and channel named by a core rule is declared (\S\ref{app:f-static}).
The rules can only reject a step:
\Rule{Step} keeps the core premise whatever $R$ is. The premises of
\Rule{Step} are the legality clause of Theorem~\ref{thm:compile-correct} for
one instruction: $H;\pos\vdash h\;\jd{core}$ is $F\models\jd{Core}_H$, checked
for both realizations of a guarded instruction (\S\ref{app:f-physical}), and
the rule premises are $F\models_H\varphi$ for every $\varphi\in R$.
Without the rule premises, the same two rules define a
\emph{core run} $H\vdash\pos\runs{K}_{\jd{core}}\pos'$
(\lean{CoreExecution}).

\FormalExample{ex:f-rules}{The rules on every frame, and the run.}
The rule program of Table~\ref{tab:f-example} has three assertions; each
is well sorted (Example~\ref{ex:f-operands}). Elaboration with the empty
environment gives the core rules
\[
\hat R=\bigl[\ \ (t=0\land\delta=0)\Rightarrow N^-(\mathrm S)=1,\qquad
 \Ntr(\mathrm S)\le1\land(\Ntr(\mathrm J)\le1\land(\Ntr(\mathrm G)\le2\land0=0)),\qquad
 1\le n_{\rm optical}\Rightarrow5\le t\ \ \bigr].
\]
\ensuremath{\rho_{\rm init}} and \ensuremath{\rho_{\rm rel}} have no variable and no $\capa$, so they
are unchanged. For \ensuremath{\rho_{\rm cap}}, the quantifier expands over
$\mathcal S_H=\mathrm S,\mathrm J,\mathrm G$: each conjunct is the
body with $s$ replaced by the site and $\capa(s)$ by the declared
capacity, and the final $0=0$ is the empty conjunction.
Table~\ref{tab:f-rules} evaluates $\hat R$ on the frames of
Table~\ref{tab:f-frames}, and Fig.~\ref{fig:f-eval} draws four of the
evaluations.

\begin{center}\begin{minipage}{\linewidth}\centering
\captionof{table}{The three rules on the frames of the feedback example.
An implication holds when its premise fails.}
\label{tab:f-rules}
\footnotesize\setlength{\tabcolsep}{5pt}
\begin{tabular}{@{}lllll@{}}
\toprule
frame & \ensuremath{\rho_{\rm init}}:\ \ $(t=0\land\delta=0)\Rightarrow N^-(\mathrm S)=1$
 & \ensuremath{\rho_{\rm cap}}:\ \ $\Ntr\le(1,1,2)$ on S, J, G
 & \ensuremath{\rho_{\rm rel}}:\ \ $1\le n_{\rm optical}\Rightarrow5\le t$ & result\\
\midrule
$F_0$ & premise $0=0\land0=0$ holds, and $1=1$ & $1\le1$, $0\le1$, $1\le2$ & premise $1\le0$ fails & $F_0\models_H R$\\
$F_1$ & premise fails: $\delta=2$ & $1\le1$, $1\le1$, $2\le2$ & premise $1\le0$ fails & $F_1\models_H R$\\
$F_2,F_3,F_4$ & premise fails: $t\ne0$ & $0\le1$, $0\le1$, $2\le2$ & premise $1\le0$ fails & holds\\
$F_5$ & premise fails & as above & $1\le1$ and $5\le5$ & holds\\
$F_6$ & premise fails & as above & $1\le1$ and $5\le7$ & holds\\
$F_7$ & premise fails & as above & $1\le2$ and $5\le11$ & holds\\
$F_8$, $F_9$ & premise fails & as above & $1\le1$ and $5\le13$, $5\le15$ & holds\\
$F_{10}$ & premise fails & as above & premise $1\le0$ fails & holds\\
$F_{11}$ & premise fails & as above & $1\le1$ and $5\le19$ & holds\\
$F_{11}^{\jd{ff}}$ & premise fails & as above & premise $1\le0$ fails & holds\\
\midrule
$\jd{frame}_H(\pos_1,h_5,2)$ & premise fails & as above & $1\le1$ holds, $5\le2$ fails & \ensuremath{\rho_{\rm rel}} fails\\
\bottomrule
\end{tabular}
\end{minipage}\end{center}

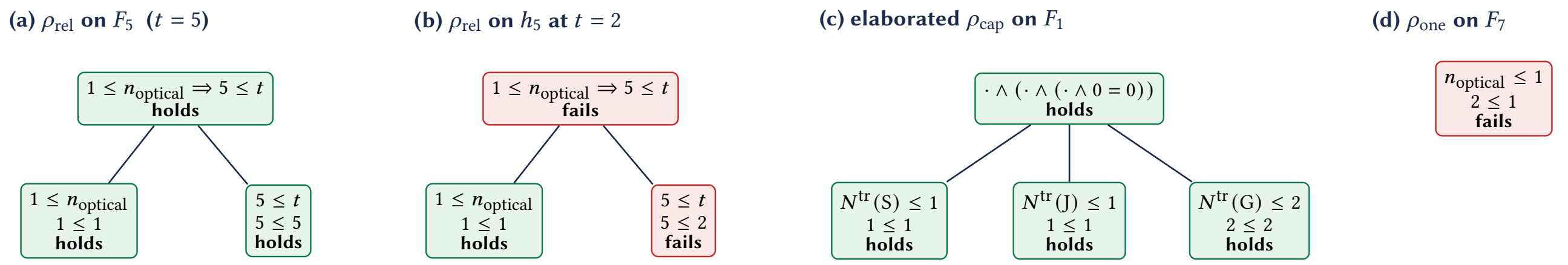
\FxFigure{%
\begin{tikzpicture}[x=1pt,y=1pt]
\path[use as bounding box] (0,7) rectangle (500,-96);
\node[fxhead] at (0,0) {(a) \ensuremath{\rho_{\rm rel}} on $F_5$\ \ ($t=5$)};
\FxEval{ra}{52}{-24}{$1\le n_{\rm optical}\Rightarrow5\le t$}{holds}{fxok}
\FxEval{rb}{22}{-62}{$1\le n_{\rm optical}$\\$1\le1$}{holds}{fxok}
\FxEval{rc}{84}{-62}{$5\le t$\\$5\le5$}{holds}{fxok}
\draw[fxwire] (ra)--(rb) (ra)--(rc);
\node[fxhead] at (126,0) {(b) \ensuremath{\rho_{\rm rel}} on $h_5$ at $t=2$};
\FxEval{sa}{178}{-24}{$1\le n_{\rm optical}\Rightarrow5\le t$}{fails}{fxbad}
\FxEval{sb}{148}{-62}{$1\le n_{\rm optical}$\\$1\le1$}{holds}{fxok}
\FxEval{sc}{210}{-62}{$5\le t$\\$5\le2$}{fails}{fxbad}
\draw[fxwire] (sa)--(sb) (sa)--(sc);
\node[fxhead] at (252,0) {(c) elaborated \ensuremath{\rho_{\rm cap}} on $F_1$};
\FxEval{ca}{330}{-24}{$\cdot\land(\cdot\land(\cdot\land0=0))$}{holds}{fxok}
\FxEval{cb}{274}{-62}{$\Ntr(\mathrm S)\le1$\\$1\le1$}{holds}{fxok}
\FxEval{cc}{330}{-62}{$\Ntr(\mathrm J)\le1$\\$1\le1$}{holds}{fxok}
\FxEval{cd}{386}{-62}{$\Ntr(\mathrm G)\le2$\\$2\le2$}{holds}{fxok}
\draw[fxwire] (ca)--(cb) (ca)--(cc) (ca)--(cd);
\node[fxhead] at (424,0) {(d) \ensuremath{\rho_{\rm one}} on $F_7$};
\FxEval{oa}{462}{-24}{$n_{\rm optical}\le1$\\$2\le1$}{fails}{fxbad}
\node[fxnote,anchor=north west] at (0,-84) {Each node shows a subformula, its instance on the
  frame, and its truth value. (b) is why the target code waits until tick 5; (d) is
  Example~\ref{ex:f-one}.};
\end{tikzpicture}}
{Rules are decided by evaluation on a frame (Lemma~\ref{lem:decide}).
The values of $n_{\rm optical}$, $t$ and $\Ntr$ are those of
Table~\ref{tab:f-frames}.}{fig:f-eval}

\noindent The table and the earlier examples give every premise of the
run:
\begin{center}\small\setlength{\tabcolsep}{4pt}
\begin{tabular}{@{}l>{\raggedright\arraybackslash}p{.86\linewidth}@{}}
\Rule{Init} & $H\vdash\pos_0:2$ (Example~\ref{ex:f-device}); $H\vdash\hat R\;\jd{ok}$,
  because $\hat R$ names only the sites S, J, G of $H$; and $F_0\models_H R$. Hence
  $H;R\vdash\pos_0\;\jd{init}$\\
\Rule{Step} & for $h_1$: $H;\pos_0\vdash h_1\;\jd{core}$ (Example~\ref{ex:f-run}) and
  $F_1\models_H R$; its two frames coincide, because $h_1$ is not guarded. Hence
  $H;R\vdash\langle\pos_0,0\rangle\steps{h_1}\langle\pos_1,2\rangle$. Likewise for
  $h_2$ to $h_{10}$ with $F_2$ to $F_{10}$\\
\Rule{Step} & for $h_{11}$: core legality needs both realizations, and the rule
  premises are $F_{11}\models_H R$ and $F_{11}^{\jd{ff}}\models_H R$. Hence
  $H;R\vdash\langle\pos_1,19\rangle\steps{h_{11}}\langle\pos_1,23\rangle$\\
\Rule{Run-Cons} & eleven times, then \Rule{Run-Nil}:\ \
  $H;R\vdash\langle\pos_0,0\rangle\runs{K}\langle\pos_1,23\rangle$
\end{tabular}
\end{center}
The last line of Table~\ref{tab:f-rules} explains the three waits. At
$\langle\pos_1,2\rangle$ the next instruction is the pulse $h_5$, but
\Rule{Step} does not apply to it: its frame has $n_{\rm optical}=1$ and
$t=2$, which falsifies \ensuremath{\rho_{\rm rel}}. \Rule{Step} does apply to
$\execkw{idle}\;1$, whose frame has no optical action, and after three
waits the clock reads 5. Lemma~\ref{lem:elaborate} on $F_1$: the formula
$\forall s{:}\mathcal S.\ \Ntr(s)\le\capa(s)$ holds on $F_1$ iff
$\sem{\Ntr(s)}^{F_1}_{[s\mapsto v]}\le\sem{\capa(s)}^{F_1}_{[s\mapsto v]}$
for $v=\mathrm S,\mathrm J,\mathrm G$, that is $1\le1$, $1\le1$ and
$2\le2$: the three comparisons of the elaborated rule
(Fig.~\ref{fig:f-eval}c).

\FormalExample{ex:f-one}{A rule that changes the target code.}
Add one assertion to the rule program,
$\rulekw{assert}\;\ensuremath{\rho_{\rm one}}:\ n_{\rm optical}\le1$, and keep the
source and the device. It is well sorted and elaborates to itself. On
the broadcast frame, $n_{\rm optical}=2$, so $F_7$ no longer satisfies
the rules (Fig.~\ref{fig:f-eval}d) and \Rule{Step} does not apply to
$h_7$. The two pulses
$\execkw{pulse}\;\irkw{R}(-\tfrac\pi2,0)\;q_0\;\execkw{via}\;k_5$ and
$\execkw{pulse}\;\irkw{R}(-\tfrac\pi2,0)\;q_1\;\execkw{via}\;k_5$ have
frames with $n_{\rm optical}=1$, which satisfy all four rules. For these
inputs the compiler returns target code with the two pulses in place of
the broadcast: twelve events that start at ticks
$0,2,3,4,5,7,11,13,15,17,20,21$ and finish at tick 25 instead of 23.
Nothing in the compiler or in the proofs below mentions this rule: both
are stated for an arbitrary $R$.

\section{Compilation rules}
\label{app:f-compile}

The compiler has nine stages (Fig.~\ref{fig:compiler-passes}). Each is
given by a function or by a judgment that relates the input of the stage
to its output. Stages~1 to~5 are functions. Stages~6 to~9 search, so
their judgments are relations: the search picks one output that the
judgment admits, or fails. The proof depends only on the judgment, not on
how the search picks.

\FormalBox{Stages 1 and 2: include resolution and domain analysis}
$\jd{incl}(P,r)$ replaces every \circkw{include} of the bundle $P$ with
root $r$ by the statements of the included unit, and fails on a missing
or cyclic include (\lean{Includes.resolveIncludes}). $\jd{dom}$ labels
each angle expression with a sign class and fails when it cannot show
that the expression is defined (\lean{Domain.analyzeProgram}). Neither
changes an operation.

\FormalBox{Stage 3: effect expansion}
$\Delta\vdash S\leadsto\Delta';\fseq e$ translates a statement to effect
instructions (\lean{expansionStep}); $\Delta\vdash O\leadsto_{\gm}\fseq e$
translates an operation under guard mark $\gm$ (\lean{expandOperation}).
\begin{mathpar}
\inferrule*[right=X-Decl]
  {S\text{ declares a register or gate}}
  {\Delta\vdash S\leadsto\Delta\oplus S;\ \epsilon}
\and
\inferrule*[right=X-Op]
  {\Delta\vdash O\leadsto_{\epsilon}\fseq e}
  {\Delta\vdash O\leadsto\Delta;\ \fseq e}
\and
\inferrule*[right=X-If]
  {\Delta\vdash a\mapsto\fseq c\\
   \Delta\vdash O\leadsto_{\scriptsize\wsaved}\fseq e}
  {\Delta\vdash\circkw{if}\,(a\,\texttt{==}\,n)\;O\leadsto\Delta;\
   \irkw{latch}\;\fseq c\,\texttt{==}\,n;\ \fseq e}
\and
\inferrule*[right=X-Barrier]
  {\Delta\vdash\fseq u\mapsto\fseq q\\ \fseq q\ne\epsilon\text{ distinct}}
  {\Delta\vdash\circkw{barrier}\;\fseq u\leadsto\Delta;\
   \irkw{barrier}\;\fseq q}
\and
\inferrule*[right=X-U]
  {\Delta\vdash u\mapsto q_1\ldots q_m}
  {\Delta\vdash\circkw{U}(\theta,\varphi,\lambda)\;u\leadsto_{\gm}
   \gm\;\irkw{U}(\theta,\varphi,\lambda)\;q_1;\ldots;
   \gm\;\irkw{U}(\theta,\varphi,\lambda)\;q_m}
\and
\inferrule*[right=X-CX]
  {\Delta\vdash u,u'\mapsto(q_1,q'_1)\ldots(q_m,q'_m)\\
   \forall j.\ q_j\ne q'_j}
  {\Delta\vdash\circkw{CX}\;u,u'\leadsto_{\gm}
   \gm\;\irkw{CX}\;q_1,q'_1;\ldots;\gm\;\irkw{CX}\;q_m,q'_m}
\and
\inferrule*[right=X-Call]
  {\Delta(a)=\circkw{gate}\;a(\fseq x)\;\fseq y\;\{\fseq B\}\\
   |\fseq\alpha|=|\fseq x|\\ |\fseq u|=|\fseq y|\\
   \Delta\vdash\fseq u\mapsto\fseq q\text{ distinct}\\
   \Delta\vdash\fseq B[\fseq\alpha/\fseq x,\fseq u/\fseq y]\leadsto_{\gm}\fseq e}
  {\Delta\vdash a(\fseq\alpha)\;\fseq u\leadsto_{\gm}\fseq e}
\and
\inferrule*[right=X-BodyBarrier]
  {\Delta\vdash\fseq u\mapsto\fseq q}
  {\Delta\vdash\circkw{barrier}\;\fseq u\leadsto_{\gm}\gm\;\irkw{barrier}\;\fseq q}
\and
\inferrule*[right=X-Measure]
  {\Delta\vdash u\mapsto q_1\ldots q_m\\ \Delta\vdash u'\mapsto c_1\ldots c_m}
  {\Delta\vdash\circkw{measure}\;u\to u'\leadsto_{\gm}
   \gm\;\irkw{measure}\;q_1\to c_1;\ldots;\gm\;\irkw{measure}\;q_m\to c_m}
\and
\inferrule*[right=X-Reset]
  {\Delta\vdash u\mapsto q_1\ldots q_m}
  {\Delta\vdash\circkw{reset}\;u\leadsto_{\gm}
   \gm\;\irkw{reset}\;q_1;\ldots;\gm\;\irkw{reset}\;q_m}
\end{mathpar}
A program is translated statement by statement, threading $\Delta$, and
the output is $P_E=\langle n_q,n_c,\fseq e\rangle$ with the register
sizes of the final $\Delta$. In \Rule{X-If} the register is read once, by
the latch, and every instruction of the expanded operation carries the
mark \wsaved. An operation with register operands is first replaced by
one instance per index, as OpenQASM prescribes; the operands of
\Rule{X-Call} are those of one instance, and a call of an \circkw{opaque}
gate has no rule, so it is rejected. \Rule{X-BodyBarrier} expands a barrier
inside a gate body under the guard mark of the call. The depth of
\Rule{X-Call} is bounded by the number of declared gates, so expansion
terminates.

\FormalBox{Stage 4: pulse selection} $\jd{sel}$ rewrites each unitary to
native pulses and keeps every other instruction
(\lean{selectInstruction}, \lean{Pulse.selectGate}):
\[
\begin{array}{@{}r@{\;}c@{\;}l@{}}
\jd{sel}(\gm\;\irkw{U}(\theta,\varphi,\lambda)\;q)&=&
 \gm\;\irkw{R}(\theta,\tfrac\pi2-\lambda)\;q;\ \
 \gm\;\irkw{R}(\pi,0)\;q;\ \
 \gm\;\irkw{R}(\pi,\tfrac{\varphi+\lambda}2)\;q\\
\jd{sel}(\gm\;\irkw{CX}\;q,q')&=&
 \gm\;\irkw{R}(\tfrac\pi2,\tfrac\pi2)\;q;\ \
 \gm\;\irkw{MS}(\tfrac\pi2)\;q,q';\ \
 \gm\;\irkw{R}(-\tfrac\pi2,0)\;q;\ \
 \gm\;\irkw{R}(-\tfrac\pi2,0)\;q';\ \
 \gm\;\irkw{R}(-\tfrac\pi2,\tfrac\pi2)\;q\\
\jd{sel}(e)&=&e\quad\text{for a measure, reset, barrier or latch}
\qquad\qquad\jd{sel}(e;\fseq e)=\jd{sel}(e);\jd{sel}(\fseq e)
\end{array}
\]

\FormalBox{Stage 5: capability binding} Let $\jd{elig}_H(i)$ be the list
of all declarations of $H$ that can play $i$: the calibrations $k$ with
$\jd{wave}(k)=\jd{wave}(w)$ for a pulse $w$, all acquisitions for a
measure, and all pumps for a reset. Binding annotates each instruction
with this list (\lean{Binding}):
\begin{mathpar}
\inferrule*[right=B-Optical]
  {n_q;n_c\vdash i\;\jd{ok}\\ i\text{ is a pulse, measure or reset}\\
   \jd{elig}_H(i)\ne\epsilon}
  {H\vdash i\leadsto i^{\,\jd{elig}_H(i)}}
\and
\inferrule*[right=B-Latch]
  {n_q;n_c\vdash i\;\jd{ok}\\ i\text{ is a latch}\\ \hwkw{controller}\in H}
  {H\vdash i\leadsto i^{\,\epsilon}}
\and
\inferrule*[right=B-Barrier]
  {n_q;n_c\vdash i\;\jd{ok}\\ i\text{ is a barrier}}
  {H\vdash i\leadsto i^{\,\epsilon}}
\end{mathpar}
Binding requires $\vdash H\;\jd{ok}$ and fails if any instruction has no
rule.

\FormalBox{Stage 6: initial placement} The stage outputs any placement
$\pos_0$ with $H;R\vdash\pos_0\;\jd{init}$ (\lean{Placed}).

\FormalBox{Stage 7: transport routing}
$H;\pos\vdash\fseq i\leadsto_{\jd{route}}A$ inserts transports and picks
one declaration for each instruction (\lean{Rules.Routes}). Let
$\jd{real}(i)$ be the set of instructions $h$ other than a broadcast with
$\jd{proj}(h)=i$: a pulse via some $k$, a readout or pump via some $a$, or the
latch or barrier itself (\lean{Rules.Real}).
\begin{mathpar}
\inferrule*[right=R-Done]{ }
  {H;\pos\vdash\epsilon\leadsto_{\jd{route}}\epsilon}
\and
\inferrule*[right=R-Move]
  {H;\pos\vdash\execkw{transport}\;\fseq m\;\jd{core}\\
   H;\nxt_H(\pos,\execkw{transport}\;\fseq m)\vdash\fseq i
     \leadsto_{\jd{route}}A}
  {H;\pos\vdash\fseq i\leadsto_{\jd{route}}\execkw{transport}\;\fseq m;\,A}
\and
\inferrule*[right=R-Emit]
  {h\in\jd{real}(i)\\ H;\pos\vdash h\;\jd{core}\\
   H;\pos\vdash\fseq i\leadsto_{\jd{route}}A}
  {H;\pos\vdash i;\fseq i\leadsto_{\jd{route}}h;\,A}
\end{mathpar}

\FormalBox{Stage 8: broadcast formation}
$H;\pos\vdash A\leadsto_{\jd{pack}}A'$ replaces runs of adjacent pulses by
declared broadcasts (\lean{Rules.Packs}). An instruction is \emph{plain}
when it is neither a wait nor a broadcast.
\begin{mathpar}
\inferrule*[right=P-Done]{ }
  {H;\pos\vdash\epsilon\leadsto_{\jd{pack}}\epsilon}
\and
\inferrule*[right=P-Keep]
  {h\text{ plain}\\ H;\pos\vdash h\;\jd{core}\\
   H;\nxt_H(\pos,h)\vdash A\leadsto_{\jd{pack}}A'}
  {H;\pos\vdash h;A\leadsto_{\jd{pack}}h;A'}
\and
\inferrule*[right=P-Join]
  {H;\pos\vdash\gm\;\execkw{broadcast}\;w_1\ldots w_m\;\execkw{via}\;
     b[k_1\ldots k_m]\;\jd{core}\\
   H;\pos\vdash A\leadsto_{\jd{pack}}A'}
  {H;\pos\vdash\gm\;\execkw{pulse}\;w_1\;\execkw{via}\;k_1;\ldots;
   \gm\;\execkw{pulse}\;w_m\;\execkw{via}\;k_m;\,A
   \ \leadsto_{\jd{pack}}\
   \gm\;\execkw{broadcast}\;w_1\ldots w_m\;\execkw{via}\;b[k_1\ldots k_m];\,A'}
\end{mathpar}
\Rule{P-Keep} alone gives $A'=A$. The compiler falls back to this choice
when stage 9 finds no timing for a packed $A'$.

\FormalBox{Stage 9: time assignment}
$H;R;\pos;t\vdash A'\leadsto_{\jd{time}}K$ gives every event its start time
and inserts waits (\lean{Rules.Times}):
\begin{mathpar}
\inferrule*[right=T-Done]{ }
  {H;R;\pos;t\vdash\epsilon\leadsto_{\jd{time}}\epsilon}
\and
\inferrule*[right=T-Emit]
  {h\text{ not a wait}\\
   H;R\vdash\langle\pos,t\rangle\steps{h}\langle\pos',t'\rangle\\
   H;R;\pos';t'\vdash A'\leadsto_{\jd{time}}K}
  {H;R;\pos;t\vdash h;A'\leadsto_{\jd{time}}h;K}
\and
\inferrule*[right=T-Wait]
  {H;R\vdash\langle\pos,t\rangle\steps{\execkw{idle}\;n}\langle\pos,t+n\rangle\\
   H;R;\pos;t+n\vdash A'\leadsto_{\jd{time}}K}
  {H;R;\pos;t\vdash A'\leadsto_{\jd{time}}\execkw{idle}\;n;\,K}
\end{mathpar}

\FormalBox{The compiler} For a source bundle $P$ with root unit $r$, a device
$H$, a rule program $R$ and search settings $\kappa$, we write
$\jd{compile}(H,R,P,\kappa)\Downarrow K$, with target code
$K=\langle\pos_0,h_1\ldots h_n\rangle$, when $R$ elaborates to $\hat R$
(the compiler checks $\hat R$, which by Lemma~\ref{lem:elaborate} decides
$\models_H R$) and there are $P_E$, $P_P=\langle n_q,n_c,\fseq i\rangle$, $A$
and $A'$ with (\lean{compileQASM}, \lean{QASMCompilation})
\[
\begin{array}{@{}l@{}}
\jd{dom}(\jd{incl}(P,r))\text{ succeeds and expands to }P_E
\qquad P_P=\jd{sel}(P_E)\qquad
\vdash H\;\jd{ok}\qquad H\vdash i\leadsto i^{\ldots}\text{ for every }i\in\fseq i\\
H;R\vdash\pos_0\;\jd{init}\qquad
H;\pos_0\vdash\fseq i\leadsto_{\jd{route}}A\qquad
H;\pos_0\vdash A\leadsto_{\jd{pack}}A'\qquad
H;R;\pos_0;0\vdash A'\leadsto_{\jd{time}}K
\end{array}
\]
Theorem~\ref{thm:compile-correct} writes $\jd{compile}(H,R,P,\kappa)\Downarrow(K,T)$:
after stage~9, the compiler also reports a timeline $T$ for $K$, which
Appendix~\ref{app:pf-objects} defines. The search settings $\kappa$ (search
bounds, a routing proposal from an external tool, the fallback router)
influence only which derivation the search finds; every output must have the
derivations above, so the proof does not depend on $\kappa$.

\begin{remark}[Rules and records]
In the compiler, each of stages 7 to 9 returns its output together with
a record of the facts that the proof needs (\lean{Routed},
\lean{Packed}, \lean{Timed}); for example, a \lean{Routed} value holds
a core run of $A$ from $\pos$, the equation $\jd{proj}(A)=\fseq i$, and the
facts that $A$ has no wait and no broadcast. The three judgments above
are inductive definitions in Lean, and we prove that an output has a
derivation exactly when it has the record (\lean{Rules.routes_iff},
\lean{Rules.packs_iff}, \lean{Rules.times_iff}). Every successful
compilation therefore consists of the derivations listed above
(\lean{Rules.compilation_derivations}).
\end{remark}
\clearpage
\section{The proof on one program}
\label{app:hoare-example}

Before Appendix~\ref{app:proof} proves Theorem~\ref{thm:compile-correct}
for every program, this appendix replays the proof on one small
program\ifqccdtutorial, Example~1 of Appendix~\ref{app:tutorial}\fi: the
feedback example of Example~\ref{ex:f-languages} without its conditional
reset, on the same device and rules. Each step applies the compilation rules
of Appendix~\ref{app:f-compile} to concrete inputs and checks, on these
inputs, the fact that the corresponding lemma of Appendix~\ref{app:proof}
establishes in general. Every rule used is defined in
Appendices~\ref{app:formal} and~\ref{app:f-compile}, and the programs after
each stage are the compiler's output for these inputs.
Section~\ref{app:hoare} then defines a Hoare logic for target code, and
Section~\ref{app:hoare-derivation} writes the same proof as one derivation in
it.

\subsection{Every stage of one compilation}
\label{app:replay-stages}

\FormalBox{Inputs}
\[
\begin{array}{@{}l@{\qquad}l@{}}
P:\ \ \circkw{qreg}\;\texttt{q[2]};\ \ \circkw{creg}\;\texttt{c[1]};\ \
 \circkw{CX}\;\texttt{q[0],q[1]};\ \
 \circkw{measure}\;\texttt{q[1]}\to\texttt{c[0]}
&\text{(source)}\\[3pt]
H:\ \ \hwkw{site}\;\mathrm S[1];\ \hwkw{site}\;\mathrm J[1];\
 \hwkw{site}\;\mathrm G[2];\ \
 \hwkw{path}\;p_0:\mathrm S\to\mathrm J\to\mathrm G\;
 \hwkw{channel}\;0\;\hwkw{waveform}\;0\;\hwkw{duration}\;2;\ \
 \hwkw{optical}\;\hwkw{channel}\;7;
&\text{(device)}\\
\phantom{H:\ \ }k_3:\hwkw{R}(\tfrac\pi2,\tfrac\pi2)\,@\mathrm G,\ d{=}2;\ \
 k_4:\hwkw{MS}(\tfrac\pi2)\,@\mathrm G,\ d{=}4;\ \
 k_5:\hwkw{R}(-\tfrac\pi2,0)\,@\mathrm G,\ d{=}2;\ \
 k_6:\hwkw{R}(-\tfrac\pi2,\tfrac\pi2)\,@\mathrm G,\ d{=}2;\\
\phantom{H:\ \ }\hwkw{broadcast}\;b_0:\hwkw{R}(-\tfrac\pi2,0)\;
 \hwkw{sites}\;\mathrm G,\ d{=}2,\ \hwkw{limit}\;2;\ \
 \hwkw{acquire}\;a_0\,@\mathrm G,\ d{=}3;\ \
 \text{all on optical channel 7}\\[3pt]
\rulekw{assert}\;\ensuremath{\rho_{\rm init}}:\ (t=0\land\delta=0)\Rightarrow N^-(\mathrm S)=1
\qquad
\rulekw{assert}\;\ensuremath{\rho_{\rm cap}}:\ \forall s{:}\mathcal S.\ \Ntr(s)\le\capa(s)
\qquad
\rulekw{assert}\;\ensuremath{\rho_{\rm rel}}:\ n_{\rm optical}>0\Rightarrow t\ge5
&\text{(rules)}
\end{array}
\]
The device also declares the calibrations $k_0$ to $k_2$, a pump and a
controller, which this program does not use\ifqccdtutorial; the
listings of Fig.~\ref{fig:tut-one-inputs} print $k_j$ as
\texttt{c}$j$\fi. $H$ satisfies (W1) to (W9),
so $\vdash H\;\jd{ok}$. Each assertion is well sorted (\Rule{A-Quant} and
\Rule{A-Var} for \ensuremath{\rho_{\rm cap}}), and elaboration gives
\[
\hat R=\ \bigl[\ (t=0\land\delta=0)\Rightarrow N^-(\mathrm S)=1,\ \ \
 \Ntr(\mathrm S)\le1\land\Ntr(\mathrm J)\le1\land\Ntr(\mathrm G)\le2,\ \ \
 n_{\rm optical}>0\Rightarrow t\ge5\ \bigr],
\]
where the trailing empty conjunction $0=0$ of Example~\ref{ex:f-rules} is
omitted. By Lemma~\ref{lem:elaborate}, $F\models\hat R$ iff $F\models_H R$,
that is, iff $F$ satisfies the three assertions.

\FormalBox{Step 1: stages 1 to 3} The source has no include and no angle,
so stages 1 and 2 return it unchanged. Expansion:
\[
\begin{array}{@{}l@{\quad}l@{}}
\text{\Rule{X-Decl}, twice}&
 \Delta=\{\texttt{q}\mapsto q_0q_1,\ \texttt{c}\mapsto c_0\},\ \text{no instruction}\\
\text{\Rule{X-Op} with \Rule{X-CX}}&
 \Delta\vdash\texttt{q[0]},\texttt{q[1]}\mapsto(q_0,q_1),\ q_0\ne q_1
 \ \ \text{gives}\ \ \irkw{CX}\;q_0,q_1\\
\text{\Rule{X-Op} with \Rule{X-Measure}}&
 \Delta\vdash\texttt{q[1]}\mapsto q_1,\ \Delta\vdash\texttt{c[0]}\mapsto c_0
 \ \ \text{gives}\ \ \irkw{measure}\;q_1\to c_0
\end{array}
\qquad
P_E=\langle2,1,\ \irkw{CX}\;q_0,q_1;\ \irkw{measure}\;q_1\to c_0\rangle
\]
Lemma~\ref{lem:expand}: for a one-branch ensemble $\sigma=(C,b,\rho)$,
\Rule{S-Unitary} and \Rule{S-Measure} give the source result
$(C[c_0{\mapsto}0],\Pi^0_{q_1}\cdot\CX\cdot\rho)\;
 (C[c_0{\mapsto}1],\Pi^1_{q_1}\cdot\CX\cdot\rho)$, and
$\sem{P_E}\sigma$ is the same two branches with the flag $b$.

\FormalBox{Step 2: stage 4} $\jd{sel}$ rewrites the \irkw{CX} and keeps
the measurement:
\[
P_P=\langle2,1,\ \ \irkw{R}(\tfrac\pi2,\tfrac\pi2)\;q_0;\ \
 \irkw{MS}(\tfrac\pi2)\;q_0,q_1;\ \ \irkw{R}(-\tfrac\pi2,0)\;q_0;\ \
 \irkw{R}(-\tfrac\pi2,0)\;q_1;\ \ \irkw{R}(-\tfrac\pi2,\tfrac\pi2)\;q_0;\ \
 \irkw{measure}\;q_1\to c_0\rangle
\]
Both instructions of $P_E$ are well formed (\Rule{Ok-MS} for the operands
of \irkw{CX}, \Rule{Ok-Measure}), so Lemma~\ref{lem:select} gives
$\sem{P_P}\sigma=\sem{P_E}\sigma$, by the second operator identity of its
proof.

\FormalBox{Step 3: stages 5 and 6} \Rule{B-Optical} applies to all six
instructions, with $\jd{elig}_H$ equal to
$k_3$, $k_4$, $k_5$, $k_5$, $k_6$ and $a_0$. The placement
$\pos_0=(\mathrm S,\mathrm G)$, that is $q_0$ at S and $q_1$ at G, is
valid by \Rule{WF-Place}: $N(\mathrm S)=1\le1$, $N(\mathrm J)=0\le1$,
$N(\mathrm G)=1\le2$. Its initial frame
$F_0=\jd{init}_H(\pos_0)$ has $t=\delta=0$, $N^-(\mathrm S)=1$,
$\Ntr=(1,0,1)$ on S, J, G, and $n_{\rm optical}=0$, so $F_0\models_H R$.
\Rule{Init} gives $H;R\vdash\pos_0\;\jd{init}$.

\FormalBox{Step 4: stage 7} With $\pos_1=(\mathrm G,\mathrm G)$:
\[
\begin{array}{@{}l@{\quad}l@{\quad}l@{}}
\text{\Rule{R-Move}}&\execkw{transport}\;q_0:p_0
 &\text{\Rule{C-Transport} at $\pos_0$: $\pos_0(q_0)=\mathrm S=\jd{src}(p_0)$;
  the next placement $\pos_1$ is valid ($N(\mathrm G)=2\le2$);}\\
 &&\text{$\Ntr=(1,1,2)\le(1,1,2)$ on S, J, G; the duration is $2>0$;
  one move, so $\jd{compat}$ and $\jd{complete}$ hold}\\
\text{\Rule{R-Emit}}&\execkw{pulse}\;\irkw{R}(\tfrac\pi2,\tfrac\pi2)\;q_0\;\execkw{via}\;k_3
 &\text{\Rule{C-Pulse} at $\pos_1$: $\jd{wave}(k_3)$ is the waveform;
  $\pos_1(q_0)=\mathrm G=\jd{site}(k_3)$}\\
\text{\Rule{R-Emit}}&\execkw{pulse}\;\irkw{MS}(\tfrac\pi2)\;q_0,q_1\;\execkw{via}\;k_4
 &\text{\Rule{C-Pulse}: $\pos_1(q_0)=\pos_1(q_1)=\mathrm G=\jd{site}(k_4)$}\\
\text{\Rule{R-Emit}}&\execkw{pulse}\;\irkw{R}(-\tfrac\pi2,0)\;q_0\;\execkw{via}\;k_5
 &\text{\Rule{C-Pulse}}\\
\text{\Rule{R-Emit}}&\execkw{pulse}\;\irkw{R}(-\tfrac\pi2,0)\;q_1\;\execkw{via}\;k_5
 &\text{\Rule{C-Pulse}}\\
\text{\Rule{R-Emit}}&\execkw{pulse}\;\irkw{R}(-\tfrac\pi2,\tfrac\pi2)\;q_0\;\execkw{via}\;k_6
 &\text{\Rule{C-Pulse}}\\
\text{\Rule{R-Emit}}&\execkw{readout}\;q_1\to c_0\;\execkw{via}\;a_0
 &\text{\Rule{C-Readout}: $\pos_1(q_1)=\mathrm G=\jd{site}(a_0)$}\\
\text{\Rule{R-Done}}
\end{array}
\]
No instruction is guarded, so $\rlz_H(\jd{ff},h)=h$ and each line is
core legal. Call the seven instructions $A$. By Lemma~\ref{lem:route},
$\jd{proj}(A)$ is the code of $P_P$ and $\sem{A}_H\sigma=\sem{P_P}\sigma$: the
transport contributes nothing to either.

\FormalBox{Step 5: stage 8} \Rule{P-Keep} keeps the transport and the
first two pulses, which are plain. \Rule{P-Join} with $m=2$ replaces the two pulses via
$k_5$ by the broadcast
\[
\execkw{broadcast}\;\irkw{R}(-\tfrac\pi2,0)\;q_0,\
 \irkw{R}(-\tfrac\pi2,0)\;q_1\;\execkw{via}\;b_0[k_5,k_5],
\]
which is core legal by \Rule{C-Broadcast}: the ions $q_0,q_1$ are distinct; each member
is legal by \Rule{C-Pulse}; the waveform, channel 7 and duration 2 of
$k_5$ are those of $b_0$; $\mathrm G\in\jd{sites}(b_0)$; and two members
at G do not exceed the limit 2. \Rule{P-Keep} keeps the last pulse and
the readout, and \Rule{P-Done} ends. Call the result $A'$ (six
instructions). By Lemma~\ref{lem:pack},
$\sem{A'}_H\sigma=\sem{A}_H\sigma$: the broadcast acts by
$\Rp(-\tfrac\pi2,0)_{q_0}\Rp(-\tfrac\pi2,0)_{q_1}$, the product of the
two pulses.

\FormalBox{Step 6: stage 9} The rule \ensuremath{\rho_{\rm rel}} rejects an optical
action before tick 5, so the derivation uses \Rule{T-Wait} three times
after the transport. Each line below is one use of \Rule{Step}; its two
frames coincide because nothing is guarded.
\begin{center}\small
\setlength{\tabcolsep}{5pt}
\begin{tabular}{@{}lllccccl@{}}
\toprule
rule & instruction $h$ & $\pos$ & $t$ & $\delta$ & $\Ntr$ on S, J, G &
$n_{\rm optical}$ & why $\jd{frame}_H(\pos,h,t)\models_H R$\\
\midrule
\Rule{T-Emit} & \execkw{transport} $q_0:p_0$ & $\pos_0$ & 0 & 2 & 1, 1, 2 & 0 &
 \ensuremath{\rho_{\rm init}}: $\delta\ne0$;\ \ \ensuremath{\rho_{\rm cap}}: $1\le1$, $1\le1$, $2\le2$;\ \ \ensuremath{\rho_{\rm rel}}: no optical action\\
\Rule{T-Wait} & \execkw{idle} 1 & $\pos_1$ & 2 & 1 & 0, 0, 2 & 0 & \ensuremath{\rho_{\rm cap}}: $2\le2$;\ \ \ensuremath{\rho_{\rm rel}}: no optical action\\
\Rule{T-Wait} & \execkw{idle} 1 & $\pos_1$ & 3 & 1 & 0, 0, 2 & 0 & as above\\
\Rule{T-Wait} & \execkw{idle} 1 & $\pos_1$ & 4 & 1 & 0, 0, 2 & 0 & as above\\
\Rule{T-Emit} & \execkw{pulse} \irkw{R} $q_0$ \execkw{via} $k_3$ & $\pos_1$ & 5 & 2 & 0, 0, 2 & 1 & \ensuremath{\rho_{\rm rel}}: $5\ge5$\\
\Rule{T-Emit} & \execkw{pulse} \irkw{MS} $q_0,q_1$ \execkw{via} $k_4$ & $\pos_1$ & 7 & 4 & 0, 0, 2 & 1 & \ensuremath{\rho_{\rm rel}}: $7\ge5$\\
\Rule{T-Emit} & \execkw{broadcast} \irkw{R} $q_0,q_1$ \execkw{via} $b_0$ & $\pos_1$ & 11 & 2 & 0, 0, 2 & 2 & \ensuremath{\rho_{\rm rel}}: $11\ge5$\\
\Rule{T-Emit} & \execkw{pulse} \irkw{R} $q_0$ \execkw{via} $k_6$ & $\pos_1$ & 13 & 2 & 0, 0, 2 & 1 & \ensuremath{\rho_{\rm rel}}: $13\ge5$\\
\Rule{T-Emit} & \execkw{readout} $q_1\to c_0$ \execkw{via} $a_0$ & $\pos_1$ & 15 & 3 & 0, 0, 2 & 1 & \ensuremath{\rho_{\rm rel}}: $15\ge5$\\
\Rule{T-Done} & & $\pos_1$ & 18\\
\bottomrule
\end{tabular}
\end{center}
After the transport the next instruction of $A'$ is an optical pulse.
\Rule{T-Emit} does not apply to it before tick 5: its \Rule{Step}
premise needs a frame with $n_{\rm optical}=1$ and $t<5$, which falsifies
\ensuremath{\rho_{\rm rel}}. Only \Rule{T-Wait} applies there. Call the nine instructions $K$. \Rule{Run-Cons}, nine
times, and \Rule{Run-Nil} give
$H;R\vdash\langle\pos_0,0\rangle\runs{K}\langle\pos_1,18\rangle$, and by
Lemma~\ref{lem:time}, $\sem{K}_H\sigma=\sem{A'}_H\sigma$.

\FormalBox{Conclusion} The sequential contract (Theorem~\ref{thm:formal}) for
this compilation reads: for every $\sigma=(C,b,\rho)$,
\[
\jd{obs}(\sem{K}_H\sigma)
 =\jd{obs}(\sem{A'}_H\sigma)=\jd{obs}(\sem{A}_H\sigma)
 =\jd{obs}(\sem{P_P}\sigma)=\jd{obs}(\sem{P_E}\sigma)
 =(C[c_0{\mapsto}0],\Pi^0_{q_1}\cdot\CX\cdot\rho)\;
  (C[c_0{\mapsto}1],\Pi^1_{q_1}\cdot\CX\cdot\rho),
\]
the result of the source, by Steps 6, 5, 4, 2 and 1 in this order;
$\jd{proj}(K)$ is the code of $P_P$; and $\vdash H\;\jd{ok}$,
$H;R\vdash\pos_0\;\jd{init}$ and the run above hold.

\subsection{Hoare logic for target code}
\label{app:hoare}

The same proof can be written as one derivation in a Hoare logic for target
code. The logic reasons on a fixed device $H$ with rules $R$ and angle
environment $\eta$; we write $\Gamma=(H,R,n_q,n_c,\eta)$. (In Lean,
\lean{Hoare.Context} takes an arbitrary state type and meaning function in
place of $\eta$ and ensembles; the logic below is its instance for the
semantics of \S\ref{app:f-semantics}, and the lemma below holds for every
instance.) A configuration is $x=(\pos,t,\sigma)$: placement, clock and
ensemble. An assertion $\Phi$, $\Psi$, $\Xi$ is a predicate on
configurations (\lean{Hoare.Assertion}); it may mention $\pos$, $t$ and
$\sigma$. One instruction acts on a configuration by
\[
\jd{step}_h(\pos,t,\sigma)=\bigl(\nxt_H(\pos,h),\ t+\delta_H(h),\
  \sem{h}_{H,\eta}^{\dagger}\sigma\bigr)
\qquad\qquad
\jd{legal}_h(\pos,t,\sigma)\ \text{ iff }\
H;R\vdash\langle\pos,t\rangle\steps{h}\langle\_,\_\rangle .
\]
A triple $\Gamma\vdash\{\Phi\}\;K\;\{\Psi\}$, where $K$ is a list of
instructions, is derived by four rules (\lean{Hoare.Derivation}):
\begin{mathpar}
\inferrule*[right=H-Empty]{ }{\Gamma\vdash\{\Phi\}\;\epsilon\;\{\Phi\}}
\and
\inferrule*[right=H-Prim]{ }
  {\Gamma\vdash\{\jd{legal}_h\land\Psi\circ\jd{step}_h\}\;h\;\{\Psi\}}
\and
\inferrule*[right=H-Seq]
  {\Gamma\vdash\{\Phi\}\;K_1\;\{\Xi\}\\ \Gamma\vdash\{\Xi\}\;K_2\;\{\Psi\}}
  {\Gamma\vdash\{\Phi\}\;K_1;K_2\;\{\Psi\}}
\and
\inferrule*[right=H-Conseq]
  {\Phi\Rightarrow\Phi'\\ \Gamma\vdash\{\Phi'\}\;K\;\{\Psi'\}\\ \Psi'\Rightarrow\Psi}
  {\Gamma\vdash\{\Phi\}\;K\;\{\Psi\}}
\end{mathpar}
\Rule{H-Prim} is the assignment axiom of Hoare
logic~\cite{hoare1969axiomatic} with one addition: the precondition
requires the step to be legal. A triple is \emph{valid},
$\Gamma\models\{\Phi\}\;K\;\{\Psi\}$, when from every configuration that
satisfies $\Phi$ the instructions $K$ run legally to their end and the final
configuration satisfies $\Psi$ (\lean{Hoare.Valid}). Target code is finite,
so this is total correctness.

\begin{lemma}[Soundness and relative completeness]
\label{lem:hoare}
$\Gamma\vdash\{\Phi\}\;K\;\{\Psi\}$ iff $\Gamma\models\{\Phi\}\;K\;\{\Psi\}$.\quad
(\lean{Hoare.sound}, \lean{Hoare.complete})
\end{lemma}
\begin{proof}
Soundness, by induction on the derivation. \Rule{H-Empty}: the empty
run. \Rule{H-Prim}: the precondition gives the legal step, and $\Psi$ holds
after it. \Rule{H-Seq}: run $K_1$ to a configuration that satisfies $\Xi$,
then $K_2$ from that configuration. \Rule{H-Conseq}: compose the
implications. Completeness: define the weakest precondition by
$\jd{wp}(\epsilon,\Psi)=\Psi$ and
$\jd{wp}(h;K,\Psi)=\jd{legal}_h\land\jd{wp}(K,\Psi)\circ\jd{step}_h$. By
induction on $K$, $\Gamma\vdash\{\jd{wp}(K,\Psi)\}\;K\;\{\Psi\}$, using
\Rule{H-Prim} and \Rule{H-Seq} (\lean{Hoare.wp_derivation}); a valid
triple has $\Phi\Rightarrow\jd{wp}(K,\Psi)$, and \Rule{H-Conseq} concludes.
Completeness is relative to the assertions being arbitrary predicates.
\end{proof}

\subsection{The same proof as a Hoare derivation}
\label{app:hoare-derivation}

Let
$W=\Rp(-\tfrac\pi2,\tfrac\pi2)_{q_0}\Rp(-\tfrac\pi2,0)_{q_1}
\Rp(-\tfrac\pi2,0)_{q_0}\MS(\tfrac\pi2)\Rp(\tfrac\pi2,\tfrac\pi2)_{q_0}$,
and write $U\cdot\sigma$ for $U$ acting on the matrix of every branch.
The annotated target code below is a derivation of the triple
\[
\Gamma\vdash\{\pos=\pos_0\land t=0\land\sigma=\sigma_0\}\ \ K\ \
\{H\vdash\pos:2\ \land\ t=18\ \land\ P\vdash\jd{obs}(\sigma_0)\Downarrow\jd{obs}(\sigma)\}
\]
for $\sigma_0=(C,b,\rho)$, the instance of Corollary~\ref{cor:hoare} for this
compilation. Each instruction with the assertion before and after it is an
instance of \Rule{H-Prim} followed by \Rule{H-Conseq}: the assertion before it
implies $\jd{legal}_h$ by the corresponding line of Step~6, and implies the
assertion after it composed with $\jd{step}_h$ by the definitions of
$\nxt_H$, $\delta_H$ and $\sem{h}_H$. \Rule{H-Seq} joins the nine triples;
the three waits share one line.
\[
\begin{array}{@{}l@{\qquad}l@{}}
\{\pos=\pos_0\land t=0\land\sigma=\sigma_0\}\\
\quad\execkw{transport}\;q_0:p_0
 &\{\pos=\pos_1\land t=2\land\sigma=\sigma_0\}\\
\quad\execkw{idle}\;1;\ \execkw{idle}\;1;\ \execkw{idle}\;1
 &\{\pos=\pos_1\land t=5\land\sigma=\sigma_0\}\\
\quad\execkw{pulse}\;\irkw{R}(\tfrac\pi2,\tfrac\pi2)\;q_0\;\execkw{via}\;k_3
 &\{\pos=\pos_1\land t=7\land\sigma=\Rp(\tfrac\pi2,\tfrac\pi2)_{q_0}\cdot\sigma_0\}\\
\quad\execkw{pulse}\;\irkw{MS}(\tfrac\pi2)\;q_0,q_1\;\execkw{via}\;k_4
 &\{\pos=\pos_1\land t=11\land\sigma=
   (\MS(\tfrac\pi2)\Rp(\tfrac\pi2,\tfrac\pi2)_{q_0})\cdot\sigma_0\}\\
\quad\execkw{broadcast}\;\irkw{R}(-\tfrac\pi2,0)\;q_0,q_1\;\execkw{via}\;b_0
 &\{\pos=\pos_1\land t=13\land\sigma=
   (\Rp(-\tfrac\pi2,0)_{q_1}\Rp(-\tfrac\pi2,0)_{q_0}
    \MS(\tfrac\pi2)\Rp(\tfrac\pi2,\tfrac\pi2)_{q_0})\cdot\sigma_0\}\\
\quad\execkw{pulse}\;\irkw{R}(-\tfrac\pi2,\tfrac\pi2)\;q_0\;\execkw{via}\;k_6
 &\{\pos=\pos_1\land t=15\land\sigma=W\cdot\sigma_0\}\\
\quad\execkw{readout}\;q_1\to c_0\;\execkw{via}\;a_0
 &\{\pos=\pos_1\land t=18\land\sigma=
   (C[c_0{\mapsto}0],b,\Pi^0_{q_1}\cdot W\cdot\rho)\;
   (C[c_0{\mapsto}1],b,\Pi^1_{q_1}\cdot W\cdot\rho)\}
\end{array}
\]
A last use of \Rule{H-Conseq} rewrites the postcondition: $\pos_1$ is
valid by \Rule{WF-Place}; $W=e^{i\pi/4}\CX$, so $W\cdot\rho=\CX\cdot\rho$, by
the operator identity in the proof of Lemma~\ref{lem:select}; and the two
branches, observed, are the result that \Rule{S-Unitary} and
\Rule{S-Measure} give the source. The derivation uses the rules
\Rule{H-Prim}, \Rule{H-Seq} and \Rule{H-Conseq} and no other fact about the
compiler: target code is correct when it has such a derivation, however it
was found.
\clearpage
\providecommand{\jd}[1]{\mathsf{#1}}
\providecommand{\sem}[1]{\llbracket#1\rrbracket}
\providecommand{\fseq}[1]{\overline{#1}}
\providecommand{\nxt}{\jd{next}}
\providecommand{\dur}{\jd{dur}}
\providecommand{\rlz}{\jd{realize}}
\providecommand{\Ntr}{N^{\mathrm{tr}}}
\providecommand{\lean}[1]{{\upshape\urlstyle{tt}\nolinkurl{#1}}}
\providecommand{\wsaved}{\irkw{when}\;\irkw{saved}}
\providecommand{\hsaved}{\execkw{when}\;\execkw{saved}}
\providecommand{\runs}[1]{\stackrel{#1}{\Longrightarrow}}
\providecommand{\steps}[1]{\stackrel{#1}{\longrightarrow}}
\newcommand{\pfres}{\jd{res}}
\newcommand{\pfresb}{\jd{res}^{\flat}}
\newcommand{\pfslots}{\jd{slots}}
\newcommand{\pfkey}{\jd{key}}
\newcommand{\pfglob}{\jd{glob}}
\newcommand{\pfconf}{\mathrel{\#}}
\newcommand{\pfready}{\jd{ready}}
\newcommand{\pfavail}{\jd{av}}
\newcommand{\pffloor}{\jd{fl}}
\newcommand{\pfhz}{\jd{hz}}
\newcommand{\pfplays}{\jd{Pl}}
\newcommand{\pferase}{\jd{erase}}
\newcommand{\pfcheck}{\jd{chain}_H}
\newcommand{\pfagree}{\sim}
\newcommand{\pfcovers}{\jd{Covers}}
\newcommand{\pfclear}{\jd{Clear}}
\newcommand{\pfchoose}{\jd{choose}}
\newcommand{\pfsort}{\jd{byStart}}
\newcommand{\pfins}{\jd{insert}}
\newcommand{\pfind}{\jd{Ind}}
\newcommand{\pfperm}{\simeq}
\newcommand{\pfobs}{\jd{obs}}
\newcommand{\pfinit}{\jd{init}}
\newcommand{\pfframe}{\jd{frame}}
\newcommand{\pfxframe}{\jd{xframe}}
\newcommand{\pfend}{\jd{end}}
\newcommand{\pfmode}{\beta}
\newcommand{\pfStep}[1]{\par\smallskip\noindent\textit{#1}\quad}
\raggedbottom

\section{Complete proof of Theorem~\ref{thm:compile-correct}}
\label{app:proof}

Theorem~\ref{thm:compile-correct} states, for every hardware description, rule
program, source program and search configuration, that target code returned by
the compiler has the meaning of its source, obeys the hardware rules, and is
charged the time of the declared hardware model.
Appendices~\ref{app:formal} and~\ref{app:f-compile} define every language,
judgment and compilation rule, and Appendix~\ref{app:hoare-example} replays the
proof on one program. This appendix gives the complete proof for every
input: it restates the theorem with every clause made precise
(\S\ref{app:pf-statement}), defines the objects of the reported timeline
(\S\ref{app:pf-objects}), and then proves every lemma in full: the sequential
contract (\S\ref{app:pf-sequential}), chain order (\S\ref{app:pf-chain}), the
schedules (\S\ref{app:pf-schedule}), commutation of independent events
(\S\ref{app:pf-commute}) and the meaning of the timeline in order of time
(\S\ref{app:pf-order}). Section~\ref{app:pf-assembly} assembles the theorem.
The proof is a human-readable account of the Lean development; each lemma
names the Lean declaration that checks it, and Table~\ref{tab:pf-map} collects
them.

\paragraph{How the proof is organized}
Why is the proof longer than a chain of pass lemmas? The compiler produces two
objects. The \emph{target code} is one sequence of instructions in which every
event starts when its predecessor ends. The \emph{timeline} reports the same
instructions with chain exchanges inserted and lets independent events overlap.
The source passes and the physical passes are proved about the first object, by
composing one contract per pass, as in CompCert~\cite{leroy2009compcert}. The
second object needs three further arguments that no pass lemma provides: the
scheduler never overlaps two events that share a resource; events that share
no resource act on the quantum and classical state in either order with the
same result; and sorting the events by start time only exchanges such
independent neighbours. Together they transfer the meaning of the sequential
code to the parallel timeline.

\paragraph{Notation}
We keep the notation of the main text and of Appendices~\ref{app:formal}
and~\ref{app:f-compile}. The target code returned by stage~9 is
$K=\langle\pos_0,h_1\ldots h_n\rangle$; $A$ and $A'$ are the outputs of
stages~7 and~8; $\jd{proj}$ is the projection to the pulse program; $R$ is the
rule program as written and $\hat R$ its elaboration. The placement after the
first $k$ instructions is
$\pos_k=\nxt_H(\pos_{k-1},h_k)$, and $t^{\rm seq}_k$ is the start time of $h_k$
in the run of stage~9. For lists, $x;\fseq x$ is a list with head $x$,
$\fseq x\,\fseq y$ is concatenation, and $\sigma\pfperm\sigma'$ says that two
ensembles contain the same branches, possibly listed in a different order.
Natural numbers start at 0, and $\max\emptyset=0$.

\subsection{The theorem, clause by clause}
\label{app:pf-statement}

\paragraph{Hypotheses}
The theorem has three hypotheses, and each states only what a procedure
returned.
\begin{itemize}[leftmargin=2.2em]
\item[(H1)] The rule program $R$ that the designer wrote elaborates to the
  core rules $\hat R=\jd{elab}_H(R)$ (\S\ref{app:f-assertions}).
\item[(H2)] The compiler, given $H$, $R$, the source bundle $P$ and the
  search configuration $\kappa$, returns a compilation,
  $\jd{compile}(H,R,P,\kappa)\Downarrow K$: the effect program $P_E$, the
  pulse program $P_P=\jd{sel}(P_E)$, written $\jd{native}(P)$ in
  Theorem~\ref{thm:compile-correct}, and the target code $K$
  (Appendix~\ref{app:f-compile}).
\item[(H3)] $T$ is a timeline reported for $K$ (Definition~\ref{def:pf-timeline}):
  steps $e_1\ldots e_m$, a timing that is \emph{parallel} or
  \emph{sequential}, start times $t_1\ldots t_m$ and a finish time
  $t_{\rm end}$. The scheduling mode $\pfmode$ is \emph{exclusive} or
  \emph{broadcast}.
\end{itemize}
Write $\delta_i$ for the duration of step $e_i$ and $\tau_i=[t_i,t_i+\delta_i)$.
For a compiled step $e_i=h_k$ let $\pos^{(i)}=\pos_{k-1}$, and for an exchange
let $\pos^{(i)}$ be the placement after the compiled steps before it.

\begin{theorem}[Verified compilation, complete statement]
\label{thm:pf-main}
Under (H1) to (H3), for every scheduling mode and either timing:
\par\smallskip\noindent\textbf{Meaning.}
\begin{itemize}[leftmargin=2.6em]
\item[(M1)] $\pferase(e_1\ldots e_m)=h_1\ldots h_n$: the steps are the target
  code with exchanges inserted.
\item[(M2)] $\jd{proj}(h_1\ldots h_n)=P_P$: deleting transports and waits and
  unpacking broadcasts gives the native program.
\item[(M3)] For every angle environment $\eta$ and every ensemble $\sigma$ there
  is exactly one $\sigma'$ with $P\vdash_\eta\pfobs(\sigma)\Downarrow\sigma'$,
  and $\pfobs(\sem{K^{T}}_{H,\eta}\sigma)\pfperm\sigma'$, where $K^{T}$ is the
  target code in order of the start times of $T$ (Definition~\ref{def:pf-order}).
\end{itemize}
\textbf{Legality.}
\begin{itemize}[leftmargin=2.6em]
\item[(L1)] $\vdash H\;\jd{ok}$, $H;R\vdash\pos_0\;\jd{init}$, and
  $H;R\vdash\langle\pos_0,0\rangle\runs{K}\langle\pos_n,t_{\rm seq}\rangle$.
\item[(L2)] For every compiled step $e_i=h_k$: $H;\pos_{k-1}\vdash h_k\;\jd{core}$,
  $\pfframe_H(\pos_{k-1},h_k,t_i)\models_H R$ and
  $\pfframe_H(\pos_{k-1},\rlz_H(\jd{ff},h_k),t_i)\models_H R$. For every exchange
  step $e_i$: $\pfxframe(\pos^{(i)},t_i,\delta_i)\models_H R$. Here
  $F\models_H R$ is satisfaction of the rules as the designer wrote them, and
  the compiler decides it by evaluating $\hat R$ (Lemma~\ref{lem:elaborate}).
\item[(L3)] The chain machine accepts the steps,
  $\pfcheck(e_1\ldots e_m)\Downarrow\chi_m$ from the declared initial chains
  $\chi_0$, every departure that it performs takes the ion at the end of its
  chain, and $\chi_m\sim\pos_n$.
\item[(L4)] For all $i<j$: $e_i\pfconf_\pfmode e_j\Rightarrow t_i+\delta_i\le t_j$.
\item[(L5)] For all $i\neq j$ and every shared transport channel $c$ that both
  steps use: $\tau_i\cap\tau_j=\emptyset$, or $w_i(c)=w_j(c)$ and
  $\tau_i=\tau_j$.
\item[(L6)] If $\jd{Uniform}(H)$, then any two moves $a,b$ of one transport
  of $K$ on one shared channel have the same motion, $\mu(a)=\mu(b)$, and the
  same duration, $d(a)=d(b)$.
\end{itemize}
\textbf{Time.}
\begin{itemize}[leftmargin=2.6em]
\item[(T1)] $t_{\rm end}=\max_i(t_i+\delta_i)\le\sum_i\delta_i$, with
  equality on the right for the sequential timing.
\item[(T2)] For every pulse $h_k=\gm\;\execkw{pulse}\;w\;\execkw{via}\;k$ of
  $K$: $k$ is a declared calibration, $\delta_H(h_k)=d(k)$, and every ion of
  $w$ is at $\jd{site}(k)$ in $\pos_{k-1}$.
\end{itemize}
\end{theorem}

The clauses refine the displayed equations of
Theorem~\ref{thm:compile-correct}: (M1) to (M3) are Eq.~\eqref{eq:meaning},
(L1) to (L6) are Eq.~\eqref{eq:legality}, and (T1), (T2) are
Eq.~\eqref{eq:time}. Clause (L1) is the sequential run of stage~9, and (L2)
checks every frame $F_i$ of Eq.~\eqref{eq:legality}, at the reported start
time $t_i$: for a compiled step, $H;\pos\vdash h\;\jd{core}$ is
$F_i\models\jd{Core}_H$, and for a guarded instruction both realizations are
checked. For an exchange, the core condition is the chain machine's (two
distinct ions of one trap, (L3)), and its frame is checked against the rules.
(L5) is the activation clause: two steps on one broadcast channel are apart
in time or form one activation. The Lean statement
\lean{verified_compilation} records (L4) and (L5) for the parallel timing and
the sequential start times for the sequential timing; Lemma~\ref{lem:pf-seq}
derives (L4), (L5) and (T1) from the latter.

\subsection{Objects of the reported timeline}
\label{app:pf-objects}

The timeline adds three kinds of objects to those of
Appendix~\ref{app:formal}: chain exchanges and the machine that replays them,
the resources that an event holds, and the two schedulers that compute start
times. We define them before they are used.

\begin{definition}[Steps]
\label{def:pf-steps}
A \emph{step} is a compiled instruction $h$ or an exchange
$\jd{exch}(s,a,b)$ of two ions $a,b$ in the trap $s$. The erasure
$\pferase(e_1\ldots e_m)$ keeps the compiled instructions, in order, and
drops the exchanges. An exchange reorders ions inside one trap and applies no
instruction: its projection is empty, $\jd{proj}(\jd{exch}(s,a,b))=\epsilon$, and it
does not change the placement. Its frame is
$\pfxframe(\pos,t,\delta)=(H,\pos,\pos,\bot,t,\delta)$, on which every count
of moves and optical actions is 0.
(\lean{Chain.Step}, \lean{Chain.erase}, \lean{Reported.exchangeFrame})
\end{definition}

\begin{definition}[Chain machine]
\label{def:pf-chain}
A linear trap orders its ions, and an ion leaves only through the end that
faces its transport segment. A \emph{chain state} $\chi=(\jd{ch},\jd{tr})$
assigns each site $s$ a list $\jd{ch}(s)$ of ions, left end first, and lists
the ions $\jd{tr}$ in transit. The device declares, for each path $p$, the end
$\jd{lv}(p)\in\{\mathrm L,\mathrm R\}$ through which it leaves its source and
the end $\jd{en}(p)$ through which it enters its destination, and for each
site $s$ the duration $\jd{xd}(s)$ of one exchange. Write
$\pfend_{\mathrm L}(\ell)$ and $\pfend_{\mathrm R}(\ell)$ for the first and last
element of a list $\ell$, and $\ell\setminus q$ for $\ell$ without $q$. The
machine has three operations, each accepted only under its condition:
\[
\begin{array}{@{}l@{\quad}l@{\quad}l@{}}
\jd{xc}(s,a,b) & \text{if }a\neq b,\ a,b\in\jd{ch}(s) &
  \jd{ch}(s):=\jd{ch}(s)\text{ with $a$ and $b$ swapped}\\
\jd{dep}(s,\sigma,q) & \text{if }\pfend_\sigma(\jd{ch}(s))=q &
  \jd{ch}(s):=\jd{ch}(s)\setminus q,\quad \jd{tr}:=q;\jd{tr}\\
\jd{arr}(s,\sigma,q) & \text{if }q\in\jd{tr}\text{ and }|\jd{ch}(s)|<\capa(s) &
  \jd{ch}(s):=\text{$q$ attached at end $\sigma$},\quad\jd{tr}:=\jd{tr}\setminus q
\end{array}
\]
The operations of a step are: $[\jd{xc}(s,a,b)]$ for an exchange;
$[\jd{dep}(\jd{src}(p),\jd{lv}(p),q);\ \jd{arr}(\jd{dst}(p),\jd{en}(p),q)]$ for a
transport of one move $q{:}p$, defined only if $\jd{src}(p)\neq\jd{dst}(p)$;
for a transport of two or more moves, defined only if their ions are distinct,
their sources are distinct and their destinations are distinct, the
concatenation of the operations of its moves; and no operation for every other
instruction. $\pfcheck(\chi,e_1\ldots e_m)$ applies the operations of the steps
in order and is undefined as soon as one is undefined or not accepted. A state
\emph{agrees} with a placement, $\chi\sim\pos$, when no ion is in
transit, no chain lists an ion twice, and $q\in\jd{ch}(s)$ iff $\pos(q)=s$.
Theorem~\ref{thm:compile-correct} writes $\pfcheck(e_1\ldots e_m)\Downarrow\chi$
for $\pfcheck(\chi_0,e_1\ldots e_m)=\chi$, where $\chi_0$ is the chain state
declared with the device.
(\lean{Chain.step}, \lean{Chain.Step.ops}, \lean{Chain.check}, \lean{Chain.Agrees})
\end{definition}

\begin{definition}[Events, resources and conflict]
\label{def:pf-events}
Each step $e$ is scheduled as an \emph{event}: a set of resources
$\pfres(e)$, a duration $\delta(e)$ and a flag $\pfglob(e)$. Resources are
ions, sites, optical channels, transport channels, classical bits and the
controller $\jd{ctl}$ that saves the condition of a guard. Write
$G_\gm=\{\jd{ctl}\}$ if $\gm=\wsaved$ and $G_\gm=\emptyset$ otherwise.
\[
\begin{array}{@{}l@{\quad}l@{}}
\text{step} & \pfres(\text{step})\\\hline
\execkw{transport}\;\fseq m &
  \bigcup_{(q:p)\in\fseq m}\bigl(\{q,\jd{src}(p),\jd{dst}(p),\jd{ch}(p)\}\cup\via(p)\cup D(p)\bigr)\\
\gm\;\execkw{pulse}\;w\;\execkw{via}\;k &
  \jd{ions}(w)\cup\{\jd{site}(k),\jd{och}(k)\}\cup G_\gm\\
\gm\;\execkw{broadcast}\;\fseq w\;\execkw{via}\;b[\fseq k] &
  \jd{ions}(\fseq w)\cup\bigcup_j\{\jd{site}(k_j),\jd{och}(k_j)\}\cup\{\jd{och}(b)\}\cup G_\gm\\
\gm\;\execkw{readout}\;q\to c\;\execkw{via}\;a &
  \{q,c,\jd{site}(a),\jd{och}(a)\}\cup G_\gm\\
\gm\;\execkw{pump}\;q\;\execkw{via}\;a & \{q,\jd{site}(a),\jd{och}(a)\}\cup G_\gm\\
\execkw{latch}\;\fseq c\,\texttt{==}\,n & \{\jd{ctl}\}\cup\fseq c\\
\gm\;\execkw{barrier}\;\fseq q & \fseq q\cup G_\gm\\
\execkw{idle}\;n & \emptyset\\
\jd{exch}(s,a,b) & \{a,b,s\}
\end{array}
\]
Here $\jd{och}$ is an optical channel and $\jd{ch}(p)$ the transport channel of
path $p$. $D(p)$ is empty unless $\jd{ch}(p)$ is mandatory; then it holds the
source site of every path with the channel and waveform of $p$, because a
mandatory channel moves every ion at those sites, so the legality of the
transport depends on which ions they hold. The duration is
$\delta(h)=\delta_H(h)$ and $\delta(\jd{exch}(s,a,b))=\jd{xd}(s)$; only an
$\execkw{idle}$ is global, since stage~9 inserts it to satisfy a timing rule
and it separates everything before it from everything after it.

In the \emph{exclusive} mode every resource is exclusive. In the
\emph{broadcast} mode a shared transport channel is not held but played:
$\pfresb(e)$ is $\pfres(e)$ without the shared transport channels, and
$\pfslots(e)$ lists, for each shared channel $c$ that a move of $e$ uses, the
pair $(c,\pfkey_e(c))$ with $\pfkey_e(c)=(W_e(c),\delta(e))$, where $W_e(c)$ is
the set of waveforms of the moves of $e$ on $c$. For a core-legal transport,
$\jd{compat}_H$ gives every move on one shared channel the same waveform, so
$W_e(c)$ has exactly one element, the waveform $w_e(c)$ that $e$ applies on
$c$; Theorem~\ref{thm:compile-correct} writes $w_i(c)$ for $w_{e_i}(c)$.
Two events \emph{conflict},
\[
e\pfconf_\pfmode e'\quad\text{iff}\quad\pfglob(e)\lor\pfglob(e')\lor
  \pfres_\pfmode(e)\cap\pfres_\pfmode(e')\neq\emptyset,
\]
with $\pfres_\pfmode=\pfres$ in the exclusive mode and $\pfresb$ in the
broadcast mode.
(\lean{Parallel.instructionResources}, \lean{Chain.stepEvent},
\lean{Shared.fromParallel}, \lean{Parallel.Conflict}, \lean{Shared.Conflict})
\end{definition}

\begin{definition}[Exclusive schedule]
\label{def:pf-sched}
A scheduler state $S=(\pfavail,\pffloor,\pfhz)$ gives each resource $r$ the
time $\pfavail(r)$ at which it becomes free (0 if never used), a floor
$\pffloor$ before which nothing may start, and the horizon $\pfhz$, the latest
finish so far. The earliest start that a set $X$ of resources permits is
$\pfready_S(X)=\max(\{\pffloor\}\cup\{\pfavail(r)\mid r\in X\})$. Scheduling an
event $e$ returns its start $t$ and a new state $S'$:
\[
\begin{array}{@{}l@{\quad}l@{}}
\text{global }e: & t=\pfhz,\qquad S'=(\pfavail,\ \pfhz+\delta,\ \pfhz+\delta)\\
\text{otherwise:} & t=\pfready_S(\pfres(e)),\qquad
  S'=(\pfavail[r\mapsto t+\delta\text{ for }r\in\pfres(e)],\ \pffloor,\
  \max(\pfhz,t+\delta))
\end{array}
\]
with $\delta=\delta(e)$. Starting from $S_0=(0,0,0)$ and scheduling
$e_1,\ldots,e_m$ in order gives the parallel start times $t_1,\ldots,t_m$; the
parallel finish is the horizon of the last state.
(\lean{Parallel.step}, \lean{Chain.timeline}, \lean{Chain.finish})
\end{definition}

\begin{definition}[Broadcast schedule]
\label{def:pf-bsched}
A broadcast channel keeps no state between transports, so the scheduler
records its \emph{activations}: a play $(c,\kappa,t^-,t^+)$ says that channel
$c$ applies key $\kappa$ from $t^-$ to $t^+$. The state is
$S=(\pfavail,\pffloor,\pfhz,\pfplays)$ with a list of plays. A play $\ell$
\emph{leaves room} for $e$ at $t$, $\pfclear_e(t,\ell)$, when for every slot
$(c,\kappa)$ of $e$ with $c=\ell.c$:
$\ell.t^+\le t$, or $t+\delta(e)\le\ell.t^-$, or $\ell.\kappa=\kappa$,
$\ell.t^-=t$ and $\ell.t^+=t+\delta(e)$. Let $r=\pfready_S(\pfresb(e))$, let
$\jd{Rel}$ be the plays with $\ell.t^+>r$, and let
$f=\max(r,\max_{\ell\in\jd{Rel}}\ell.t^+)$. Then $\pfchoose(S,e)$ is the least
element of the set
\[
\{f\}\cup\bigl\{t\in\{r\}\cup\{\ell.t^-,\ell.t^+\mid\ell\in\jd{Rel}\}\ \bigm|\
  t\ge r\ \land\ \forall\ell\in\jd{Rel}.\ \pfclear_e(t,\ell)\bigr\}.
\]
Scheduling a global event is as in Definition~\ref{def:pf-sched}; a
non-global event starts at $t=\pfchoose(S,e)$, occupies $\pfresb(e)$ until
$t+\delta$, and both kinds add the plays $(c,\kappa,t,t+\delta)$ for its slots
$(c,\kappa)$ to $\pfplays$.
(\lean{Shared.choose}, \lean{Shared.step}, \lean{Shared.chainTimeline})
\end{definition}

\begin{definition}[Sequential timing]
\label{def:pf-seqtime}
The sequential start times are $t_1=0$ and $t_{i+1}=t_i+\delta_i$; the
sequential finish is $\sum_i\delta_i$.
(\lean{Reported.sequentialStarts}, \lean{Chain.sequentialFinish})
\end{definition}

\begin{definition}[Reported timeline]
\label{def:pf-timeline}
A timeline reported for $K$ from the initial chains $\chi_0$ is a tuple
$T=(e_1\ldots e_m,\ \chi_m,\ \text{timing},\ t_1\ldots t_m,\ t_{\rm end})$
such that
\begin{itemize}[leftmargin=2.6em]
\item[(i)] $\pferase(e_1\ldots e_m)=h_1\ldots h_n$;
\item[(ii)] $\pfcheck(\chi_0,e_1\ldots e_m)=\chi_m$ and (iii)
  $\chi_m\sim\pos_n$;
\item[(iv)] the start times and the finish are those of the timing: of
  Definition~\ref{def:pf-sched} (exclusive mode) or~\ref{def:pf-bsched}
  (broadcast mode) for the parallel timing, of
  Definition~\ref{def:pf-seqtime} for the sequential one;
\item[(v)] every compiled step is \emph{legal at its start time}: for
  $e_i=h_k$, $H;\pos_{k-1}\vdash h_k\;\jd{core}$ and the two frames of
  $h_k$ at $t_i$ satisfy $R$, as in \Rule{Step} with the clock set to $t_i$;
\item[(vi)] every exchange step satisfies $\pfxframe(\pos^{(i)},t_i,\delta_i)\models_H R$.
\end{itemize}
(\lean{Reported.Timeline}, \lean{Parallel.TimedRun}, \lean{Reported.ExchangesLegal})
\end{definition}

\noindent How does the compiler obtain such a tuple? It inserts, before every
transport, the exchange that brings each departing ion to the end of its
chain, replays the result on the chain machine after checking that $\chi_0$
agrees with $\pos_0$, computes the parallel start times, and rechecks the rules
on every frame at those times; if a rule fails it uses the sequential start
times, rechecked in the same way; if that fails too, it reports no time.
Lemma~\ref{lem:pf-driver} proves that whatever it reports satisfies (i)
to~(vi). The theorem quantifies over every tuple with these properties, so
how the insertion and the search choose has no effect on the proof.

\begin{definition}[Target code in order of time]
\label{def:pf-order}
Let $x_i=(e_i,t_i)$. The list $\pfsort(x_1\ldots x_m)$ sorts it by start time
by insertion, $\pfsort(\epsilon)=\epsilon$ and
$\pfsort(x;\fseq x)=\pfins(x,\pfsort(\fseq x))$, where $\pfins(x,\fseq y)$
places $x$ before the first element of $\fseq y$ whose start is not smaller;
steps with equal start times keep their program order. $K^{T}$ is the erasure
of the steps of $\pfsort(x_1\ldots x_m)$, and $\sem{K^{T}}_{H,\eta}$ runs it with
the semantics of target code (\S\ref{app:f-semantics}).
(\lean{TimeOrder.byStart}, \lean{TimeOrder.ordered})
\end{definition}

\begin{definition}[Uniform motion]
\label{def:pf-uniform}
The device description gives every path $p$ its list of hop-by-hop motion
labels $\mu(p)$, such as an axis direction or a loop step (in Lean, a
\emph{wiring}). For a move $a=q{:}p$ write $\mu(a)=\mu(p)$ and $d(a)=d(p)$.
$\jd{Uniform}(H)$ holds when any two paths $p,p'$ with
$\jd{ch}(p)=\jd{ch}(p')$ shared and $\jd{wf}(p)=\jd{wf}(p')$ have
$\mu(p)=\mu(p')$ and $d(p)=d(p')$ (Section~\ref{sec:broadcast}).
(\lean{Broadcast.WiringUniform})
\end{definition}

\subsection{The sequential contract}
\label{app:pf-sequential}

We first prove the success contract of Theorem~\ref{thm:formal} with every
step written out. It gives (M2), (L1) and the meaning of $K$ in program
order, on which (M3) rests. Three facts about the semantics are used
throughout.

\begin{lemma}[Lifting]
\label{lem:pf-lift}
For all lists $X,Y$ of instructions and ensembles $\sigma_1,\sigma_2$:
$\sem{X}(\sigma_1\sigma_2)=\sem{X}\sigma_1\,\sem{X}\sigma_2$ and
$\sem{X\,Y}=\sem{Y}\circ\sem{X}$. If $\sigma_1\pfperm\sigma_2$ then
$\sem{X}\sigma_1\pfperm\sem{X}\sigma_2$. The same holds for target code and
for $\pfobs$. (\lean{interpretPulseCode_append},
\lean{Independence.interpretPulseCode_perm})
\end{lemma}
\begin{proof}
The lifting $f^\dagger$ of a branch function maps each branch and concatenates
the results, so it distributes over concatenation and maps a permutation of
its input to a permutation of its output. The first two equations follow by
induction on $X$ from $\sem{\epsilon}\sigma=\sigma$ and
$\sem{x;X}\sigma=\sem{X}(\sem{x}^\dagger\sigma)$, and the third by induction on
$X$ from the second property of $f^\dagger$. $\pfobs$ is applied pointwise.
\end{proof}

\begin{lemma}[The flag]
\label{lem:pf-flag}
Every instruction other than a latch keeps the flag: every branch of
$\sem{x}(C,b,\rho)$ has flag $b$. Consequently, if $X=\gm x_1\ldots\gm x_r$
with $\gm=\wsaved$ and no $x_j$ a latch, then
$\sem{X}(C,\jd{ff},\rho)=(C,\jd{ff},\rho)$ and
$\sem{X}(C,\jd{tt},\rho)=\sem{X^\circ}(C,\jd{tt},\rho)$, where $X^\circ$ is $X$
without its marks. (\lean{guarded_comp})
\end{lemma}
\begin{proof}
The first claim is read off the denotations of \S\ref{app:f-semantics}: only
the latch writes the second component of a branch. For the second, induct on
$r$. With flag $\jd{ff}$ the guard of $\gm x_1$ returns the branch unchanged,
and the induction hypothesis applies to the rest. With flag $\jd{tt}$ the
guard applies $\sem{x_1}$; by the first claim every resulting branch still has
flag $\jd{tt}$, so the induction hypothesis applies to each of them, and
Lemma~\ref{lem:pf-lift} reassembles the ensemble.
\end{proof}

\begin{lemma}[Determinism of the source]
\label{lem:pf-det}
If both
$\Delta\vdash\langle\fseq S,\sigma\rangle\Downarrow\langle\Delta_1,\sigma_1\rangle$
and $\Delta\vdash\langle\fseq S,\sigma\rangle\Downarrow\langle\Delta_2,\sigma_2\rangle$
hold, then $\Delta_1=\Delta_2$ and $\sigma_1=\sigma_2$.
(\lean{StatementsExecute.deterministic})
\end{lemma}
\begin{proof}
By induction on the first derivation. Operand resolution
$\Delta\vdash u\mapsto\fseq q$ and the action $\Delta\vdash O:f$ of a unitary
operation are functions of $\Delta$ and the syntax: the latter by induction on
the depth of gate calls, which declarations bound. Hence in \Rule{S-Decl},
\Rule{S-Unitary}, \Rule{S-Measure} and \Rule{S-Reset} the conclusion is
determined by the premises. In \Rule{S-If} each branch satisfies exactly one
of the two disjuncts, because the register value either equals $n$ or does
not, and by the induction hypothesis the result $\sigma_j$ of each branch is
unique. \Rule{S-Seq} composes two unique results.
\end{proof}

\begin{lemma}[Stages 1 and 2]
\label{lem:pf-incl}
\label{lem:incl}
If $\jd{incl}(P,r)=P_R$, then $P\vdash\sigma\Downarrow\sigma'$ iff
$P_R\vdash\sigma\Downarrow\sigma'$. If $\jd{dom}(P_R)=P_D$, then erasing the
labels of $P_D$ gives $P_R$.
(\lean{resolveIncludes_executes_iff}, \lean{analyzeProgram_erase})
\end{lemma}
\begin{proof}
The meaning of a bundle is the meaning of its statements with every
\circkw{include} replaced by the statements of the included unit
(\S\ref{app:f-semantics}). $\jd{incl}$ computes exactly this list: by
induction on the include tree, which is finite because $\jd{incl}$ fails on a
cycle, each include is replaced by the resolved statements of its unit and
every other statement is kept in place. Both sides are therefore executions of
the same statement list. $\jd{dom}$ attaches a label to each angle expression
and changes nothing else, by induction on statements and expressions.
\end{proof}

\begin{lemma}[Stage 3: expansion simulates the source]
\label{lem:pf-expand}
\label{lem:expand}
If $\Delta\vdash\fseq S\leadsto\Delta';\fseq e$ (statement by statement,
threading $\Delta$), then for every ensemble $\sigma$:
$\Delta\vdash\langle\fseq S,\pfobs(\sigma)\rangle\Downarrow
\langle\Delta',\pfobs(\sem{\fseq e}\sigma)\rangle$.
(\lean{expandProgram_source_executes})
\end{lemma}
\begin{proof}
We first prove the claim for one operation $O$ expanded with the empty mark:
if $\Delta\vdash O\leadsto_\epsilon\fseq e$, then
$\Delta\vdash\langle O,\pfobs(\sigma)\rangle\Downarrow\langle\Delta,\pfobs(\sem{\fseq e}\sigma)\rangle$.
By Lemma~\ref{lem:pf-lift} it suffices to take $\sigma$ a single branch
$(C,b,\rho)$. By cases on the expansion rule:
\begin{itemize}[leftmargin=1.6em]
\item \Rule{X-U}, \Rule{X-CX}: the instructions apply the operator of
  \circkw{U} or \circkw{CX} at the resolved ions $q_1,\ldots,q_m$ in index
  order, and do not touch $c$ or $b$. \Rule{S-Unitary} applies the composition
  $f$ of the same operators in the same order to the matrix; observation drops
  $b$ on both sides.
\item \Rule{X-Measure}: the instructions apply $\jd{meas}_{q_1,c_1},\ldots,
  \jd{meas}_{q_m,c_m}$ in this order, which is the composition in
  \Rule{S-Measure}; each keeps the flag, so observing after the instructions
  equals applying the flag-free functions to the observed branch. \Rule{X-Reset}
  is the same with $\jd{rst}$.
\item \Rule{X-Call}: by induction on the depth of the call. The body is
  expanded with the arguments substituted, its operations are unitary (or
  barriers, which are the identity), and by the induction hypothesis each acts
  as the action that \Rule{S-Unitary} assigns to it; the action of the call is
  by definition the composition of these actions in body order.
\end{itemize}
Now induct on $\fseq S$. For the empty list both sides are the identity.
For $S;\fseq S'$ with $\Delta\vdash S\leadsto\Delta_1;\fseq e_1$ and
$\Delta_1\vdash\fseq S'\leadsto\Delta';\fseq e_2$, we show
$\Delta\vdash\langle S,\pfobs(\sigma)\rangle\Downarrow\langle\Delta_1,\pfobs(\sem{\fseq e_1}\sigma)\rangle$
by cases on the rule; the induction hypothesis for $\fseq S'$, applied to the
ensemble $\sem{\fseq e_1}\sigma$, and $\sem{\fseq e_1\fseq e_2}=\sem{\fseq e_2}\circ\sem{\fseq e_1}$
(Lemma~\ref{lem:pf-lift}) then give the claim by \Rule{S-Seq}.
\begin{itemize}[leftmargin=1.6em]
\item \Rule{X-Decl}: $\fseq e_1=\epsilon$ and \Rule{S-Decl} records the same
  declaration and leaves the ensemble unchanged.
\item \Rule{X-Op}: the claim for operations.
\item \Rule{X-Barrier}: both sides are the identity.
\item \Rule{X-If}: $\fseq e_1=\irkw{latch}\;\fseq c\,\texttt{==}\,n;\ \fseq e_O$
  where $\Delta\vdash O\leadsto_{\wsaved}\fseq e_O$, so every instruction of
  $\fseq e_O$ carries the mark and none is a latch. Both \Rule{S-If} and the
  instructions act branch by branch, so by Lemma~\ref{lem:pf-lift} take one
  branch $(C,b,\rho)$ with observed branch $(C,\rho)$. The latch gives
  $(C,v,\rho)$ with $v=(\jd{val}_C(\fseq c)=n)$, which is the test of
  \Rule{S-If} on $(C,\rho)$. If $v=\jd{ff}$, Lemma~\ref{lem:pf-flag} returns
  $(C,\jd{ff},\rho)$, observed $(C,\rho)$: the second disjunct of \Rule{S-If}.
  If $v=\jd{tt}$, Lemma~\ref{lem:pf-flag} gives $\sem{\fseq e_O^\circ}(C,\jd{tt},\rho)$,
  and $\fseq e_O^\circ$ is the expansion of $O$ with the empty mark, so by the
  claim for operations its observation is the result of $O$ on $(C,\rho)$: the
  first disjunct. The register is read once, by the latch, before $O$ runs,
  exactly as \Rule{S-If} reads it once; this remains true when $O$ overwrites
  the register (Example~\ref{ex:f-latch}).\qedhere
\end{itemize}
\end{proof}

\begin{lemma}[Stage 4: pulse selection]
\label{lem:pf-select}
\label{lem:select}
If every instruction of $P_E$ has well-formed operands, then
$\sem{\jd{sel}(P_E)}\sigma=\sem{P_E}\sigma$ for every $\sigma$.
(\lean{selectProgram_correct}, \lean{selectGate_correct}, \lean{cxPulse_density})
\end{lemma}
\begin{proof}
$\jd{sel}$ works instruction by instruction and keeps every non-unitary
instruction, so by Lemma~\ref{lem:pf-lift} it suffices to show, for one branch,
that the pulses of a gate act as the gate. All pulses of $\jd{sel}(x)$ carry the
mark $\gm$ of $x$ and none is a latch, so by Lemma~\ref{lem:pf-flag} they act
as the identity where $x$ does (mark set, flag false), and otherwise as their
unmarked composition. It remains to show that the composition of the pulses
equals the gate up to a global phase: a phase cancels in
$U\cdot\rho=U\rho U^\dagger$, since $(e^{i\alpha}U)\rho(e^{i\alpha}U)^\dagger=U\rho U^\dagger$.

Write $R_x(\alpha)=e^{-i\alpha X/2}$, $R_y(\alpha)=e^{-i\alpha Y/2}$,
$R_z(\alpha)=e^{-i\alpha Z/2}$. By definition
$\Rp(\theta,\varphi)=e^{-i\theta(\cos\varphi X+\sin\varphi Y)/2}$ and
$\MS(\theta)=e^{-i\theta X_1X_2/2}$. Two conjugation rules hold: for all $\psi$,
$R_z(\psi)XR_z(-\psi)=\cos\psi\,X+\sin\psi\,Y$, and
$R_y(\beta)XR_y(-\beta)=\cos\beta\,X-\sin\beta\,Z$. (Both sides of each satisfy
the same linear differential equation in the angle, with the same value at 0;
for instance $\frac{d}{d\psi}$ of the left side of the first is
$-\tfrac i2[Z,\cdot\,]$ applied to it, and $-\tfrac i2[Z,X]=Y$,
$-\tfrac i2[Z,Y]=-X$.) Since $Ve^{A}V^\dagger=e^{VAV^\dagger}$ for every unitary
$V$, the first rule gives $\Rp(\theta,\varphi)=R_z(\varphi)R_x(\theta)R_z(-\varphi)$.

\emph{The U gate.} $\jd{sel}(\irkw{U}(\theta,\varphi,\lambda)\,q)$ applies
$\Rp(\theta,\tfrac\pi2-\lambda)$, then $\Rp(\pi,0)$, then
$\Rp(\pi,\tfrac{\varphi+\lambda}2)$. With $\Rp(\pi,a)=-i(\cos a\,X+\sin a\,Y)$
and $YX=-iZ$,
\[
\Rp(\pi,a)\Rp(\pi,0)=-(\cos a\,X+\sin a\,Y)X=-(\cos a\,I-i\sin a\,Z)=-R_z(2a).
\]
With $a=\tfrac{\varphi+\lambda}2$ and
$\Rp(\theta,\tfrac\pi2-\lambda)=R_z(\tfrac\pi2-\lambda)R_x(\theta)R_z(\lambda-\tfrac\pi2)
=R_z(-\lambda)R_y(\theta)R_z(\lambda)$, using
$R_z(\tfrac\pi2)R_x(\theta)R_z(-\tfrac\pi2)=\Rp(\theta,\tfrac\pi2)=R_y(\theta)$, the product is
\[
-R_z(\varphi+\lambda)R_z(-\lambda)R_y(\theta)R_z(\lambda)=-R_z(\varphi)R_y(\theta)R_z(\lambda)
=-\mathrm U(\theta,\varphi,\lambda),
\]
where the last step multiplies out the three $2\times2$ matrices, which gives
the matrix of \S\ref{app:f-semantics} (the SU(2) form of \circkw{U} in the
OpenQASM~2 specification, \lean{qasmU}).

\emph{The CX gate.} $\jd{sel}(\irkw{CX}\;q,q')$ applies, with $q$ as ion 1
and $q'$ as ion 2, $R_y(\tfrac\pi2)_1=\Rp(\tfrac\pi2,\tfrac\pi2)_1$, then
$\MS(\tfrac\pi2)=e^{-i\frac\pi4X_1X_2}$, then $R_x(-\tfrac\pi2)_1$,
$R_x(-\tfrac\pi2)_2$ and $R_y(-\tfrac\pi2)_1$. The product is
\[
W=R_y(-\tfrac\pi2)_1R_x(-\tfrac\pi2)_2R_x(-\tfrac\pi2)_1\,e^{-i\frac\pi4X_1X_2}\,R_y(\tfrac\pi2)_1 .
\]
$R_x(\cdot)_1$ and $R_x(\cdot)_2$ commute with $X_1X_2$; move them to the right
of the exponential and insert $R_y(\tfrac\pi2)_1R_y(-\tfrac\pi2)_1=I$:
\[
W=\bigl(R_y(-\tfrac\pi2)_1e^{-i\frac\pi4X_1X_2}R_y(\tfrac\pi2)_1\bigr)
  \bigl(R_y(-\tfrac\pi2)_1R_x(-\tfrac\pi2)_1R_y(\tfrac\pi2)_1\bigr)R_x(-\tfrac\pi2)_2 .
\]
By the second conjugation rule with $\beta=-\tfrac\pi2$,
$R_y(-\tfrac\pi2)XR_y(\tfrac\pi2)=Z$, so
$W=e^{-i\frac\pi4Z_1X_2}\,e^{i\frac\pi4Z_1}\,e^{i\frac\pi4X_2}$. The operators
$Z_1X_2$, $Z_1$ and $X_2$ commute pairwise, hence
$W=e^{-i\frac\pi4(Z_1X_2-Z_1-X_2)}=e^{i\pi/4}e^{-i\pi Q}$ with
$Q=\tfrac14(I-Z_1-X_2+Z_1X_2)=\tfrac{I-Z_1}2\otimes\tfrac{I-X_2}2$. $Q$ is the
projector onto $\ket1\otimes\ket-$, so $e^{-i\pi Q}=I-2Q$, and
$I-2Q=\ket0\!\langle0|\otimes I+\ket1\!\langle1|\otimes X=\CX$. Thus
$W=e^{i\pi/4}\CX$.

Both identities are between operators on one or two ions. Embedded in the
register they become $A\otimes I$ and $e^{i\alpha}A\otimes I$ on the other
ions, so they hold for every matrix of the register, including entangled ones.
The two-ion embedding needs $q\neq q'$, which well-formed operands give
(\Rule{Ok-MS}).
\end{proof}

\begin{lemma}[Stages 5 and 6]
\label{lem:pf-bind}
\label{lem:bind}
Binding changes only annotations, each annotation is the list
$\jd{elig}_H(i)$, and binding succeeds only if $\vdash H\;\jd{ok}$. Placement
returns $\pos_0$ with $H;R\vdash\pos_0\;\jd{init}$ and leaves the program
unchanged. (\lean{Binding.stage_correct}, \lean{Placed.stage_correct})
\end{lemma}
\begin{proof}
Each of \Rule{B-Optical}, \Rule{B-Latch} and \Rule{B-Barrier} outputs its
input instruction with an annotation and nothing else, and binding requires
$\vdash H\;\jd{ok}$. Stage~6 outputs a placement admitted by its judgment,
which is \Rule{Init}.
\end{proof}

\begin{lemma}[A core-legal instruction performs its projection]
\label{lem:pf-instr}
\label{lem:instr}
If $H;\pos\vdash h\;\jd{core}$ for some placement $\pos$, then
$\sem{h}^\dagger_{H}\sigma=\sem{\jd{proj}(h)}\sigma$ for every $\sigma$.
Consequently, if every instruction of a list $Y$ is core legal at some
placement, not necessarily the one that $Y$ reaches, then
$\sem{Y}_H\sigma=\sem{\jd{proj}(Y)}\sigma$.
(\lean{hardwareInstruction_refines}, \lean{TimeOrder.interpretHardwareCode_project})
\end{lemma}
\begin{proof}
By Lemma~\ref{lem:pf-lift} take one branch $(C,b,\rho)$; recall
$\sem{h}_H(C,b,\rho)=\jd{act}(\rlz_H(b,h))(C,b,\rho)$. By cases on $h$.
\begin{itemize}[leftmargin=1.6em]
\item Transport, wait: $\jd{act}$ is the identity and $\jd{proj}(h)=\epsilon$.
\item $\gm\;\execkw{pulse}\;w\;\execkw{via}\;k$: if $\gm=\wsaved$ and
  $b=\jd{ff}$, $\rlz_H$ gives a wait, the identity, and the guard of
  $\jd{proj}(h)=\gm\,w$ returns the branch unchanged. Otherwise both sides apply the
  matrix of $w$ on its ions. Readout and pump are the same with
  $\jd{act}=\sem{\irkw{measure}\;q\to c'}$ and $\sem{\irkw{reset}\;q}$, which
  are the denotations of their projections.
\item Latch, barrier: $\jd{proj}(h)=h$ and $\jd{act}$ is its denotation.
\item $\gm\;\execkw{broadcast}\;w_1\ldots w_m\;\execkw{via}\;b'[\fseq k]$: the
  flag-false case is as for a pulse. Otherwise $\jd{act}$ applies
  $\prod_{q}\Rp(\theta_{b'},\varphi_{b'})_q$ over the ions of $\fseq w$, with the
  angles declared by $b'$, and $\jd{proj}(h)=\gm w_1;\ldots;\gm w_m$ applies the
  pulses $w_j$ in order. By \Rule{C-Broadcast}, $\jd{wave}(w_j)=\jd{wave}(b')$,
  so every $w_j$ is $\irkw{R}(\theta_{b'},\varphi_{b'})$ on its ion, and the
  ions are distinct. One-ion operators on distinct ions commute, so the product
  is well defined and equals the composition in the listed order.
\end{itemize}
The second claim follows by induction on $Y$ with
$\jd{proj}(h;Y)=\jd{proj}(h)\,\jd{proj}(Y)$ and Lemma~\ref{lem:pf-lift}; it uses only that each
instruction is core legal at some placement.
\end{proof}

\begin{lemma}[Stage 7: routing]
\label{lem:pf-route}
\label{lem:route}
If $H;\pos\vdash\fseq i\leadsto_{\jd{route}}A$ and $H\vdash\pos:n_q$, then
$\jd{proj}(A)=\fseq i$, $A$ has no wait and no broadcast, $A$ is a core run from
$\pos$, every placement that $A$ passes through is valid, and
$\sem{A}_H\sigma=\sem{\fseq i}\sigma$.
(\lean{Rules.Routes.facts}, \lean{Routed.stage_correct}, \lean{CoreExecution.prefix_invariant})
\end{lemma}
\begin{proof}
By induction on the derivation. \Rule{R-Done}: $A=\epsilon$. \Rule{R-Move}:
$A=\execkw{transport}\;\fseq m;A'$ where the transport is core legal at $\pos$
and $A'$ is routed from $\pos'=\nxt_H(\pos,\execkw{transport}\;\fseq m)$.
\Rule{C-Transport} has the premise $H\vdash\pos':n_q$, so $\pos'$ is valid and
the induction hypothesis applies; $\jd{proj}$ of a transport is $\epsilon$, so
$\jd{proj}(A)=\jd{proj}(A')=\fseq i$. \Rule{R-Emit}: $A=h;A'$ with $h\in\jd{real}(i)$
core legal at $\pos$; $h$ is not a transport or a broadcast, so it does not
move an ion and $A'$ is routed from $\pos$; $\jd{proj}(h)=i$ by the definition of
$\jd{real}$, so $\jd{proj}(A)=i;\fseq i'=\fseq i$. In every case the head is core legal
at the placement where it fires and no wait or broadcast is added. The meaning
follows from Lemma~\ref{lem:pf-instr}: $\sem{A}_H\sigma=\sem{\jd{proj}(A)}\sigma=\sem{\fseq i}\sigma$.
\end{proof}

\begin{lemma}[Stage 8: broadcast formation]
\label{lem:pf-pack}
\label{lem:pack}
If $H;\pos\vdash A\leadsto_{\jd{pack}}A'$, where $A$ is the output of routing
from $\pos$, then $A'$ is a core run from $\pos$ without waits, $A'$ and $A$ end
in the same placement, expanding each broadcast of $A'$ into its pulses gives
$A$, $\jd{proj}(A')=\jd{proj}(A)$, and $\sem{A'}_H\sigma=\sem{A}_H\sigma$.
(\lean{Rules.Packs.facts}, \lean{Packed.stage_correct})
\end{lemma}
\begin{proof}
By induction on the derivation. \Rule{P-Keep} keeps a plain instruction that
is core legal at $\pos$ and continues from $\nxt_H(\pos,h)$. \Rule{P-Join}
replaces $\gm\,\execkw{pulse}\;w_j\;\execkw{via}\;k_j$ for $j=1..m$ by one
broadcast that is core legal at $\pos$; pulses and broadcasts move no ion, so
the remainder continues from $\pos$ in both programs, and
$\jd{proj}(\text{broadcast})=\gm w_1;\ldots;\gm w_m=\jd{proj}(\text{pulse}_1)\ldots\jd{proj}(\text{pulse}_m)$.
Neither rule adds a wait. The meanings agree by Lemma~\ref{lem:pf-instr}, applied
to $A'$ and to $A$ (Lemma~\ref{lem:pf-route}):
$\sem{A'}_H\sigma=\sem{\jd{proj}(A')}\sigma=\sem{\jd{proj}(A)}\sigma=\sem{A}_H\sigma$.
\end{proof}

\begin{lemma}[Stage 9: time assignment]
\label{lem:pf-time}
\label{lem:time}
If $H;R;\pos;t\vdash A'\leadsto_{\jd{time}}K$, then erasing the waits of $K$
gives $A'$, $H;R\vdash\langle\pos,t\rangle\runs{K}\langle\pos',t'\rangle$ for
the placement $\pos'$ and time $t'$ that $K$ reaches, and
$\sem{K}_H\sigma=\sem{A'}_H\sigma$.
(\lean{Rules.Times.facts}, \lean{Timed.stage_correct},
\lean{interpretHardwareCode_eraseIdles})
\end{lemma}
\begin{proof}
By induction on the derivation. \Rule{T-Emit} contributes one instruction of
$A'$ with a derivation of \Rule{Step}; \Rule{T-Wait} contributes one wait with
a derivation of \Rule{Step}. \Rule{Run-Cons} composes them and \Rule{T-Done}
ends with \Rule{Run-Nil}. A wait acts as the identity on every branch, and
the realization of a guarded instruction depends only on the flag, which waits
do not change; hence deleting the waits does not change the meaning.
\end{proof}

\begin{theorem}[The sequential contract]
\label{thm:pf-seq}
\label{thm:formal}
Under (H2): $\vdash H\;\jd{ok}$; $H;R\vdash\pos_0\;\jd{init}$;
$H;R\vdash\langle\pos_0,0\rangle\runs{K}\langle\pos_n,t_{\rm seq}\rangle$;
$\jd{proj}(K)=P_P$; and for every $\eta$ and $\sigma$,
$P\vdash_\eta\pfobs(\sigma)\Downarrow\pfobs(\sem{K}_{H,\eta}\sigma)$, which is
the only result of $P$ from $\pfobs(\sigma)$.
(\lean{compileQASM_sound}, \lean{QASMCompilation.source_correct})
\end{theorem}
\begin{proof}
The first two facts are Lemma~\ref{lem:pf-bind} and the run is
Lemma~\ref{lem:pf-time} from $\langle\pos_0,0\rangle$. Projection: waits
project to $\epsilon$, so $\jd{proj}(K)=\jd{proj}(A')$; $\jd{proj}(A')=\jd{proj}(A)$ by
Lemma~\ref{lem:pf-pack}; $\jd{proj}(A)=P_P$ by Lemma~\ref{lem:pf-route}, since
binding does not change the instructions (Lemma~\ref{lem:pf-bind}). Meaning:
\[
\sem{K}_H\sigma\overset{\ref{lem:pf-time}}{=}\sem{A'}_H\sigma
\overset{\ref{lem:pf-pack}}{=}\sem{A}_H\sigma
\overset{\ref{lem:pf-route}}{=}\sem{P_P}\sigma
\overset{\ref{lem:pf-select}}{=}\sem{P_E}\sigma .
\]
Lemma~\ref{lem:pf-select} applies because expansion produces well-formed
operands (it resolves every operand against the declared registers, and
\Rule{X-CX} requires distinct ions). Lemma~\ref{lem:pf-expand} gives
$P_R\vdash\pfobs(\sigma)\Downarrow\pfobs(\sem{P_E}\sigma)$, Lemma~\ref{lem:pf-incl}
carries it to $P$ (domain analysis does not change the program), and
Lemma~\ref{lem:pf-det} gives uniqueness.
\end{proof}

\subsection{Chain order}
\label{app:pf-chain}

The site-level legality of $K$ does not say that an ion can leave its trap:
it may sit in the middle of a chain. The timeline therefore inserts exchanges,
and the chain machine checks that every departure takes an end ion. This
section proves that an accepted replay keeps the chains equal to the
placement of $K$.

\begin{lemma}[An exchange keeps agreement]
\label{lem:pf-xc}
If $\chi\sim\pos$ and $\jd{xc}(s,a,b)$ is accepted in $\chi$, giving
$\chi'$, then $\chi'\sim\pos$. (\lean{Chain.agrees_exchange})
\end{lemma}
\begin{proof}
Only $\jd{ch}(s)$ changes, to its image under the transposition $\sigma$ of $a$
and $b$. Since $a,b\in\jd{ch}(s)$, $\sigma$ maps the set of ions of $\jd{ch}(s)$
onto itself, so membership in every chain is unchanged and so is $\pos$. Since
$\sigma$ is injective, the image of a list without repetitions has none. No
ion enters transit.
\end{proof}

\begin{lemma}[A move keeps agreement]
\label{lem:pf-move}
Let $p$ be a path with $\jd{src}(p)\neq\jd{dst}(p)$. If $\chi\sim\pos$ and the
operations
\[
\jd{dep}(\jd{src}(p),\jd{lv}(p),q);\ \jd{arr}(\jd{dst}(p),\jd{en}(p),q)
\]
are accepted from $\chi$, giving $\chi'$, then
$\chi'\sim\nxt_H(\pos,\execkw{transport}\;q{:}p)$.
(\lean{Chain.agrees_move})
\end{lemma}
\begin{proof}
Write $s=\jd{src}(p)$, $d=\jd{dst}(p)$ and $\pos'$ for the new placement. The
departure is accepted, so $q$ is an end of $\jd{ch}(s)$, hence $q\in\jd{ch}(s)$
and $\pos(q)=s$ by agreement; $\jd{tr}$ was empty and becomes $[q]$. The arrival
is accepted ($q\in\jd{tr}$), attaches $q$ to $\jd{ch}(d)$ and empties $\jd{tr}$.
No repetition: $\jd{ch}(s)\setminus q$ is a sublist of a list without
repetitions, and $q\notin\jd{ch}(d)$ before the arrival, since otherwise
$\pos(q)=d\neq s$. Membership: $q$ is now only in $\jd{ch}(d)$, because it
occurred once in $\jd{ch}(s)$ and was removed, and $\pos'(q)=d$; every other ion
keeps its chain and $\pos'(r)=\pos(r)$.
\end{proof}

\begin{lemma}[A step keeps agreement]
\label{lem:pf-step}
If $\chi\sim\pos$ and the operations of a step $e$ are accepted from
$\chi$, giving $\chi'$, then $\chi'\sim\pos'$, where $\pos'=\pos$ for an
exchange and $\pos'=\nxt_H(\pos,h)$ for a compiled step $h$. Hence, by
induction, $\pfcheck(\chi_0,e_1\ldots e_m)=\chi_m$ and $\chi_0\sim\pos_0$
imply $\chi_m\sim\pos_n$.
(\lean{Chain.advance_agrees}, \lean{Chain.check_agrees}, \lean{Chain.placement_finalPlacement})
\end{lemma}
\begin{proof}
An exchange is Lemma~\ref{lem:pf-xc}, and a transport of one move is
Lemma~\ref{lem:pf-move}. For a transport of moves $m_1\ldots m_r$, $r\ge2$, the
ions are distinct, so $\nxt_H(\pos,\execkw{transport}\;m_1\ldots m_r)$ equals
moving $m_1$ first and then the rest; induction on $r$ with
Lemma~\ref{lem:pf-move} at each move gives the claim. Every other instruction
has no operation and does not change the placement. For the consequence,
induct on the steps: the placement after $e_1\ldots e_i$ is that of the
compiled steps among them, because exchanges do not change it, and by (M1)
this is $\pos_n$ after all $m$ steps.
\end{proof}

\noindent In a batch, sources and destinations are distinct as well, so each
departing ion is at the end of its chain at the start of the event: an
earlier arrival in the same batch can only cover the end of a chain that no
later move of the batch leaves.

\begin{lemma}[Departures are physical]
\label{lem:pf-depart}
In an accepted replay every $\jd{dep}(s,\sigma,q)$ removes the ion at end
$\sigma$ of $\jd{ch}(s)$. (\lean{Chain.depart_end})
\end{lemma}
\begin{proof}
This is the condition under which $\jd{dep}$ is accepted.
\end{proof}

\begin{lemma}[Inserting exchanges keeps the code]
\label{lem:pf-refine}
The steps computed by the compiler satisfy $\pferase(e_1\ldots e_m)=h_1\ldots h_n$.
(\lean{Chain.erase_refine})
\end{lemma}
\begin{proof}
Before each $h_k$ the compiler inserts only exchange steps, which erasure
drops, and then $h_k$ itself; if the replay of that block fails it appends
$h_{k+1}\ldots h_n$ unchanged. By induction on $n$ the erasure is
$h_1\ldots h_n$ in both cases. (A failed replay is then rejected by the final
check, so no timeline is reported.)
\end{proof}

\subsection{The schedules}
\label{app:pf-schedule}

Two events that share a resource must not overlap, and two events on one
broadcast channel must not play different waveforms at once. We prove these
properties for every list of events, by one invariant of the scheduler state.

\begin{definition}[Coverage]
A state $S$ \emph{covers} a scheduled event $(e,t)$ with $\delta=\delta(e)$,
written $\pfcovers(S,(e,t))$, when (a)~$t+\delta\le\pfhz$;
(b)~$\pfglob(e)\Rightarrow t+\delta\le\pffloor$; (c)~for every
$r\in\pfres_\pfmode(e)$, $t+\delta\le\max(\pffloor,\pfavail(r))$; and, in the
broadcast mode, (d)~$(c,\kappa,t,t+\delta)\in\pfplays$ for every slot
$(c,\kappa)$ of $e$. (\lean{Parallel.Covers}, \lean{Shared.Covers})
\end{definition}

\begin{lemma}[The choice of a start on broadcast channels]
\label{lem:pf-choose}
For every broadcast state $S$ and event $e$, with $r=\pfready_S(\pfresb(e))$
and $t=\pfchoose(S,e)$: (i)~$t\ge r$; (ii)~$\pfclear_e(t,\ell)$ for every
$\ell\in\pfplays$; (iii)~$t\le\max(r,\max_{\ell\in\jd{Rel}}\ell.t^+)$.
(\lean{Shared.choose_spec})
\end{lemma}
\begin{proof}
Let $Q(t)$ say that $t\ge r$ and $\pfclear_e(t,\ell)$ for every $\ell\in\jd{Rel}$.
The fallback $f$ satisfies $Q$: $f\ge r$, and every $\ell\in\jd{Rel}$ has
$\ell.t^+\le f$, which is the first disjunct of $\pfclear$. Every candidate that
the filter keeps satisfies $Q$ by the filter. The least element of a nonempty
finite set of numbers is one of them, so $Q(t)$, which gives (i) and (ii) for
the plays in $\jd{Rel}$. A play $\ell\notin\jd{Rel}$ has $\ell.t^+\le r\le t$, the
first disjunct again. (iii) holds because $f$ belongs to the set.
\end{proof}

\begin{lemma}[Coverage is established and kept]
\label{lem:pf-covers}
Let $(t,S')$ be the result of scheduling $e$ from $S$, in either mode.
\begin{enumerate}
\item $\pfcovers(S',(e,t))$.
\item If $\pfcovers(S,x)$ then $\pfcovers(S',x)$.
\item If $\pfcovers(S,(e',t'))$ and $e'\pfconf_\pfmode e$, then $t'+\delta(e')\le t$.
\item In the broadcast mode, if $\pfcovers(S,(e',t'))$, then for every slot
  $(c,\kappa')$ of $e'$ and $(c,\kappa)$ of $e$ on one channel $c$:
  $t'+\delta(e')\le t$, or $t+\delta(e)\le t'$, or $\kappa'=\kappa$, $t'=t$ and
  $\delta(e')=\delta(e)$.
\end{enumerate}
(\lean{Parallel.covers_step_self}, \lean{Parallel.covers_step_mono},
\lean{Parallel.step_respects}, and their \lean{Shared} versions)
\end{lemma}
\begin{proof}
Write $\delta=\delta(e)$, $\delta'=\delta(e')$, and $r$ for the ready time of
$e$; a non-global event starts at $t=r$ in the exclusive mode and at
$t=\pfchoose(S,e)\ge r$ (Lemma~\ref{lem:pf-choose}) in the broadcast mode.
\pfStep{1.}If $e$ is global, $t=\pfhz$ and $\pfhz'=\pffloor'=\pfhz+\delta=t+\delta$,
which gives (a) and (b), and (c) since $\max(\pffloor',\cdot)\ge\pffloor'$.
Otherwise $\pfhz'=\max(\pfhz,t+\delta)\ge t+\delta$, (b) is vacuous, and
$\pfavail'(x)=t+\delta$ for $x\in\pfres_\pfmode(e)$. In the broadcast mode the
plays of the slots of $e$ at $t$ are added, which is (d).
\pfStep{2.}Let $x=(e',t')$. If $e$ is global, $\pfhz'=\pffloor'=\pfhz+\delta\ge\pfhz\ge t'+\delta'$
by (a), and (a) to (c) follow. Otherwise $\pfhz'\ge\pfhz$ and $\pffloor'=\pffloor$;
for $x'\in\pfres_\pfmode(e')$, if $x'\in\pfres_\pfmode(e)$ then
$\pfavail'(x')=t+\delta\ge r\ge\max(\pffloor,\pfavail(x'))\ge t'+\delta'$ by (c),
and if not, $\pfavail'(x')=\pfavail(x')$. Plays are only added, so (d) is kept.
\pfStep{3.}If $e$ is global, $t=\pfhz\ge t'+\delta'$ by (a). If $e'$ is global,
$t'+\delta'\le\pffloor\le r\le t$ by (b). Otherwise they share a resource $x'$,
and $t'+\delta'\le\max(\pffloor,\pfavail(x'))\le r\le t$ by (c).
\pfStep{4.}If $e$ is global, $t=\pfhz\ge t'+\delta'$. Otherwise the play
$\ell=(c,\kappa',t',t'+\delta')$ is in $\pfplays$ by (d), and
$\pfclear_e(t,\ell)$ by Lemma~\ref{lem:pf-choose}(ii) gives
$t'+\delta'\le t$, or $t+\delta\le t'$, or $\kappa'=\kappa$, $t'=t$ and
$t'+\delta'=t+\delta$, hence $\delta'=\delta$.
\end{proof}

\begin{lemma}[Exclusion and activations]
\label{lem:pf-exclusive}
For the parallel start times of either mode and all $i<j$:
$e_i\pfconf_\pfmode e_j\Rightarrow t_i+\delta_i\le t_j$. In the broadcast mode,
for all $i\neq j$ and every shared channel $c$ used by both, $\tau_i$ and
$\tau_j$ are disjoint, or $\pfkey_{e_i}(c)=\pfkey_{e_j}(c)$, $t_i=t_j$ and
$\delta_i=\delta_j$.
(\lean{Parallel.run_exclusive}, \lean{Chain.timeline_exclusive},
\lean{Shared.run_exclusive}, \lean{Shared.chain_timeline_exclusive})
\end{lemma}
\begin{proof}
Let $S_k$ be the state after scheduling $e_1\ldots e_k$. By
Lemma~\ref{lem:pf-covers}(1), $S_i$ covers $(e_i,t_i)$; by (2), applied to
$e_{i+1},\ldots,e_{j-1}$, so does $S_{j-1}$; and $e_j$ is scheduled from
$S_{j-1}$, so (3) gives the first claim and (4) the second, for $i<j$. The
second claim is symmetric in $i$ and $j$. Disjointness of
$\tau_i=[t_i,t_i+\delta_i)$ and $\tau_j$ is $t_i+\delta_i\le t_j$ or
$t_j+\delta_j\le t_i$.
\end{proof}

\noindent The key of a slot records the waveforms played: by \Rule{C-Transport}
($\jd{compat}_H$), the moves of one event on one shared channel use one
waveform, so $W_e(c)$ has one element, and equal keys mean the same waveform
for the same duration.

\begin{lemma}[Bound and latest end]
\label{lem:pf-bound}
For the parallel timing of either mode, the finish is
$\max_i(t_i+\delta_i)$ and is at most $\sum_i\delta_i$.
(\lean{ReportedTime.finish_latest_end}, \lean{Parallel.final_horizon_le},
\lean{Shared.final_horizon_le}, \lean{Chain.finish_le_sequential})
\end{lemma}
\begin{proof}
\emph{Latest end.} Scheduling $e$ from $S$ at start $t$ gives
$\pfhz'=\max(\pfhz,t+\delta(e))$: for a non-global event by definition, and for
a global one because $t=\pfhz$. Since $S_0$ has $\pfhz=0$, induction gives
$\pfhz_m=\max(0,\max_i(t_i+\delta_i))=\max_i(t_i+\delta_i)$.
\emph{Bound.} Call $S$ tight when $\pffloor\le\pfhz$, $\pfavail(r)\le\pfhz$ for
every $r$, and every play ends by $\pfhz$. $S_0$ is tight. In a tight state
every ready time is at most $\pfhz$, and so is $\pfchoose(S,e)$, by
Lemma~\ref{lem:pf-choose}(iii). Hence a non-global event has $t\le\pfhz$ and
$\pfhz'=\max(\pfhz,t+\delta)\le\pfhz+\delta$; a global one has
$\pfhz'=\pfhz+\delta$. The new state is tight: every value it writes is at most
$\pfhz'$. By induction $\pfhz_m\le\sum_i\delta_i$.
\end{proof}

\begin{lemma}[The sequential timing]
\label{lem:pf-seq}
For the sequential start times, $t_i+\delta_i\le t_j$ for all $i<j$, the finish
is $\sum_i\delta_i=\max_i(t_i+\delta_i)$, and therefore (L4), (L5) and (T1) hold.
(\lean{TimeOrder.prefixSums_ge}, \lean{Chain.sequentialFinish})
\end{lemma}
\begin{proof}
$t_j=t_i+\delta_i+\sum_{i<l<j}\delta_l\ge t_i+\delta_i$. The finish is by
definition $\sum_i\delta_i=t_m+\delta_m$, and every $t_i+\delta_i\le t_m+\delta_m$
by the first claim. Every two steps are disjoint in time, which gives (L4) and
the first disjunct of (L5).
\end{proof}

\begin{lemma}[What the compiler reports is a timeline]
\label{lem:pf-driver}
Every tuple that the compiler reports satisfies (i) to (vi) of
Definition~\ref{def:pf-timeline}.
(\lean{Reported.timeline}, \lean{Chain.refineChecked}, \lean{Parallel.timedRun_of_rulesAt},
\lean{Reported.exchangesLegal_of_rulesAt})
\end{lemma}
\begin{proof}
(i) is Lemma~\ref{lem:pf-refine}. (ii) is checked before a timeline is
reported, and so is $\chi_0\sim\pos_0$; Lemma~\ref{lem:pf-step} then
gives (iii). (iv) holds by construction. For (v), let $e_i=h_k$. By (L1), the
run of $K$ contains a \Rule{Step} for $h_k$ at $\pos_{k-1}$, whose premise
$H;\pos_{k-1}\vdash h_k\;\jd{core}$ mentions no time; it therefore holds at any
start time. The compiler evaluates every core rule of $\hat R$ on the two
frames of $h_k$ at $t_i$, and on every exchange frame at its start time, before
it reports $T$; by Lemma~\ref{lem:decide} a true evaluation means that the core
rule holds, and by Lemma~\ref{lem:elaborate} the frame then satisfies $R$.
This gives (v) and (vi). The placements of these frames follow program order,
exactly as in the run of $K$, because both are computed by $\nxt_H$ over
$h_1\ldots h_n$.
\end{proof}

\subsection{Events on disjoint resources commute}
\label{app:pf-commute}

Overlapping events must act on the state as if they ran in either order.
This section proves it from the matrices of the primitives; nothing about
overlapping events is assumed. We work with pulse-level instructions, the
projections of the steps.

\begin{definition}[Pulse-level resources and independence]
The resources of a pulse-level instruction are
\[
\begin{array}{@{}l@{\qquad}l@{\qquad}l@{}}
\pfres(\gm\,w)=\jd{ions}(w)\cup G_\gm &
\pfres(\gm\;\irkw{measure}\;q\to c)=\{q,c\}\cup G_\gm &
\pfres(\gm\;\irkw{reset}\;q)=\{q\}\cup G_\gm\\
\pfres(\irkw{latch}\;\fseq c\,\texttt{==}\,n)=\{\jd{ctl}\}\cup\fseq c &
\pfres(\gm\;\irkw{barrier}\;\fseq q)=\emptyset
\end{array}
\]
Two instructions $x,y$ are
\emph{independent}, $\pfind(x,y)$, when $\pfres(x)\cap\pfres(y)=\emptyset$.
(\lean{Independence.resources}, \lean{Independence.Independent})
\end{definition}

\begin{lemma}[Local operations on disjoint ions commute]
\label{lem:pf-local}
Call a map $\Lambda$ on register matrices \emph{local} to a set $Q$ of ions when
$\Lambda(\rho)=\sum_k A_k\rho A_k^\dagger$ with every $A_k$ of the form
$B_k\otimes I$, $B_k$ acting on the ions of $Q$ and $I$ on the others. If
$\Lambda$ is local to $Q$ and $\Lambda'$ to $Q'$ with $Q\cap Q'=\emptyset$, then
$\Lambda\circ\Lambda'=\Lambda'\circ\Lambda$. The action of a pulse, of a measurement
projector and the reset channel are local to their ions.
(\lean{Independence.Local.commute}, \lean{Independence.LocalAction.comm},
\lean{Independence.local_pulse}, \lean{Independence.local_reset})
\end{lemma}
\begin{proof}
Operators of the forms $B\otimes I$ and $I\otimes C$ on disjoint ions commute:
both products are $B\otimes C$ after reordering the tensor factors. If
$\Lambda(\rho)=\sum_kA_k\rho A_k^\dagger$ and $\Lambda'(\rho)=\sum_lC_l\rho C_l^\dagger$,
then $\Lambda(\Lambda'(\rho))=\sum_{k,l}A_kC_l\rho C_l^\dagger A_k^\dagger=
\sum_{l,k}C_lA_k\rho A_k^\dagger C_l^\dagger=\Lambda'(\Lambda(\rho))$. A pulse acts by
$\rho\mapsto W\rho W^\dagger$ with $W$ its matrix on its ions, a measurement
outcome by $\Pi^v_q\rho\Pi^v_q$, and the reset by
$K^0_q\rho K^{0\dagger}_q+K^1_q\rho K^{1\dagger}_q$.
\end{proof}

\begin{lemma}[Independent instructions commute]
\label{lem:pf-comm}
If $\pfind(x,y)$ then
$\sem{y}^\dagger(\sem{x}^\dagger\sigma)\pfperm\sem{x}^\dagger(\sem{y}^\dagger\sigma)$
for every $\sigma$.
(\lean{Independence.branch_comm}, \lean{Independence.pulse_comm})
\end{lemma}
\begin{proof}
By Lemma~\ref{lem:pf-lift} it suffices to take $\sigma=\beta=(C,b,\rho)$. Every
instruction other than a latch has the form
$\sem{x}=\gm_x\,?\,\bigl(\beta\mapsto o_1(\beta)\ldots o_r(\beta)\bigr)$, a guard
around a list of \emph{outcomes}. An outcome $o$ optionally writes one bit
$\jd{wr}(o)$ and applies a map $\Lambda_o$ to the matrix; it keeps the flag. A
pulse has one outcome with $\Lambda_o(\rho)=W\rho W^\dagger$; a measurement $q\to c'$
has two, writing $c'\mapsto v$ with $\Lambda_o(\rho)=\Pi^v_q\rho\Pi^v_q$ for $v=0,1$;
a reset has one, the reset channel; a barrier has one, the identity. Every
written bit and every ion on which $\Lambda_o$ acts is a resource of $x$. Three
cases.

\emph{Neither is a latch.} By Lemma~\ref{lem:pf-flag} every branch keeps flag
$b$. If $x$ is disabled on $\beta$ (marked and $b=\jd{ff}$), it is disabled on
every branch of $\sem{y}\beta$ as well, so both sides equal $\sem{y}\beta$; the
same if $y$ is disabled. Otherwise both are enabled on every branch, and
\[
\sem{y}^\dagger(\sem{x}\beta)=\bigl[o'(o(\beta))\bigr]_{o\in\jd{out}(x),\,o'\in\jd{out}(y)},\qquad
\sem{x}^\dagger(\sem{y}\beta)=\bigl[o(o'(\beta))\bigr]_{o'\in\jd{out}(y),\,o\in\jd{out}(x)},
\]
the first listed with $o$ outermost and the second with $o'$ outermost. For
each pair, $o'(o(\beta))=o(o'(\beta))$: the maps commute by
Lemma~\ref{lem:pf-local}, since their ions are disjoint resources; the writes
go to distinct bits, since written bits are resources, and writes to distinct
bits commute; neither changes the flag. The two lists therefore contain the
same branches in transposed order. (Two measurements are the case where the
order really differs.)

\emph{$x$ is the latch $\irkw{latch}\;\fseq c\,\texttt{==}\,n$ and $y$ is not.}
$\jd{ctl}\in\pfres(x)$, so $y$ is unmarked; and $\fseq c\subseteq\pfres(x)$,
so no outcome of $y$ writes a bit of $\fseq c$. The latch sets
$b:=(\jd{val}_C(\fseq c)=n)$ and changes nothing else. An outcome $o$ of $y$
leaves the bits $\fseq c$ and the flag unchanged, so the value that the latch
reads is the same before and after $o$, and $o$ does not read the flag. Hence
$o(\jd{latch}(\beta))=\jd{latch}(o(\beta))$ for each outcome, in the same list
order. The case where $y$ is the latch is symmetric.

\emph{Both are latches.} Both hold $\jd{ctl}$, so they are not independent.
\end{proof}

\begin{lemma}[Blocks of independent instructions commute]
\label{lem:pf-blocks}
If $\pfind(x,y)$ for every $x$ in a list $X$ and $y$ in a list $Y$, then
$\sem{Y}(\sem{X}\sigma)\pfperm\sem{X}(\sem{Y}\sigma)$.
(\lean{Independence.code_comm})
\end{lemma}
\begin{proof}
First one instruction $x$ past a list $Y$, by induction on $Y$: move $x$ past
the head of $Y$ by Lemma~\ref{lem:pf-comm}, then past the rest by the
induction hypothesis, using that $\sem{Y'}$ maps $\pfperm$-related ensembles to
$\pfperm$-related ensembles (Lemma~\ref{lem:pf-lift}) and that $\pfperm$ is
transitive. Then a list $X$ past $Y$, by induction on $X$ in the same way.
\end{proof}

\begin{lemma}[Non-conflicting steps have independent projections]
\label{lem:pf-proj}
Every resource of an instruction of $\jd{proj}(h)$ is in $\pfres(h)$, and none is a
transport channel. Hence, in either mode, if two steps do not conflict then
every instruction of the projection of one is independent of every
instruction of the projection of the other.
(\lean{Independence.project_resources}, \lean{Independence.resources_not_transport},
\lean{TimeOrder.steps_independent}, \lean{TimeOrder.steps_independent_broadcast})
\end{lemma}
\begin{proof}
By cases on $h$, comparing the table of Definition~\ref{def:pf-events} with
the pulse-level resources: a transport and a wait project to $\epsilon$; a
pulse, a readout and a pump keep their ions, their bit and their guard
resource; a broadcast projects to pulses on its ions with its guard; a latch
and a barrier project to themselves, and a barrier uses nothing at the pulse
level. No pulse-level instruction holds a transport channel. An exchange
projects to $\epsilon$. If $e,e'$ do not conflict in mode $\pfmode$, then
$\pfres_\pfmode(e)\cap\pfres_\pfmode(e')=\emptyset$; the pulse-level resources
of their projections lie in $\pfres(e)$ and $\pfres(e')$ and contain no
transport channel, so they lie in $\pfres_\pfmode(e)$ and $\pfres_\pfmode(e')$
and are disjoint.
\end{proof}

\subsection{The timeline in order of time}
\label{app:pf-order}

Sorting the steps by start time reorders them; we show that it only exchanges
neighbours that do not conflict, and then that such exchanges preserve the
meaning.

\begin{lemma}[Sorting is a sequence of exchanges]
\label{lem:pf-sort}
Let $I$ be a relation on items. If for all $i<j$, $t_j<t_i$ implies
$I(x_i,x_j)$, then $\pfsort(x_1\ldots x_m)$ is obtained from $x_1\ldots x_m$ by
finitely many exchanges of adjacent items $\ldots u\,v\ldots\to\ldots v\,u\ldots$
with $I(u,v)$, and it is sorted by start time.
(\lean{TimeOrder.commutes_byStart}, \lean{TimeOrder.byStart_sorted})
\end{lemma}
\begin{proof}
By induction on $m$. For $x_1;\fseq x$, the induction hypothesis turns $\fseq x$
into $\fseq y=\pfsort(\fseq x)$ by exchanges, and putting $x_1$ in front keeps
them valid. $\pfins(x_1,\fseq y)$ moves $x_1$ to the right past the initial
items $y$ of $\fseq y$ with start smaller than $t_1$, one exchange each. Each
such $y$ is some $x_j$ with $j>1$ and $t_j<t_1$, so $I(x_1,y)$ holds by the
hypothesis. Sortedness: $\pfins$ puts $x_1$ before the first item that does not
start earlier, so a sorted list stays sorted.
\end{proof}

\begin{lemma}[Only non-conflicting steps are exchanged]
\label{lem:pf-swaps}
For the reported start times, $\pfsort(x_1\ldots x_m)$ is obtained from
$x_1\ldots x_m$ by exchanges of adjacent steps that do not conflict
($\lnot\,e\pfconf_\pfmode e'$). For the sequential timing no exchange is needed.
(\lean{TimeOrder.time_order}, \lean{TimeOrder.time_order_broadcast},
\lean{TimeOrder.time_order_sequential})
\end{lemma}
\begin{proof}
Parallel timing: if $i<j$ and $t_j<t_i$, the steps cannot conflict, since
otherwise Lemma~\ref{lem:pf-exclusive} gives $t_j\ge t_i+\delta_i\ge t_i$.
Lemma~\ref{lem:pf-sort} applies with $I=\lnot\pfconf_\pfmode$. Sequential timing:
$t_i\le t_j$ for $i<j$ (Lemma~\ref{lem:pf-seq}), so the hypothesis of
Lemma~\ref{lem:pf-sort} holds with $I$ empty.
\end{proof}

\noindent Two remarks. Two steps on one broadcast channel do not conflict in
the broadcast mode and may be exchanged; their projections are empty, since
both are transports or exchanges, so Lemma~\ref{lem:pf-proj} still applies.
And because the exchanged pairs never share a resource, the steps that hold
any given resource keep their program order in $\pfsort$
(\lean{TimeOrder.resource_order}): the frame on which a step was checked lists,
for everything it holds, exactly the events that precede it in time.

\begin{lemma}[The order of time has the meaning of the program order]
\label{lem:pf-meaning}
For every $\eta$ and $\sigma$:
$\sem{K^{T}}_{H,\eta}\sigma\pfperm\sem{K}_{H,\eta}\sigma$.
(\lean{TimeOrder.perm_of_commutes}, \lean{TimeOrder.time_order_sound},
\lean{TimeOrder.time_order_sound_broadcast}, \lean{TimeOrder.time_order_sound_sequential})
\end{lemma}
\begin{proof}
Every instruction of $K^{T}$ is an instruction of $K$, and every instruction
of $K$ is core legal at the placement where it fires (L1). By
Lemma~\ref{lem:pf-instr}, which needs core legality at \emph{some} placement,
$\sem{K^{T}}_H\sigma=\sem{\jd{proj}(K^{T})}\sigma$ and $\sem{K}_H\sigma=\sem{\jd{proj}(K)}\sigma$.
Give each step $e$ the action $\alpha_e=\sem{\jd{proj}(e)}$ on ensembles; an
exchange has the identity action. Running the steps of a list in order
composes their actions, and since $\jd{proj}$ of a list is the concatenation of the
projections and exchanges project to $\epsilon$, running the steps of
$\pfsort(x_1\ldots x_m)$ gives $\sem{\jd{proj}(K^{T})}\sigma$ and running
$e_1\ldots e_m$ gives $\sem{\jd{proj}(K)}\sigma$ (by (M1)). By
Lemma~\ref{lem:pf-swaps} the first list arises from the second by exchanges of
adjacent non-conflicting steps $u,v$. For one exchange, $\alpha_v\circ\alpha_u$
and $\alpha_u\circ\alpha_v$ give $\pfperm$-related results by
Lemmas~\ref{lem:pf-proj} and~\ref{lem:pf-blocks}; the actions of the steps
after the exchanged pair preserve $\pfperm$ (Lemma~\ref{lem:pf-lift}); and the
steps before it are unchanged. By transitivity of $\pfperm$ over the finitely
many exchanges, $\sem{\jd{proj}(K^{T})}\sigma\pfperm\sem{\jd{proj}(K)}\sigma$.
\end{proof}

\subsection{Assembling the proof}
\label{app:pf-assembly}

\begin{proof}[Proof of Theorem~\ref{thm:pf-main}, and so of Theorem~\ref{thm:compile-correct}]
Fix (H1) to (H3). We take each clause in turn.

\emph{(M1)} is property (i) of the timeline (Lemma~\ref{lem:pf-refine}).
\emph{(M2)} is the projection part of Theorem~\ref{thm:pf-seq}.
\emph{(M3)}: let $\sigma'=\pfobs(\sem{K}_{H,\eta}\sigma)$. By
Theorem~\ref{thm:pf-seq}, $P\vdash_\eta\pfobs(\sigma)\Downarrow\sigma'$ and it is
the only such result. By Lemma~\ref{lem:pf-meaning}
$\sem{K^{T}}_{H,\eta}\sigma\pfperm\sem{K}_{H,\eta}\sigma$, and $\pfobs$, applied
branch by branch, preserves $\pfperm$, so
$\pfobs(\sem{K^{T}}_{H,\eta}\sigma)\pfperm\sigma'$.

\emph{(L1)} is Theorem~\ref{thm:pf-seq}. \emph{(L2)}: the conditions on
compiled steps and exchange steps are (v) and (vi) of the timeline
(Lemma~\ref{lem:pf-driver}), where the compiler decides $F\models_H R$ by
evaluating the elaborated rules $\hat R$ of (H1)
(Lemmas~\ref{lem:elaborate} and~\ref{lem:decide}).
\emph{(L3)} is (ii), (iii) and Lemma~\ref{lem:pf-depart}.
\emph{(L4)} and \emph{(L5)}: for the parallel timing, Lemma~\ref{lem:pf-exclusive}
gives (L4), and for two steps on one shared channel $c$ that are not disjoint
in time it gives $\pfkey_{e_i}(c)=\pfkey_{e_j}(c)$, $t_i=t_j$ and
$\delta_i=\delta_j$. Equal keys mean $W_{e_i}(c)=W_{e_j}(c)$, so
$w_i(c)=w_j(c)$ (Definition~\ref{def:pf-events}), and equal starts and
durations mean $\tau_i=\tau_j$. In the exclusive mode a shared transport
channel is a resource of every transport that uses it, so two steps on it
conflict and are disjoint in time by (L4), which is the first disjunct of
(L5). For the sequential timing, Lemma~\ref{lem:pf-seq}.
\emph{(L6)}: let two moves $q{:}p$ and $q'{:}p'$ of one transport of $K$ use
one shared channel. If $q=q'$ they are the same move, because the ions of a
transport are distinct (\Rule{C-Transport}). Otherwise the transport is core
legal (L1), so $\jd{compat}_H$ gives $\jd{wf}(p)=\jd{wf}(p')$, and
$\jd{Uniform}(H)$ (Definition~\ref{def:pf-uniform}) gives $\mu(p)=\mu(p')$ and
$d(p)=d(p')$, that is, $\mu(a)=\mu(b)$ and $d(a)=d(b)$ for the two moves.
(\lean{Broadcast.transport_same_motion}, \lean{Broadcast.run_same_motion})

\emph{(T1)}: by (iv) the finish is that of the timing; for the parallel timing
Lemma~\ref{lem:pf-bound} gives
$t_{\rm end}=\max_i(t_i+\delta_i)\le\sum_i\delta_i$, and for the sequential
timing Lemma~\ref{lem:pf-seq} gives equality.
\emph{(T2)}: let $h_k=\gm\;\execkw{pulse}\;w\;\execkw{via}\;k$. By (L1),
$H;\pos_{k-1}\vdash h_k\;\jd{core}$, so \Rule{C-Pulse} holds at $\pos_{k-1}$:
$k\in H$ and every ion of $w$ is at $\jd{site}(k)$; and
$\delta_H(h_k)=d(k)$ by the definition of $\delta_H$
(\lean{ReportedTime.pulse_duration_at_site}, \lean{ReportedTime.gate_durations}).
For a broadcast, \Rule{C-Broadcast} gives $d(k_j)=d(b)=\delta_H(h_k)$ for
each member, so a broadcast is charged the declared duration of each of its
calibrations.
\end{proof}

\noindent The sequential contract can also be stated in the Hoare logic of
Section~\ref{app:hoare}, with $\Gamma=(H,R,n_q,n_c,\eta)$.

\begin{corollary}[Hoare form]
\label{cor:hoare}
Under (H1) and (H2), for every $\eta$ and $\sigma_0$:
\[
\Gamma\vdash
\{\pos=\pos_0\land t=0\land\sigma=\sigma_0\}\ \ K\ \
\{H\vdash\pos:n_q\ \land\ t=t_{\rm seq}\ \land\
  P\vdash_\eta\pfobs(\sigma_0)\Downarrow\pfobs(\sigma)\}.
\]
(\lean{QASMCompilation.source_hoare}, \lean{compileQASM_hoare_sound})
\end{corollary}
\begin{proof}
By (L1), from every configuration that satisfies the precondition, $K$ runs
legally to $\langle\pos_n,t_{\rm seq}\rangle$ with ensemble $\sem{K}_{H,\eta}\sigma_0$,
so the triple whose postcondition fixes this final configuration is valid; by
Lemma~\ref{lem:hoare} it has a derivation (\lean{Hoare.hoare_of_run}).
\Rule{H-Conseq} weakens the postcondition: $\pos_n$ is valid because every
\Rule{Step} keeps the placement valid, and the source result is
Theorem~\ref{thm:pf-seq}.
\end{proof}

\paragraph{What the proof assumes}
The proof uses the ideal semantics of the primitives declared in
\S\ref{app:f-semantics}: each pulse acts by its matrix, a readout by the two
projectors, a pump by the reset channel, and a transport, a wait and an
exchange act as the identity on the computation. It assumes nothing about
overlapping events: Lemma~\ref{lem:pf-meaning} derives their meaning from the
primitives' matrices. Lean's kernel checks every lemma above; the dependencies
of the headline declarations use only the axioms \lean{propext},
\lean{Classical.choice} and \lean{Quot.sound} (\S\ref{sec:trust}). The theorem
is a soundness theorem: it constrains every result that the compiler returns
and does not claim that compilation succeeds.

\begin{table}[h]
\caption{The lemmas of this appendix and the Lean declarations that check them.}
\label{tab:pf-map}
\small\centering
\begin{tabular}{@{}l>{\raggedright\arraybackslash}p{.72\linewidth}@{}}
\toprule
statement & Lean declaration\\
\midrule
Theorem~\ref{thm:pf-main} & \lean{verified_compilation}\\
Theorem~\ref{thm:pf-seq} & \lean{compileQASM_sound}, \lean{QASMCompilation.source_correct}\\
Lemmas~\ref{lem:pf-lift}, \ref{lem:pf-flag}, \ref{lem:pf-det} &
  \lean{interpretPulseCode_append}, \lean{guarded_comp}, \lean{StatementsExecute.deterministic}\\
Lemmas~\ref{lem:pf-incl} to~\ref{lem:pf-select} &
  \lean{resolveIncludes_executes_iff}, \lean{expandProgram_source_executes}, \lean{selectProgram_correct}\\
Lemmas~\ref{lem:pf-bind} to~\ref{lem:pf-time} &
  \lean{Binding.stage_correct}, \lean{hardwareInstruction_refines}, \lean{Routed.stage_correct},
  \lean{Packed.stage_correct}, \lean{Timed.stage_correct}\\
Lemmas~\ref{lem:pf-xc} to~\ref{lem:pf-refine} &
  \lean{Chain.agrees_exchange}, \lean{Chain.agrees_move}, \lean{Chain.check_agrees},
  \lean{Chain.depart_end}, \lean{Chain.erase_refine}\\
Lemmas~\ref{lem:pf-choose} to~\ref{lem:pf-exclusive} &
  \lean{Shared.choose_spec}, \lean{Parallel.step_respects}, \lean{Parallel.run_exclusive},
  \lean{Shared.run_exclusive}\\
Lemmas~\ref{lem:pf-bound}, \ref{lem:pf-seq}, \ref{lem:pf-driver} &
  \lean{ReportedTime.finish_latest_end}, \lean{Chain.finish_le_sequential}, \lean{Reported.timeline}\\
Lemmas~\ref{lem:pf-local} to~\ref{lem:pf-proj} &
  \lean{Independence.Local.commute}, \lean{Independence.pulse_comm}, \lean{Independence.code_comm},
  \lean{Independence.project_resources}\\
Lemmas~\ref{lem:pf-sort} to~\ref{lem:pf-meaning} &
  \lean{TimeOrder.commutes_byStart}, \lean{TimeOrder.time_order}, \lean{TimeOrder.reported_time_order_sound}\\
(L6), (T2) & \lean{Broadcast.run_same_motion}, \lean{ReportedTime.gate_durations}\\
\bottomrule
\end{tabular}
\end{table}
\clearpage
  \fi
  \ifqccdreference
    \section{Complete language syntax}
\label{app:syntax}

This appendix is reference material: the complete grammar of every language,
of which Appendix~\ref{app:f-syntax} gives the abstract syntax that the proof
manipulates. Hardware is blue, OpenQASM purple, intermediate programs teal,
rules green and target code orange. The concrete hardware and rule spellings
are presentations of the fields of the input files;
Section~\ref{app:syntax-mapping} states which readers and printers are
proved. The verified source is an OpenQASM~2.0 abstract-syntax bundle.

\subsection{Names, expressions and sequences}
\label{app:syntax-names}

Let $a,s,p,k,b$ be names, $n$ a natural number, $z$ an integer, $q_i$ an ion
identity and $c_i$ a classical address. Names are finite identifier strings,
and file names are quoted strings. Source decimals denote exact rationals.
The following recursive sequence notation abbreviates BNF, not a new program
construct:
\[
\begin{array}{rcl}
\mathit{Seq}(X)&::=&\epsilon\mid X\;\mathit{Seq}(X)\\
\mathit{List}(X)&::=&\epsilon\mid X\;\mathit{Tail}(X)\\
\mathit{Tail}(X)&::=&\epsilon\mid ,X\;\mathit{Tail}(X).
\end{array}
\]
Expressions retain their complete trees, including unsimplified arithmetic:
\begingroup\small
\[
\begin{array}{rcl}
\alpha&::=&z/n\mid\circkw{pi}\mid a\mid-\alpha\mid f(\alpha)\\
&&{}\mid(\alpha+\alpha)\mid(\alpha-\alpha)\mid(\alpha*\alpha)\\
&&{}\mid(\alpha/\alpha)\mid(\alpha\mathbin{\texttt{\char94}}\alpha)\\
f&::=&\circkw{sin}\mid\circkw{cos}\mid\circkw{tan}\mid\circkw{exp}
       \mid\circkw{ln}\mid\circkw{sqrt}.
\end{array}
\]
\endgroup
The rational constructor requires $n>0$. Integers abbreviate denominator one.
Source text also permits ordinary decimal literals. Definedness is
established by Domain Analysis before effects are expanded. Its analysis is
conservative: for example, it proves $\pi/7$ defined without floating point,
but may reject another mathematically defined expression.

\subsection{Physical hardware programs}
\label{app:syntax-hardware}

The complete hardware declaration vocabulary is:
\begingroup\small
\[
\begin{array}{@{}r@{\;}c@{\;}l@{}}
H&::=&\hwkw{hardware}\;a;\;\mathit{Seq}(D)\\
D&::=&\hwkw{site}\;s[n];\\
&&{}\mid\hwkw{path}\;p:\mathit{Route}\;\hwkw{channel}\;n\\
&&\quad\hwkw{waveform}\;n\;\hwkw{duration}\;n;\\
&&{}\mid\hwkw{share}\;\hwkw{channel}\;n\;\mathit{Mode};\\
&&{}\mid\hwkw{optical}\;\hwkw{channel}\;n;\\
&&{}\mid\hwkw{calibration}\;k:W\;@s\;\mathit{Opt};\\
&&{}\mid\hwkw{broadcast}\;b:\hwkw{R}(\alpha,\alpha)\\
&&\quad\hwkw{sites}\;\mathit{List}(s)\;\mathit{Opt}\;\hwkw{limit}\;n;\\
&&{}\mid\hwkw{acquire}\;a\;@s\;\mathit{Opt};\\
&&{}\mid\hwkw{pump}\;p\;@s\;\mathit{Opt};\\
&&{}\mid\hwkw{controller}\;\hwkw{duration}\;n;\\
\mathit{Opt}&::=&\hwkw{channel}\;n\;\hwkw{duration}\;n\\
\mathit{Route}&::=&s\to s\;\mathit{RouteTail}\\
\mathit{RouteTail}&::=&\epsilon\mid{}\to s\;\mathit{RouteTail}\\
\mathit{Mode}&::=&\hwkw{mandatory}\mid\hwkw{optional}\\
W&::=&\hwkw{R}(\alpha,\alpha)\mid\hwkw{MS}(\alpha).
\end{array}
\]
\endgroup
No declaration fixes initial ion placement. A path is directed, and its interior
sites are occupied in transit. Its duration is explicit. Shared transport
channels require a common waveform. A mandatory channel also moves all ions
at every matching source, including spectators. Optical and transport channel
numbers belong to separate namespaces.

A calibration is one exact waveform at one site with a declared channel and
positive duration. A broadcast profile specifies an addressed R waveform,
allowed sites, duration and maximum addressed ions per site. Acquisition and
pumping implement measurement and reset. The optional controller gives the
latency of sampling a register comparison. Omitting one of these capabilities
does not implicitly grant it.

For example, calibration c4 declares MS($\pi/2$) at G on optical channel 7
with duration 4. It permits that physical pulse after $q_0$ has shuttled from
S to join $q_1$ at G. Appendix~\ref{app:native-hardware} prints every
declaration for this machine.

\subsection{Standard OpenQASM source}
\label{app:syntax-circuit}

The source constructors follow OpenQASM~2.0. Write $u$ for an indexed bit or
whole register, $a$ for a formal gate operand, and $v$ for either in its
appropriate context:
\begingroup\small
\[
\begin{array}{@{}r@{\;}c@{\;}l@{}}
C&::=&\circkw{OPENQASM}\;2.0;\;\mathit{Seq}(S)\\
u&::=&a\mid a[n]\\
S&::=&\circkw{include}\;\mathit{string};\\
&&{}\mid\circkw{qreg}\;a[n];\mid\circkw{creg}\;a[n];\\
&&{}\mid\circkw{gate}\;a\;\mathit{Params}\;\mathit{List}(a)
       \;\{\mathit{Seq}(B)\}\\
&&{}\mid\circkw{opaque}\;a\;\mathit{Params}\;\mathit{List}(a);\\
&&{}\mid O;\mid\circkw{if}(a\texttt{==}n)\;O;\\
&&{}\mid\circkw{barrier}\;\mathit{List}(u);\\
\mathit{Params}&::=&\epsilon\mid(\mathit{List}(a))\\
\mathit{Args}&::=&\epsilon\mid(\mathit{List}(\alpha))\\
O&::=&V(u)\mid\circkw{measure}\;u\to u\mid\circkw{reset}\;u\\
V(v)&::=&\circkw{U}(\alpha,\alpha,\alpha)\;v\\
&&{}\mid\circkw{CX}\;v,v\mid a\;\mathit{Args}\;\mathit{List}(v)\\
B&::=&V(a);\mid\circkw{barrier}\;\mathit{List}(a);.
\end{array}
\]
\endgroup
Gate-definition braces are part of standard OpenQASM. They are the only source
body construct and cannot contain another definition. Hardware, all executable
IRs and target programs are flat. A condition guards one operation, never an
invented nested instruction block. An opaque declaration is legal syntax, but
calling an opaque operation without an implementation is rejected.

A bundle supplies finite lists of named source units and include links, with a
chosen root identifier. An included unit may omit its version header:
\[
\begin{array}{rcl}
\mathit{Unit}&::=&\langle a,\mathit{Header},\mathit{Seq}(S)\rangle\\
\mathit{Header}&::=&\epsilon\mid\circkw{OPENQASM}\;2.0;\\
\mathit{Link}&::=&\langle a,\mathit{string},a\rangle\\
\mathit{Bundle}&::=&\langle\mathit{List}(\mathit{Unit}),
                         \mathit{List}(\mathit{Link})\rangle.
\end{array}
\]
The root must have the 2.0 header. A link maps its caller and include string to
a target unit. Include Resolution retains each statement's unit, position and
include provenance. Effect Expansion subsequently records the resolved
statement index and expansion path on every scalar occurrence.

Registers can be used sequentially: \texttt{\circkw{U}(pi/7,0,0) q[0];}
followed by \texttt{\circkw{CX} q[0],q[1];} is legal. There is no disjoint-operand
restriction between successive operations. Written order and every barrier's
operands and position are retained. Register-wide measurement pairs equal-width
quantum and classical registers in index order.

\subsection{Intermediate programs at all nine stages}
\label{app:syntax-ir}

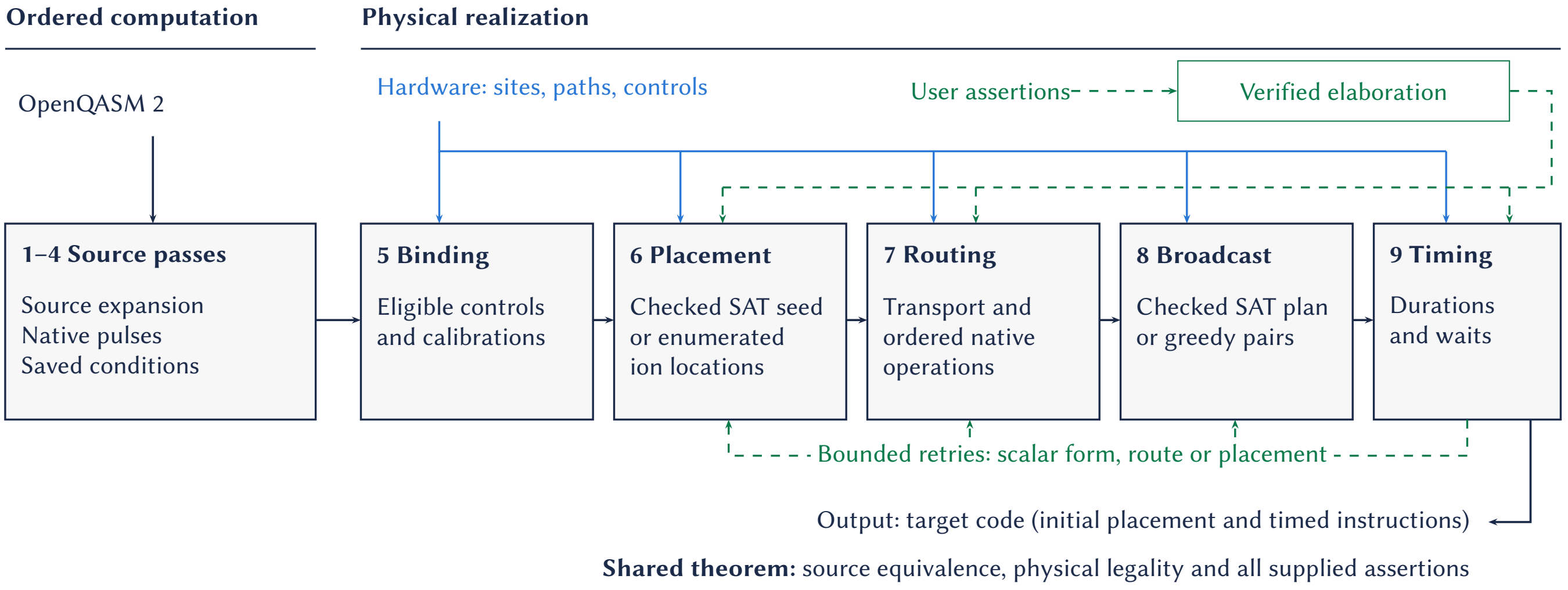
\begin{figure*}[t]
\centering
\resizebox{\linewidth}{!}{
\begin{tikzpicture}[x=1pt,y=1pt,
  font=\sffamily\fontsize{8.5}{10}\selectfont,
  txt/.style={inner sep=0pt,text=qink,align=left},
  head/.style={txt,font=\sffamily\bfseries\fontsize{8.5}{10}\selectfont},
  flow/.style={-{Stealth[length=3pt]},qink,line width=.7pt},
  hw/.style={-{Stealth[length=3pt]},qblue,line width=.65pt},
  rule/.style={-{Stealth[length=3pt]},qgreen,dashed,line width=.65pt}]
  \path[use as bounding box] (0,0) rectangle (516,-192);
  \node[head,anchor=north west] at (0,0) {Ordered computation};
  \node[head,anchor=north west] at (118,0) {Physical realization};
  \draw[qink,line width=.6pt] (0,-14)--(103,-14) (118,-14)--(516,-14);
  \node[txt,anchor=north west] at (4,-29) {OpenQASM 2};
  \draw[flow] (49,-43)--(49,-72);
  \node[txt,anchor=north west,text=qblue] at (123,-23)
    {Hardware: sites, paths, controls};
  \draw[qblue,line width=.65pt] (144,-38)--(144,-48)--(478,-48);
  \foreach \x in {144,224,308,392,478} \draw[hw] (\x,-48)--(\x,-72);
  \node[txt,anchor=west,text=qgreen] (assertions) at (300,-28) {User assertions};
  \node[draw=qgreen,fill=white,minimum width=110pt,minimum height=20pt,
    inner sep=2pt,text=qgreen] (elaboration) at (444,-28) {Verified elaboration};
  \draw[rule] (assertions.east)--(elaboration.west);
  \draw[qgreen,dashed,line width=.65pt]
    (elaboration.east)--(513,-28)--(513,-60)--(238,-60);
  \foreach \x in {238,322,499} \draw[rule] (\x,-60)--(\x,-72);
  \foreach \x/\w in {0/103,118/77,202/77,286/77,370/77,454/62}
    \draw[qink,fill=black!3,line width=.65pt] (\x,-72) rectangle +(\w,-65);
  \foreach \x/\dx in {103/15,195/7,279/7,363/7,447/7}
    \draw[flow] (\x,-104)--+(\dx,0);
  \node[head,anchor=north west] at (5,-79) {1--4 Source passes};
  \node[txt,anchor=north west] at (5,-96) {Source expansion\\Native pulses\\Saved conditions};
  \node[head,anchor=north west] at (123,-79) {5 Binding};
  \node[txt,anchor=north west] at (123,-96) {Eligible controls\\and calibrations};
  \node[head,anchor=north west] at (207,-79) {6 Placement};
  \node[txt,anchor=north west] at (207,-96) {Checked SAT seed\\or enumerated\\ion locations};
  \node[head,anchor=north west] at (291,-79) {7 Routing};
  \node[txt,anchor=north west] at (291,-96) {Transport and\\ordered native\\operations};
  \node[head,anchor=north west] at (375,-79) {8 Broadcast};
  \node[txt,anchor=north west] at (375,-96) {Checked SAT plan\\or greedy pairs};
  \node[head,anchor=north west] at (459,-79) {9 Timing};
  \node[txt,anchor=north west] at (459,-96) {Durations\\and waits};
  \draw[rule] (485,-137)--(485,-149)--(240,-149)--(240,-137);
  \draw[rule] (320,-149)--(320,-137);
  \draw[rule] (408,-149)--(408,-137);
  \node[txt,text=qgreen,fill=white,inner sep=2pt] at (354,-149)
    {Bounded retries: scalar form, route or placement};
  \draw[flow] (506,-137)--(506,-171)--(492,-171);
  \node[txt,anchor=east] at (486,-171)
    {Output: target code (initial placement and timed instructions)};
  \node[txt,anchor=east] at (486,-187)
    {\textbf{Shared theorem:} source equivalence, physical legality and all supplied assertions};
\end{tikzpicture}}
\caption{Compiler organization. Solid arrows show program flow and hardware
inputs; dashed arrows show assertion inputs and bounded retries. Stages
1--4 establish the ordered computation, and stages 5--9 construct its
physical realization. Timing checks the full rule set for both active and
idle realizations of guarded events. Optional SAT seeds and pair plans at
stages 6 and 8 are checked against the actual compilation; pass proofs are
reused. All successful outputs share the same correctness theorem.}
\label{fig:worked-pipeline}
\Description{Nine compiler stages establish source effects, then choose
controls, placement, transport, broadcast and timing under the supplied rules.}
\end{figure*}

Resolved Source is a sequence of located source statements, with includes
expanded. Let $S_R$ be the source-statement grammar without its include
alternative, and $o_R$ a unit/statement location with its include chain:
\[
\begin{array}{rcl}
P_R&::=&\mathit{Seq}(S_R^{o_R})\\
\alpha_D&::=&\langle\alpha,\sigma\rangle\\
\sigma&::=&-\mid0\mid+\mid?\\
P_D&::=&C[\alpha_D/\alpha].
\end{array}
\]
The last production uses every source constructor, substituting an annotated
angle at each root angle position. Successful Domain Analysis has no includes
and preserves all statements and expression trees. Its four sign annotations
mean negative, zero, positive and unknown. A separate theorem establishes
definedness. Erasing annotations recovers the resolved source with its 2.0
header. The include-origin table is retained separately. Neither stage changes
the source operations into target code.

The next stages use flat effect and pulse syntax. Each atom carries an origin
$o=(\text{statement},\text{expansion path})$. Superscripts below denote this
retained metadata, not additional runtime instructions:
\begingroup\small
\[
\begin{array}{@{}r@{\;}c@{\;}l@{}}
A_E&::=&\irkw{U}(\alpha,\alpha,\alpha)\;q_i
          \mid\irkw{CX}\;q_i,q_j\\
&&{}\mid\irkw{measure}\;q_i\to c_j\mid\irkw{reset}\;q_i\\
&&{}\mid\irkw{barrier}\;\mathit{List}(q_i)\\
A_P&::=&\irkw{R}(\alpha,\alpha)\;q_i
          \mid\irkw{MS}(\alpha)\;q_i,q_j\\
&&{}\mid\irkw{measure}\;q_i\to c_j\mid\irkw{reset}\;q_i\\
&&{}\mid\irkw{barrier}\;\mathit{List}(q_i)\\
L&::=&\irkw{latch}\;\irkw{saved}\leftarrow
       \irkw{little\_endian}(\mathit{List}(c_i))=n\\
I_X&::=&A_X^o;\mid\irkw{when}\;\irkw{saved}\;A_X^o;\mid L^o;\\
P_X&::=&\langle n_q,n_c,\mathit{Seq}(I_X)\rangle\quad(X=E,P).
\end{array}
\]
\endgroup
Effect Expansion outputs $P_E$. Pulse Selection outputs $P_P$, replacing U/CX
with complete ordered R/MS words and retaining measurement, reset, barriers and
saved guards exactly. During execution, a latch samples the comparison once
before the expanded effects of a conditional operation. No effect inside
that expansion changes the saved flag.

Capability Binding annotates every pulse instruction with its full candidate
list, $I_B::=I_P^{[k_1,\ldots,k_m]}$. Pulse, acquisition and pump references
have separate tables. Barriers have no candidate, and latches require the
controller. A Bound Program is $\langle n_q,n_c,\mathit{Seq}(I_B)\rangle$.
Initial Placement outputs $\langle\pos_0,\mathit{Seq}(I_P)\rangle$: it adds
a legal placement and leaves the pulse program unchanged. The binding table
is retained separately by the compilation record.

Transport Routing outputs $\langle\pos_0,\mathit{Seq}(I_H)\rangle$, using the
physical syntax below, without broadcasts or idles. Routing selects reachable
declarations from the bound candidates. Broadcast Formation has
the same underlying syntax but permits broadcast instructions and still no
idles. Time Assignment permits positive idles and establishes time-dependent
legality. Distinct IR types retain these contracts. Sharing constructors does
not conflate their obligations. For example, a routed MS at G follows the
actual S-to-G shuttle, and a timed version can insert three waits before its pulse.

\subsection{Target code}
\label{app:syntax-execution}

The complete grammar of target code is flat:
\begingroup\small
\[
\begin{array}{@{}r@{\;}c@{\;}l@{}}
K&::=&\execkw{init}\;\mathit{List}(q_i@s);\;\mathit{Seq}(I_H)\\
I_H&::=&O_H;\mid\execkw{when}\;\execkw{saved}\;O_H;\\
&&{}\mid\execkw{simd}\;\execkw{transport}\;\mathit{List}(M);\\
&&{}\mid\execkw{latch}\;\execkw{saved}\leftarrow\\
&&\qquad\execkw{little\_endian}(\mathit{List}(c_i))=n;\\
&&{}\mid\execkw{idle}\;n;\mid\execkw{barrier}\;\mathit{List}(q_i);\\
O_H&::=&\execkw{pulse}\;A\;@s\;\execkw{via}\;k\\
&&{}\mid\execkw{broadcast}\;\mathit{List}(A@s)\;\execkw{via}\;b\\
&&{}\mid\execkw{readout}\;q_i\to c_j\;@s\;\execkw{via}\;a\\
&&{}\mid\execkw{pump}\;q_i\;@s\;\execkw{via}\;p\\
A&::=&\execkw{R}(\alpha,\alpha)\;q_i\mid\execkw{MS}(\alpha)\;q_i,q_j\\
M&::=&q_i:s\to s[p].
\end{array}
\]
\endgroup
The typed instruction additionally retains each source origin. A broadcast
retains an ordered list of scalar-calibration references, one per member.
The display abbreviates that list by the profile name. A barrier retains its
source guard as metadata, but both branches realize the same zero-time
boundary. These fields remain in the intermediate report and preservation
theorems even when omitted from a printed line.

Calibration references determine physical sites and durations. Init is the
compiler's chosen placement, with one entry per ion. A move batch updates all
destinations simultaneously. Its duration is the maximum declared duration of
its paths. A false saved guard realizes an idle of the active event's duration.
Transport is unconditional. Thus measurement outcomes do not change the route
or elapsed time. Source gates never appear as hardware opcodes.

\subsection{Concrete and quantified rule grammars}
\label{app:syntax-rules}

A rule file names its observations by the field names below; the main text
and Appendix~\ref{app:formal} write them in mathematical notation, and the
listings of the appendices print them as \texttt{N-}, \texttt{N+},
\texttt{Ntr}, \texttt{cap}, \texttt{n\_optical}, \texttt{t} and
\texttt{delta}:
\begingroup\small
\[
\begin{array}{@{}r@{\;}c@{\;}l@{}}
\mathcal R&::=&\epsilon\mid\rulekw{assert}\;a:r;\;\mathcal R\\
\mathit{ph}&::=&\rulekw{before}\mid\rulekw{during}\mid\rulekw{after}\\
e&::=&n\mid\rulekw{start}\mid\rulekw{end}\mid\rulekw{duration}\\
&&{}\mid\rulekw{occ}_{\mathit{ph}}(s)\mid\rulekw{moveCount}\\
&&{}\mid\rulekw{pulseCount}\mid\rulekw{Rcount}\mid\rulekw{MScount}\\
&&{}\mid\rulekw{pulses}(s)\mid\rulekw{readoutCount}\\
&&{}\mid\rulekw{pumpCount}\mid\rulekw{latchCount}\\
&&{}\mid\rulekw{pathUse}(p)\mid\rulekw{channelUse}(c)\\
&&{}\mid\rulekw{channelWaves}(c)\\
&&{}\mid(e+e)\mid(e\mathbin{\dot{-}}e)\\
r&::=&e\le e\mid e=e\mid(r\land r)\mid(r\lor r)\\
&&{}\mid\lnot r\mid(r\Rightarrow r).
\end{array}
\]
\endgroup
All expressions are natural-valued, with $x\mathbin{\dot{-}}y=\max(x-y,0)$.
The field names and the notation of the main text correspond as follows:
\begin{center}\small\setlength{\tabcolsep}{4pt}
\begin{tabular}{@{}llllll@{}}
\toprule
field & notation & field & notation & field & notation\\
\midrule
$\rulekw{start}$ & $t$ & $\rulekw{occ}_{\rulekw{before}}(s)$ & $N^-(s)$ & $\rulekw{moveCount}$ & $n_{\rm move}$\\
$\rulekw{duration}$ & $\delta$ & $\rulekw{occ}_{\rulekw{after}}(s)$ & $N^+(s)$ & $\rulekw{pulseCount}$ & $n_{\rm optical}$\\
$\rulekw{end}$ & $t+\delta$ & $\rulekw{occ}_{\rulekw{during}}(s)$ & $\Ntr(s)$ & $\rulekw{Rcount}$, $\rulekw{MScount}$ & $n_{\rm R}$, $n_{\rm MS}$\\
$\rulekw{pathUse}(p)$ & $M(p)$ & $\rulekw{pulses}(s)$ & $P(s)$ & $\rulekw{readoutCount}$ & $n_{\rm readout}$\\
$\rulekw{channelUse}(c)$ & $M(c)$ & capacity of $s$ & $\capa(s)$ & $\rulekw{pumpCount}$ & $n_{\rm pump}$\\
$\rulekw{channelWaves}(c)$ & $W(c)$ & & & $\rulekw{latchCount}$ & $n_{\rm latch}$\\
\bottomrule
\end{tabular}
\end{center}
The optical count $n_{\rm optical}$ includes acquisition and pumping. R/MS
counts distinguish the unitary families. A two-ion R broadcast contributes two
optical actions, while an MS pulse contributes one. Initialization has zero
times and counts. Rules also apply to barriers, waits and disabled-event
idles, not just optical commands.

The quantified source language adds $\top$, $\bot$, site-capacity lookup,
sorted references, and $\forall x{:}D.\,r$ and $\exists x{:}D.\,r$, for
$D\in\{\mathcal S,\mathcal P,\mathcal C_T\}$. Site observations accept
site references, path traffic accepts path references, and channel observations
accept transport-channel references. Names bind lexically, including shadowing.
References are checked before finite expansion; each quantifier body is checked
even when its domain is empty. Capacity lookup lowers to the declared constant,
universal quantification to conjunction and existential quantification to
disjunction. The implementation retains named declarations and category labels
at the input boundary; all lowered formulas enter the same rule set.
Appendix~\ref{app:rule-language} illustrates the concrete event observations.

\subsection{Representation, readers and printers}
\label{app:syntax-mapping}

Names resolve to table indices in their own namespaces. The program model
contains placement, exact expressions, guards, origins, ordered operands and
calibration references, independently of typography. The paper's hardware
declarations supply empty legacy source-operation lists in the compatibility
geometry. The backend determines physical capabilities from calibrations, and
it does not use those legacy permissions to grant a pulse.

Theorem~\ref{thm:compile-correct} starts from the decoded bundle, hardware
and rules. Three facts about the readers and printers are proved in Lean,
outside that theorem. The device file and the rule-program file are read by
strict readers: a text is accepted exactly when it is the printed form of the
value returned, so no member of an accepted file is ignored
(\lean{Strict.hardwareFile_file}, \lean{Strict.rulesProgram_file}). OpenQASM
reading inverts printing: for every printable program, reading its printed text
returns that program (\lean{parse_printProgram}). The printed time report
carries the reported start times and finish (\lean{printed_time}). The
listings of these appendices are displays of the compiler's values in the
notation of this appendix, not input files. A displayed $\pi/2$ and an
algebraically equal expression with a different tree need not match the same
calibration. The theorem is about successful compilations: grammar
membership does not promise that conservative domain analysis, exact binding
or bounded search accepts every feasible program.

    \section{Additional hardware and intermediate-language detail}
\label{sec:native-languages}

The main text connects declarations, assertions and target code on one
running example, the device S, J, G. Fig.~\ref{fig:devices} shows how the same declaration model
represents four other layout abstractions. The examples illustrate
connectivity; their capacities, control resources and durations are not
specifications of the published devices.

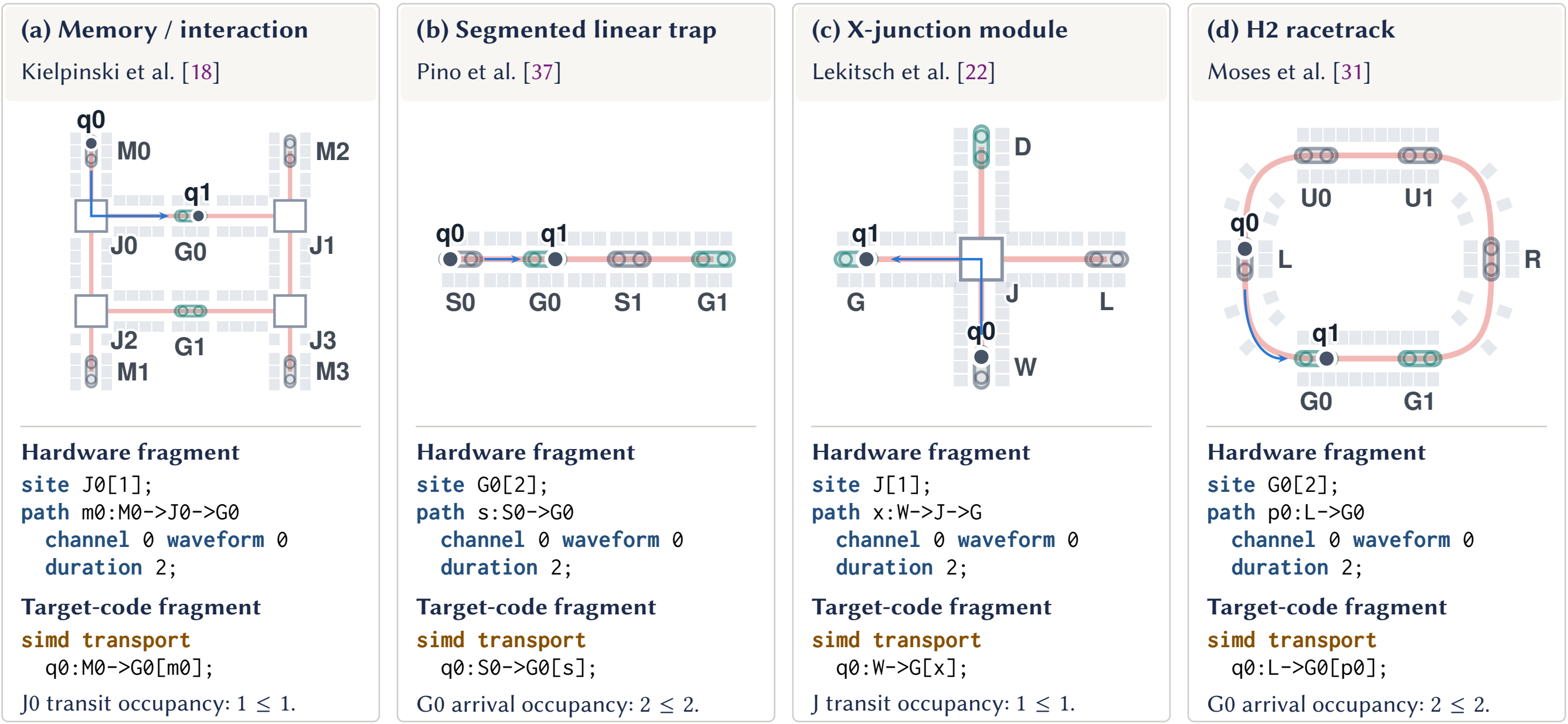
\begin{figure*}[t]
\centering
\resizebox{\linewidth}{!}{
\providecommand{\QCCDLiteratureCite}[1]{\cite{#1}}
\begin{tikzpicture}[x=1pt,y=1pt,
  card/.style={draw=qrail,fill=white,rounded corners=2pt,line width=.5pt},
  txt/.style={inner sep=0pt,align=left,text=qink},
  head/.style={txt,anchor=north west,font=\sffamily\bfseries\fontsize{8.5}{10}\selectfont},
  sub/.style={txt,anchor=north west,font=\sffamily\fontsize{8}{9.5}\selectfont},
  code/.style={txt,anchor=north west,text=black,
    font=\ttfamily\fontsize{8}{9}\selectfont}]
  \path[use as bounding box] (0,0) rectangle (514,-236);
  \foreach \x in {0,130,260,390}{
    \draw[card] (\x,0) rectangle +(124,-236);
    \fill[qpaper,rounded corners=2pt] (\x+1,-1) rectangle +(122,-31);
    \draw[qrail,line width=.5pt] (\x+6,-139)--+(112,0);
  }
  \node[head] at (6,-5) {(a) Memory / interaction};
  \node[sub] at (6,-19) {Kielpinski et al.~\QCCDLiteratureCite{kielpinski2002architecture}};
  \node[head] at (136,-5) {(b) Segmented linear trap};
  \node[sub] at (136,-19) {Pino et al.~\QCCDLiteratureCite{pino2021demonstration}};
  \node[head] at (266,-5) {(c) X-junction module};
  \node[sub] at (266,-19) {Lekitsch et al.~\QCCDLiteratureCite{lekitsch2017blueprint}};
  \node[head] at (396,-5) {(d) H2 racetrack};
  \node[sub] at (396,-19) {Moses et al.~\QCCDLiteratureCite{moses2023race}};
  \foreach \x/\scene in {62/kmw-fragment,192/pino-fragment,322/lekitsch-fragment,452/moses-fragment}{
    \node[inner sep=0pt] at (\x,-84)
      {\resizebox{108pt}{!}{\QCCDScene{\scene}}};
  }
  \foreach \x in {6,136,266,396}{
    \node[sub,font=\sffamily\bfseries\fontsize{8}{9}\selectfont]
      at (\x,-144) {Hardware fragment};
    \node[sub,font=\sffamily\bfseries\fontsize{8}{9}\selectfont]
      at (\x,-195) {Target-code fragment};
  }
  \node[code] at (6,-155) {
    \hwkw{site} J0[1];\\
    \hwkw{path} m0:M0->J0->G0\\
    \quad\hwkw{channel} 0 \hwkw{waveform} 0\\
    \quad\hwkw{duration} 2;};
  \node[code] at (136,-155) {
    \hwkw{site} G0[2];\\
    \hwkw{path} s:S0->G0\\
    \quad\hwkw{channel} 0 \hwkw{waveform} 0\\
    \quad\hwkw{duration} 2;};
  \node[code] at (266,-155) {
    \hwkw{site} J[1];\\
    \hwkw{path} x:W->J->G\\
    \quad\hwkw{channel} 0 \hwkw{waveform} 0\\
    \quad\hwkw{duration} 2;};
  \node[code] at (396,-155) {
    \hwkw{site} G0[2];\\
    \hwkw{path} p0:L->G0\\
    \quad\hwkw{channel} 0 \hwkw{waveform} 0\\
    \quad\hwkw{duration} 2;};
  \node[code] at (6,-206) {
    \execkw{simd} \execkw{transport}\\
    \quad q0:M0->G0[m0];};
  \node[code] at (136,-206) {
    \execkw{simd} \execkw{transport}\\
    \quad q0:S0->G0[s];};
  \node[code] at (266,-206) {
    \execkw{simd} \execkw{transport}\\
    \quad q0:W->G[x];};
  \node[code] at (396,-206) {
    \execkw{simd} \execkw{transport}\\
    \quad q0:L->G0[p0];};
  \node[sub] at (6,-227) {J0 transit occupancy: $1\leq1$.};
  \node[sub] at (136,-227) {G0 arrival occupancy: $2\leq2$.};
  \node[sub] at (266,-227) {J transit occupancy: $1\leq1$.};
  \node[sub] at (396,-227) {G0 arrival occupancy: $2\leq2$.};
\end{tikzpicture}}
\caption{Topology abstractions of four published designs: (a) the
memory--interaction network of Kielpinski et al.~\cite{kielpinski2002architecture},
(b) the segmented linear trap of Pino et al.~\cite{pino2021demonstration},
(c) an X-junction cell from Lekitsch et al.~\cite{lekitsch2017blueprint}, and
(d) the H2 racetrack~\protect\dglyph{racetrack-h2} of Moses et al.~\cite{moses2023race}.
Each hardware fragment declares the drawn path, and its execution fragment
moves $q_0$ to $q_1$ from the shown placement. Capacities, channels, waveforms
and durations are illustrative.}
\label{fig:devices}
\end{figure*}

R is an equatorial rotation. The two primitive unitaries are
\[
\begin{aligned}
R(\theta,\phi)&=\exp[-i\theta(\cos\phi X+\sin\phi Y)/2],\\
MS(\theta)&=\exp[-i\theta(X\otimes X)/2].
\end{aligned}
\]
The five-pulse CX decomposition has the same density-matrix action as the
source gate, including when its operands are entangled with other qubits.
Table~\ref{tab:irs} summarizes the nine intermediate programs whose
independently defined semantics support this preservation argument.

\begin{table*}[t]
\centering
\caption{Nine intermediate programs. $X^*$ denotes a finite statement sequence.
Programs carry register sizes and origin metadata. Annotations are not runtime commands.}
\label{tab:irs}
\small\renewcommand{\arraystretch}{1.15}
\resizebox{\linewidth}{!}{%
\begin{tabular}{@{}p{114pt}p{214pt}p{164pt}@{}}
\toprule
Rewrite output & Program syntax & Concrete change / obligation\\
\midrule
1. Resolved Source & $S^{o*}$: located QASM statements, no includes & Expand includes and retain origins.\\
2. Analyzed Source & $S^{o*}[\langle\alpha,\mathit{sign}\rangle/\alpha]$ & Retain $\pi/7$ with definedness evidence.\\
3. Effect Program & $I_E^*$, $E::=\irkw{U}\mid\irkw{CX}\mid\irkw{measure}\mid\irkw{reset}\mid\irkw{barrier}$ & Expand operands with one saved guard per condition.\\
4. Pulse Program $P_P$ & $I_P^*$, $P::=\irkw{R}\mid\irkw{MS}\mid\irkw{measure}\mid\irkw{reset}\mid\irkw{barrier}$ & CX becomes a five-pulse R/MS word.\\
5. Bound Program $P_B$ & $(I_P^{[k_1,\ldots,k_m]})^*$ & Annotate all eligible calibrations at G.\\
6. Placed Program & $\langle\pos_0,I_P^*\rangle$ & Choose legal initial $q_0@S,q_1@G$.\\
7. Routed Program $P_R$ & $\langle\pos_0,T^*\rangle$, $T::=\execkw{transport}\ m\mid F$ & Insert $q_0:S\to G$ preserving pulse prefix.\\
8. Packed Program $P_K$ & $\langle\pos_0,K^*\rangle$, $K::=T\mid\execkw{broadcast}\ b,\vec q$ & Adjacent equal R pulses share profile \texttt{b0}.\\
9. Timed Program $P_T$ & $\langle\pos_0,V^*\rangle$, $V::=K\mid\execkw{idle}\ \delta$ & Add three unit idles for event legality.\\
\bottomrule
\end{tabular}}
\par\smallskip
\begin{minipage}{\textwidth}\small
$I_X::=X^o;\mid\irkw{when}\ \irkw{saved}\ X^o;\mid
\irkw{latch}\ \irkw{saved}\leftarrow(\vec c=n);$ for $X=E,P$.
Operands and angles follow their source/pulse constructors.
$F$ is a calibrated pulse, acquisition, pump, latch or barrier.
Non-latch effects, including broadcasts, may retain a saved guard.
$m$ lists ion/path pairs, $b$ names a broadcast profile and $\delta>0$.
Resolved statements carry source locations $o$. Effect/pulse origins also
record expansion paths. The compilation record retains the binding table
after placement. Each syntax admits only its stage's operations.
Separate contracts establish legality and simulation.
\end{minipage}
\end{table*}

\begin{figure*}[t]
\centering
\resizebox{\linewidth}{!}{
\begin{tikzpicture}[x=1pt,y=1pt,font=\sffamily\fontsize{8}{10}\selectfont,
  repair note/.style={anchor=north,align=center,text width=158pt,inner sep=0pt}]
  \path[use as bounding box] (0,0) rectangle (514,-242);
  \node[anchor=north,align=center,text width=514pt,inner sep=0pt]
    at (257,0) {\hwkw{capacity}$(\mathrm S)=\hwkw{capacity}(\mathrm G)=3$
      \qquad $N^{\rm tr}(\mathrm G)\leq2$ \qquad two ticks per move};
  \draw[qrail] (169,-24)--(169,-237) (344,-24)--(344,-237);

  \node[repair note,font=\sffamily\bfseries\fontsize{9}{11}\selectfont]
    at (81,-26) {(a) A move waiting cannot repair};
  \node[repair note] at (81,-51) {Tick 16: $q_2@\mathrm S$, $q_0,q_1@\mathrm G$};
  \node[inner sep=0pt] at (81,-109) {\resizebox{150pt}{!}{%
    \begin{tikzpicture}[x=\QCCDUnit,y=-\QCCDUnit,transform shape]
      \QCCDGeometry{80}{20}
      \QCCDDeclareSite{S}{0}{0}{0}{qhwTrap}{3}{qhwTrap}
      \QCCDDeclareSite{G}{120}{0}{0}{qhwTrap}{3}{qhwTrap}
      \QCCDRail{S}{G}\QCCDDrawSite{S}\QCCDDrawSite{G}
      \QCCDSiteLabel{S}\QCCDSiteLabel{G}
      \QCCDIon{q2}{S}{0}\QCCDIon{q0}{G}{0}\QCCDIon{q1}{G}{1}
      \draw[hop red] ($(S)+(0,-42)$) to[bend left=8] ($(G)+(0,-42)$);
      \draw[qred,line width=1.6pt] (54,-52)--(66,-40) (54,-40)--(66,-52);
      \pgfresetboundingbox\path[use as bounding box] (-40,-61) rectangle (166,54);
    \end{tikzpicture}}};
  \node[repair note] at (81,-158) {$q_2:\mathrm S\to\mathrm G$ would make\\$N^{\rm tr}(\mathrm G)=3$.};
  \node[repair note,text=qred,font=\sffamily\bfseries\fontsize{8.5}{10}\selectfont]
    at (81,-189) {Prune this edge before\\extending the route.};
  \node[repair note] at (81,-220) {Former candidate: 3 moves,\\36 ticks. Invalid under the rule.};

  \node[repair note,font=\sffamily\bfseries\fontsize{9}{11}\selectfont]
    at (256,-26) {(b) First make room};
  \node[repair note] at (256,-51) {Move $q_0$ out during $[16,18)$};
  \node[inner sep=0pt] at (256,-109) {\resizebox{150pt}{!}{%
    \begin{tikzpicture}[x=\QCCDUnit,y=-\QCCDUnit,transform shape]
      \QCCDGeometry{80}{20}
      \QCCDDeclareSite{S}{0}{0}{0}{qhwTrap}{3}{qhwTrap}
      \QCCDDeclareSite{G}{120}{0}{0}{qhwTrap}{3}{qhwTrap}
      \QCCDRail{S}{G}\QCCDDrawSite{S}\QCCDDrawSite{G}
      \QCCDSiteLabel{S}\QCCDSiteLabel{G}
      \QCCDIon{q0}{S}{0}\QCCDIon{q2}{S}{2}\QCCDIon{q1}{G}{1}
      \draw[hop blue] ($(G)+(0,-42)$) to[bend right=8] ($(S)+(0,-42)$);
      \pgfresetboundingbox\path[use as bounding box] (-40,-61) rectangle (166,54);
    \end{tikzpicture}}};
  \node[repair note] at (256,-158) {\textcolor{black}{\texttt{\execkw{simd} \execkw{transport}}}\\
    \textcolor{black}{\texttt{q0:G->S[p1];}}\\State after the move: $[S,G,S]$.};
  \node[repair note,text=qgreen,font=\sffamily\bfseries\fontsize{8.5}{10}\selectfont]
    at (256,-189) {G occupancy: two during,\\one after.};
  \node[repair note] at (256,-220) {Physical capacity and the\\programmable rule both hold.};

  \node[repair note,font=\sffamily\bfseries\fontsize{9}{11}\selectfont]
    at (431,-26) {(c) Admit the next operand};
  \node[repair note] at (431,-51) {Move $q_2$ in during $[18,20)$};
  \node[inner sep=0pt] at (431,-109) {\resizebox{150pt}{!}{%
    \begin{tikzpicture}[x=\QCCDUnit,y=-\QCCDUnit,transform shape]
      \QCCDGeometry{80}{20}
      \QCCDDeclareSite{S}{0}{0}{0}{qhwTrap}{3}{qhwTrap}
      \QCCDDeclareSite{G}{120}{0}{0}{qhwTrap}{3}{qhwTrap}
      \QCCDRail{S}{G}\QCCDDrawSite{S}\QCCDDrawSite{G}
      \QCCDSiteLabel{S}\QCCDSiteLabel{G}
      \QCCDIon{q0}{S}{0}\QCCDIon{q1}{G}{1}\QCCDIon{q2}{G}{2}
      \draw[hop orange] ($(S)+(0,-42)$) to[bend left=8] ($(G)+(0,-42)$);
      \pgfresetboundingbox\path[use as bounding box] (-40,-61) rectangle (166,54);
    \end{tikzpicture}}};
  \node[repair note] at (431,-158) {\textcolor{black}{\texttt{\execkw{simd} \execkw{transport}}}\\
    \textcolor{black}{\texttt{q2:S->G[p0];}}\\State after the move: $[S,G,G]$.};
  \node[repair note,text=qgreen,font=\sffamily\bfseries\fontsize{8.5}{10}\selectfont]
    at (431,-189) {Next \execkw{MS} pulse acts\\on $q_1,q_2$ at G.};
  \node[repair note] at (431,-220) {Complete repaired target code:\\6 moves, 42 ticks.};
\end{tikzpicture}}
\caption{A stricter rule changes routing on unchanged three-slot hardware.
Path \texttt{p0} runs S$\to$G; \texttt{p1} runs G$\to$S. At tick 16,
direct admission of $q_2$ violates the two-ion occupancy rule. Evicting $q_0$
before admitting $q_2$ prepares the next MS pulse legally. The complete
execution takes six moves and 42 ticks. The former three-move, 36-tick
candidate violates the rule. Panels show the critical fragment.}
\label{fig:routing-repair}
\end{figure*}
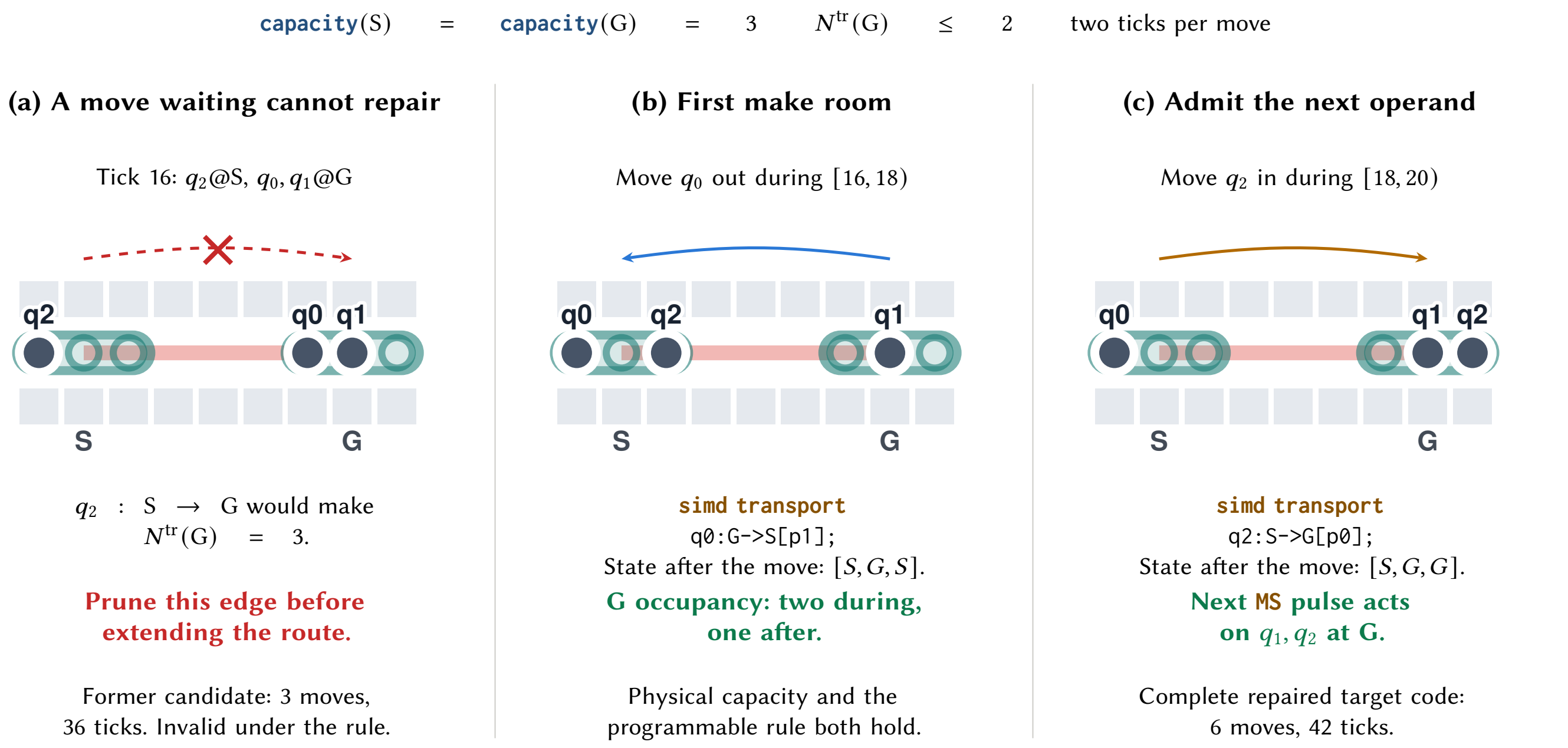

    \section{Complete constraint catalogue}
\label{app:rules}
\raggedbottom

These conditions cover declarations, operands, geometry,
capabilities and event rules. They are established by the compilation
passes and their theorems. User formulas strengthen the mandatory physical
conditions and cannot disable them. Each entry isolates one condition with a
concrete trap or ion example.

\begin{figure*}[t]
\centering
\resizebox{\textwidth}{!}{
\begin{tikzpicture}[x=1pt,y=1pt,font=\sffamily\fontsize{8.5}{10}\selectfont]
\node[anchor=north west,text=qink,font=\sffamily\bfseries\fontsize{9}{11}\selectfont] at (0,170)
  {(a) Route to an interaction site};
\node[anchor=north west,inner sep=0pt] at (0,152) {\resizebox{165pt}{!}{%
  \begin{tikzpicture}[x=\QCCDUnit,y=-\QCCDUnit,transform shape]
    \QCCDGeometry{72}{18}
    \QCCDDeclareSite{S}{0}{0}{0}{qhwTrap}{2}{qhwTrap}
    \QCCDDeclareSite{G}{144}{0}{0}{qhwTrap}{2}{qhwTrap}
    \QCCDDeclareSite{P}{72}{72}{0}{qhwTrap}{1}{qhwTrap}
    \QCCDDeclareJunction{J}{72}{0}{qhwGrid}
    \QCCDRail{S}{J}\QCCDRail{J}{G}\QCCDRail{J}{P}
    \QCCDDrawSite{S}\QCCDDrawSite{G}\QCCDDrawSite{P}\QCCDDrawJunction{J}
    \QCCDSiteLabel{S}\QCCDSiteLabel{G}\QCCDSiteLabel{P}\QCCDJunctionLabel{J}
    \QCCDIon{q0}{S}{0}\QCCDIon{q1}{G}{0}
    \draw[hop blue] ($(S)+(0,-30)$) to[bend left=10] node[above,font=\sffamily\fontsize{10}{11}\selectfont]{$p_0$} ($(G)+(0,-30)$);
  \end{tikzpicture}}};
\node[anchor=north west,align=left,text width=165pt] at (0,0)
  {$q_0:S\!\to\!J\!\to\!G$, then $\execkw{MS}(\pi/2)$ on $q_0,q_1$.\\
   $G$: $1\!\to\!2$ ions. $J$: one passing ion.};

\node[anchor=north west,text=qink,font=\sffamily\bfseries\fontsize{9}{11}\selectfont] at (177,170)
  {(b) Endpoints alone can miss overflow};
\node[anchor=north west,inner sep=0pt] at (181,135) {\resizebox{165pt}{!}{%
  \begin{tikzpicture}[x=\QCCDUnit,y=-\QCCDUnit,transform shape]
    \QCCDGeometry{72}{9}
    \QCCDDeclareSite{S}{0}{0}{0}{qhwTrap}{1}{qhwTrap}
    \QCCDDeclareSite{G}{72}{0}{0}{qhwTrap}{1}{qhwTrap}
    \QCCDDeclareSite{P}{144}{0}{0}{qhwTrap}{1}{qhwTrap}
    \QCCDRail{S}{G}\QCCDRail{G}{P}
    \QCCDDrawSite{S}\QCCDDrawSite{G}\QCCDDrawSite{P}
    \QCCDSiteLabel{S}\QCCDSiteLabel{G}\QCCDSiteLabel{P}
    \QCCDIon{q0}{S}{0}\QCCDIon{q1}{G}{0}
    \draw[hop blue] ($(S)+(0,-28)$) to[bend left=12] ($(G)+(0,-28)$);
    \draw[hop orange] ($(G)+(0,-28)$) to[bend left=12] ($(P)+(0,-28)$);
  \end{tikzpicture}}};
\node[anchor=north west,align=left,text width=165pt] at (177,0)
  {At $G$: before $=1$, after $=1$, reserved $=2$.\\
   Capacity 1 requires $q_1$ to leave first.};

\node[anchor=north west,text=qink,font=\sffamily\bfseries\fontsize{9}{11}\selectfont] at (354,170)
  {(c) Shared transport moves spectators};
\node[anchor=north west,inner sep=0pt] at (374,152) {\resizebox{100pt}{!}{%
  \begin{tikzpicture}[x=\QCCDUnit,y=-\QCCDUnit,transform shape]
    \QCCDGeometry{72}{9}
    \QCCDDeclareSite{S}{0}{0}{0}{qhwTrap}{2}{qhwTrap}
    \QCCDDeclareSite{G}{72}{0}{0}{qhwTrap}{2}{qhwTrap}
    \QCCDDeclareSite{P}{0}{72}{0}{qhwTrap}{2}{qhwTrap}
    \QCCDDeclareSite{T}{72}{72}{0}{qhwTrap}{2}{qhwTrap}
    \QCCDRail{S}{G}\QCCDRail{P}{T}
    \QCCDDrawSite{S}\QCCDDrawSite{G}\QCCDDrawSite{P}\QCCDDrawSite{T}
    \QCCDSiteLabel{S}\QCCDSiteLabel{G}\QCCDSiteLabel{P}\QCCDSiteLabel{T}
    \QCCDIon{q0}{S}{0}\QCCDIon{q1}{P}{0}
    \draw[hop blue] ($(S)+(0,-28)$) to[bend left=12] ($(G)+(0,-28)$);
    \draw[hop orange] ($(P)+(0,-28)$) to[bend left=12] ($(T)+(0,-28)$);
  \end{tikzpicture}}};
\node[anchor=north west,align=left,text width=165pt] at (354,0)
  {Both paths: channel 0, waveform 7.\\
   Mandatory means $q_0$ \emph{and} $q_1$ move.};
\end{tikzpicture}}
\caption{Concrete cases reused throughout the catalogue. (a) A declared
S--J--G path brings $q_0$ to $q_1$ before an MS pulse. (b) A departing ion and
a different arriving ion both count during a conservative transport frame.
(c) Mandatory shared transport includes a spectator on a second path. All
hardware glyphs use the editable website-derived drawing primitives.}
\label{fig:rule-atlas}
\end{figure*}
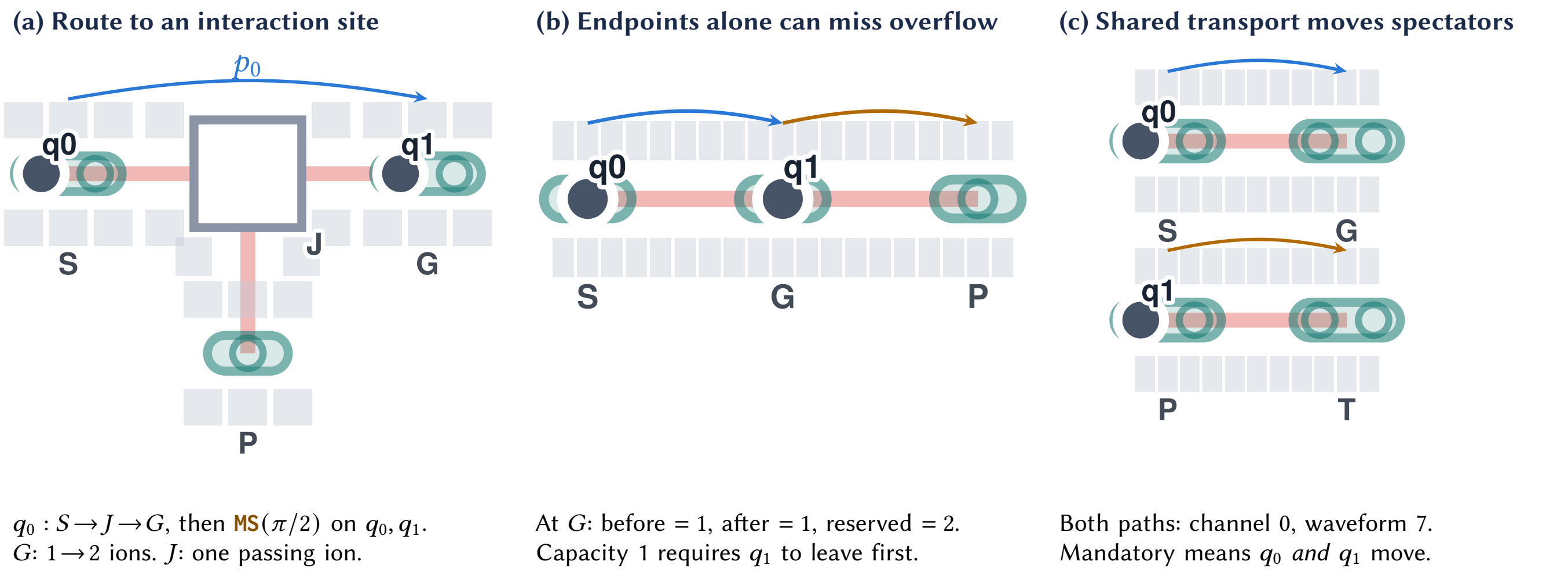

The atlas uses $S[2]$, $G[2]$, $P[1]$ and $J[1]$, with
$p_0:S\to J\to G$. The path has channel 0, waveform 7 and duration 2.
G declares the needed physical R/MS calibrations. Starting with $q_0@S,q_1@G$,
a legal shuttle produces $q_0@G,q_1@G$ without changing their logical
identities. The two-zone example in Appendix~\ref{app:native} removes J and P
and gives a complete executable instance.

\subsection{Hardware declarations: H1--H12}
\label{app:rule-hardware}

\par\smallskip
\begingroup\small
\setlength{\tabcolsep}{3pt}\renewcommand{\arraystretch}{1.13}
\noindent\begin{tabular}{@{}p{.09\columnwidth}p{.86\columnwidth}@{}}
\toprule
ID & Condition and concrete example\\
\midrule
H1 & Sites are nonempty and their names distinct. S and G pass, but two sites both
named G cannot identify where $q_0$ is located.\\
H2 & Every site has positive capacity. $J[1]$ permits one passing ion, whereas
$J[0]$ is invalid. Compatibility inputs' legacy operation lists must have no
duplicates. They grant no physical capability.\\
H3 & Path endpoints exist and differ. Interior sites exist, are distinct and
exclude both endpoints. $S\to J\to G$ passes, but $S\to J\to S$ fails.\\
H4 & Every path has exactly one positive duration. The declared $p_0$ takes
2 ticks. A missing or zero duration is invalid.\\
H5 & Shared transport channels are declared once and name at least one path.
Channel 0 may drive $p_0$, but an otherwise unused shared channel 9 fails.\\
H6 & A mandatory channel/waveform selects at most one path at a given source.
Two channel-0/waveform-7 paths leaving S would make the destination of $q_0$
ambiguous and are invalid.\\
\bottomrule
\end{tabular}
\par\smallskip
\noindent\begin{tabular}{@{}p{.09\columnwidth}p{.86\columnwidth}@{}}
\toprule
ID & Condition and concrete example (continued)\\
\midrule
H7 & Optical channel identifiers are distinct. One declaration of optical
channel 7 permits the R/MS calibrations at G. Duplicating it fails.\\
H8 & Each R/MS calibration uses a declared optical channel, positive duration
and a site with capacity at least its arity. MS at $G[2]$ passes, but MS at $J[1]$
fails even before ions are placed.\\
H9 & A broadcast profile has a declared channel, positive duration and address
limit, and a nonempty, duplicate-free list of existing positive-capacity sites.
An R profile for G with limit 2 can address $q_0,q_1$.\\
H10 & Each acquisition names an existing positive-capacity site, declared optical
channel and positive duration. Reading $q_0$ at G in 3 ticks is a capability.
A declaration at nonexistent U is invalid.\\
H11 & Pump declarations have the same site/channel/duration requirements.
Resetting $q_0$ at G requires a declared pump, for example duration 4.\\
H12 & A present controller has positive sampling latency. Duration 1 supports
a condition on $c_0,c_1$. No controller means no latch capability.\\
\bottomrule
\end{tabular}
\endgroup\par\smallskip

Multiple matching scalar calibrations are permitted and remain distinct table
entries. This lets binding retain several physical choices for the same pulse.
Transport channel 0 and optical channel 0 would still be distinct namespaces.

\subsection{Source programs: C1--C8}
\label{app:rule-circuit}

\par\smallskip
\begingroup\small
\setlength{\tabcolsep}{3pt}\renewcommand{\arraystretch}{1.13}
\noindent\begin{tabular}{@{}p{.09\columnwidth}p{.86\columnwidth}@{}}
\toprule
ID & Condition and concrete example\\
\midrule
C1 & The root is version 2.0 and included units are version-compatible. Bundle
identifiers and caller/path links are unique, references resolve and expansion
is acyclic. A library defining the operation on $q_0$ must be supplied.\\
C2 & Declarations and names satisfy the source environment's uniqueness,
kind and positive-register-width requirements. Declaring \texttt{q[2]} allocates
$q_0,q_1$, but using an undeclared \texttt{r[0]} fails.\\
C3 & Gate formals are distinct, qubit-formal lists nonempty, parameters bound,
body names declared and call arities matched. A one-parameter, one-qubit
definition called on $q_0$ receives one angle and one operand. An opaque call
without a body fails.\\
C4 & Every angle is accepted by the sound domain analysis. $\pi/7$ is
defined, while a zero denominator fails. No approximation is substituted for an
unsupported operating point on $q_0$.\\
C5 & Indexed operands are in bounds, register expansion has compatible widths,
and each scalar CX has distinct operands. \texttt{\circkw{CX} q[0],q[1];} passes, whereas
\texttt{\circkw{CX} q[0],q[0];} fails. Successive operations may reuse $q_0$.\\
C6 & Measurement pairs two scalar operands or equal-width registers of the
correct quantum/classical kinds. Measuring width-two q into width-two c expands
to $q_0\to c_0,q_1\to c_1$. Scalar/register mixing fails.\\
C7 & Reset addresses declared ions. A conditional reads one declared complete
classical register. Its latch has nonempty, distinct, in-range bit addresses.
An out-of-range integer comparison is simply false when no store matches it.\\
C8 & Barriers have nonempty, distinct, in-range resolved operands. A barrier on
$q_0,q_1$ retains both ions and its exact position. Duplicate $q_0$ fails.
Expansion never sorts or reschedules the source operations.\\
\bottomrule
\end{tabular}
\endgroup\par\smallskip

These conditions impose no fixed H/X/CX basis and no requirement that adjacent
operations act on disjoint ions. U/CX plus expanded source definitions provide
the unitary source language. Measurement, reset and conditions have separate
quantum/classical semantics.

\subsection{Placement and initialization: P1--P4}
\label{app:rule-placement}

\par\smallskip
\begingroup\small
\setlength{\tabcolsep}{3pt}\renewcommand{\arraystretch}{1.13}
\noindent\begin{tabular}{@{}p{.09\columnwidth}p{.86\columnwidth}@{}}
\toprule
P1 & Placement has exactly one site for each logical ion. For two qubits,
$(S,G)$ locates $q_0,q_1$, whereas $(S)$ leaves $q_1$ unplaced. The index of an entry
is the ion identity and never changes under transport.\\
P2 & Every assigned site exists and its occupancy is within capacity.
$(S,G)$ passes for $S[1],G[2]$, but $(S,S)$ fails.\\
P3 & Every site, path and channel reference in a rule expression exists
in its declared domain.
$\Ntr(\mathrm J)\le1$ refers to J.
The same formula for an undeclared U fails.\\
P4 & Every supplied formula holds at the initial frame: unchanged placement,
$t=\delta=0$ and zero event counts. A rule requiring one ion in S
therefore constrains the compiler's placement of $q_0,q_1$.\\
\bottomrule
\end{tabular}
\endgroup\par\smallskip

\subsection{Shuttling and shared controls: T1--T7}
\label{app:rule-transport}

\par\smallskip
\begingroup\small
\setlength{\tabcolsep}{3pt}\renewcommand{\arraystretch}{1.13}
\noindent\begin{tabular}{@{}p{.09\columnwidth}p{.86\columnwidth}@{}}
\toprule
T1 & A transport batch is nonempty and moves each ion at most once.
Moving $q_0$ once along $p_0$ passes, but assigning two paths to $q_0$ fails.\\
T2 & Every move names an existing path and an in-range ion currently at that
path's source. After $q_0:S\to G$, using that same forward path again fails.\\
T3 & Distinct simultaneous moves never traverse a link in opposite directions.
Moving $q_0:S\to G$ together with $q_1:G\to S$ fails.\\
T4 & Distinct simultaneous paths have disjoint interior sites. Two routes
through the same J fail even if endpoint capacities would permit them.\\
\bottomrule
\end{tabular}
\endgroup\par\smallskip
\begingroup\small
\setlength{\tabcolsep}{3pt}\renewcommand{\arraystretch}{1.13}
\noindent\begin{tabular}{@{}p{.09\columnwidth}p{.86\columnwidth}@{}}
\toprule
\multicolumn{2}{@{}l@{}}{\textit{Shared transport conditions (continued)}}\\
\midrule
T5 & A shared transport channel uses one waveform per batch. Two channel-0
moves with waveform 7 agree. Waveforms 7 and 8 conflict.\\
T6 & A mandatory channel includes every ion at every matching path source.
In Fig.~\ref{fig:rule-atlas}(c), moving $q_0:S\to G$ also requires
$q_1:P\to T$, even if the next optical pulse uses only $q_0$.\\
T7 & Placements before and after the batch are valid, and conservative
transit count $\Ntr$ fits every site's capacity. In (b), G contains departing
$q_1$ and arriving $q_0$: $\Ntr(\mathrm G)=2$ exceeds capacity 1.\\
\bottomrule
\end{tabular}
\endgroup\par\smallskip

The maximum path duration is the duration of the whole simultaneous batch.
The routing invariant follows the actual destination updates, including every
spectator. It is not a claim that an ion teleports to a site mentioned by a
later pulse.

\subsection{Physical effects and timing: E1--E10}
\label{app:rule-frames}

\par\smallskip
\begingroup\small
\setlength{\tabcolsep}{3pt}\renewcommand{\arraystretch}{1.13}
\noindent\begin{tabular}{@{}p{.09\columnwidth}p{.86\columnwidth}@{}}
\toprule
E1 & Scalar pulses have valid operands and an existing exact matching
calibration. R addresses one ion and MS addresses two distinct ions.
MS on $q_0,q_0$ fails.\\
E2 & Every pulse operand occupies its selected calibration site.
MS on $q_0,q_1$ at G becomes available after the S-to-G shuttle,
not while $q_0$ remains at S.\\
E3 & An optical broadcast has at least two members, one scalar reference per
member, and no repeated ion. Addressing $q_0,q_1$ once each passes this test.\\
E4 & Every broadcast member is scalar-legal and matches the profile's exact R
parameters, channel, duration and allowed sites. R($-\pi/2,0$) at G cannot
join a profile for R($\pi/2,0$). MS is not a broadcast member.\\
E5 & Per-site addresses respect the profile's limit. The two addressed ions
at G require limit at least 2, even if G has capacity 3. Formation combines
only adjacent independent pulses with the same saved guard.\\
\bottomrule
\end{tabular}
\endgroup\par\smallskip
\begingroup\small
\setlength{\tabcolsep}{3pt}\renewcommand{\arraystretch}{1.13}
\noindent\begin{tabular}{@{}p{.09\columnwidth}p{.86\columnwidth}@{}}
\toprule
\multicolumn{2}{@{}l@{}}{\textit{Acquisition, pumping and control conditions}}\\
\midrule
E6 & Readout addresses an in-range ion and classical bit at a declared
acquisition site. $q_0\to c_0$ at G passes with acquisition $a0$.
The same instruction while $q_0@S$ fails.\\
E7 & Pumping addresses an in-range ion at its declared pump site.
Pumping $q_0@G$ requires $p0$ at G. An R calibration alone is insufficient.\\
E8 & Latch addresses a nonempty duplicate-free classical list and a controller.
Readout may later overwrite $c_0$, but cannot overwrite the saved comparison
used by a following guarded operation on $q_1$.\\
E9 & Physical durations are positive, except barriers, whose duration is zero.
A false guarded 3-tick readout becomes a 3-tick idle. Every event preserves
valid placement, and clocks advance by the declared duration.\\
E10 & Every user formula holds on every scheduled event, including barriers,
inserted waits and both realizations of a guarded event. A condition requiring
$n_{\rm optical}=1$ would reject the inactive idle, even if its fired
readout at G passes.\\
\bottomrule
\end{tabular}
\endgroup\par\smallskip

The compiler additionally preserves the exact ordered pulse/effect projection:
transport and waits erase, and a broadcast expands to its stored scalar
occurrences. Origins, barriers, classical destinations and guards are retained.
This is a refinement invariant, not an extra user-configurable hardware rule.

\section{Expressing and extending policies}
\label{app:rule-language}

A rule is a formula over one physical event frame. The frame supplies the
device, placements before and after the event, its actual instruction, start
time and duration. For a guarded instruction, legality includes two frames:
the fired instruction and the equal-duration idle. Formula evaluation agrees
with its independent propositional meaning by induction on the formula.

\subsection{Every observation on a concrete frame}
\label{app:rule-observations}

For the atlas move $q_0:S\to J\to G$, take initial placement $(S,G)$,
start 0 and duration 2, using primitive $p$ and channel $c$. A subsequent
scalar R on $q_0@G$ starts at 5 and lasts 2. The observations are:

\par\smallskip
\begingroup\small
\setlength{\tabcolsep}{3pt}\renewcommand{\arraystretch}{1.12}
\noindent\begin{tabular}{@{}p{.53\columnwidth}rr@{}}
\toprule
Observation & Shuttle & R pulse\\
\midrule
$t,\ t+\delta,\ \delta$ & $0,2,2$ & $5,7,2$\\
$N^-(\mathrm S)$ & 1 & 0\\
$N^+(\mathrm G)$ & 2 & 2\\
$\Ntr(\mathrm J)$ & 1 & 0\\
$n_{\rm move}$ & 1 & 0\\
$n_{\rm optical},n_{\rm R}$ & $0,0$ & $1,1$\\
$n_{\rm MS}$ & 0 & 0\\
$P(\mathrm G)$ & 0 & 1\\
$n_{\rm readout}$ & 0 & 0\\
$n_{\rm pump}$ & 0 & 0\\
$n_{\rm latch}$ & 0 & 0\\
$M(p),M(c),W(c)$ & $1,1,1$ & $0,0,0$\\
$\capa(\mathrm S),\capa(\mathrm G)$ & $2,2$ & $2,2$\\
\bottomrule
\end{tabular}
\endgroup\par\smallskip

$N^-$ and $N^+$ count resident ions. $\Ntr$ counts the union of old and new
residents plus passing ions at interior sites. It therefore
counts both different ions at G in Fig.~\ref{fig:rule-atlas}(b), rather than
taking the maximum endpoint count. An R broadcast on $q_0,q_1@G$ has
$n_{\rm R}=n_{\rm optical}=P(\mathrm G)=2$.
An MS on those ions has $n_{\rm MS}=n_{\rm optical}=1$.
Readout and pump each contribute one optical action and one dedicated count.
A latch has $n_{\rm latch}=1$ and $n_{\rm optical}=0$. Idles and barriers
have zero counts.

\subsection{Six concrete policies}
\label{app:rule-policies}

The complete formula grammar supports, for example:
\begingroup\small
\[
\begin{array}{@{}l@{}}
\text{P-G: }\Ntr(\mathrm G)\le2\\
\text{P-M: }n_{\rm move}\le1\\
\text{P-R: }1\le n_{\rm optical}\Rightarrow5\le t\\
\text{P-B: }P(\mathrm G)\le2\\
\text{P-A: }1\le n_{\rm readout}\Rightarrow3\le\delta\\
\text{P-I: }(t=0\land\delta=0)
             \Rightarrow N^-(\mathrm S)=1.
\end{array}
\]
\endgroup
P-G constrains both resident and passing ions at G. P-M rejects the two-path
mandatory movement in Fig.~\ref{fig:rule-atlas}(c). It cannot override that
channel's compulsory spectator participation. P-R delays the first R on
$q_0@G$ until tick 5. P-B permits the two-ion R broadcast but forbids three
addressed pulses there in one event. P-A constrains the declared acquisition
duration, not the logical source measurement's cost. P-I requires one ion at
S in the initial frame. It also applies to any zero-duration event at tick 0,
because the language has no separate initialization-test observation.

In the two-zone example the shuttle ends at tick 2. Three legal unit waits
satisfy P-R before the first optical event at tick 5. A false guarded pulse
has $n_{\rm optical}=0$, so P-R is vacuous on its idle realization, while
occupancy and duration formulas still apply.

\subsection{Expressibility and extensions}
\label{app:rule-extension}

A new Boolean combination of the listed observations reuses the same rule
soundness theorem. For example, adding
$n_{\rm R}\le1\lor N^+(\mathrm G)\le2$
requires no new compiler proof. The rule must hold on the concrete G occupancy
at every event. The name of the rule alone supplies no such guarantee.

The language does not inspect arbitrary history, classical values, analog
waveforms or noise. A cooldown condition depending on time since the last
acquisition needs an explicit history state and a new observation. The current
start-time field cannot express this condition. Such an extension requires
a defined transition, an observation interpretation and an extended
inductive preservation argument. New physical primitives additionally require
independent logical semantics and a refinement proof. These boundaries keep
flexibility precise rather than treating every possible physical constraint
as already formalized.

\section{Full design comparison}
\label{app:design-comparison}
Table~\ref{tab:eval-summary} lists every design of RQ3 and RQ5 with its
device properties, wins and median times per circuit interaction graph.

\begin{table*}[t]
\caption{Design comparisons of Sections~\ref{sec:eval-selection} and~\ref{sec:eval-families}. Time: median modeled time relative to the fastest compared design for the same program. Wins: programs a design is fastest for (ties included). Device columns: ion-holding sites, junctions and gate-capable share of capacity at 96 ions. Moves and chain exchanges per two-qubit gate and the transport share of operation time are medians. Teal: median slowest-to-fastest ratio. Bold, underline: best, second best; --: no checked compilation.}
\label{tab:eval-summary}
\centering\footnotesize\setlength{\tabcolsep}{4.4pt}\renewcommand{\arraystretch}{0.90}
\begin{tabular}{@{}lrrrrrrrrrrrrr@{}}
\toprule
 & \multicolumn{3}{c}{Device at 96 ions} & \multicolumn{2}{c}{All programs} & \multicolumn{5}{c}{Time by circuit interaction graph} & \multicolumn{2}{c}{Per 2-qubit gate} & Transp.\\
\cmidrule(lr){2-4}\cmidrule(lr){5-6}\cmidrule(lr){7-11}\cmidrule(lr){12-13}
QCCD design & Sites & Junct. & Gate\,\% & Wins & Time & \multicolumn{1}{c}{\pglyph{chain}\,Chain} & \multicolumn{1}{c}{\pglyph{star}\,Star} & \multicolumn{1}{c}{\pglyph{sparse}\,Sparse} & \multicolumn{1}{c}{\pglyph{dense}\,Dense} & \multicolumn{1}{c}{\pglyph{local}\,Local} & Moves & Exch. & time\,\%\\
\midrule
\multicolumn{14}{c}{\cellcolor{qpaper}\emph{Trap size and layout at equal capacity (Section~\ref{sec:eval-selection}), relative to the fastest design of the class; 75 programs: 27, 13, 24, 8, 3 by interaction graph}}\\
\dglyph{line-c4}~Line, 4-ion traps & 24 & 0 & 100 & 12 & \underline{1.08} & \underline{1.04} & 1.65 & 2.19 & \textbf{1.04} & -- & 0.34 & 0.12 & 13\\
\dglyph{line-c6}~Line, 6-ion traps & 16 & 0 & 100 & 6 & 1.11 & 1.04 & 1.41 & 1.67 & 1.16 & -- & 0.25 & 0.10 & 11\\
\dglyph{line-c8}~Line, 8-ion traps & 12 & 0 & 100 & \underline{17} & 1.12 & \textbf{1.01} & \underline{1.31} & 1.21 & 1.16 & \textbf{1.00} & 0.20 & 0.08 & 9\\
\dglyph{line-c12}~Line, 12-ion traps & 8 & 0 & 100 & 10 & 1.13 & 1.14 & 1.32 & 1.10 & \underline{1.07} & \underline{1.01} & 0.13 & 0.06 & 7\\
\dglyph{line-c16}~Line, 16-ion traps & 6 & 0 & 100 & 0 & 1.44 & 1.49 & 1.58 & 1.34 & 1.54 & 1.35 & 0.09 & 0.05 & 5\\
\dglyph{line-c24}~Line, 24-ion traps & 4 & 0 & 100 & 1 & 2.13 & 2.28 & 1.95 & 1.84 & 2.18 & 1.95 & 0.06 & 0.02 & 2\\
\dglyph{line-c32}~Line, 32-ion traps & 3 & 0 & 100 & 0 & 2.75 & 2.97 & 2.62 & 2.07 & 3.16 & 2.63 & 0.04 & 0.02 & 1\\
\dglyph{line-c48}~Line, 48-ion traps & 2 & 0 & 100 & 0 & 4.04 & 4.62 & 3.51 & 3.62 & 5.11 & 4.18 & 0.01 & 0.00 & 0\\
\dglyph{ring-c12}~Ring, 12-ion traps & 8 & 0 & 100 & 12 & \textbf{1.08} & 1.11 & \textbf{1.00} & \underline{1.10} & 1.08 & \underline{1.01} & 0.13 & 0.07 & 5\\
\dglyph{ladder-c12}~Ladder, 12-ion traps & 8 & 0 & 100 & \textbf{20} & 1.21 & 1.38 & 1.51 & \textbf{1.00} & 1.16 & 1.10 & 0.17 & 0.08 & 11\\
\addlinespace[1pt]
\textcolor{black}{\textbf{Slowest $\div$ fastest design}} & & & & & \textcolor{black}{\textbf{3.44$\times$}} & \textcolor{black}{\textbf{4.44$\times$}} & \textcolor{black}{\textbf{3.14$\times$}} & \textcolor{black}{\textbf{2.80$\times$}} & \textcolor{black}{\textbf{4.04$\times$}} & \textcolor{black}{\textbf{3.86$\times$}} & & &\\
\midrule
\multicolumn{14}{c}{\cellcolor{qpaper}\emph{Published design families (Section~\ref{sec:eval-families}), relative to the fastest of these and the generic shapes below; 65 programs: 22, 12, 20, 8, 3 by interaction graph}}\\
\dglyph{line-moveless}~Line, 7-ion traps & 14 & 0 & 100 & \underline{13} & \textbf{1.14} & \textbf{1.01} & 1.40 & 1.12 & \textbf{1.19} & 1.16 & 0.27 & 0.15 & 13\\
\dglyph{ring-cyclone-c12}~Ring, 12-ion traps~\cite{khan2026cyclone} & 8 & 4 & 100 & 1 & 1.21 & 1.26 & 1.16 & 1.17 & 1.20 & \underline{1.04} & 0.13 & 0.08 & 9\\
\dglyph{ring-cyclone-c4}~Ring, 4-ion traps~\cite{khan2026cyclone} & 24 & 4 & 100 & 2 & 1.29 & 1.29 & 1.35 & 1.23 & 1.40 & 1.52 & 0.43 & 0.28 & 17\\
\dglyph{line-l6}~Line, six traps~\cite{murali2020architecting} & 6 & 5 & 100 & 2 & 1.38 & 1.41 & 1.42 & 1.24 & 1.38 & 1.42 & 0.22 & 0.10 & 10\\
\dglyph{grid-g2x3}~Six traps on a bus~\cite{murali2020architecting} & 6 & 3 & 100 & 4 & 1.40 & 1.70 & 1.10 & 1.22 & 1.42 & 1.61 & 0.13 & 0.08 & 13\\
\dglyph{line-chain2}~Line, 2-ion sites & 48 & 0 & 100 & 7 & 1.90 & 1.45 & 4.36 & 1.97 & 1.97 & 4.15 & 1.87 & 0.97 & 33\\
\dglyph{grid-c2}~Grid, 2-ion traps & 49 & 30 & 100 & 1 & 3.16 & 4.51 & 2.99 & 2.63 & 3.10 & 3.53 & 1.73 & 0.91 & 46\\
\dglyph{wise}~Switched-sort grid (WISE)~\cite{malinowski2023wise} & 48 & 16 & 100 & 0 & 3.22 & 3.39 & 3.30 & 2.85 & 3.31 & 4.19 & 1.86 & 0.95 & 44\\
\dglyph{dualloop-coupled}~Coupled dual loop & 48 & 0 & 50 & 0 & 5.25 & 7.03 & 7.15 & 3.85 & 4.17 & 5.91 & 4.50 & 2.33 & 66\\
\dglyph{ladder}~Ladder with highways & 48 & 0 & 50 & 0 & 5.63 & 7.52 & 9.70 & 4.20 & 4.70 & 5.50 & 5.71 & 3.55 & 65\\
\dglyph{ring-docks}~Ring with docks & 48 & 0 & 17 & 0 & 5.88 & 6.79 & 8.02 & 4.58 & 4.64 & 5.87 & 11.84 & 8.33 & 84\\
\dglyph{single-trap}~Single trap & 1 & 0 & 100 & 0 & 6.12 & 8.48 & 3.54 & 4.64 & 6.82 & 9.24 & 0.00 & 0.00 & 0\\
\dglyph{grid-zoned}~Grid of zoned traps & 72 & 9 & 50 & 0 & 6.55 & 8.07 & 6.73 & 6.45 & 5.21 & 6.62 & 16.44 & 2.88 & 72\\
\dglyph{grid-pz}~Grid with processing zone~\cite{schoenberger2025multizone} & 99 & 57 & 2 & 0 & 8.58 & 10.91 & 13.86 & 5.38 & 5.73 & 6.39 & 14.16 & 0.70 & 88\\
\dglyph{racetrack-h2}~Race track, gate zones~\cite{moses2023race} & 78 & 0 & 8 & 0 & 14.79 & 29.08 & 45.40 & 9.19 & 13.05 & 10.12 & 62.17 & 13.67 & 85\\
\multicolumn{14}{c}{\cellcolor{qpaper}\emph{Generic shapes: gate traps of one size, only the layout differs}}\\
\dglyph{line-c12}~Line, 12-ion traps & 8 & 0 & 100 & 6 & \underline{1.15} & 1.18 & 1.25 & 1.14 & 1.32 & \textbf{1.00} & 0.14 & 0.07 & 7\\
\dglyph{line-c6}~Line, 6-ion traps & 16 & 0 & 100 & 5 & 1.17 & \underline{1.05} & 1.44 & 1.15 & 1.21 & 1.33 & 0.32 & 0.20 & 14\\
\dglyph{tree-c12}~Tree, 12-ion traps & 8 & 5 & 100 & 0 & 1.17 & 1.46 & 1.04 & \underline{1.11} & 1.34 & 1.12 & 0.12 & 0.07 & 10\\
\dglyph{ladder-c12}~Ladder, 12-ion traps & 8 & 0 & 100 & \textbf{18} & 1.19 & 1.33 & 1.18 & \textbf{1.00} & 1.29 & 1.11 & 0.15 & 0.06 & 8\\
\dglyph{bus-c12}~Bus, 12-ion traps & 8 & 4 & 100 & 4 & 1.19 & 1.43 & \textbf{1.02} & 1.14 & 1.37 & 1.11 & 0.12 & 0.07 & 9\\
\dglyph{ring-c6}~Ring, 6-ion traps & 16 & 4 & 100 & 0 & 1.21 & 1.23 & 1.21 & 1.20 & \underline{1.20} & 1.24 & 0.31 & 0.20 & 13\\
\dglyph{ringcross-c6}~Ring with cross, 6-ion traps & 16 & 5 & 100 & 0 & 1.28 & 1.52 & 1.18 & 1.28 & 1.38 & 1.21 & 0.23 & 0.12 & 18\\
\dglyph{tree-c6}~Tree, 6-ion traps & 16 & 9 & 100 & 0 & 1.33 & 1.74 & 1.06 & 1.26 & 1.40 & 1.44 & 0.21 & 0.11 & 19\\
\dglyph{bus-c6}~Bus, 6-ion traps & 16 & 8 & 100 & 4 & 1.35 & 1.68 & \underline{1.04} & 1.27 & 1.35 & 1.33 & 0.23 & 0.12 & 19\\
\dglyph{mesh-c6}~Mesh, 6-ion traps & 16 & 0 & 100 & 2 & 1.40 & 1.67 & 1.59 & 1.23 & 1.31 & 1.40 & 0.34 & 0.16 & 21\\
\addlinespace[1pt]
\textcolor{black}{\textbf{Slowest $\div$ fastest generic shape}} & & & & & \textcolor{black}{\textbf{1.46$\times$}} & \textcolor{black}{\textbf{1.67$\times$}} & \textcolor{black}{\textbf{1.56$\times$}} & \textcolor{black}{\textbf{1.36$\times$}} & \textcolor{black}{\textbf{1.48$\times$}} & \textcolor{black}{\textbf{1.43$\times$}} & & &\\
\bottomrule
\end{tabular}
\end{table*}

  \fi
  \clearpage
  \typeout{QCCD-APPENDIX-PAGES: \number\numexpr\value{page}-\value{appendixstartpage}\relax}
  \typeout{QCCD-CONTENT-PAGES: \number\numexpr\value{bodycontentpages}+\value{page}-\value{appendixstartpage}\relax}
\else
  \typeout{QCCD-APPENDIX-PAGES: 0}
  \typeout{QCCD-CONTENT-PAGES: \arabic{bodycontentpages}}
\fi
\end{document}